\documentclass[11pt,british]{article}
\usepackage{lmodern}

\usepackage[T1]{fontenc}
\usepackage[latin9]{inputenc}
\usepackage{geometry}
\usepackage{xcolor}
\usepackage{verbatim}
\usepackage{prettyref}
\usepackage{refstyle}
\usepackage{amsmath}
\usepackage{amsthm}
\usepackage{amssymb}
\usepackage{cancel}
\usepackage{graphicx}

\makeatletter

\AtBeginDocument{\providecommand\Figref[1]{\ref{Fig:#1}}}
\AtBeginDocument{\providecommand\Subsecref[1]{\ref{Subsec:#1}}}
\AtBeginDocument{\providecommand\Thmref[1]{\ref{Thm:#1}}}
\AtBeginDocument{\providecommand\Defref[1]{\ref{Def:#1}}}
\AtBeginDocument{\providecommand\Eqref[1]{\ref{Eq:#1}}}
\AtBeginDocument{\providecommand\Secref[1]{\ref{Sec:#1}}}
\AtBeginDocument{\providecommand\Lemref[1]{\ref{Lem:#1}}}
\AtBeginDocument{\providecommand\secref[1]{\ref{sec:#1}}}
\AtBeginDocument{\providecommand\Corref[1]{\ref{Cor:#1}}}
\AtBeginDocument{\providecommand\Claimref[1]{\ref{Claim:#1}}}
\RS@ifundefined{subsecref}
  {\newref{subsec}{name = \RSsectxt}}
  {}
\RS@ifundefined{thmref}
  {\def\RSthmtxt{theorem~}\newref{thm}{name = \RSthmtxt}}
  {}
\RS@ifundefined{lemref}
  {\def\RSlemtxt{lemma~}\newref{lem}{name = \RSlemtxt}}
  {}

\theoremstyle{plain}
\newtheorem{thm}{\protect\theoremname}
\theoremstyle{definition}
\newtheorem*{defn*}{\protect\definitionname}
\theoremstyle{definition}
\newtheorem{defn}[thm]{\protect\definitionname}
\theoremstyle{plain}
\newtheorem{prop}[thm]{\protect\propositionname}
\theoremstyle{plain}
\newtheorem{lem}[thm]{\protect\lemmaname}
\theoremstyle{plain}
\newtheorem{fact}[thm]{\protect\factname}
\theoremstyle{plain}
\newtheorem{cor}[thm]{\protect\corollaryname}
\theoremstyle{definition}
\newtheorem{example}[thm]{\protect\examplename}
\theoremstyle{plain}
\newtheorem*{lem*}{\protect\lemmaname}
\theoremstyle{plain}
\newtheorem*{prop*}{\protect\propositionname}
\theoremstyle{remark}
\newtheorem{rem}[thm]{\protect\remarkname}
\theoremstyle{remark}
\newtheorem{claim}[thm]{\protect\claimname}

\usepackage{tikz}
\usetikzlibrary{calc}
\usepackage{prettyref}

\usepackage{multicol}

\usepackage{qtree}

\usepackage{hyperref}

\hypersetup{
colorlinks=true,
urlcolor=cyan,
linkcolor=black,
citecolor=blue,
}

\usepackage[super]{nth}

\newref{thm}{name=theorem~,Name=Theorem~,names=theorems~,Names=Theorems~}
\newref{def}{name=definition~,Name=Definition~,names=definitions~,Names=Definitions~}
\newref{cor}{name=corollary~,Name=Corollary~,names=corollaries~,Names=Corollaries~}
\newref{lem}{name=lemma~,Name=Lemma~,names=lemmas~,Names=Lemmas~}
\newref{claim}{name=claim~,Name=Claim~,names=claims~,Names=Claims~}
\newref{sec}{name=section~,Name=Section~,names=sections~,Names=Sections~}
\newref{subsec}{name=subsection~,Name=Subsection~,names=subsections~,Names=Subsections~}

\usepackage{color}
\definecolor{purple}{RGB}{120,20,120}

\graphicspath{{./unifiedWCF/},{./unifiedWCF/intro}}

\@ifundefined{showcaptionsetup}{}{%
 \PassOptionsToPackage{caption=false}{subfig}}
\usepackage{subfig}
\makeatother

\usepackage{babel}
\usepackage[style=phys]{biblatex}
\providecommand{\claimname}{Claim}
\providecommand{\corollaryname}{Corollary}
\providecommand{\definitionname}{Definition}
\providecommand{\examplename}{Example}
\providecommand{\factname}{Fact}
\providecommand{\lemmaname}{Lemma}
\providecommand{\propositionname}{Proposition}
\providecommand{\remarkname}{Remark}
\providecommand{\theoremname}{Theorem}

\begin{document}
\title{Quantum Weak Coin Flipping}
\author{Atul Singh Arora\thanks{responsible for all mistakes}, Jérémie Roland,
Stephan Weis\thanks{currently at the Universidade de Coimbra, Portugal}\\
\emph{~}\\
\emph{Université libre de Bruxelles}}
\date{November 5, 2018\\
(v2, November 2019)}

\maketitle
\vspace{-5mm}

\begin{abstract}
We investigate weak coin flipping, a fundamental cryptographic primitive
where two distrustful parties need to remotely establish a shared
random bit. A cheating player can try to bias the output bit towards
a preferred value. For weak coin flipping the players have known opposite
preferred values. A weak coin-flipping protocol has a bias $\epsilon$
if neither player can force the outcome towards their preferred value
with probability more than $\frac{1}{2}+\epsilon$. While it is known
that all classical protocols have $\epsilon=\frac{1}{2}$, Mochon
showed in 2007 \cite{Mochon07} that quantumly weak coin flipping
can be achieved with arbitrarily small bias (near perfect) but the
best known explicit protocol has bias $1/6$ (also due to Mochon,
2005 \cite{Mochon05}). We propose a framework to construct new explicit
protocols achieving biases below $1/6$. In particular, we construct
explicit unitaries for protocols with bias approaching $1/10$. To
go below, we introduce what we call the Elliptic Monotone Align (EMA)
algorithm which, together with the framework, allows us to numerically
construct protocols with arbitrarily small biases.
\end{abstract}
\vspace{1mm}

\tableofcontents{}\vfill{}

\pagebreak\textcolor{purple}{}
\global\long\def\supp{\text{supp}}%
\textcolor{purple}{}
\global\long\def\diag{\text{diag}}%

\pagenumbering{arabic}
\setcounter{page}{1}

\section{Introduction}

We investigate coin flipping, a fundamental cryptographic primitive
where two distrustful parties need to remotely generate a shared unbiased
random bit. A cheating player can try to bias the output bit towards
a preferred value. For weak coin flipping the players have known opposite
preferred values. A weak coin flipping (WCF) protocol has a bias $\epsilon$
if neither player can force the outcome towards his/her preferred
value with probability more than $\frac{1}{2}+\epsilon$. For strong
coin-flipping there are no a priori preferred values and the bias
is defined similarly. Restricting to classical resources, neither
weak nor strong coin flipping is possible under information-theoretic
security, as there always exists a player \cite{Cleve86} who can
force any outcome with probability $1$. However, in a quantum world,
strong coin flipping protocols with bias strictly less than $\frac{1}{2}$
have been shown and the best known explicit protocol has bias $\frac{1}{4}$
\cite{Ambainis04b}. Nevertheless, Kitaev gave a lower bound of $\frac{1}{\sqrt{2}}-\frac{1}{2}$
for the bias of any quantum strong coin flipping, so an unbiased protocol
is not possible.

As for weak coin flipping, the current best known explicit protocol\textemdash the
Dip Dip Boom protocol\textemdash is due to Mochon \cite{Mochon05}
and has bias $1/6$. In a breakthrough result, he even proved the
existence of a quantum weak coin-flipping protocol with arbitrarily
low bias $\epsilon>0$, hence showing that near-perfect weak coin
flipping is theoretically possible \cite{Mochon07}. This fundamental
result for quantum cryptography, unfortunately, was proved non-constructively,
by elaborate successive reductions (80 pages) of the protocol to different
versions of so-called point games, a formalism introduced by Kitaev
\cite{Kitaev03} in order to study coin flipping. Consequently, the
structure of the protocol whose existence is proved is lost. A systematic
verification of this by independent researchers recently led to a
simplified proof \cite{ACG+14} (\emph{only} 50 pages) but eleven
years later, an explicit weak coin-flipping protocol is still unknown,
despite various expert approaches ranging from the distillation of
a protocol using the proof of existence to numerical search \cite{NST14,Nayak2015}.
Further, weak coin flipping provides, via black-box reductions, optimal
protocols for strong coin flipping \parencite{CK09}, bit commitment
\parencite{CK11} and a variant of oblivious transfer \parencite{Chailloux2013a}
(fundamental cryptographic primitives). It is also used to implement
other cryptographic tasks such as leader election \parencite{Ganz2009}
and dice rolling \parencite{Aharon2010}. 

We construct a framework that allows us to convert simple point games
(i.e. corresponding to known protocols) into explicit quantum protocols
defined in terms of unitaries and projectors. We use the said framework
to convert a bias $1/10$ point game into its corresponding explicit
protocol making it the first improvement of its kind in the last thirteen
years since Mochon's Dip Dip Boom protocol (bias $1/6$) \cite{Mochon05}.

Our second contribution, the Elliptic Monotone Align (EMA) algorithm,
is a numerical algorithm which can provably find the unitaries required
for implementing protocols with arbitrary biases, including the ones
with $\epsilon\to0$. 

\subsection{State of the Art | Kitaev's Formalisms and Mochon's Games}

Let us start with noting two features of weak coin flipping. First,
note that we can say, without loss of generality, that if the bit
is zero it means Alice won and if the bit is one it means Bob won.
Why is that? In weak coin flipping we know both players have known
preferences. Alice wants zero and Bob wants one\footnote{Alice wanting a zero and Bob wanting a one is just an uninteresting
relabelling.} since if they both wanted the same outcome bit, there would be no
need to flip a coin. If a player gets what they want, we say they
won. Second, we note that there are four situations which can arise
in a weak coin flipping scenario of which three are of interest. Let
us denote by HH the situation where both Alice and Bob are honest,
i.e. follow the protocol. In this situation we want the protocol to
be such that both Alice and Bob (a) win with equal probability and
(b) are in agreement with each other. In the situation HC where Alice
is honest and Bob is cheating, the protocol must protect Alice from
a cheating Bob. In this situation, a cheating Bob tries to convince
an honest Alice that he has won. His probability of succeeding by
using his best cheating strategy is denoted by $P_{B}^{*}$ where
the star/asterisk refers to a cheating player and the subscript denotes
the outcome he desires to enforce on the honest player. The CH situation
where Bob is honest and Alice is cheating naturally points us to the
corresponding definition of $P_{A}^{*}$. The situation CC where both
players are cheating is not of interest to us as nothing can be said
which depends on the protocol. This is because nobody is following
the protocol.

A trivial example of a weak coin flipping protocol is where Alice
flips a coin and reveals the outcome to Bob over the phone. A cheating
Alice can simply lie and always win against an honest Bob which means
$P_{A}^{*}=1$. On the other hand, a cheating Bob can not do anything
to convince Alice that he has won, unless it happens by random chance
on the coin flip. This corresponds to $P_{B}^{*}=\frac{1}{2}$. The
bias of the protocol is $\max[P_{A}^{*},P_{B}^{*}]-\frac{1}{2}$ which
for this naïve protocol amounts to $\frac{1}{2}$, the worst possible.
Manifestly, constructing protocols where one player is protected is
nearly trivial. Constructing protocols where neither player is able
to cheat (against an honest player) is the real challenge.

\begin{figure}
\centering{}\includegraphics[viewport=0bp -20bp 177bp 297bp,clip,width=4cm]{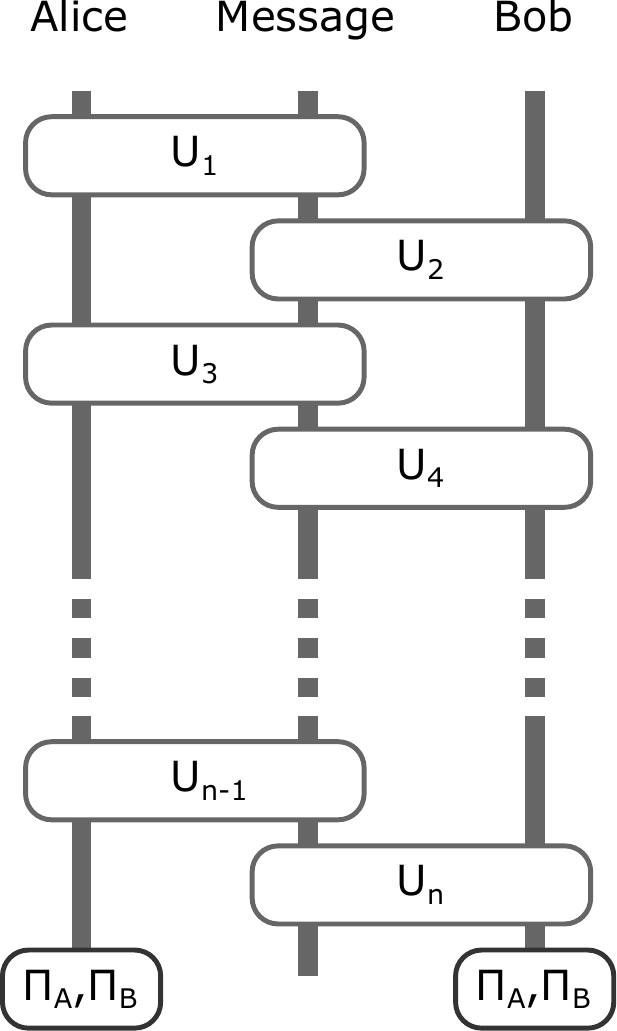}\bigskip{}
\caption{General structure of a Weak Coin Flipping protocol.\label{fig:General-structure-of}}
\end{figure}

Given a WCF protocol it is not a priori clear how the best success
probability of a cheating player, denoted by $P_{A/B}^{*}$, should
be computed as the strategy space can be dauntingly large. It turns
out that all quantum WCF protocols can be defined using the exchange
of a message register interleaved with the players applying the unitaries
$U_{i}$ locally (see \Figref{General-structure-of}) until a final
measurement, say $\Pi_{A}$ denoting Alice won and $\Pi_{B}$ denoting
Bob won, is made in the end. Computing $P_{A}^{*}$ in this case reduces
to a semi-definite program (SDP) in $\rho$: maximise $P_{A}^{*}=\text{tr}(\Pi_{A}\rho)$
given the constraint that the honest player follows the protocol.
Similarly for computing $P_{B}^{*}$ one can define another SDP. Using
SDP duality one can turn this maximization problem over cheating strategies
into a minimization problem over dual variables $Z_{A/B}$. Any dual
feasible assignment then provides an upper bound on the cheating probabilities
$P_{A/B}^{*}$. SDPs are usually easy to handle but in this case,
there are two SDPs, and we must optimise both simultaneously (see
\Subsecref{WCF-protocol-as-an-SDP-and-its-Dual}). Note that here
we assume the protocol is known and we are trying to find bounds on
$P_{A}^{*}$ and $P_{B}^{*}$. However, our goal is to find good protocols.
So what we would like is a framework which allows us to do both, construct
protocols and find the associated $P_{A}^{*}$ and $P_{B}^{*}$. Kitaev
gave us such a framework.

He converted this problem about matrices ($Z$s, $\rho$s and $U$s)
into a problem about points on a plane, which Mochon called Kitaev's
Time Dependent Point Game (TDPG) framework. In this framework, one
is concerned with a sequence of frames\textemdash the positive quadrant
of the plane with some points and their probability weights\textemdash which
must start with a fixed frame and end with a frame that has only one
point. The fixed starting frame consists of two points at $[0,1]$
and $[1,0]$ with weight $1/2$. The end frame must be a single point,
say at $[\beta,\alpha]$, with weight $1$. The objective of the protocol
designer is to get this end point as close to the origin as possible
by transitioning through intermediate frames (see \Figref{Point-game-corresponding})
by following certain rules (we describe these shortly). The magic
of this formalism, roughly stated, is that if one abides by these
rules then corresponding to every such valid sequence of frames, there
exists a WCF protocol with $P_{A}^{*}=\alpha$, $P_{B}^{*}=\beta$
(see \Subsecref{TDPG-with-valid-transitions}). 

We now describe these rules. Consider a given frame and focus on a
set of points that fall on this vertical (or horizontal) line. Let
the $y$ coordinate (or $x$ coordinate) of the $i$th point be given
by $z_{g_{i}}$ and the weight be given by $p_{g_{i}}$. Let $z_{h_{i}}$
and $p_{h_{i}}$ denote the corresponding quantity in the subsequent
frame. Then, the following conditions must hold 
\begin{enumerate}
\item the probabilities are conserved, viz. $\sum_{i}p_{g_{i}}=\sum_{i}p_{h_{i}}$
\begin{enumerate}
\item for all $\lambda>0$ 
\begin{equation}
\sum_{i}\frac{\lambda z_{g_{i}}}{\lambda+z_{g_{i}}}p_{g_{i}}\le\sum_{i}\frac{\lambda z_{h_{i}}}{\lambda+z_{h_{i}}}p_{h_{i}}.\label{eq:scalarCondition}
\end{equation}
\end{enumerate}
\end{enumerate}
Note that from one frame to the next, one can either make a horizontal
transition or a vertical transition. By combining these sequentially
one can obtain the desired form of the final frame, i.e. a single
point. The aforesaid rule and the points in the frames arise from
the dual variables $Z_{A/B}$ . Just as the state $\rho$ evolves
through the protocol, so do the dual variables $Z_{A/B}$. The points
and their weights in the TDPG are exactly the eigenvalue pairs of
$Z_{A/B}$ with the probability weight assigned to them by the honest
state $\left|\psi\right\rangle $ at a given point in the protocol
($\left|\psi\right\rangle $ and $\rho$ are closely related). The
aforementioned rules are related to the dual constraints. Given an
explicit WCF protocol and a feasible assignment for the dual variables
witnessing a given bias, it is straightforward to construct the TDPG.
However, going backwards, constructing the WCF dual from a TDPG is
highly non-trivial and no general construction is known.

Our main contribution is precisely to this part. We construct a framework
which allows for a ready conversion of simple TDPGs into explicit
protocols, and once supplemented with the EMA algorithm, it can convert
any TDPG into its corresponding protocol. This is relevant because
Mochon's breakthrough result was to define a family of games\footnote{Mochon describes his games in Kitaev's Time Independent Point Game
(TIPG) framework but it is straightforward to go back from a TIPG
to a TDPG.} with bias $\epsilon=\frac{1}{4k+2}$ where $k$ encodes the number
of points that are involved in the non-trivial step (for $k=1$ it
reduces to a version of the Dip Dip Boom (bias $1/6$) protocol) which
means, effectively, we can numerically construct quantum weak coin
flipping protocols with arbitrarily small bias (see Section 5 of either
\parencite{ACG+14,Mochon07}).

As this point game formalism is the cornerstone of the analysis, we
simplify the rules further and then apply them to construct a simple
example game. Later, we convert this example game into an explicit
protocol using our framework. If we restrict ourselves to transitions
involving only one initial and one final point, the second condition
reduces to $z_{g}\le z_{h}$ (we suppressed the subscript). This is
called a \emph{raise}. It means that we can always increase the coordinate
of a single point. What about going from one initial point to many
final points (note that the points before and after must lie along
either a horizontal or a vertical line)? The second condition in this
case becomes $1/z_{g}\ge\left\langle 1/z_{h}\right\rangle $, that
is the harmonic mean of the final points must be greater than or equal
to that of the initial point, where $\left\langle f(z_{h})\right\rangle :=\left(\sum_{i}f(z_{h_{i}})p_{h_{i}}\right)/\left(\sum_{j}p_{h_{j}}\right)$.
This is called a \emph{split}. Finally, we can ask: What happens upon
merging many points into a single point? The second condition becomes
$\left\langle z_{g}\right\rangle \le z_{h}$, that is the final position
must not be smaller than the average initial position (where $\left\langle f(z_{g})\right\rangle $
is analogously defined). This is called a \emph{merge}. While these
three transitions/moves do not exhaust the set of moves, they are
enough to construct games that almost achieve the bias $1/6$. Let
us construct a simple game as an example. We start with the initial
frame and raise the point $[1,0]$ along the vertical to $[1,1]$
(see \Figref{Point-game-corresponding}). We know this move is allowed
as it is just a raise. Next we merge the points $[0,1]$ with $[1,1]$
using a horizontal merge. The $x$-coordinate of the resulting point
can at best be $\frac{1}{2}.0+\frac{1}{2}.1=\frac{1}{2}$ where we
used the fact that both points have weight $1/2$. Thus we end up
with a single point at $[\frac{1}{2},1]$ with all the weight. Kitaev's
framework tells us that there must exist a protocol which yields $P_{A}^{*}=1$
while $P_{B}^{*}=\frac{1}{2}$. This, however, is the phone protocol
that we started our discussion with! It is a neat consistency check
but it yields a trivial bias. This is because we did not use the split.
If we use a split once, we can, by simply matching the weights, already
obtain a game with $P_{A}^{*}=P_{B}^{*}=\frac{1}{\sqrt{2}}$. Protocols
corresponding to this bias were found by various researchers \cites{SR02}{NS03}{KerenidisNayak04}
long before this framework was known. In fact, the bias of the said
weak coin flipping protocol, $\epsilon=\frac{1}{\sqrt{2}}-\frac{1}{2}$,
was exactly the lower bound for \emph{strong }coin flipping. It was
an exciting time (we imagine) as the technique used to bound strong
coin flipping fails for weak coin flipping. The matter was not resolved
for a while. This protocol held the record for being the best known
weak coin flipping protocol until Mochon progressively showed that
if we use multiple splits wisely at the beginning followed by a raise,
one simply needs to use merges thereafter to obtain a game with bias
almost $1/6$, which corresponds to his Dip Dip Boom protocol. The
Dip Dip Boom protocol, is actually a family of protocols which in
the limit of infinite rounds of communication yields bias $1/6$.
Going lower, therefore, is not a straight forward extension and we
need to use moves which can not be decomposed into the three basic
ones, splits, merges and raises. Our contribution is to find ways
of constructing the unitaries corresponding to these moves.

\subsection{A Framework | First Contribution}

We first describe our framework for converting a TDPG into an explicit
protocol. We start by defining a `canonical form' for any given frame
of a TDPG. This allows one to write the WCF dual variables, $Z$s,
and the honest state $\left|\psi\right\rangle $ associated with each
frame of the TDPG. We define a sequence of quantum operations, unitaries
and projections, which allow Alice and Bob to transition from the
initial frame to the final frame. It turns out that there is only
one non-trivial quantum operation in the sequence which we leave partially
specified for the moment. This means that we know that the unitary
should send the honest initial state to the honest final state. However
the action of the unitary on the orthogonal space, which intuitively
is what would bestow on it the cheating prevention/detection capability,
is obtained as an interesting constraint. Using the SDP formalism
we write the constraints at each step of the sequence on the $Z$s
and show that they are indeed satisfied (see \Thmref{TEFconstraint}
for a full statement of the following, \Subsecref{The-Framework}
for its proof and the description of the framework).
\begin{thm}[TEF constraint (simplified)]
 \begin{sloppy}If a unitary matrix $U$ acting on the space $\text{span}\{\left|g_{1}\right\rangle ,\left|g_{2}\right\rangle \dots,\left|h_{1}\right\rangle ,\left|h_{2}\right\rangle \dots\}$
satisfying the constraints 
\[
U\left|v\right\rangle =\left|w\right\rangle ,
\]
\begin{equation}
\sum_{i}x_{h_{i}}\left|h_{i}\right\rangle \left\langle h_{i}\right|-\sum_{i}x_{g_{i}}E_{h}U\left|g_{i}\right\rangle \left\langle g_{i}\right|U^{\dagger}E_{h}\ge0\label{eq:constraint}
\end{equation}
can be found for every move/transition (see \Defref{transition} and
\Defref{EBMlineTransition}) of a TDPG then an explicit protocol with
the corresponding bias can be obtained using the TDPG\textendash to\textendash Explicit\textendash protocol
Framework (TEF), where $\{\left|g_{i}\right\rangle \},\{\left|h_{i}\right\rangle \}$
are orthonormal vectors and if the transition is horizontal
\begin{itemize}
\item the initial points have $x_{g_{i}}$ as their $x$-coordinate and
$p_{g_{i}}$ as their corresponding probability weight,
\item the final points have, similarly, $x_{h_{i}}$ as their $x$-coordinate
and $p_{h_{i}}$ as their corresponding probability weight
\item $E_{h}$ is a projection onto the $\text{span}\left\{ \left|h_{i}\right\rangle \right\} $
space, 
\item $\left|v\right\rangle =\sum_{i}\sqrt{p_{g_{i}}}\left|g_{i}\right\rangle /\sqrt{\sum p_{g_{i}}},$
$\left|w\right\rangle =\sum_{i}\sqrt{p_{h_{i}}}\left|h_{i}\right\rangle /\sqrt{\sum p_{h_{i}}}$
\end{itemize}
and if the transition is vertical, the $x_{g_{i}}$ and $x_{h_{i}}$
become the $y$-coordinates $y_{g_{i}}$ and $y_{h_{i}}$ with everything
else unchanged.\label{thm:TEFconstraint-inf}\end{sloppy}
\end{thm}

\begin{figure}
\begin{centering}
~~
\par\end{centering}
\centering{}~~\\
\includegraphics[viewport=0bp -20bp 356bp 89bp,width=8cm]{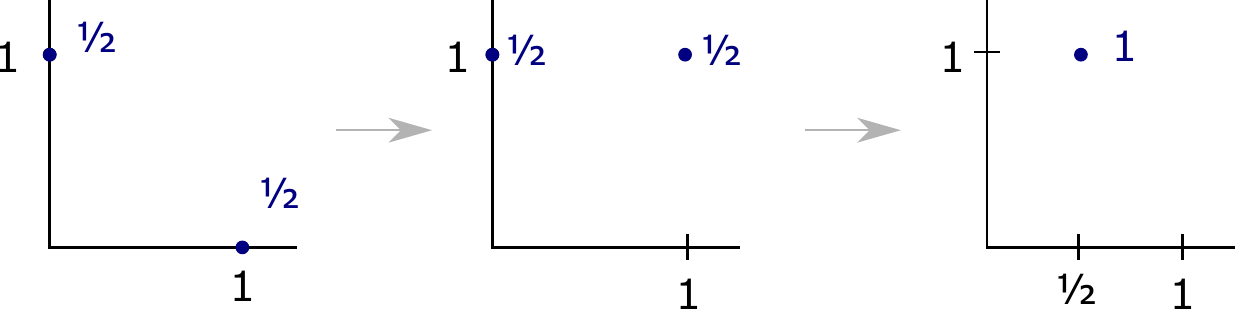}\caption{Point game corresponding to the weak coin flipping over the phone
protocol.\label{fig:Point-game-corresponding}}
\end{figure}

{Note that the TDPG already specifies the coordinates $x_{h_{i}},x_{g_{i}}$
and the probabilities $p_{h_{i}},p_{g_{i}}$ which satisfy, \Eqref{scalarCondition},
the scalar condition. Our task therefore reduces to finding the correct
$U$ which satisfies the aforesaid matrix constraints. It is this
general problem that is solved by our EMA algorithm which we describe
later. 

Given such a unitary $U$ acting on the space $\text{span}\{\left|g_{1}\right\rangle ,\left|g_{2}\right\rangle \dots$
$\left|h_{1}\right\rangle ,\left|h_{2}\right\rangle \dots\}$ one
can construct a unitary, $U_{AM}^{(2)}$, acting non-trivially on
the space $\text{span}\{\left|g_{1}g_{1}\right\rangle _{AM},\left|g_{2}g_{2}\right\rangle _{AM}\dots,\left|h_{1}h_{1}\right\rangle _{AM},\left|h_{2}h_{2}\right\rangle _{AM},\allowbreak\dots\}$
by mapping $\left|g_{i}\right\rangle \to\left|g_{i}g_{i}\right\rangle _{AM}$,
$\left|h_{i}\right\rangle \to\left|h_{i}h_{i}\right\rangle _{AM}$
and as identity otherwise. We now informally describe how to convert
a TDPG into an explicit protocol. It suffices to show what a transition
from a given frame to the next frame corresponds to in terms of the
protocol. In this discussion, we refer to them as the initial frame
and the final frame. Assume that the corresponding non-trivial $U_{AM}^{(2)}$
is known. As we saw, a given transition would either be horizontal
or vertical. We assume it is horizontal without loss of generality\footnote{Mochon's point games have a repeating structure he calls a ``ladder''.
Corresponding to each $k$ he constructs a family of point games parametrised
by the number of points in this ladder. The game approaches the bias
$\epsilon=(4k+2)^{-1}$ as the number of points is increased (the
value is reached in the limit of infinite points). Consequently, we
consider a finite set of points in the transition.}. We label the points that do not participate in this horizontal transition,
i.e. remain unchanged in both frames, by $k_{1},k_{2}\dots$ in both
frames. The points in the initial frame involved in this transition
are labelled $g_{1},g_{2}\dots$ and the ones in the final frame are
labelled $h_{1},h_{2}\dots$. All the points are now labelled. We
denote the coordinates of the final points by $x_{h_{1}},x_{h_{2}}\dots$
and the probability weights by $p_{h_{1}},p_{h_{2}}\dots$. We similarly
define $x_{g_{i}},p_{g_{i}}$ and $x_{k_{i}},p_{k_{i}}$. The Hilbert
space of interest is given by $\mathcal{H}:=\text{span}\{\left|k_{1}\right\rangle ,\left|k_{2}\right\rangle \dots,\left|g_{1}\right\rangle ,\left|g_{2}\right\rangle \dots,\left|h_{1}\right\rangle ,\left|h_{2}\right\rangle \dots,\left|m\right\rangle \}$
where each vector is assumed orthonormal ($\left|m\right\rangle $
is just an idle state in which the message register is assumed to
be initially and returned to finally). We assume that Alice's register,
Bob's register and the message register each have dimension at least
as large as $\dim(\mathcal{H})$. The state (by state in this discussion,
we mean the honest state) corresponding to the initial frame is assumed
to have the form 
\[
\left|\psi_{(1)}\right\rangle =\left(\sum_{i}\sqrt{p_{g_{i}}}\left|g_{i}g_{i}\right\rangle _{AB}+\sum_{i}\sqrt{p_{k_{i}}}\left|k_{i}k_{i}\right\rangle _{AB}\right)\otimes\left|m\right\rangle _{M}.
\]
} 

\textbf{Bob}: Assume Bob has the message register. He applies the
conditional swap $U_{BM}^{\text{SWP}\{\vec{g},m\}}$ where $U_{BM}^{\text{SWP}\{\vec{g},m\}}$
swaps conditionally on both registers being in the subspace $\text{span}\{\left|g_{1}\right\rangle ,\left|g_{2}\right\rangle \dots,\left|m\right\rangle \}$.
The state after this operation is
\[
\left|\psi_{(2)}\right\rangle =\sum_{i}\sqrt{p_{g_{i}}}\left|g_{i}g_{i}\right\rangle _{AM}\otimes\left|m\right\rangle _{B}+\sum_{i}\sqrt{p_{k_{i}}}\left|k_{i}k_{i}\right\rangle _{AB}\otimes\left|m\right\rangle _{M}.
\]
He then sends the message register to Alice.

\textbf{Alice}: Alice applies the non-trivial unitary $U_{AM}^{(2)}$
on her local register and the message register. She then measures
$\{E^{(2)},\mathbb{I}-E^{(2)}\}$ where $E^{(2)}:=\left(\sum\left|h_{i}\right\rangle \left\langle h_{i}\right|+\sum\left|k_{i}\right\rangle \left\langle k_{i}\right|\right)_{A}\otimes\mathbb{I}_{M}$.
The state at this point is 
\[
\left|\psi_{(3)}\right\rangle =\sum_{i}\sqrt{p_{h_{i}}}\left|h_{i}h_{i}\right\rangle _{AM}\otimes\left|m\right\rangle _{B}+\sum_{i}\sqrt{p_{k_{i}}}\left|k_{i}k_{i}\right\rangle _{AB}\otimes\left|m\right\rangle _{M}.
\]
If the outcome corresponds to the latter, she declares herself to
be the winner. Otherwise she sends the message register back to Bob.

\textbf{Bob}: Bob again applies a conditional swap $U_{BM}^{\text{SWP}\{\vec{h},m\}}$
followed by a measurement corresponding to $\{E^{(3)},\mathbb{I}-E^{(3)}\}$
where $E^{(3)}:=\left(\sum_{i}\left|h_{i}\right\rangle \left\langle h_{i}\right|+\sum_{i}\left|k_{i}\right\rangle \left\langle k_{i}\right|\right)_{B}\otimes\mathbb{I}_{M}$.
The final state is 
\[
\left|\psi_{(4)}\right\rangle =\left(\sum_{i}\sqrt{p_{h_{i}}}\left|h_{i}h_{i}\right\rangle _{AB}+\sum_{i}\sqrt{p_{k_{i}}}\left|k_{i}k_{i}\right\rangle _{AB}\right)\otimes\left|m\right\rangle _{M}.
\]
If the outcome corresponds to $\mathbb{I}-E^{(3)}$, Bob declares
himself the winner. 

As the final state is in the same form as the initial state, one
can progressively build the sequence corresponding to the complete
protocol. Once the entire sequence is known, one must reverse the
order of all the operations to obtain the final protocol. Note that
the message register is initially decoupled, it then gets entangled,
and finally it emerges decoupled again. This simplifies the analysis
(and also entails that one need not keep the message register coherent
for the duration of the protocol; keeping it coherent for each round
individually is sufficient).

Let us try to apply this procedure to our example game (see \Figref{Point-game-corresponding}).
We label the points in the first frame as $g_{1}$ and $g_{2}$. The
state is given by $\frac{1}{\sqrt{2}}\left(\left|g_{1}g_{1}\right\rangle _{AB}+\left|g_{2}g_{2}\right\rangle _{AB}\right)\otimes\left|m\right\rangle _{M}$.
(This should make it clear that the order is reversed here because
we want to end with an EPR like state so that when Alice and Bob make
a measurement, they agree on a random bit.) We simply claim for the
moment that raising does not require Alice and Bob to do anything.
This means that we can consider the second frame with the same labels.
We now apply the merge transition by using the aforesaid recipe, where
Bob applies a swap, sends the message register to Alice, she applies
$U_{AM}^{(2)}$ and the projector, returns the message register to
Bob and he applies the final swap and measurement. We continue to
assume we are given the correct $U_{AM}^{(2)}$ that implements the
merge step. The state one obtains after the application of these unitaries
turns out to be $\left|h_{1}h_{2}\right\rangle _{AB}\otimes\left|m\right\rangle _{M}$.
(This looks like the state we should start with, completely unentangled.
This is intuitively why the actual protocol is a reversed version
of what we have.) Our procedure can be applied to any point game,
granted the non-trivial unitary $U^{(2)}$ can be found. The central
issue is that there is no general recipe known for constructing $U^{(2)}$s.

To address this we can prove that what we call the Blinkered Unitary
satisfies the required constraints for both the split and merge moves
(see \Subsecref{BlinkeredUnitary}). It is defined as
\begin{equation}
U_{\text{blink}}=\left|w\right\rangle \left\langle v\right|+\left|v\right\rangle \left\langle w\right|+\sum_{i}\left|v_{i}\right\rangle \left\langle v_{i}\right|+\sum_{i}\left|w_{i}\right\rangle \left\langle w_{i}\right|\label{eq:Ublink}
\end{equation}
 where $\left|v\right\rangle $, $\left\{ \left|v_{i}\right\rangle \right\} $
and $\left|w\right\rangle $, $\left\{ \left|w_{i}\right\rangle \right\} $
are orthonormal vectors spanning the $\left\{ \left|g_{i}\right\rangle \right\} $
and $\left\{ \left|h_{i}\right\rangle \right\} $ space respectively.
With these the former best protocol (bias $1/6$) can already be derived
from its TDPG, in a manner analogous to the one used for the example
game. This was not known (to the best of our knowledge), even though
the protocol itself was separately known and analysed. We next study
the family of bias $1/10$ TDPGs and isolate the precise moves required
to implement it (see \Subsecref{Bias1by10}). Let $n_{g}\to n_{h}$
denote a move from $n_{g}$ initial points to $n_{h}$ final points.
While the bias $1/6$ games used a $2\to1$ merge as its key move,
the bias $1/10$ games use a combination of $3\to2$ and $2\to2$
moves (these can not be produced by a combination of merges and splits,
as was pointed out earlier). We give analytic expressions for these
unitaries and show that they satisfy the required constraints (see
\Subsecref{Bias1by10protocol}). In particular, we show that for $3\to2$
moves with $x_{g_{1}}<x_{g_{2}}<x_{g_{3}}$ and $x_{h_{1}}<x_{h_{2}}$
\begin{equation}
U_{3\to2}=\left|w\right\rangle \left\langle v\right|+\left|w_{1}\right\rangle \left\langle v_{1}'\right|+\left|v_{2}'\right\rangle \left\langle v_{2}'\right|+\left|v_{1}'\right\rangle \left\langle w_{1}\right|+\left|v\right\rangle \left\langle w\right|\label{eq:U3to2}
\end{equation}
satisfies the required constraints (under some further technical conditions
which are satisfied by the $1/10$ games of interest), where 
\begin{align*}
\left|v\right\rangle  & =\frac{\sqrt{p_{g_{1}}}\left|g_{1}\right\rangle +\sqrt{p_{g_{2}}}\left|g_{2}\right\rangle +\sqrt{p_{g_{3}}}\left|g_{3}\right\rangle }{N_{g}},\\
\left|v_{1}\right\rangle  & =\frac{\sqrt{p_{g_{3}}}\left|g_{2}\right\rangle -\sqrt{p_{g_{2}}}\left|g_{3}\right\rangle }{N_{v_{1}}},\\
\left|v_{2}\right\rangle  & =\frac{-\frac{\left(p_{g_{2}}+p_{g_{3}}\right)}{\sqrt{p_{g_{1}}}}\left|g_{1}\right\rangle +\sqrt{p_{g_{2}}}\left|g_{2}\right\rangle +\sqrt{p_{g_{3}}}\left|g_{3}\right\rangle }{N_{v_{2}}}
\end{align*}
and 
\[
\left|w\right\rangle =\frac{\sqrt{p_{h_{1}}}\left|h_{1}\right\rangle +\sqrt{p_{h_{2}}}\left|h_{2}\right\rangle }{N_{h}},\left|w_{1}\right\rangle =\frac{\sqrt{p_{h_{2}}}\left|h_{1}\right\rangle -\sqrt{p_{h_{1}}}\left|h_{2}\right\rangle }{N_{h}}
\]
are normalised vectors (this fixes the normalisation factors) which
we use to define 
\[
\left|v_{1}'\right\rangle =\cos\theta\left|v_{1}\right\rangle +\sin\theta\left|v_{2}\right\rangle ,\left|v_{2}'\right\rangle =\sin\theta\left|v_{1}\right\rangle -\cos\theta\left|v_{2}\right\rangle 
\]
where $\cos\theta$ is obtained by solving 
\begin{align*}
\frac{\sqrt{p_{h_{1}}p_{h_{2}}}}{N_{h}^{2}}\left(x_{h_{1}}-x_{h_{2}}\right)-\cos\theta\frac{\sqrt{p_{g_{2}}p_{g_{3}}}}{N_{g}N_{v_{1}}}\left(x_{g_{2}}-x_{g_{3}}\right)\\
-\sin\theta\left\langle x_{g}\right\rangle \frac{N_{g}}{N_{v_{2}}} & =0
\end{align*}
and choosing the solution which is closer to $1$. Similarly we give
an explicit unitary corresponding to the second move, i.e. the $2\to2$
move. For the second move, i.e. the $2\to2$ move with $x_{g_{1}}<x_{g_{2}}$
and $x_{h_{1}}<x_{h_{2}}$, we show that
\[
U_{2\to2}=\left|w\right\rangle \left\langle v\right|+\left(\alpha\left|v\right\rangle +\beta\left|w_{1}\right\rangle \right)\left\langle v_{1}\right|+\left|v\right\rangle \left\langle w\right|+\left(\beta\left|v\right\rangle -\alpha\left|w_{1}\right\rangle \right)\left\langle w_{1}\right|
\]
satisfies the required constraints (again, under further technical
conditions which are satisfied by the $1/10$ games of interest) where
\begin{align*}
\left|v\right\rangle  & =\frac{1}{N_{g}}\left(\sqrt{p_{g_{1}}}\left|g_{1}\right\rangle +\sqrt{p_{g_{2}}}\left|g_{2}\right\rangle \right),\\
\left|v_{1}\right\rangle  & =\frac{1}{N_{g}}\left(\sqrt{p_{g_{2}}}\left|g_{1}\right\rangle -\sqrt{p_{g_{1}}}\left|g_{2}\right\rangle \right)
\end{align*}
and 
\begin{align*}
\left|w\right\rangle  & =\frac{1}{N_{h}}\left(\sqrt{p_{h_{1}}}\left|h_{1}\right\rangle +\sqrt{p_{h_{2}}}\left|h_{2}\right\rangle \right)\\
\left|w_{1}\right\rangle  & =\frac{1}{N_{h}}\left(\sqrt{p_{h_{2}}}\left|h_{1}\right\rangle -\sqrt{p_{h_{1}}}\left|h_{2}\right\rangle \right).
\end{align*}
Further, $\alpha,\beta\in\mathbb{R}$ are such that $\alpha^{2}+\beta^{2}=1$
and 
\[
\beta=\sqrt{\frac{p_{h_{1}}p_{h_{2}}}{p_{g_{1}}p_{g_{2}}}}\frac{(x_{h_{1}}-x_{h_{2}})}{(x_{g_{1}}-x_{g_{2}})}.
\]
This lets us, in effect, convert Mochon's family of bias $1/10$ games
into explicit protocols, finally breaking the $1/6$ barrier. Mochon's
games achieving lower biases correspond to larger unitary matrices.
Consequently, this approach based on guessing the correct form of
the solution becomes untenable.

\subsection{EMA Algorithm | Second Contribution\label{subsec:EMA-Algorithm}}

\begin{figure}
\centering{}\includegraphics[viewport=0bp -20bp 554bp 301bp,width=10.5cm]{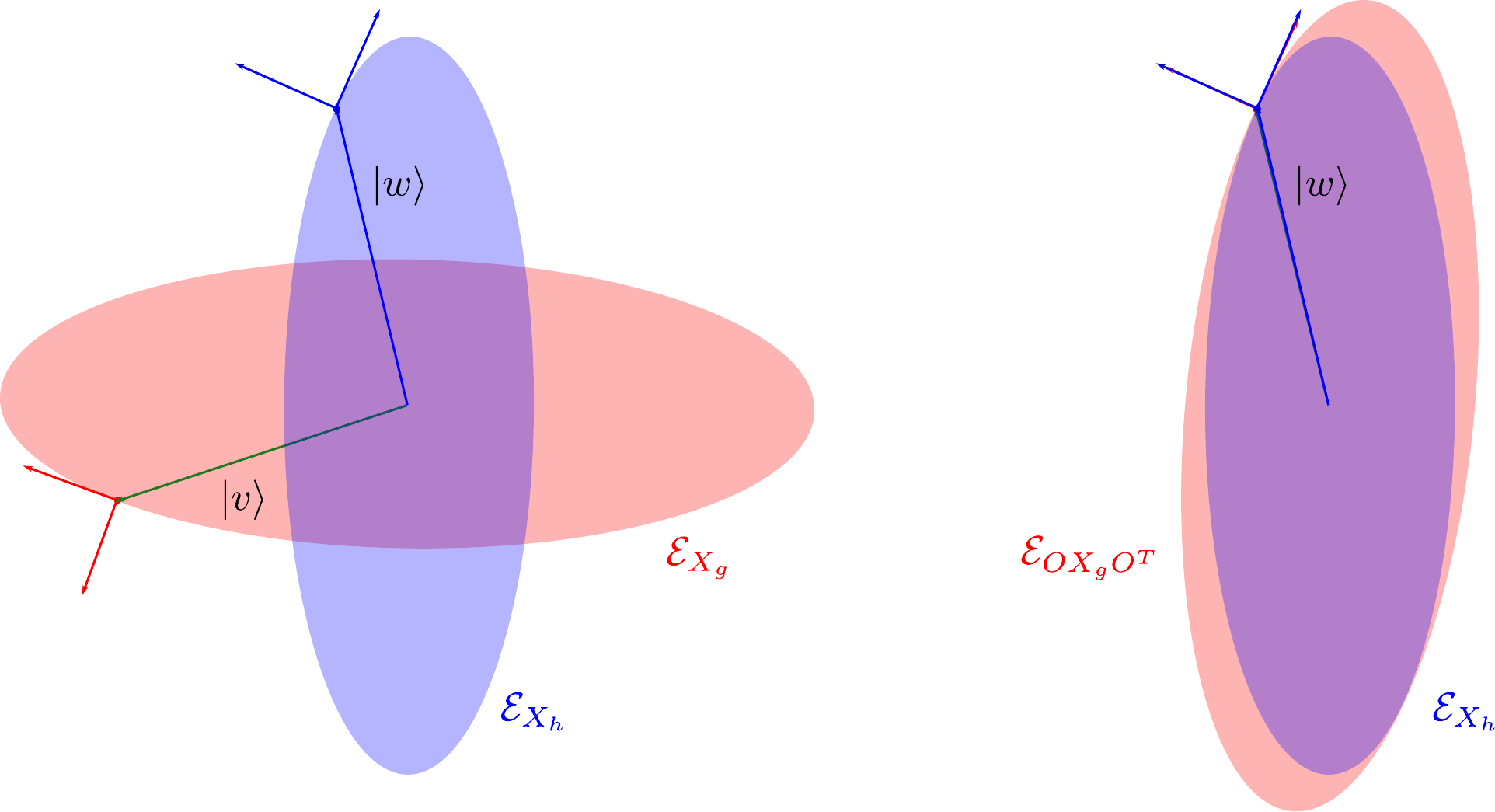}\caption{On the left the ellipsoids correspond to the diagonal matrices $X_{g}$
and $X_{h}$. The vectors $\left|w\right\rangle $ and $\left|v\right\rangle $
indicate only the direction. On the right, the larger ellipsoid is
now rotated to corresponding to $OX_{g}O^{T}$. The point of contact
is along the vector $\left|w\right\rangle =O\left|v\right\rangle $.\label{fig:Visual-aid-EMA}}
\end{figure}

To go lower than $1/10$ we use our Elliptic Monotone Align (EMA)
algorithm which we now describe. Note that if we neglect the projector
in \Eqref{constraint}, we can express it as $X_{h}\ge UX_{g}U^{\dagger}$
where $X_{h},X_{g}$ are diagonal matrices with positive entries (justified
in \Subsecref{CPFandCOF}). Surprisingly, it is possible to show that
we can restrict ourselves to orthogonal matrices without loss of generality
(see \Subsecref{EBMtoEBRMtoCOF}). Once we restrict to real numbers,
it is easy to see that the set of vectors $\mathcal{E}_{X_{h}}:=\left\{ \left|u\right\rangle |\left\langle u\right|X_{h}\left|u\right\rangle =1\right\} $
describe the boundary of an ellipsoid as $\sum_{i}u_{i}^{2}/(x_{h_{i}}^{-1})=1$
(note $x_{h_{i}}$ is fixed here and $u_{i}$ is the variable). Similarly
$\mathcal{E}_{OX_{g}O^{T}}$ represents a rotated ellipsoid where
$O$ is orthogonal (see \Figref{Visual-aid-EMA}). Note that larger
the $x_{h_{i}}$ (or $x_{g_{i}}$) higher is the curvature of the
ellipsoid along the associated direction. It is not hard to see the
aforesaid inequality, geometrically, as the $\mathcal{E}_{X_{h}}$
ellipsoid being contained inside the $\mathcal{E}_{OX_{g}O^{T}}$
ellipsoid (the order gets reversed; see \Subsecref{inequalityAsEllipsoids}).

Recall that the orthogonal matrix also has the property $O\left|v\right\rangle =\left|w\right\rangle $.
Imagine that in addition, we have $\left\langle w\right|X_{h}\left|w\right\rangle =\left\langle v\right|X_{g}\left|v\right\rangle $
which in terms of the point game means that the average is preserved
(as was the case for merge). In terms of the ellipsoids, it means
that the ellipsoids touch along the $\left|w\right\rangle $ direction.
More precisely, the point $\left|c\right\rangle :=\left|w\right\rangle /\sqrt{\left\langle w\right|X_{h}\left|w\right\rangle }$
belongs to both $\mathcal{E}_{X_{h}}$ and $\mathcal{E}_{OX_{g}O^{T}}$.
Since the inequality tells us the smaller $h$ ellipsoid is contained
inside the larger $g$ ellipsoid, and we now know that they touch
at the point $\left|c\right\rangle $, we conclude that their normals
evaluated at $\left|c\right\rangle $ must be equal. Further, we can
conclude that the inner ellipsoid must be more curved than the outer
ellipsoid.

Mark the point $\left|c\right\rangle $ on the $\mathcal{E}_{OX_{g}O^{T}}$
ellipsoid. Now imagine rotating the $\mathcal{E}_{X_{g}}$ ellipsoid
to the $\mathcal{E}_{OX_{g}O^{T}}$ ellipsoid. The normal at the marked
point must be mapped to the normal of $\mathcal{E}_{X_{h}}$ at $\left|c\right\rangle $.
It turns out that to evaluate the normals $\left|n_{h}\right\rangle $
on $\mathcal{E}_{X_{h}}$ at $\left|c\right\rangle $ and $\left|n_{g}\right\rangle $
on $\mathcal{E}_{X_{g}}$ at the marked point, one only needs to know
$X_{h},X_{g},\left|v\right\rangle $ and $\left|w\right\rangle $.
Complete knowledge of $O$ is not required and yet we can be sure
that $O\left|n_{g}\right\rangle =\left|n_{h}\right\rangle $ which
means $O$ must have a term $\left|n_{h}\right\rangle \left\langle n_{g}\right|$.
In fact, one can even evaluate the curvature from the aforesaid quantities.
It so turns out that when this condition is expressed precisely, it
becomes an instance of the same problem we started with one less dimension
allowing us to iteratively find $O$, which so far we had only assumed
to exist. This, however, only works under our assumption that $\left\langle w\right|X_{h}\left|w\right\rangle =\left\langle v\right|X_{g}\left|v\right\rangle $.
This is not always the case which we address next.

A monotone function $f$ is defined to be a function which has the
property ``$x\ge y$ $\implies$ $f(x)\ge f(y)$''. An operator
monotone function\footnote{Note that the monotone function $f(x)=x^{2}$ is not an operator monotone.
This is a counter-example: $\left[\begin{array}{cc}
2 & 1\\
1 & 1
\end{array}\right]\ge\left[\begin{array}{cc}
1 & 1\\
1 & 1
\end{array}\right]$.} is obtained from a generalisation of the aforesaid property to matrices,
which in our notation can be expressed as ``$X_{h}\ge OX_{g}O^{T}$
$\implies$ $f(X_{h})\ge Of(X_{g})O^{T}$''. In mathematics, it is
known that for a certain class of operator monotone functions $f$,
$f^{-1}$ is also an operator monotone. Using these results in conjunction
with results from Aharonov et al. \cite{ACG+14} we conclude that
one can show that there is always an operator monotone $f$ such that
$\left\langle w\right|f(X_{h})\left|w\right\rangle =\left\langle v\right|f(X_{g})\left|v\right\rangle $.
(This result also admits a beautiful geometric interpretation. It
means that to establish $\mathcal{E}_{X_{h}}$ is inside $\mathcal{E}_{OX_{g}O^{T}}$,
which essentially means we look at all different directions and make
sure the $h$ ellipsoid is inside the $g$ ellipsoid, we can instead
look along a single direction $\left|w\right\rangle $ and check that
all the different ellipsoids $\mathcal{E}_{f(X_{h})}$ are inside
the corresponding $\mathcal{E}_{Of(X_{g})O^{T}}$ ellipsoids along
just this direction, for every operator monotone $f$ in the class
indicated earlier.) Since the orthogonal matrix which solves the initial
problem also solves the one mapped by $f$, we can use our technique
on the latter to proceed. This shows how and why our EMA algorithm
works. Let us summarise the algorithm into an informal statement.
\begin{defn*}[EMA Algorithm (informal)]
 Given a transition from a TDPG the algorithm proceeds in three phases. 
\begin{enumerate}
\item Initialise
\begin{itemize}
\item Tightening procedure: Bring the final points close to zero until the
corresponding ellipsoids start to touch. 
\item Spectral domain, matrices: Find the spectrum of the matrices which
represent the ellipsoid. Evaluate the smallest matrix size $n$ needed
to represent the problem using ellipsoids.
\item Bootstrapping: Using the aforesaid, define $\left(X_{h}^{(n)},X_{g}^{(n)},\left|w^{(n)}\right\rangle ,\left|v^{(n)}\right\rangle \right):=\text{\ensuremath{\underbar{X}}}^{(n)}$
where the superscript denotes the size of the matrix and vectors.
\end{itemize}
\item Iterate (neglecting special cases) 

Input: $\text{\ensuremath{\underbar{X}}}^{(k)}$

Output: $\text{\ensuremath{\underbar{X}}}^{(k-1)}$, the vector $\left|u_{h}^{(k)}\right\rangle $
and the orthogonal matrices $\bar{O}_{g}^{(k)},\bar{O}_{h}^{(k)}$

Procedure:
\begin{itemize}
\item Tightening procedure: Similar to the one above, shrink the outer ellipsoid
until it touches the inner ellipsoid.
\item Honest align: Use operator monotone functions to make the ellipsoids
touch along the $\left|w\right\rangle $ direction.
\item Evaluate the Reverse Weingarten Map: Evaluate the curvatures and the
normal (which fixes $\left|u_{h}^{(k)}\right\rangle $) along the
$\left|w\right\rangle $ direction.
\item Finite Method: Use the curvatures to specify $\text{\ensuremath{\underbar{X}}}^{(k-1)}$
and find the orthogonal matrices $\bar{O}_{g}^{(k)},\bar{O}_{h}^{(k)}$.
\end{itemize}
\item Reconstruction

Evaluate $O^{(n)}$ recursively using $O^{(k)}=\bar{O}_{g}^{(k)}\left(\left|u_{h}^{(k)}\right\rangle \left\langle u_{h}^{(k)}\right|+O^{(k-1)}\right)\bar{O}_{h}^{(k)}$.
\end{enumerate}
\end{defn*}

\begin{thm}[Correctness of the EMA Algorithm (informal)]
 Given a transition of a TDPG, the EMA Algorithm always finds a $U$
such that the constraints in \Thmref{TEFconstraint-inf} are satisfied.
\end{thm}

See \Subsecref{EMA-Algorithm} for the complete algorithm and proof
of the theorem; in particular \Defref{EMAalgorithm} and \Thmref{EMAcorrectnessFormal}
for the corresponding formal statements. Results from a preliminary
numerical implementation of the EMA algorithm are discussed in \Secref{Conclusion}.

Despite the apparent simplicity of the main argument there were many
difficulties we had to address in order to prove the aforesaid statement.
We had to extend the results about operator monotone functions to
be able to use them for performing the tightening step as indicated
and for being certain that the solution unitary/orthogonal matrix
stays unchanged under these transformations. We also extended some
results related to different representations of the aforesaid transitions
as these situations arise in the tightening procedure (see \Subsecref{EMA-Lemmas-Generalisations}).
Finding an easy method for evaluating the curvatures\textemdash the
reverse Weingarten map\textemdash was key as has been noted (see \Subsecref{Weingarten}).
The trickiest part of the algorithm, which we have not mentioned here
in the introduction, was handling the cases where one of the tangent
directions of an ellipsoid has an infinite curvature. For concreteness,
imagine an ellipse which under an operator monotone gets mapped to
a line segment. The tip of the line segment, if viewed as a limit
of an ellipse, has an infinite curvature. In these cases, our finite
analysis breaks down as the normal is no longer well defined. For
the moves used by Mochon in his $1/18$ game for instance that we
tried to numerically solve using this algorithm, this infinite case
does not appear. However, to solve the split move using the algorithm
(instead of the blinkered unitaries) the infinite case does appear.
In either case, our algorithm can handle these infinite cases using
what we call the Wiggle-v method (see \Subsecref{lemmas-for-Wiggle-v}
and \Subsecref{EMA-The-Algorithm-itself}).

The implication is that we can now numerically convert known games
with arbitrarily small bias into complete protocols. One remaining
question is the effect of noise. In the current analysis two idealising
assumptions have been made. First, the EMA algorithm assumes one can
solve certain problems with arbitrary precision classically, such
as finding the roots of polynomials and diagonalising matrices. Second,
in Mochon/Kitaev's point game formalisms, one assumes that the unitaries
are known and applied exactly. Neither of these will hold practically,
therefore, the effect of noise on the bias of the protocol must be
quantified, which we leave as an open problem for further work.

The remaining document starts with stating the previously known formal
results and motivating their proofs (\Secref{Prior-Art}). Thereafter
the document is divided into two parts. The first part establishes
the TDPG-to-Explicit-protocol Framework, TEF, (\Secref{TEF}) and
then discusses its application to Mochon's games (\Secref{Games-and-Protocols})
including the one with bias $1/10$. The second part starts with laying
the groundwork for viewing the problem in terms of ellipsoids (\Secref{Canonical-Forms-Revisited}),
summarises some results about ellipsoids (\Secref{Ellipsoids}) and
finally discusses the Elliptic Monotone Algorithm itself (\Secref{EMAalgorithm}).
It ends by stating some observations made through a preliminary numerical
implementation of the said algorithm and briefly discussing related
open problems (\Secref{Conclusion}).


\vfill{}

\pagebreak{}

\section{Prior Art\label{sec:Prior-Art}}

\textcolor{purple}{We start with stating the results up to a certain
point from Aharonov et al's \cite{ACG+14} paper (which completely
formalises Mochon's results \cite{Mochon07} and simplifies one of
the results proved in its appendix; we build on this improved proof
in our work). We will motivate the statements as we go along. It is
unlikely that the next section will make perfect sense unless one
has already read Aharonov's et al's and/or Mochon's article. The ideas
should get clear in the following sections as we use them to construct
explicit protocols. Logically, however, we have tried to keep everything
consistent (even though the presentation thereof may not be optimal).}

\textcolor{purple}{We define $\mathbb{R}_{\ge}:=[0,\infty)$, $\mathbb{R}_{>}:=(0,\infty)$
and similarly $\mathbb{R}_{\le}:=(-\infty,0],\mathbb{R}_{<}:=(-\infty,0)$.
A note about the colour scheme. We use purple for intuitive and non-technical
discussions and, in this section, blue for statements/corrections
that we have added to the results from Aharonov et al's article.}

\subsection{WCF protocol as an SDP and its Dual\label{subsec:WCF-protocol-as-an-SDP-and-its-Dual}}

\textcolor{purple}{Any weak coin flipping protocol can be expressed
in the following general form (we do not prove this claim here; see
\cite{Ambainis04b}).}

\begin{defn}[WCF protocol with bias $\epsilon$]
 For $n$ even, an $n$-message WCF protocol between two players,
Alice and Bob, is described by
\begin{itemize}
\item three Hilbert spaces with $\mathcal{A},\,\mathcal{B}$ corresponding
to Alice and Bob's private workspaces (Bob does not have any access
to $\mathcal{A}$ and Alice to $\mathcal{B}$), and a message space
$\mathcal{M}$;
\item an initial product state $\left|\psi_{0}\right\rangle =\left|\psi_{A,0}\right\rangle \otimes\left|\psi_{M,0}\right\rangle \otimes\left|\psi_{B,0}\right\rangle \in\mathcal{A}\otimes\mathcal{M}\otimes\mathcal{B}$;
\item a set of $n$ unitaries $\{U_{1},\dots U_{n}\}$ acting on $\mathcal{A}\otimes\mathcal{M}\otimes\mathcal{B}$
with $U_{i}=U_{A,i}\otimes\mathbb{I}_{\mathcal{B}}$ for $i$ odd
and $U_{i}=\mathbb{I}_{\mathcal{A}}\otimes U_{B,i}$ for $i$ even;
\item a set of honest states $\{\left|\psi_{i}\right\rangle :i\in[n]\}$
defined by $\left|\psi_{i}\right\rangle =U_{i}U_{i-1}\dots U_{1}\left|\psi_{0}\right\rangle $;
\item a set of $n$ projectors $\{E_{1},\dots E_{n}\}$ acting on $\mathcal{A}\otimes\mathcal{M}\otimes\mathcal{B}$
with $E_{i}=E_{A,i}\otimes\mathbb{I}_{\mathcal{B}}$ for $i$ odd,
and $E_{i}=\mathbb{I}_{\mathcal{A}}\otimes E_{B,i}$ for $i$ even,
such that $E_{i}\left|\psi_{i}\right\rangle =\left|\psi_{i}\right\rangle $;
\item two final positive operator valued measure (POVM) $\{\Pi_{A}^{(0)},\Pi_{A}^{(1)}\}$
acting on $\mathcal{A}$ and $\{\Pi_{B}^{(0)},\Pi_{B}^{(1)}\}$ acting
on $\mathcal{B}$.
\end{itemize}
The WCF protocol proceeds as follows:
\begin{itemize}
\item In the beginning, Alice holds $\left|\psi_{A,0}\right\rangle \left|\psi_{M,0}\right\rangle $
and Bob $\left|\psi_{B,0}\right\rangle $.
\item For $i=1$ to $n$:
\begin{itemize}
\item If $i$ is odd, Alice applies $U_{i}$ and measures the resulting
state with the POVM $\{E_{i},\mathbb{I}-E_{i}\}$. On the first outcome,
Alice sends the message qubits to Bob; on the second outcome, she
ends the protocol by outputting ``0'', i.e, Alice declares herself
to be the winner.
\item If $i$ is even, Bob applies $U_{i}$ and measures the resulting state
with the POVM $\{E_{i},\mathbb{I}-E_{i}\}$. On the first outcome,
Bob sends the message qubits to Alice; on the second outcome, he ends
the protocol by outputting ``1'', i.e., Bob declares himself to
be the winner.
\item Alice and Bob measure their part of the state with the final POVM
and output the outcome of their measurements. Alice wins on outcome
``0'' and Bob on outcome ``1''.
\end{itemize}
\end{itemize}
The WCF protocol has the following properties:
\begin{itemize}
\item Correctness: When both players are honest, Alice and Bob's outcomes
are always the same: $\Pi_{A}^{(0)}\otimes\mathbb{I}_{\mathcal{M}}\otimes\Pi_{B}^{(1)}\left|\psi_{n}\right\rangle =\Pi_{A}^{(1)}\otimes\mathbb{I}_{\mathcal{M}}\otimes\Pi_{B}^{(0)}\left|\psi_{n}\right\rangle =0$.
\item Balanced: When both players are honest, they win with probability
$1/2$: \\
$P_{A}=\left|\Pi_{A}^{(0)}\otimes\mathbb{I}_{\mathcal{M}}\otimes\Pi_{B}^{(0)}\left|\psi_{n}\right\rangle \right|^{2}=\frac{1}{2}$
and $P_{B}=\left|\Pi_{A}^{(1)}\otimes\mathbb{I}_{\mathcal{M}}\otimes\Pi_{B}^{(1)}\left|\psi_{n}\right\rangle \right|^{2}=\frac{1}{2}.$
\item $\epsilon$ biased: When Alice is honest, the probability that both
players agree on Bob winning is $P_{B}^{*}\le\frac{1}{2}+\epsilon$.
And conversely, if Bob is honest, the probability that both players
agree on Alice winning is $P_{A}^{*}\le\frac{1}{2}+\epsilon$.
\end{itemize}
\end{defn}

\textcolor{purple}{To be able to define the bias, we need $P_{A}^{*}$
and $P_{B}^{*}$ which correspond to the best possible cheating strategy
of the opponent. The primal semi-definite program (SDP) formalises
this statement.}

\begin{thm}[Primal]
~\\
$P_{B}^{*}=\max\text{Tr}((\Pi_{A}^{(1)}\otimes\mathbb{I}_{\mathcal{M}})\rho_{AM,n})$
over all $\rho_{AM,i}$ satisfying the constraints
\begin{itemize}
\item $\text{Tr}_{\mathcal{M}}(\rho_{AM,0})=\text{Tr}_{\mathcal{MB}}(\left|\psi_{0}\right\rangle \left\langle \psi_{0}\right|)=\left|\psi_{A,0}\right\rangle \left\langle \psi_{A,0}\right|$;
\item for $i$ odd, $\text{Tr}_{\mathcal{M}}(\rho_{AM,i})=\text{Tr}_{\mathcal{M}}(E_{i}U_{i}\rho_{AM,i-1}U_{i}^{\dagger}E_{i})$;
\item for $i$ even, $\text{Tr}_{\mathcal{M}}(\rho_{AM,i})=\text{Tr}_{\mathcal{M}}(\rho_{AM,i-1}).$
\end{itemize}
$P_{A}^{*}=\max\text{Tr}((\mathbb{I}_{\mathcal{M}}\otimes\Pi_{B}^{(0)})\rho_{MB,n})$
over all $\rho_{BM,i}$ satisfying the constraints
\begin{itemize}
\item $\text{Tr}_{\mathcal{M}}(\rho_{MB,0})=\text{Tr}_{\mathcal{AM}}(\left|\psi_{0}\right\rangle \left\langle \psi_{0}\right|)=\left|\psi_{B,0}\right\rangle \left\langle \psi_{B,0}\right|$;
\item for $i$ even, $\text{Tr}_{\mathcal{M}}(\rho_{MB,i})=\text{Tr}_{\mathcal{M}}(E_{i}U_{i}\rho_{MB,i-1}U_{i}^{\dagger}E_{i});$
\item for $i$ odd, $\text{Tr}_{\mathcal{M}}(\rho_{MB,i})=\text{Tr}_{\mathcal{M}}(\rho_{MB,i-1})$.
\end{itemize}
\end{thm}

\textcolor{purple}{A feasible solution to an optimisation problem
is one which satisfies the constraints but is not necessarily optimal.
A feasible solution to the primal problems gives a lower bound on
$P_{A}^{*}$ and $P_{B}^{*}$. If we consider the duals instead, it
is known that, a feasible solution gives an upper bound on $P_{A}^{*}$
and $P_{B}^{*}$. This certifies how good the protocol is.}

\begin{thm}[Dual]
\label{thm:dual}~\\
$P_{B}^{*}=\min\text{Tr}(Z_{A,0}\left|\psi_{A,0}\right\rangle \left\langle \psi_{A,0}\right|)$
over all $Z_{A,i}$ under the constraints 
\begin{enumerate}
\item $\forall i,$ $Z_{A,i}\ge0$;
\item for $i$ odd, $Z_{A,i-1}\otimes\mathbb{I}_{\mathcal{M}}\ge U_{A,i}^{\dagger}E_{A,i}(Z_{A,i}\otimes\mathbb{I}_{\mathcal{M}})E_{A,i}U_{A,i};$
\item for $i$ even, $Z_{A,i-1}=Z_{A,i}$;
\item $Z_{A,n}=\Pi_{A}^{(1)}$.
\end{enumerate}
$P_{A}^{*}=\min\text{Tr}(Z_{B,0}\left|\psi_{B,0}\right\rangle \left\langle \psi_{B,0}\right|)$
over all $Z_{B,i}$ under the constraints
\begin{enumerate}
\item $\forall i,$ $Z_{B,i}\ge0$;
\item for $i$ even, $\mathbb{I}_{\mathcal{M}}\otimes Z_{B,i-1}\ge U_{B,i}^{\dagger}E_{B,i}(\mathbb{I}_{\mathcal{M}}\otimes Z_{B,i})E_{B,i}U_{B,i};$
\item for $i$ odd, $Z_{B,i-1}=Z_{B,i}$;
\item $Z_{B,n}=\prod_{B}^{(0)}$.
\end{enumerate}
We add one more constraint to the above dual SDPs.
\begin{enumerate}
\item[5.]  $\left|\psi_{A,0}\right\rangle $ is an eigenvector of $Z_{A,0}$
with eigenvalue $\alpha>0$ and $\left|\psi_{B,0}\right\rangle $
is an eigenvector of $Z_{B,0}$ with eigenvalue $\beta>0$.
\end{enumerate}
\end{thm}

\begin{defn}[dual feasible points]
 We call \emph{dual feasible points} any two sets of matrices $\{Z_{A,0},\dots,Z_{A,n}\}$
and $\{Z_{B,0},\dots,Z_{B,n}\}$ that satisfy the corresponding conditions
1 to 5 as listed in \Thmref{dual}.
\end{defn}

\textcolor{purple}{It turns out that strong duality holds for the
primal problems which means that there is a cheating strategy for
Alice and Bob matching the upper bound on $P_{A}^{*}$ and $P_{B}^{*}$
respectively.}

\begin{prop}
$P_{A}^{*}=\inf\alpha$ and $P_{B}^{*}=\inf\beta$ where the infimum
is over all dual feasible points and $\beta,\alpha$ are defined in
constraint 5 of the definition of the dual feasible points.
\end{prop}

\subsection{(Time Dependent) Point Games with EBM transitions\label{subsec:TDPG-with-EBM-transitions}}

\textcolor{purple}{The basic idea here is to remove all inessential
information, that is the basis information, from the two aforesaid
dual problems. Kitaev's genius was to achieve this by considering,
at a given step, the dual variable $Z_{A},Z_{B}$ as observables with
$\left|\psi\right\rangle $ governing the probability. This combines
the evolution of the certificates on cheating probabilities with the
evolution of the honest state\textemdash the state obtained when both
players follow the protocol (nobody cheats). Originally, using a similar
manoeuvre, Kitaev settled solvability of the quantum strong coin flipping
problem by giving a bound on $\epsilon$. To make this insight precise,
first ``prob'' is defined.}
\begin{defn}[prob]
 Consider $Z\ge0$ and let $\Pi^{[z]}$ represent the projector on
the eigenspace of eigenvalue $z\in\text{sp}(Z)$. We have $Z=\sum_{z}z\Pi^{[z]}$.
Let $\left|\psi\right\rangle $ be a (not necessarily normalised)
vector. We define the function with finite support $\text{prob}[Z,\psi]:[0,\infty)\to[0,\infty)$
as 
\[
\text{prob}[Z,\psi](z)=\begin{cases}
\left\langle \psi\right|\Pi^{[z]}\left|\psi\right\rangle  & \text{if }z\in\text{sp}(Z)\\
0 & \text{else}.
\end{cases}
\]
If $Z=Z_{A}\otimes\mathbb{I}_{\mathcal{M}}\otimes Z_{B}$, using the
same notation, we define the $2-$variate function with finite support
$\text{prob}[Z_{A},Z_{B},\psi]:[0,\infty)\times[0,\infty)\to[0,\infty)$
as 
\[
\text{prob}[Z_{A},Z_{B},\psi](z_{A},z_{B})=\begin{cases}
\left\langle \psi\right|\Pi^{[z_{A}]}\otimes\mathbb{I}_{\mathcal{M}}\otimes\Pi^{[z_{B}]}\left|\psi\right\rangle  & \text{if }(z_{A},z_{B})\in\text{sp}(Z_{A})\times\text{sp}(Z_{B}),\\
0 & \text{else}.
\end{cases}
\]
\label{def:prob}
\end{defn}

\textcolor{purple}{Think of the aforesaid 2\textendash variate function
as assigning a weight on each point of the plane. Going from one such
configuration to another is what we would intuitively refer to as
a ``move'' for the moment. Notice that at an odd step $i$, the
dual variable $Z_{B}$ doesn't change while $Z_{A}$ does (see \Thmref{dual}).
The constraint equation in this step is $Z_{A,i-1}\otimes\mathbb{I}_{M}\ge U_{i}^{\dagger}\left(Z_{A,i}\otimes\mathbb{I}_{M}\right)U_{i}$.
The honest state can be expressed as $\left|\psi_{i}\right\rangle =U_{i}\left|\psi_{i-1}\right\rangle $
but this acts on the complete $\mathcal{A}\otimes\mathcal{M}\otimes\mathcal{B}$
space. Applying the aforesaid method of removing the basis information
using the $\text{prob}$ method, and appending the fixed $Z_{B,i-1}=Z_{B,i}$,
we conclude that $\text{prob}(Z_{A,i-1}\otimes\mathbb{I}_{M}\otimes Z_{B,i},\left|\psi_{i-1}\right\rangle )\to\text{prob}(Z_{A,i}\otimes\mathbb{I}_{M}\otimes Z_{B,i},\left|\psi_{i}\right\rangle )$
should constitute an ``allowed move'' as it is simply re-expressing
the dual SDP in a basis independent form. For the dual, we are assuming
the protocol is given to us, i.e. $U_{i}$ (unitary operations), $\Pi_{A/B}$
(measurements) and $\left|\psi_{0}\right\rangle $ (initial state)
are specified, and we have to find the appropriate $Z$s. However,
when we discuss the notion of an ``allowed move'' we are moving
towards a framework which will free us from discussing a specific
protocol. This motivates the following definitions.}\begin{defn}[EBM line transition]
 Let $g,h:[0,\infty)\to[0,\infty)$ be two functions with finite
supports. The line transition $g\to h$ is EBM if there exist two
matrices $0\le G\le H$ and a (not necessarily normalised) vector
$\left|\psi\right\rangle $ such that $g=\text{prob}[G,\psi]$ and
$h=\text{prob}[H,\psi]$.\label{def:EBMlineTransition}
\end{defn}

\begin{defn}[EBM transition]
 Let $p,q:[0,\infty)\times[0,\infty)\to[0,\infty)$ be two functions
with finite supports. The transition $p\to q$ is an 
\begin{itemize}
\item EBM horizontal transition if for all $y\in[0,\infty),$ $p(.,y)\to q(.,y)$
is an EBM line transition, and 
\item EBM vertical transition if for all $x\in[0,\infty),$ $p(x,.)\to q(x,.)$
is an EBM line transition.
\end{itemize}
\end{defn}

\textcolor{purple}{It turns out that when one writes the dual, the
order of the constraints gets inverted, i.e. the condition associated
with the final measurements and states appears first and the condition
associated with the initial state appears in the end. We expect the
final state to be like an EPR state and, intuitively, expect two points
(in terms of the 2\textendash variate function as described earlier)
to be associated with it. This makes it plausible that we will start
with two points in the dual when it is expressed in the aforementioned
basis independent way. The initial state of the protocol is unentangled.
This we expect should correspond to a single point. This helps us
accept that we end with a single point in the basis independent expression
of the dual. The rules for moving these points must be related to
the dual constraints. We already formalised these conditions into
EBM transitions. The notation 
\[
[x_{g},y_{g}](x,y)=\begin{cases}
1 & x_{g}=x\text{ and }y_{g}=y\\
0 & \text{else}
\end{cases}
\]
will be useful for formalising the complete description into what
Mochon dubbed an ``Expressible by Matrices'' (Time Dependent) point
game.}

\begin{defn}[EBM point game]
 An EBM point game is a sequence of functions $\{p_{0},p_{1},\dots,p_{n}\}$
with finite support such that 
\begin{itemize}
\item $p_{0}=1/2[0,1]+1/2[1,0]$;
\item for all even $i$, $p_{i}\to p_{i+1}$ is an EBM vertical transition;
\item for all odd $i$, $p_{i}\to p_{i+1}$ is an EBM horizontal transition;
\item $p_{n}=1[\beta,\alpha]$ for some $\alpha,\beta\in[0,1]$. We call
$[\beta,\alpha]$ the final point of the EBM point game.
\end{itemize}
\end{defn}

Since we started with a WCF protocol, considered its dual and re-expressed
it as a TDPG (which is just a basis independent representation), the
following should not come as a surprise.
\begin{prop}[WCF $\implies$ EBM point game]
Given a WCF protocol with cheating probabilities $P_{A}^{*}$ and
$P_{B}^{*}$, along with a positive real number $\delta>0$, there
exists an EBM point game with final point $[P_{B}^{*}+\delta,P_{A}^{*}+\delta]$.
\end{prop}

\textcolor{purple}{What is slightly more non-trivial is that given
this TDPG one can construct a WCF protocol. This means that by using
only ``allowed moves'' one can be sure that there exists a corresponding
sequence of unitaries $U_{i}$, the measurements $\Pi_{A/B}$ and
the initial state $\left|\psi_{0}\right\rangle $ complemented by
the dual variables $Z_{A,i}$ and $Z_{B,i}$ which certify the bias
corresponding to the coordinates of the final point in the point game.
This establishes the equivalence between TDPGs and WCF protocols.
The precise statement is as follows.}
\begin{thm}[EBM to protocol]
 Given an EBM point game with final point $[\beta,\alpha]$, there
exists a WCF protocol and two dual feasible points proving that the
optimal cheating probabilities are $P_{A}^{*}\le\alpha$ and $P_{B}^{*}\le\beta$.
\end{thm}

\textcolor{purple}{Our first contribution is related to this part.
We construct protocols from EBM point games in a slightly different
way which results in two important improvements. The first improvement
makes the protocol more practical as the message register gets decoupled
from Alice and Bob's registers after each round. (In Mochon/Aharonov
et al's version the message register is highly entangled and stays
that way until the very end.) The second improvement is due to the
addition of a cheat detection measurement at every round (similar
to Mochon's improved Dip Dip Boom protocol) which allows us to consider
certain matrices with infinite eigenvalues in a well defined way.
These pave the path for converting the bias $1/10$ point game (due
to Mochon; will be introduced later) into a protocol.}

\subsection{(Time Dependent) Point Games with valid transitions\label{subsec:TDPG-with-valid-transitions}}

\textcolor{purple}{While the problem has been simplified by the removal
of the basis information, it is still hard to know which transitions
are allowed, i.e. are EBM transitions. This is because finding the
matrices certifying that a transition is EBM is not easy. The goal
of this section is to find another criterion for establishing that
a transition is EBM. This criterion is at the heart of coin flipping.
It would turn out that the set of EBM functions (closely related to
EBM transitions) form a closed convex cone. The dual of this cone
happens to be the set operator monotone functions (as described earlier
in \Subsecref{EMA-Algorithm}, these are a generalisation of monotone
functions, $x\ge y\implies f(x)\ge f(y)$, to $X\ge Y\implies f(X)\ge f(Y)$
where $X$ and $Y$ are now matrices). These functions have a very
nice and simple characterisation. This is what leads to the key simplification
for WCF. To be able to harness this, one can use the known fact that
for a closed convex cone, the dual of the dual is the original cone
itself (also called bi-dual). So this dual of operator monotone functions,
i.e. the bi-dual of the cone of EBM functions, equals the cone of
EBM functions. The dual of operator monotone functions has an easy
description because operator monotone functions have an easy description.
Combining these, one obtains an easy characterisation of EBM functions
which allows one to construct interesting WCF protocols. The catch
is that we establish that the two cones, the cone of EBM functions
and the dual of the cone of operator monotone functions, are the same
but given a point in the second cone we do not have a recipe for finding
the matrices certifying it is an EBM. Without the matrices we can
not implement the protocol even though we know the matrices must exist
as the cones are the same. These notions are now formalised.}

\subsubsection{Formalising the equivalence between transitions and functions}

\textcolor{purple}{Working with functions instead of transitions will
be rather useful as will be evident from the next subsection.}

\begin{defn}[$K$, EBM functions]
 A function $a:\mathbb{R}_{\ge0}\to\mathbb{R}$ with finite support
is an \emph{EBM function} if the line transition $a^{-}\to a^{+}$
is EBM, where $a^{+}:\mathbb{R}_{\ge0}\to\mathbb{R}_{\ge0}$ and $a^{-}:\mathbb{R}_{\ge0}\to\mathbb{R}_{\ge0}$
denote, respectively, the positive and the negative part of $a$ ($a=a^{+}-a^{-})$. 

We denote by $K$ the set of EBM functions.\label{def:EBMfunctions}
\end{defn}

\begin{defn}[{$K_{\Lambda},$ EBM functions on $[0,\Lambda]$}]
 For any finite $\Lambda$, a function $a:[0,\Lambda)\to\mathbb{R}$
with finite support is an \emph{EBM function with support on $[0,\Lambda]$}
if the line transition $a^{-}\to a^{+}$ is EBM with its spectrum
in $[0,\Lambda]$, where $a^{-}:[0,\Lambda)\to\mathbb{R}_{\ge0}$
and $a^{+}:[0,\Lambda)\to\mathbb{R}_{\ge0}$ denote, respectively,
the positive and the negative part of $a$. 

We denote the set of EBM functions with support on $[0,\Lambda]$
by $K_{\Lambda}$.\label{def:lambdaEBMfunctions}
\end{defn}

\textcolor{purple}{It is evident that if the functions $g,h$ denoting
the transition $g\to h$ have no common support then the function
description uniquely captures the said transition. In this section
we restrict to such transitions and therefore use them interchangeably.
In later sections we revisit this notion.}

\textcolor{purple}{To be able to talk about different characterisations
of EBM functions it is useful to abstract it (the characterisation)
into a property $\mathcal{P}$ which the function must satisfy. Using
this we can define games which use these $\mathcal{P}$ functions.
This is done to be able to handle subtleties which arise in proving
that the set of EBM functions is the same as the set of $\mathcal{P}$
functions for specific $\mathcal{P}$s. }

\begin{defn}[Horizontal and vertical $\mathcal{P}$-functions]
 A $\mathcal{P}$-function $a:\mathbb{R}_{\ge0}\to\mathbb{R}$ is
a function with finite support that has the property $\mathcal{P}$.

A function $t:\mathbb{R}_{\ge0}\times\mathbb{R}_{\ge0}\to\mathbb{R}$
is a
\begin{itemize}
\item \emph{horizontal $\mathcal{P}$-function} if for all $y\ge0$, $t(.,y)$
is a $\mathcal{P}$-function;
\item \emph{vertical $\mathcal{P}$-function} if for all $x\ge0,$ $t(x,.)$
is a $\mathcal{P}$-function.
\end{itemize}
\end{defn}

\begin{defn}[point games with $\mathcal{P}$-functions]
 A point game with $\mathcal{P}$-functions is a set $\{t_{1},\dots,t_{n}\}$
of $n$ $\mathcal{P}$-functions alternatively horizontal and vertical
such that 
\begin{itemize}
\item $\frac{1}{2}[0,1]+\frac{1}{2}[1,0]+\sum_{i=1}^{n}t_{i}=[\beta,\alpha];$
\item $\forall j\in\{1,\dots,n\}$, $\frac{1}{2}[0,1]+\frac{1}{2}[1,0]+\sum_{i=1}^{j}t_{i}\ge0.$
\end{itemize}
We call $[\beta,\alpha]$ the final point of the game.
\end{defn}

\textcolor{purple}{We note the following before looking at $\mathcal{P}$
functions in more detail.}\begin{lem}[point game with EBM functions $\implies$ point game with EBM transitions]
 Given a point game with $n$ EBM functions and final point $[\beta,\alpha]$
we can construct a point game with $n$ EBM transitions and final
point $[\beta,\alpha]$.
\end{lem}

\subsubsection{Operator monotone functions and valid functions}

\textcolor{purple}{This discussion is essential to understand our
second contribution. The set of EBM functions forms a convex cone.
To see this we recall the definition of a convex cone.}\begin{defn}[convex cone]
 A set $C$ in a vector space $V$ is a cone if for all $x\in C$
and for all $\lambda>0$, $\lambda x\in C$. It is convex if for all
$x,y\in C$, $x+y\in C$.
\end{defn}

\textcolor{purple}{Noting that the state $\left|\psi\right\rangle $
in the definition of an EBM function (which in turn invokes an EBM
transition) is unnormalised the set of EBM functions is easily seen
to form a cone. By taking a direct sum one can establish convexity
as well. The vector space of interest here is given by the span of
the basis $\left\{ [x_{g}]\right\} _{x_{g}\in[0,\infty)}$ where $[x_{g}](x)=\delta_{x_{g},x}$.
The notation is similar to the one introduced earlier. We use it shortly.}

\begin{lem}
$K$ is a convex cone. Also, for any $\Lambda\in(0,\infty)$, $K_{\Lambda}$
is a convex cone.
\end{lem}

\textcolor{purple}{To establish an alternative characterisation of
the cone of EBM functions we need to define what is called a dual
cone.}

\begin{defn}[dual cone]
 Let $C$ be a cone in a normed vector space $V$. We denote by $V'$
the space of continuous linear functionals from $V$ to $\mathbb{R}$.
The dual cone of a set $C\subseteq V$ is 
\[
C^{*}=\left\{ \Phi\in V'|\forall a\in C,\Phi(a)\ge0\right\} .
\]
\end{defn}

\textcolor{purple}{For our purpose linear functionals can be thought
of simply as functions which map objects in the cone to a non-negative
real number with the added property of being linear in its argument.}

\textcolor{purple}{We now formally define operator monotone functions.}\begin{defn}[operator monotone functions]
 A function $f:\mathbb{R}_{\ge0}\to\mathbb{R}$ is operator monotone
if for all $0\le X\le Y$ we have $f(X)\le f(Y)$.\label{def:operatorMonotone}
\end{defn}

\begin{defn}[{operator monotone functions on $[0,\Lambda]$}]
 A function $f:[0,\Lambda]\to\mathbb{R}$ is operator monotone on
$[0,\Lambda]$ if for all $0\le X\le Y$ with spectrum in $[0,\Lambda]$
we have $f(X)\le f(Y)$.\label{def:operatorMonotoneLambda}
\end{defn}

\textcolor{purple}{The pivotal result of this (sub)section is the
equivalence between the cone of operator monotone functions and the
dual cone of EBM functions.}\begin{lem}
$\Phi\in K^{*}$ if and only if $f_{\Phi}$ is operator monotone in
$[0,\infty]$. Also, for any $\Lambda\in(0,\infty)$, $\Phi\in K_{\Lambda}^{*}$
if and only if $f_{\Phi}$ is operator monotone on $[0,\Lambda]$.
\end{lem}

(NB: We need to use the bijection between $\Phi$ (a linear functional
from $V\to\mathbb{R}$) and a function on reals (from $\mathbb{R}\to\mathbb{R}$)
given by the identification $f_{\Phi}(x)=\Phi([x])$ to make such
a statement)

\textcolor{purple}{The proof of this crucial result is not too hard
(almost trivial in one direction) and follows from the respective
definitions with some work for unpacking. What makes this connection
interesting is the following beautiful characterisation of operator
monotone functions introduced by Löwner (in 1934, see \cite{Bhatia2013}).}\begin{lem}[characterisation of operator monotone functions]
 Any operator monotone function $f:\mathbb{R}_{\ge0}\to\mathbb{R}$
can be written as 
\[
f(x)=c_{0}+c_{1}x+\int_{0}^{\infty}\frac{\lambda x}{\lambda+x}d\omega(\lambda)
\]
for a measure $\omega$ satisfying $\int_{0}^{\infty}\frac{\lambda}{1+\lambda}d\omega(\lambda)<\infty$.
\label{lem:opMonChar}
\end{lem}

\begin{lem}[{characterisation of operator monotone functions on $[0,\Lambda]$}]
 Any operator monotone function $f:[0,\Lambda]\to\mathbb{R}$ can
be written as 
\[
f(x)=c_{0}+c_{1}x+\int\frac{\lambda x}{\lambda+x}d\omega(\lambda)
\]
with the integral ranging over $\lambda\in(-\infty,-\Lambda)\cup(0,\infty)$
\textcolor{blue}{satisfying $\int\frac{\lambda}{1+\lambda}d\omega(\lambda)<\infty$}
where $\omega$ is a measure.\label{lem:opMonCharLambda}
\end{lem}

\textcolor{purple}{As will become clear when we discuss the dual of
the cone of operator monotones, it suffices to consider operator monotones
of the form $\lambda x/(\lambda+x)$ (which basically is because $\omega$
is a measure). }

\textcolor{purple}{So far the statements from Aharonov et al's paper
were made in the same order as they had originally appeared. We now
re-order these a little with an eye on our end-goal (as opposed to
the one of Mochon/Aharonov). It is known that the bi-dual of a cone
is the closure of the cone we started with. }
\begin{fact}
Let $C\subseteq V$ be a convex cone, then $C^{**}=\text{cl}(C)$
where $C^{*}$ is the dual cone of $C$.\footnote{See {[}Boyd and Vandenberghe 2004{]} for proofs of these facts.}
\end{fact}

\textcolor{purple}{The astute reader would have guessed where we are
going with this discussion. We define, from hindsight, the bi-dual
of EBM functions to be the cone of valid functions. Since the dual
of EBM functions has an easy characterisation, the bi-dual also has
an easy characterisation which is why we are interested in it.}
\begin{defn}[$\Lambda$ valid functions]
\textcolor{blue}{ A function $a:[0,\Lambda]\to\mathbb{R}$ with finite
support on $[0,\Lambda]$ is $\Lambda$ valid if $a\in K_{\Lambda}^{**}$.\label{def:lambdaValid}}
\end{defn}

\textcolor{purple}{To be able to use the aforementioned fact we note
that the cone of interest, the cone of EBM functions, is closed. As
one can imagine, proving this is easier if the matrices involved have
a bounded spectrum. We consider only these for now. This means that
the cone of valid functions is the same as the cone of EBM functions.
We state these precisely below.}
\begin{lem}
For $\Lambda\in(0,\infty)$, $K_{\Lambda}$ is closed (which implies
$K_{\Lambda}^{**}=K_{\Lambda}$).
\end{lem}

\begin{cor}
For $\Lambda\in(0,\infty)$, $K_{\Lambda}=\{a\in V|\forall\Phi\in K_{\Lambda}^{*},\Phi(a)\ge0\}.$
\end{cor}

\begin{cor}[{EBM on $[0,\Lambda]$ is equivalent to $\Lambda$ valid}]
 A function $a:[0,\Lambda]\to\mathbb{R}$ with finite support on
$[0,\Lambda]$ is EBM on $[0,\Lambda]$ if and only if $\sum_{x}a(x)=0$,
$\sum_{x}xa(x)\ge0$ and $\forall\lambda\in(-\infty,-\Lambda]\cup(0,\infty)$,
$\sum_{x}\frac{\lambda x}{\lambda+x}a(x)\ge0$.\label{cor:EBMandLambdaValid}
\end{cor}

\textcolor{purple}{In the last statement, the characterisation of
operator monotone functions was used which we introduced earlier.
Note that all the statements made here assume that the matrices used
in EBM functions have a finite spectrum. Our EMA algorithm heavily
relies on this part of the analysis which is due to Aharonov et al. }

\textcolor{purple}{It is worth pointing out that Mochon outlines this
scheme used by Aharonov et al. but himself proceeds by using matrix
perturbation theory for proving a similar result.}

\subsubsection{Strictly valid functions are EBM functions}

\textcolor{purple}{To be able to simplify the conditions one needs
to check, it is useful to relax the condition on the spectrum of the
matrices involved. This is evident from range of $\lambda$ one needs
to use in the characterisation of operator monotone functions (compare
\Lemref{opMonCharLambda} and \Lemref{opMonChar}). }

\textcolor{purple}{It is easy to describe the interior of the dual
of a cone. It is also possible to relate the interior with the closure
of the cone, but in finite dimensions. This reasoning fails for infinite
dimensions. They still serve as motivation for the definition of valid
and strictly valid functions. }

\begin{fact}
Let $C$ be a convex set, then $\text{int}(C)=\text{int}(\text{cl}(C))$.
\end{fact}

\begin{sloppy}
\begin{fact}
Let $C$ be a cone in the finite-dimensional vector space $V$, then
$\text{int}(C^{*})=\left\{ \Phi\in V'|\forall a\in C-\{0\},\Phi(a)>0\right\} $.
\end{fact}

\end{sloppy}

\begin{defn}[valid function]
 A function $a:\mathbb{R}_{\ge0}\to\mathbb{R}$ with finite support
is valid if for every operator monotone function $f:\mathbb{R}_{\ge0}\to\mathbb{R}$
we have $\sum_{x\in\text{supp}(h)}f(x)a(x)\ge0$.
\end{defn}

\begin{defn}[strictly valid function]
 A function $a:\mathbb{R}_{\ge0}\to\mathbb{R}$ with finite support
is strictly valid if for every non-constant operator monotone function
$f:\mathbb{R}_{\ge0}\to\mathbb{R}$ we have $\sum_{x\in\text{supp}(a)}f(x)a(x)>0$.
\end{defn}

\textcolor{purple}{One can use the characterisation of operator monotone
functions to explicitly characterise the set of valid and strictly
valid functions.}
\begin{lem}
Let $a:\mathbb{R}_{\ge0}\to\mathbb{R}$ be a function with finite
support such that $\sum_{x}a(x)=0$. The function a is a strictly
valid function if and only if for all $\lambda>0$, $\sum_{x}\frac{-a(x)}{\lambda+x}>0$
\textcolor{blue}{and $\sum_{x}x.a(x)>0$. }\\
\textcolor{blue}{(We added the last condition else the merge (discussed
later) becomes a strictly valid function but it can be shown that
no bounded matrices exist for which it is EBM.)}

The function $a$ is valid if and only if for all $\lambda>0$, $\sum_{x}\frac{-a(x)}{\lambda+x}\ge0$
\textcolor{blue}{and $\sum_{x}x.a(x)\ge0$}.
\end{lem}

\textcolor{purple}{The set of strictly valid functions can be shown
to also be $\Lambda$ valid for some finite $\Lambda$. This means
that it would also be EBM on $[0,\Lambda]$ which in turn means it
would be an EBM function. We hence have the following.}
\begin{lem}
Any strictly valid function is an EBM function.\label{lem:strictlyValidIsEBM}
\end{lem}

\subsubsection{From valid functions to EBM functions}

\textcolor{purple}{If we construct a point game with valid functions
we can convert it into a game with EBM functions with an arbitrarily
small overhead on the bias. The trick is to raise the coordinates
of all the final points (ones with positive weight) a little at each
step, to convert a valid function into a strictly valid function. }
\begin{thm}[valid to EBM]
 Given a point game with $2m$ valid functions and final point $[\beta,\alpha]$
and any $\epsilon>0$, we can construct a point game with $2m$ EBM
functions and final point $[\beta+\epsilon,\alpha+\epsilon]$.
\end{thm}

\begin{lem}
Fix $\epsilon>0$. Given a point game with $2m$ valid functions and
final point $[\beta,\alpha]$ we can construct a point game with $2m$
strictly valid functions and final point $[\beta+\epsilon,\alpha+\epsilon]$.\label{lem:ValidToStrictlyValidGame}
\end{lem}

\subsubsection{Examples of valid line transitions}

\textcolor{purple}{We go back to transitions to discuss some simple
valid and strictly valid line transitions which are defined similar
to the corresponding functions.}
\begin{defn}[Valid and strictly valid line transitions]
 Let $g,h:\mathbb{R}_{\ge0}\to\mathbb{R}$ be two functions with
finite support. The transition $g\to h$ is valid (resp., strictly
valid) if the function $h-g$ is valid (resp., strictly valid).
\end{defn}

\textcolor{purple}{We focus our attention to the simplest cases. The
first is to increase the coordinate of a point. The second considers
the case of merging two points into one. The third is about splitting
a single point into two.}
\begin{example}[Point raise]
 $p[x_{g}]\to p[x_{h}]$ with $x_{h}\ge x_{g}$.
\end{example}

\begin{example}[Point merge]
 $p_{g_{1}}[x_{g_{1}}]+p_{g_{2}}[x_{g_{2}}]\to(p_{g_{1}}+p_{g_{2}})[x_{h}]$
with $x_{h}\ge\frac{p_{g_{1}}x_{g_{1}}+p_{g_{2}}x_{g_{2}}}{p_{g_{1}}+p_{g_{2}}}$.
\end{example}

\begin{example}[Point split]
 $p_{g}[x_{g}]\to p_{h_{1}}[x_{h_{1}}]+p_{h_{2}}[x_{h_{2}}]$ with
$p_{g}=p_{h_{1}}+p_{h_{2}}$ and $\frac{p_{g}}{x_{g}}\ge\frac{p_{h_{1}}}{x_{h_{1}}}+\frac{p_{h_{2}}}{x_{h_{2}}}$.
\end{example}

\textcolor{purple}{The merge and split can be generalised to many
points and be shown to have the same form. }

\subsection{TIPGs.}

\textcolor{purple}{Mochon's Dip Dip Boom protocol, the one with bias
$1/6$, can be expressed already as a (time dependent) point game.
However, it is possible to simplify the point game formalism even
further and it is in this simplified formalism Mochon constructs his
family of point games that achieve an arbitrarily small bias. Instead
of worrying about the entire sequence of horizontal and vertical transitions,
one can focus on just two functions as described below. }
\begin{defn}[TIPG]
 A TIPG is a valid horizontal function $a$ and a valid vertical
function $b$ such that 
\[
a+b=1[\beta,\alpha]-\frac{1}{2}[0,1]-\frac{1}{2}[1,0]
\]
for some $\alpha,\beta>1/2$. We call the point $[\beta,\alpha]$
the final point of the game.
\end{defn}

\textcolor{purple}{The main difference here is that we do not worry
about the sequence in which one must apply the transitions to obtain
the final configuration. This justifies the name TIPG which stands
for a Time Independent Point Game. It is not too hard to see that
if we have a valid point game we can combine the horizontal functions
and the vertical functions to obtain $a$ and $b$. It is a little
counter-intuitive in fact to learn that one can convert a TIPG into
a valid (time dependent) point game with an arbitrarily small cost
on the bias. It is counter-intuitive because it is not clear that
one can flesh out a time ordered sequence as one can, and in fact
does for Mochon's point games, run into causal loops that is you expect
a point to be present to create another point which in turn is required
to produce the first point. The trick that is used to fix this problem
is known as the ``catalyst state''. One deposits a little bit of
weight wherever there is negative weight for $a$, for instance, and
then one can implement a scaled down round of $a$ and $b$. The scaling
is proportional to the weight that is placed to start with. Repeating
this procedure multiple times yields the required final state along
with the ``catalyst state'' which stays unchanged. Absorbing the
catalyst state leads to a small increase in the bias. The number of
rounds increases with how small one wants this increase in bias to
be.}
\begin{thm}[TIPG to valid point games]
 Given a TIPG with a valid horizontal function $a$ and a valid vertical
function $b$ such that $a+b=1[\beta,\alpha]-\frac{1}{2}[0,1]-\frac{1}{2}[1,0]$,
we can construct, for all $\epsilon>0$, a valid point game with final
point $[\beta+\epsilon,\alpha+\epsilon]$ where the number of transitions
depends on $\epsilon$.
\end{thm}

\textcolor{purple}{It is important to state that the conversion from
a TIPG to a valid (time dependent) point game, TDPG, is easy and explicit.}

\textcolor{purple}{A word about resource usage. The size (dimension)
of the physical system we use depends on the number of points involved
in the point games linearly. The number of rounds on the other hand
needs to be calculated with more care as it depends on the choice
of the catalyst state. These calculations with respect to Mochon's
game and in general have been performed by Aharonov et al. in their
article and we do not discuss it here. }

\textcolor{purple}{We have stated enough results to be able to commence
the discussion of our work.\label{endPriorArt}}

\clearpage{}

\part{Bias $1/10$}

\section{TDPG $\to$ Explicit Protocol, Framework (TEF)\label{sec:TEF}}

\textcolor{purple}{We strongly recommend that the reader looks at
the third section titled ``The illustrated guide to point games''
from Mochon's \cite{Mochon07} article, if they have not already,
before proceeding.}

\subsection{Motivation and Conventions}

\textcolor{purple}{We wish to construct a protocol such that its dual
matches a given TDPG. The main difference in our construction, compared
to the one used by Aharonov et al. and Mochon, is the introduction
of a message register that decouples after each round and of suitably
adapted projectors. Consequently, the non-trivial constraint that
the dual matrices must satisfy would be similar to, but not quite
the same as, the EBM condition.}

\textcolor{purple}{Keep \Defref{prob} in mind. Intuitively, the most
natural way of constructing $Z$s and a $\left|\psi\right\rangle $
given an arbitrary frame (think of a TDPG as a sequence of frames)
is to construct an entangled state that encodes the weight and define
$Z$s to contain the coordinates corresponding to the weight. Let
us make this idea more precise.}

\begin{figure}[h]
\begin{centering}
\subfloat[Frame of a TDPG\label{fig:TDPGframe}]{\begin{centering}
\includegraphics[height=6cm]{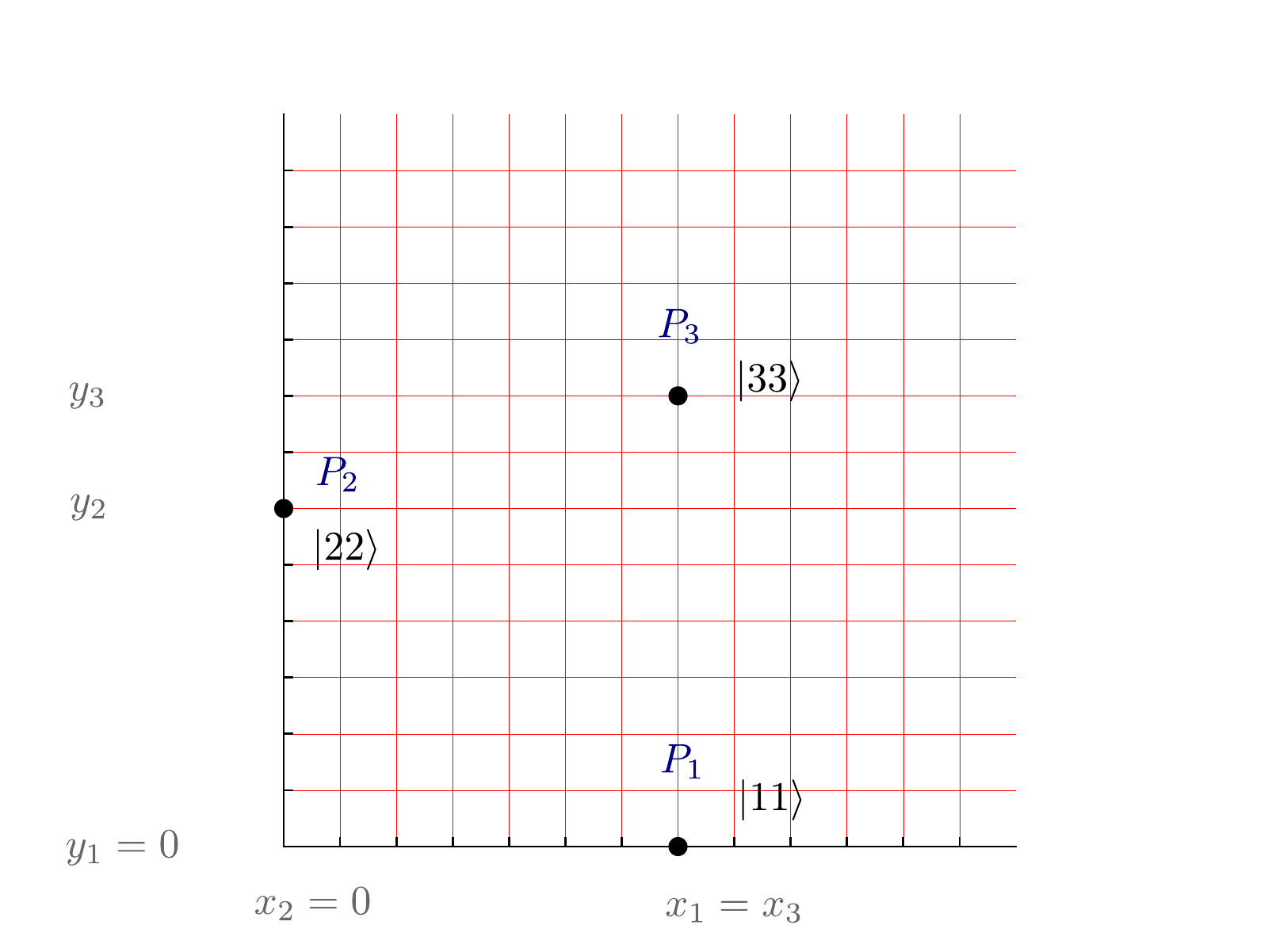}
\par\end{centering}
}
\par\end{centering}
\begin{centering}
\subfloat[The points which are unchanged from one frame to another are labelled
by $\{k_{i}\}$. Among the points that change, the initial ones are
labelled by $\{g_{i}\}$ and the final ones by $\{h_{i}\}$.\label{fig:TDPGillustrating_kgh}]{\begin{centering}
\includegraphics[height=6cm]{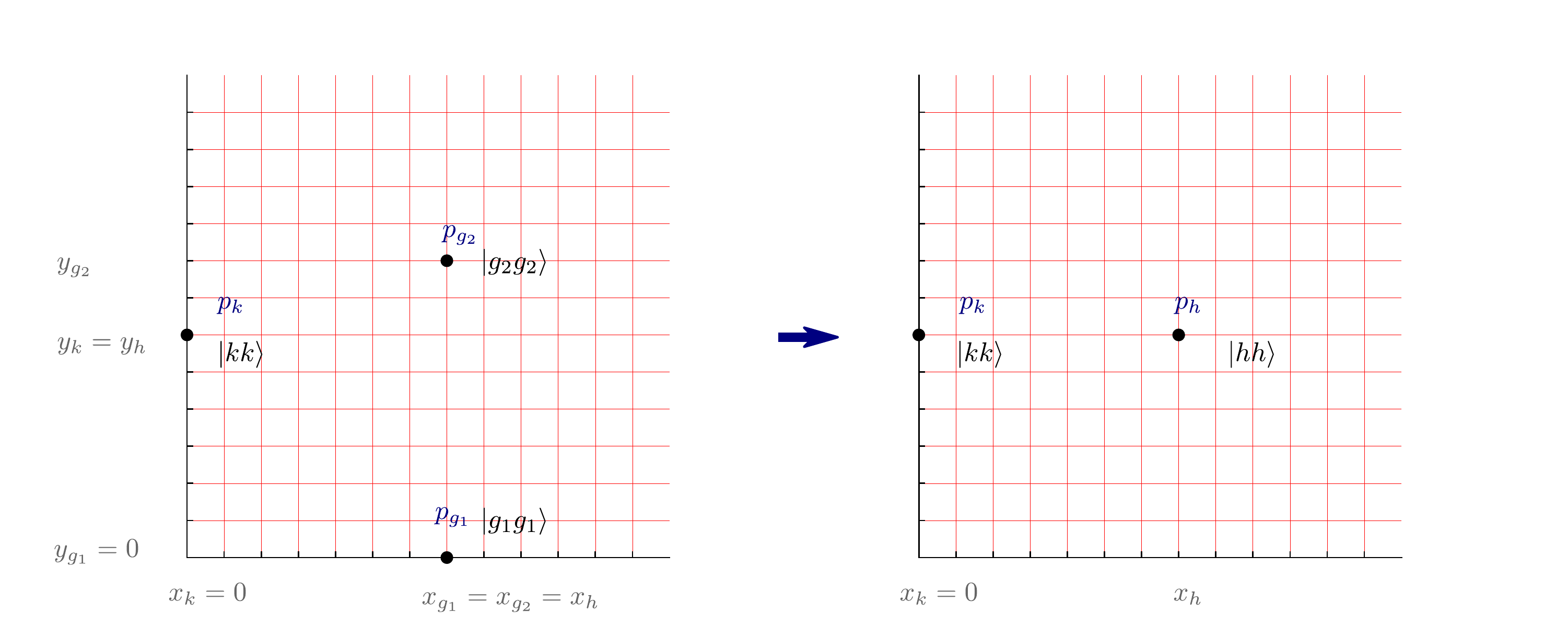}
\par\end{centering}
}
\par\end{centering}
\caption{Illustrations for the Canonical Form}
\end{figure}

\begin{defn*}[Canonical Form]
 The tuple $(\left|\psi\right\rangle ,Z^{A},Z^{B})$ is said to be
in the Canonical Form with respect to a set of points in a frame of
a TDPG if (see \Figref{TDPGframe}) $\left|\psi\right\rangle =\sum_{i}\sqrt{P_{i}}\left|ii\right\rangle _{AB}\otimes\left|.\right\rangle _{M}$,
$Z^{A}=\left(\sum x_{i}\left|i\right\rangle \left\langle i\right|_{A}\right)\otimes\left|.\right\rangle \left\langle .\right|_{M}$
and $Z^{B}=\left(\sum y_{i}\left|i\right\rangle \left\langle i\right|_{B}\right)\otimes\left|.\right\rangle \left\langle .\right|_{M}$
where $\left|.\right\rangle _{M}$ represent extra uncoupled registers
which might be present.
\end{defn*}

\textcolor{purple}{It is easy to see that the `label' $\left|ii\right\rangle $
correspond to a point with coordinates $x_{i},y_{i}$ and weight $P_{i}$
in the frame (see \Defref{prob}). It is tempting to imagine that
we systematically construct, from each frame of a TDPG, a canonical
form of $\left|\psi\right\rangle s$ and $Z$s. The unitaries can
be deduced from the evolution of $\left|\psi\right\rangle $. This
approach has two problems, (1) it does not manifestly mean that the
unitaries would be decomposable into moves by Alice and Bob who communicate
only through the message register and (2) the constraints imposed
consecutive $Z$s, of the form $Z_{n-1}\otimes\mathbb{I}\ge U_{n}^{\dagger}\left(Z_{n}\otimes\mathbb{I}\right)U_{n}$,
are not satisfied in general. This construction ensures these issues
are dissolved. }

\textcolor{purple}{The framework will output variables in the reverse
time convention indexed as, for example, $\left|\psi_{(i)}\right\rangle $,
$Z_{(i)}$, $U_{(i)}$. The variables at the $i^{\text{th}}$ step
of the protocol (which follows the forward time convention) would
be given by $\left|\psi_{i}\right\rangle =\left|\psi_{(N-i)}\right\rangle $,
$Z_{i}=Z_{(N-i)}$ and $U_{i}=U_{(N-i)}^{\dagger}$. Note that the
results so obtain extend naturally to the case where $U_{i}$ may
not be unitary and contains projections.}

\subsubsection*{\textcolor{purple}{Basic Moves Work Out of the Box}}

\textcolor{purple}{Recall the three basic moves of a TDPG were given
by}
\begin{enumerate}
\item \textcolor{purple}{Raise: $p_{1}[x,y]\to p_{1}[x',y]$ s.t. $x'\ge x$.}
\item \textcolor{purple}{Merge: $p_{1}[x_{1},y]+p_{2}[x_{2},y]\to p_{1}+p_{2}\left[\frac{p_{1}x_{1}+p_{2}x_{2}}{p_{1}+p_{2}},y\right]$}
\item \textcolor{purple}{Split: $\left(p_{1}+p_{2}\right)\left[\left(\frac{p_{1}w_{1}+p_{2}w_{2}}{p_{1}+p_{2}}\right)^{-1},y\right]\to p_{1}[x_{1},y]+p_{2}[x_{2},y]$
where $w_{1}=1/p_{1}$ and $w_{2}=1/p_{2}$.}
\end{enumerate}
\textcolor{purple}{We construct the explicit Unitaries that implement
these moves which in turn (when generalised to $n$ points) are enough
to construct the former best known protocol from its TDPG. Note, however,
that these moves do not exhaust the set of moves and more advanced
moves will be constructed to go beyond the $1/6$ limit.}

\subsection{The Framework\label{subsec:The-Framework}}

\subsection*{Intuition}

\textcolor{purple}{Imagine a canonical description is given. Let the
labels on the points one wants to transform be indexed by $\{g_{i}\}$
and let us also assume that one wishes to apply an $x$-transition
(meaning Alice performs the non-trivial step). Let the labels of the
points that one wishes to leave untouched be given by $\{k_{i}\}$
(see \Figref{TDPGillustrating_kgh}). We can write the state as}

\textcolor{purple}{
\[
\left|\psi_{(1)}\right\rangle =\left(\sum_{i}\sqrt{p_{g_{i}}}\left|g_{i}g_{i}\right\rangle _{AB}+\sum_{i}\sqrt{p_{k_{i}}}\left|k_{i}k_{i}\right\rangle _{AB}\right)\otimes\left|m\right\rangle _{M}.
\]
We want Bob to send his part of $\left|g_{i}\right\rangle $ states
to Alice through the message register. One way is that he conditionally
swaps to obtain the following
\[
\left|\psi_{(2)}\right\rangle =\sum_{i}\sqrt{p_{g_{i}}}\left|g_{i}g_{i}\right\rangle _{AM}\otimes\left|m\right\rangle _{B}+\sum_{i}\sqrt{p_{k_{i}}}\left|k_{i}k_{i}\right\rangle _{AB}\otimes\left|m\right\rangle _{M}.
\]
This should at most force all the points to align along the $y$-axis
but no non-trivial constraint should arise (speaking with hindsight).
Let $\{h_{i}\}$ be the labels of the new points after the transformation.
We assume that $h$ and $g$ index orthonormal vectors. Alice can
update the probabilities and labels by locally performing a unitary
to obtain
\[
\left|\psi_{(3)}\right\rangle =\sum_{i}\sqrt{p_{h_{i}}}\left|h_{i}h_{i}\right\rangle _{AM}\otimes\left|m\right\rangle _{B}+\sum_{i}\sqrt{p_{k_{i}}}\left|k_{i}k_{i}\right\rangle _{AB}\otimes\left|m\right\rangle _{M}.
\]
It is precisely this step which yields the non-trivial constraint.
Bob must now accept this by `unswapping' to get 
\[
\left|\psi_{(4)}\right\rangle =\left(\sum_{i}\sqrt{p_{h_{i}}}\left|h_{i}h_{i}\right\rangle _{AB}+\sum_{i}\sqrt{p_{k_{i}}}\left|k_{i}k_{i}\right\rangle _{AB}\right)\otimes\left|m\right\rangle _{M}
\]
which leaves Bob's $Z$ in essentially the standard form (we will
see). Remember that in the actual protocol the sequence will get reversed
as described above. }

\textcolor{purple}{Note that we add a few extra frames to the final
TDPG to go from a given frame to the next of the initial TDPG. This
is irrelevant as the bias stays the same but we mention it to avoid
confusion.}

\subsection*{Formal Description and Proofs}

\begin{enumerate}
\item \textbf{First frame.} 
\begin{align*}
\left|\psi_{(1)}\right\rangle  & =\left(\sum_{i}\sqrt{p_{g_{i}}}\left|g_{i}g_{i}\right\rangle _{AB}+\sum_{i}\sqrt{p_{k_{i}}}\left|k_{i}k_{i}\right\rangle _{AB}\right)\otimes\left|m\right\rangle _{M}\\
Z_{(1)}^{A} & =\sum_{i}x_{g_{i}}\left|g_{i}\right\rangle \left\langle g_{i}\right|_{A}+\sum_{i}x_{k_{i}}\left|k_{i}\right\rangle \left\langle k_{i}\right|_{A}\\
Z_{(1)}^{B} & =\sum_{i}y_{g_{i}}\left|g_{i}\right\rangle \left\langle g_{i}\right|_{B}+\sum_{i}y_{k_{i}}\left|k_{i}\right\rangle \left\langle k_{i}\right|_{B}.
\end{align*}
\begin{proof}
Follows from the assumption of starting with a Canonical Form.
\end{proof}
\item \textbf{Bob sends to Alice.} With $y\ge\text{max}\{y_{g_{i}}\}$ the
following is a valid choice
\begin{align*}
\left|\psi_{(2)}\right\rangle  & =\sum_{i}\sqrt{p_{g_{i}}}\left|g_{i}g_{i}\right\rangle _{AM}\otimes\left|m\right\rangle _{B}+\sum_{i}\sqrt{p_{k_{i}}}\left|k_{i}k_{i}\right\rangle _{AB}\otimes\left|m\right\rangle _{M}\\
U^{(1)} & =U_{BM}^{\text{SWP}\{\vec{g},m\}}\\
Z_{(2)}^{A} & =Z_{(1)}^{A}\\
Z_{(2)}^{B} & =y\mathbb{I}_{B}^{\{\vec{g},m\}}+\sum_{i}y_{k_{i}}\left|k_{i}\right\rangle \left\langle k_{i}\right|_{B}.
\end{align*}
\begin{proof}
We have to prove: (1) $\left|\psi_{(2)}\right\rangle =U^{(1)}\left|\psi_{(1)}\right\rangle $
and (2) $U^{(1)\dagger}\left(Z_{(2)}^{B}\otimes\mathbb{I}_{M}\right)U^{(1)}\ge\left(Z_{(1)}^{B}\otimes\mathbb{I}_{M}\right)$.\\
(1) It follows trivially from the defining action of $U^{(1)}$.\\
(2) For convenience, let momentarily $U=U^{(1)}$ and note that $U^{\dagger}=U$
so that we can write
\begin{align*}
 & U\left(Z_{(2)}^{B}\otimes\mathbb{I}_{M}\right)U\\
 & =y\left(U\left(\mathbb{I}_{B}^{\{\vec{g},m\}}\otimes\mathbb{I}_{M}^{\{\vec{g},m\}}\right)U+U\underbrace{\left(\mathbb{I}_{B}^{\{\vec{g},m\}}\otimes\mathbb{I}_{M}^{\{\vec{k},\vec{h}\}}\right)}_{\text{outside }U\text{'s action space}}U\right)+U\underbrace{\left(\sum y_{k_{i}}\left|k_{i}\right\rangle \left\langle k_{i}\right|\otimes\mathbb{I}\right)}_{\text{outside }U\text{'s action space}}U\\
 & =Z_{(2)}\otimes\mathbb{I}_{M}\ge Z_{(1)}\otimes\mathbb{I}_{M}
\end{align*}
so long as $y\ge y_{g_{i}}$ which is guaranteed by the choice of
$y$.
\end{proof}
\item \textbf{Alice's non-trivial step. }We claim that the following is
a valid choice, 
\begin{align*}
\left|\psi_{(3)}\right\rangle  & =\sum_{i}\sqrt{p_{h_{i}}}\left|h_{i}h_{i}\right\rangle _{AM}\otimes\left|m\right\rangle _{B}+\sum_{i}\sqrt{p_{k_{i}}}\left|k_{i}k_{i}\right\rangle _{AB}\otimes\left|m\right\rangle _{M}\\
E^{(2)}U^{(2)} & =E^{(2)}\left(\left|w\right\rangle \left\langle v\right|+\text{other terms acting on span\{}\left|h_{i}h_{i}\right\rangle ,\left|g_{i}g_{i}\right\rangle \}\right)_{AM}\\
Z_{(3)}^{A} & =\sum_{i}x_{h_{i}}\left|h_{i}\right\rangle \left\langle h_{i}\right|+\sum_{i}x_{k_{i}}\left|k_{i}\right\rangle \left\langle k_{i}\right|\\
Z_{(3)}^{B} & =Z_{(2)}^{B}
\end{align*}
where 
\[
\left|v\right\rangle =\frac{\sum_{i}\sqrt{p_{g_{i}}}\left|g_{i}g_{i}\right\rangle }{\sqrt{\sum_{i}p_{g_{i}}}},\,\left|w\right\rangle =\frac{\sum_{i}\sqrt{p_{h_{i}}}\left|h_{i}h_{i}\right\rangle }{\sqrt{\sum_{i}p_{h_{i}}}},\,E^{(2)}=\left(\sum\left|h_{i}\right\rangle \left\langle h_{i}\right|_{A}+\sum\left|k_{i}\right\rangle \left\langle k_{i}\right|_{A}\right)\otimes\mathbb{I}_{M}
\]
 subject to the condition 
\begin{equation}
\sum x_{h_{i}}\left|h_{i}h_{i}\right\rangle \left\langle h_{i}h_{i}\right|\ge\sum x_{g_{i}}E^{(2)}U^{(2)}\left|g_{i}g_{i}\right\rangle \left\langle g_{i}g_{i}\right|U^{(2)\dagger}E^{(2)}\label{eq:MainConstraintInequality}
\end{equation}
and of course the conservation of probability, viz. $\sum p_{g_{i}}=\sum p_{h_{i}}$.\begin{proof}
We must show that (1) $E^{(2)}\left|\psi_{(3)}\right\rangle =U^{(2)}\left|\psi_{(2)}\right\rangle $
and (2) 
\[
Z_{(3)}^{A}\otimes\mathbb{I}_{M}\ge E^{(2)}U^{(2)}\left(Z_{(2)}^{A}\otimes\mathbb{I}_{M}\right)U^{(2)\dagger}E^{(2)}.
\]
(1) Observing $E^{(2)}\left|\psi_{(3)}\right\rangle =\left|\psi_{(3)}\right\rangle $
the statement holds almost trivially by construction of $U^{(2)}$.\\
(2) Consider the space $\mathcal{H}=\text{span}\left\{ \left|g_{1}g_{1}\right\rangle ,\left|g_{2}g_{2}\right\rangle \dots,\left|h_{1}h_{1}\right\rangle ,\left|h_{2},h_{2}\right\rangle \dots\right\} $.
We separate all expressions (they are nearly diagonal) into the $\mathcal{H}$
space (which gets non-diagonal) and the rest. We start with the RHS,
\[
Z_{(2)}^{A}\otimes\mathbb{I}_{M}=\underbrace{\sum x_{g_{i}}\left|g_{i}g_{i}\right\rangle \left\langle g_{i}g_{i}\right|}_{\text{I}}+\sum x_{g_{i}}\left|g_{i}\right\rangle \left\langle g_{i}\right|\otimes\left(\mathbb{I}-\left|g_{i}\right\rangle \left\langle g_{i}\right|\right)+\sum x_{k_{i}}\left|k_{i}\right\rangle \left\langle k_{i}\right|\otimes\mathbb{I},
\]
where only term I is in the operator space spanned by $\mathcal{H}$.
Note that all the terms are still diagonal. Next consider the LHS,
without the $U$s, 
\begin{align*}
Z_{(3)}^{A}\otimes\mathbb{I}_{M} & =\underbrace{\sum x_{h_{i}}\left|h_{i}h_{i}\right\rangle \left\langle h_{i}h_{i}\right|}_{\text{I}}+\sum x_{h_{i}}\left|h_{i}\right\rangle \left\langle h_{i}\right|\otimes\left(\mathbb{I}-\left|h_{i}\right\rangle \left\langle h_{i}\right|\right)+\sum x_{k_{i}}\left|k_{i}\right\rangle \left\langle k_{i}\right|\otimes\mathbb{I},
\end{align*}
which also has only term I in the $\mathcal{H}$ operator space. Consequently,
only on these will $U$ have a non-trivial action. Let us first evaluate
the non-$\mathcal{H}$ part where we only need to apply the projector.
The result after separating equations where possible is
\begin{align*}
\sum x_{h_{i}}\left|h_{i}\right\rangle \left\langle h_{i}\right|\otimes\left(\mathbb{I}-\left|h_{i}\right\rangle \left\langle h_{i}\right|\right) & \ge0\\
\sum(x_{k_{i}}-x_{k_{i}})\left|k_{i}\right\rangle \left\langle k_{i}\right|\otimes\mathbb{I} & \ge0
\end{align*}
which essentially only implies
\[
x_{h_{i}}\ge0.
\]
Finally the non-trivial part yields 
\[
\sum x_{h_{i}}\left|h_{i}h_{i}\right\rangle \left\langle h_{i}h_{i}\right|\ge\sum x_{g_{i}}EU\left|g_{i}g_{i}\right\rangle \left\langle g_{i}g_{i}\right|U^{\dagger}E
\]
which completes the proof.
\end{proof}
\item \textbf{Bob accepts Alice's change.} The following is valid. 
\begin{align*}
\left|\psi_{(4)}\right\rangle  & =\left(\sum_{i}\sqrt{p_{h_{i}}}\left|h_{i}h_{i}\right\rangle _{AB}+\sum_{i}\sqrt{p_{k_{i}}}\left|k_{i}k_{i}\right\rangle _{AB}\right)\otimes\left|m\right\rangle _{M}\\
E^{(3)}U^{(3)} & =E^{(3)}U_{BM}^{\text{SWP}\{\vec{h},m\}}\\
Z_{(4)}^{A} & =Z_{(3)}^{A}\\
Z_{(4)}^{B} & =y\sum_{i}\left|h_{i}\right\rangle \left\langle h_{i}\right|+\sum_{i}y_{k_{i}}\left|k_{i}\right\rangle \left\langle k_{i}\right|_{B}
\end{align*}
where $E^{(3)}=\left(\sum\left|h_{i}\right\rangle \left\langle h_{i}\right|+\sum\left|k_{i}\right\rangle \left\langle k_{i}\right|\right)_{B}\otimes\mathbb{I}_{M}$.
\begin{proof}
We have to prove: (1) $E^{(3)}\left|\psi_{(4)}\right\rangle =U^{(3)}\left|\psi_{(3)}\right\rangle $
and (2) 
\[
Z_{(4)}^{B}\otimes\mathbb{I}_{M}\ge E^{(3)}U^{(3)}\left(Z_{(3)}^{B}\otimes\mathbb{I}_{M}\right)U^{(3)\dagger}E^{(3)}.
\]
(1) This can be proven again, by a direct application of $U^{\dagger}E$
on $\left|\psi_{(4)}\right\rangle $ (where $E$ is defined to be
$E^{(3)}$ and $U$ to be $U^{(3)}$ for the proof).\\
(2) Note that 
\begin{align*}
EU\left(\mathbb{I}_{B}^{\{\vec{g},m\}}\otimes\mathbb{I}_{M}^{\{\vec{h},\vec{g},\vec{k},m\}}\right)U^{\dagger}E & =EU\left(\mathbb{I}_{B}^{\{m\}}\otimes\mathbb{I}_{M}^{\{\vec{h},\vec{g},\vec{k},m\}}\right)U^{\dagger}E+E\left(\mathbb{I}_{B}^{\{\vec{g}\}}\otimes\mathbb{I}_{M}^{\{\vec{h},\vec{g},\vec{k},m\}}\right)E\\
 & =EU\left(\mathbb{I}_{B}^{\{m\}}\otimes\mathbb{I}_{M}^{\{\vec{h},m\}}\right)U^{\dagger}E\\
 & =\sum\left|h_{i}\right\rangle \left\langle h_{i}\right|\otimes\mathbb{I}_{M}^{\{m\}}.
\end{align*}
Since the other term in $Z_{3}^{B}\otimes\mathbb{I}$ is anyway in
the non-action space of $U$ it follows that 
\[
EU(Z_{3}^{B}\otimes\mathbb{I})U^{\dagger}E=y\sum\left|h_{i}\right\rangle \left\langle h_{i}\right|\otimes\mathbb{I}_{M}^{\{m\}}+\sum y_{k_{i}}\left|k_{i}\right\rangle \left\langle k_{i}\right|\otimes\mathbb{I}_{M}.
\]
It only remains to show that $Z_{(4)}^{B}\otimes\mathbb{I}_{M}\ge E^{(3)}U^{(3)}\left(Z_{(3)}^{B}\otimes\mathbb{I}_{M}\right)U^{(3)\dagger}E^{(3)}$
which it obviously is because $y\sum\left|h_{i}\right\rangle \left\langle h_{i}\right|\otimes\mathbb{I}_{M}\ge y\sum\left|h_{i}\right\rangle \left\langle h_{i}\right|\otimes\mathbb{I}_{M}^{\{m\}}$
and the $y_{k_{i}}$ term is common.
\end{proof}
\end{enumerate}
\textcolor{purple}{We can summarise the condition of interest as follows,
the proof of which is a trivial consequence of the aforesaid.}
\begin{thm}
For an $x$-transition (where Alice performs the non-trivial step)
\[
\sum_{i=1}^{n_{k}}p_{k_{i}}[x_{k_{i}}]+\sum_{i=1}^{n_{g}}p_{g_{i}}[x_{g_{i}}]\to\sum_{i=1}^{n_{h}}p_{h_{i}}[x_{h_{i}}]+\sum_{i=1}^{n_{k}}p_{k_{i}}[x_{k_{i}}]
\]
 to be implementable under the TDPG-to-Explicit-protocol Framework
(TEF) one must find a $U^{(2)}$ that satisfies the inequality
\begin{equation}
\sum_{i=1}^{n_{h}}x_{h_{i}}\left|h_{i}h_{i}\right\rangle \left\langle h_{i}h_{i}\right|_{AM}\ge\sum_{i=1}^{n_{g}}x_{g_{i}}E_{h}^{(2)}U^{(2)}\left|g_{i}g_{i}\right\rangle \left\langle g_{i}g_{i}\right|_{AM}U^{(2)\dagger}E_{h}^{(2)}\label{eq:ProtocolConstraintEquation}
\end{equation}
and the honest action constraint
\[
U^{(2)}\left|v\right\rangle =\left|w\right\rangle 
\]
where $\left|h_{i}\right\rangle $ and $\left|g_{i}\right\rangle $
are orthonormal basis vectors, 
\[
\left|v\right\rangle =\mathcal{N}\left(\sum\sqrt{p_{g_{i}}}\left|g_{i}g_{i}\right\rangle _{AM}\right)
\]
 and 
\[
\left|w\right\rangle =\mathcal{N}\left(\sum\sqrt{p_{h_{i}}}\left|h_{i}h_{i}\right\rangle _{AM}\right)
\]
 for $\mathcal{N}(\left|\psi\right\rangle )=\left|\psi\right\rangle /\sqrt{\left\langle \psi|\psi\right\rangle }$,
$E_{h}=\left(\sum_{i=1}^{n_{h}}\left|h_{i}\right\rangle \left\langle h_{i}\right|_{A}+\sum\left|k_{i}\right\rangle \left\langle k_{i}\right|_{A}\right)\otimes\mathbb{I}_{M}$
with $U^{(2)}$'s non-trivial action restricted to $\text{span}\left\{ \{\left|g_{i}g_{i}\right\rangle _{AM}\},\{\left|h_{i}h_{i}\right\rangle _{AM}\}\right\} $
(note $\left|k_{i}\right\rangle $ corresponds to the points that
are left unchanged in the transition).\label{thm:TEFconstraint}
\end{thm}

\subsection{Important Special Case: The Blinkered Unitary\label{subsec:BlinkeredUnitary}}

\textcolor{purple}{So far we have not specified the non-trivial $U^{(2)}$
(which we call $U$ from now) beyond requiring it to have a certain
action on the honest state.} We now define an important class of $U$, we call the Blinkered Unitary,
as 
\[
U=\left|w\right\rangle \left\langle v\right|+\left|v\right\rangle \left\langle w\right|+\sum\left|v_{i}\right\rangle \left\langle v_{i}\right|+\sum\left|w_{i}\right\rangle \left\langle w_{i}\right|+\mathbb{I}^{\text{outside }\text{\ensuremath{\mathcal{H}}}}
\]
and can even drop the last term as we are restricting our analysis
to the $\mathcal{H}$ operator space, where $\left|v\right\rangle ,\{\left|v_{i}\right\rangle \}$
form a complete orthonormal basis and so do $\left|w\right\rangle ,\{\left|w_{i}\right\rangle \}$
wrt $\text{span}\{\left|g_{i}g_{i}\right\rangle \}$ and $\text{span}\{\left|v_{i}v_{i}\right\rangle \}$
respectively. The blinkered unitary can be used to implement the two
non-trivial operations of the set of basic moves.
\begin{itemize}
\item Merge: $g_{1},g_{2}\to h_{1}$\\
We can construct from the very definitions\\
\[
\left|v\right\rangle =\frac{\sqrt{p_{g_{1}}}\left|g_{1}g_{1}\right\rangle +\sqrt{p_{g_{2}}}\left|g_{2}g_{2}\right\rangle }{N},\,\left|v_{1}\right\rangle =\frac{\sqrt{p_{g_{2}}}\left|g_{1}g_{1}\right\rangle -\sqrt{p_{g_{1}}}\left|g_{2}g_{2}\right\rangle }{N},\,\left|w\right\rangle =\left|h_{1}h_{1}\right\rangle 
\]
with $N=\sqrt{p_{g_{1}}+p_{g_{2}}}$ and even 
\[
U=\left|w\right\rangle \left\langle v\right|+\left|v\right\rangle \left\langle w\right|+\left|v_{1}\right\rangle \left\langle v_{1}\right|(=U^{\dagger}).
\]
We would need 
\[
EU\left|g_{1}g_{1}\right\rangle =\frac{\sqrt{p_{g_{1}}}\left|w\right\rangle }{N},\,EU\left|g_{2}g_{2}\right\rangle =\frac{\sqrt{p_{g_{2}}}\left|w\right\rangle }{N}
\]
because the constraint was (substituting for $m$ and $n$)
\[
x_{h}\left|h_{1}h_{1}\right\rangle \left\langle h_{1}h_{1}\right|\ge\sum x_{g_{i}}EU\left|g_{i}g_{i}\right\rangle \left\langle g_{i}g_{i}\right|U^{\dagger}E
\]
which becomes 
\[
x_{h}\ge\frac{p_{g_{1}}x_{g_{1}}+p_{g_{2}}x_{g_{2}}}{N^{22}}.
\]
\textcolor{purple}{This is precisely the merge condition Mochon derives.
This can be readily generalised to an $m\to1$ point merge condition
by simply constructing appropriate vectors (which we leave for the
appendix).}
\item Split: $g_{1}\to h_{1},h_{2}$\\
\[
\left|v\right\rangle =\left|g_{1}g_{1}\right\rangle ,\,\left|w\right\rangle =\frac{\sqrt{p_{h_{1}}}\left|h_{1}h_{1}\right\rangle +\sqrt{p_{h_{2}}}\left|h_{2}h_{2}\right\rangle }{N},\,\left|w_{1}\right\rangle =\frac{\sqrt{p_{h_{2}}}\left|h_{1}h_{1}\right\rangle -\sqrt{p_{h_{1}}}\left|h_{2}h_{2}\right\rangle }{N}
\]
with $N=\sqrt{p_{h_{1}}+p_{h_{2}}}$ and 
\[
U=\left|v\right\rangle \left\langle w\right|+\left|w\right\rangle \left\langle v\right|+\left|w_{1}\right\rangle \left\langle w_{1}\right|=U^{\dagger}.
\]
We evaluate $EU\left|g_{1}g_{1}\right\rangle =\left|w\right\rangle $
which upon being plugged into the constraint yields
\[
x_{h_{1}}\left|h_{1}h_{1}\right\rangle \left\langle h_{1}h_{1}\right|+x_{h_{2}}\left|h_{2}h_{2}\right\rangle \left\langle h_{2}h_{2}\right|-x_{g_{1}}\left|w\right\rangle \left\langle w\right|\ge0.
\]
This yields the matrix equation 
\begin{align*}
\left[\begin{array}{cc}
x_{h_{1}}\\
 & x_{h_{2}}
\end{array}\right]-\frac{x_{g_{1}}}{N^{2}}\left[\begin{array}{cc}
p_{h_{1}} & \sqrt{p_{h_{1}}p_{h_{2}}}\\
\sqrt{p_{h_{1}}p_{h_{2}}} & p_{h_{2}}
\end{array}\right] & \ge0\\
\mathbb{I} & \ge\frac{x_{g_{1}}}{N^{2}}\left[\begin{array}{cc}
\frac{p_{h_{1}}}{x_{h_{1}}} & \sqrt{\frac{p_{h_{1}}}{x_{h_{1}}}\frac{p_{h_{2}}}{x_{h_{2}}}}\\
\sqrt{\frac{p_{h_{1}}}{x_{h_{1}}}\frac{p_{h_{2}}}{x_{h_{2}}}} & \frac{p_{h_{2}}}{x_{h_{2}}}
\end{array}\right]\\
\frac{x_{g_{1}}}{N^{2}}\left(\frac{p_{h_{1}}}{x_{h_{1}}}+\frac{p_{h_{2}}}{x_{h_{2}}}\right) & \le1
\end{align*}
where in the second step we used the fact that $F-M\ge0$ implies
$\mathbb{I}-\sqrt{F}^{-1}M\sqrt{F}^{-1}\ge0$ (if $F>0$) and the
last step is obtained by writing the matrix in the previous step as
$\left|\psi\right\rangle \left\langle \psi\right|$ followed by demanding
$1\ge\left\langle \psi|\psi\right\rangle $.\\
\textcolor{purple}{The last statement is the same constraint for a
split as the one derived by Mochon. This also readily generalises
to the case of $1\to N$ splits which again we defer to the appendix.}

It would not be surprising to learn/prove that the class of unitaries
with these properties is much more general than the Blinkered Unitaries.
\item General $m\to n$: $g_{1},g_{2}\dots g_{m}\to h_{1},h_{2}\dots h_{n}$\\
It is not too hard to show that in general one obtains the constraint
\[
\frac{1}{\left\langle x_{g}\right\rangle }\ge\left\langle \frac{1}{x_{h}}\right\rangle 
\]
or more explicitly, 
\[
\frac{1}{\sum_{i=1}^{m}p_{g_{i}}x_{g_{i}}}\ge\sum_{i=1}^{n}p_{h_{i}}.\frac{1}{x_{h_{i}}},
\]
using the appropriate blinkered unitary (which also we show in the
appendix).
\end{itemize}

\textcolor{purple}{This class of unitary is enough to convert the
$1/6$ game into an explicit protocol. However, for games given by
Mochon that go beyond $1/6$ this class falls short. One way of seeing
this is that the general $m\to n$ blinkered transition effectively
behaves like an $m\to1$ merge followed by a $1\to n$ split, which
are a set of moves that are insufficient to break the $1/6$ limit
(at least using Mochon's games).}

\section{Games and Protocols\label{sec:Games-and-Protocols}}

\textcolor{purple}{We now describe two games, the bias $1/6$ game
and the bias $1/10$ game, from the family of games constructed by
Mochon to show that arbitrarily small bias is achievable. Mochon parametrises
his games by $k$ which determines the number of points involved in
the non-trivial step. The bias he obtains is $\epsilon=1/(4k+2)$.
We consider games with $k=1$ and $k=2$, yielding the aforementioned
bias.}

\subsection{Mochon's Approach}

\subsubsection{Assignments}

\textcolor{purple}{Recall that a function 
\[
\sum_{z\in\{x_{1},x_{2},\dots x_{n}\}}p(z)[z]
\]
is valid if 
\[
\sum_{z\in\{x_{1},x_{2},\dots x_{n}\}}\left(\frac{-1}{\lambda+z}\right)p(z)\ge0,\,\sum_{z\in\{x_{1},x_{2},\dots x_{n}\}}zp(z)\ge0,\,\sum_{z\in\{x_{1}\dots x_{n}\}}p(z)=0
\]
for all $\lambda>0$ where $x_{i}\ge0$. Checking if a generic assignment
for $p$ satisfies these infinite constraints in not always easy.
Mochon had used a constructive approach here and we  build on to it.
Let us state these results with some precision (proven in the appendix,
well most) where as above $n$ numbers are assumed to be represented
by $x_{i}$ and each $x_{i}\ge0$.}
\begin{lem*}[Mochon's Denominator]
 $\sum_{i=1}^{n}\frac{1}{\prod_{j\neq i}(x_{j}-x_{i})}=0$ for $n\ge2$.
\end{lem*}
\begin{lem*}[Mochon's f-assignment Lemma]
 $\sum_{i=1}^{n}\frac{f(x_{i})}{\prod_{j\neq i}(x_{j}-x_{i})}=0$
where $f(x_{i})$ is a polynomial of order $k\le n-2$.
\end{lem*}
\begin{defn*}[Mochon's TIPG assignment]
 Given a set of $n$ points $0<x_{1}<x_{2}\dots<x_{n}$, a polynomial
$f(x)$ with order $k$ at most $n-2$ and $f(-\lambda)\ge0$ for
all $\lambda\ge0$, the probability weights for a TIPG assignment
is $p(x_{i})=-\frac{f(x_{i})}{\prod_{j\neq i}(x_{j}-x_{i})}$.
\end{defn*}
\textcolor{purple}{Mochon was able to show that `Mochon's TIPG assignment'
makes for a valid function (in the TIPG formalism), given by 
\[
\sum_{i=1}^{n}p(x_{i})[x_{i},y]
\]
where the notion of validity has been extended to a pair of points.
As we will see soon, the power of this construction lies in the fact
that we can easily construct polynomials that have roots at arbitrary
locations. This allows us to create interesting repeating structures
called ladders (due to Mochon) which we can terminate using these
polynomials to obtain a game with a finite set of points. These ladders
play a pivotal role in achieving smaller biases and the ability to
obtain finite ladders is essential for being able to obtain a physical
process that would yield the said bias.}

\textcolor{purple}{We now build a little on Mochon's notation and
results.}
\begin{defn*}[Mochon's TDPG assignment]
 Given Mochon's TIPG assignment, let $\{i\}$ be the set of indices
for which $p(x_{i})<0$ and $\{k\}$ be the remaining indices with
respect to $\{1,2,\dots n\}$. The TDPG assignment (in accordance
with the notation used in TEF) is given as 
\begin{align*}
\{x_{g_{1}},x_{g_{2}}\dots\} & =\{x_{i}\}\\
\{p_{g_{1}},p_{g_{2}}\dots\} & =\{-p(x_{i})\}\\
\{x_{h_{1}},x_{h_{2}}\dots\} & =\{x_{k}\}\\
\{p_{h_{1}},p_{h_{2}}\dots\} & =\{p(x_{k})\}.
\end{align*}
\end{defn*}
\textcolor{purple}{With these in place we make some observations about
how initial and final averages behave under such an assignment.}
\begin{prop*}
$N_{h}^{2}=N_{g}^{2}$ where $N_{g}^{2}=\sum p_{g_{i}}$ and $N_{h}^{2}=\sum p_{h_{i}}$
for Mochon's TDPG assignment.
\end{prop*}
\begin{proof}
We have to show that $N_{h}^{2}-N_{g}^{2}=\sum p_{h_{i}}-\sum p_{g_{i}}=0$
which is the same as showing $\sum_{i=1}^{n}p(x_{i})=0$ which holds
because we just showed that $\sum_{i=1}^{n}f(x_{i})/\prod_{j\neq i}(x_{j}-x_{i})=0$
(Mochon's f-assignment Lemma).
\end{proof}
\begin{prop*}
$\left\langle x_{h}\right\rangle -\left\langle x_{g}\right\rangle =0$
for a Mochon's TDPG assignment with $k\le n-3$ where $\left\langle x_{h}\right\rangle =\frac{1}{N_{h}^{2}}\sum p_{h_{i}}x_{h_{i}}$
and $\left\langle x_{g}\right\rangle =\frac{1}{N_{g}^{2}}\sum p_{g_{i}}x_{g_{i}}$.
\end{prop*}
\begin{proof}
This is a direct consequence of Mochon's f-assignment lemma. Let $h$
be the $n-3$ order polynomial defined by Mochon's TDPG assignment
so that $\left\langle x_{h}\right\rangle -\left\langle x_{g}\right\rangle \propto\sum p_{h_{i}}x_{i}-\sum p_{g_{i}}x_{g_{i}}=\sum_{i=1}^{n}p(x_{i})x_{i}=\sum_{i=1}^{n}\frac{h(x_{i})x_{i}}{\prod_{j\neq i}(x_{j}-x_{i})}=\sum_{i=1}^{n}\frac{f(x_{i})}{\prod_{j\neq i}(x_{j}-x_{i})}=0$
because $f$ is an $n-2$ order polynomial.
\end{proof}
\begin{lem*}
We have $\sum_{i=1}^{n}\frac{x_{i}^{n-1}}{\prod_{j\neq i}(x_{j}-x_{i})}=(-1)^{n-1}$
for $n\ge2$ (proof in \secref{Mochons-Assignments} of the Appendix).
\end{lem*}
\begin{prop*}
$\left\langle x_{h}\right\rangle -\left\langle x_{g}\right\rangle =\frac{1}{N_{h}^{2}}=\frac{1}{N_{g}^{2}}$
for a Mochon's TDPG assignment with $k=n-2$ and coefficient of $x^{n-2}$
being $\pm1$ in $f(x)$. As above, here $\left\langle x_{h}\right\rangle =\frac{1}{N_{h}^{2}}\sum p_{h_{i}}x_{h_{i}}$
and $\left\langle x_{g}\right\rangle =\frac{1}{N_{g}^{2}}\sum p_{g_{i}}x_{g_{i}}$.
\end{prop*}
\textcolor{purple}{We will see that typically $N_{h}=N_{g}$ are quite
large and the average only slightly increase, if at all. We are now
in a position to discuss Mochon's games.}

\subsubsection{Typical Game Structure}

We assume an equally spaced $n$-point lattice given by $x_{j}=x_{0}+j\delta x$
where $\delta x=\delta y$ is small and $x_{0}$ would essentially
give a bound on $P_{B}^{*}$ which will be determined by following
the constraints; similarly $y_{j}=y_{0}+j\delta y$ and we also define
$\Gamma_{k+1}=y_{n-k}=x_{n-k}$. Let $P(x_{j})$ be the probability
weight associated with the point $[x_{j},0]$ s.t. 
\[
\sum_{j=1}^{n}P(x_{j})=\frac{1}{2},\,\sum_{j=1}^{n}\frac{P(x_{j})}{x_{j}}=\frac{1}{2}.
\]
Similarly with the point $[0,y_{j}]$ we associate $P(y_{j})$ where
$y_{j}=x_{j}$ as we also assume that $x_{0}=y_{0}$. These choices
explicitly impose symmetry between Alice and Bob which in turn entails
that we have to do only half the analysis.

\subsection{Bias $1/6$ \label{subsec:Bias1by6}}

\subsubsection{Game}

\begin{figure}
\begin{centering}
\includegraphics[width=14cm]{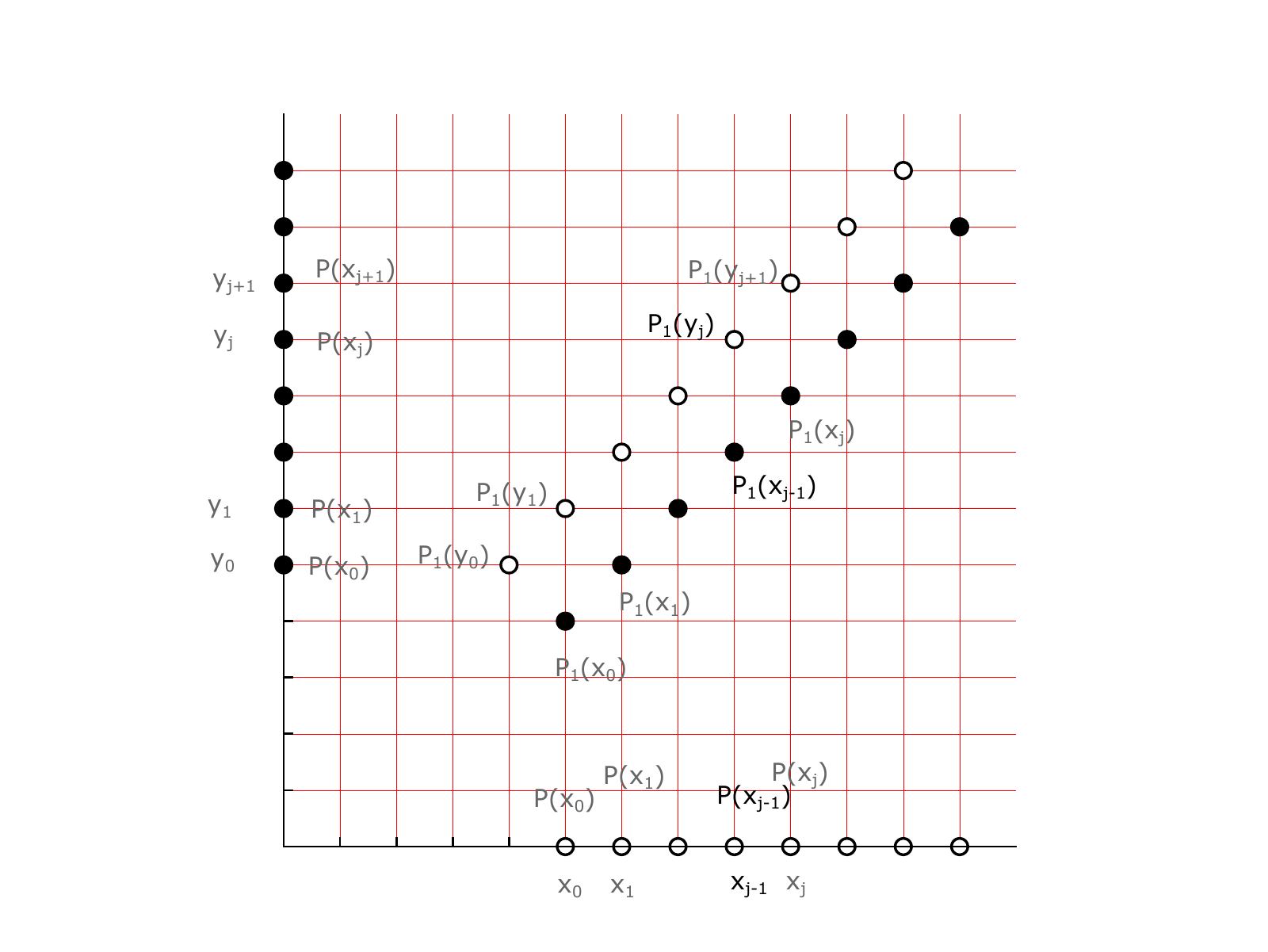}
\par\end{centering}
\caption{Building a TDPG/TIPG using merge moves\label{fig:Building-a-TDPG}}
\end{figure}

With reference to \Figref{Building-a-TDPG} we need to satisfy $P(x_{j-1})+P_{1}(y_{j})=P_{1}(x_{j-1})$
which is probability conservation and $P_{1}(y_{j})y_{j}\le P_{1}(x_{j-1})y_{j-2}$
which is the merge condition. Both of these are automatically satisfied
if we make a Mochon's denominator based assignment as follows
\begin{align*}
0 & \leftrightarrow x_{g_{1}}\\
y_{j} & \leftrightarrow x_{g_{2}}\\
y_{j-2} & \leftrightarrow x_{h_{1}}
\end{align*}
\begin{align*}
P(x_{j-1}) & \leftrightarrow p_{g_{1}}=\frac{c(x_{j-1})}{y_{j}y_{j-2}}\\
P_{1}(y_{j}) & \leftrightarrow p_{g_{2}}=\frac{c(x_{j-1})}{(y_{j}-y_{j-2})(y_{j})}=\frac{c(x_{j-1})}{2y_{j}\delta y}\\
P_{1}(x_{j-1}) & \leftrightarrow p_{h_{1}}=\frac{c(x_{j-1})}{(y_{j}-y_{j-2})(y_{j-2})}=\frac{c(x_{j-1})}{2y_{j-2}\delta y}
\end{align*}
where the function $c(x_{j-1})$ must be chosen so that $P_{1}(y_{j})=P_{1}(x_{j})$
which entails 
\[
\frac{c(x_{j-1})}{\cancel{2}y_{j}\cancel{\delta y}}=\frac{c(x_{j})}{\cancel{2}y_{j-1}\cancel{\delta y}}
\]
and that in turn is solved by $c(x_{j})=\frac{c_{0}\delta x}{x_{j}}$
where we used $x_{j}=y_{j}$, $\delta x=\delta y$ (and added a $\delta x$
as it helps approximating $\sum P(x_{j})$ by an integral). Plugging
this back we have 
\[
P_{1}(x_{j})=\frac{c_{0}}{2x_{j}x_{j-1}},\,P(x_{j})=\frac{c_{0}\delta x}{x_{j-1}x_{j}x_{j+1}}.
\]

Since they involve a sum we do this in the limit $\delta x\to0$ and
$\Gamma\to\infty$ to avoid dealing with summing a series.
\[
\sum_{j=0}^{n}P(x_{j})=\frac{1}{2}\to c_{0}\int_{x_{0}}^{\Gamma}\frac{dx}{x^{3}}=\frac{c_{0}}{(-2)}\left[\frac{1}{\Gamma^{2}}-\frac{1}{x_{0}^{2}}\right]=\frac{1}{2}
\]
which entails $c_{0}=x_{0}^{2}$. The next condition yields $x_{0}$
\[
\sum_{j=0}^{n}\frac{P(x_{j})}{x_{j}}=\frac{1}{2}\to x_{0}^{2}\int_{x_{0}}^{\Gamma}\frac{dx}{x^{4}}=\frac{x_{0}^{2}}{(-3)}\left[\frac{1}{\Gamma^{3}}-\frac{1}{x_{0}^{3}}\right]=\frac{1}{3x_{0}}=\frac{1}{2}
\]
which means 
\[
x_{0}=\frac{2}{3}\implies\epsilon=\frac{1}{6}.
\]
Of course a more careful analysis must be done to show these things
exactly. Aside from the integration step one must also set $c_{0}(x)=(\Gamma_{n+1}-x)$
in order to terminate the ladder which turns the terminating step
on the ladder into a raise. At the moment, however, we satisfy ourselves
with this and move on to the more interesting $1/10$ game. These
issues have been dealt with in general (see \cite{ACG+14}).

\subsubsection{Protocol}

Although we could only claim that one can construct the protocol once
the unitaries are known, the basic idea is that one starts with a
split, then a raise by Alice and Bob, followed by a merge by Bob,
then a merge by Alice and so on until only two points remain. Bob
can also start as the description is symmetric. These two can then
be raised to the same location and merged. The coordinates of these
points tend to $[\frac{2}{3},\frac{2}{3}]$ as calculated above. The
only creative part left would be the choice of labels that make the
description neater from the point of view of the explicit protocol.

\subsection{Bias $1/10$ Game\label{subsec:Bias1by10game}}

\begin{figure}
\begin{centering}
\includegraphics[width=14cm]{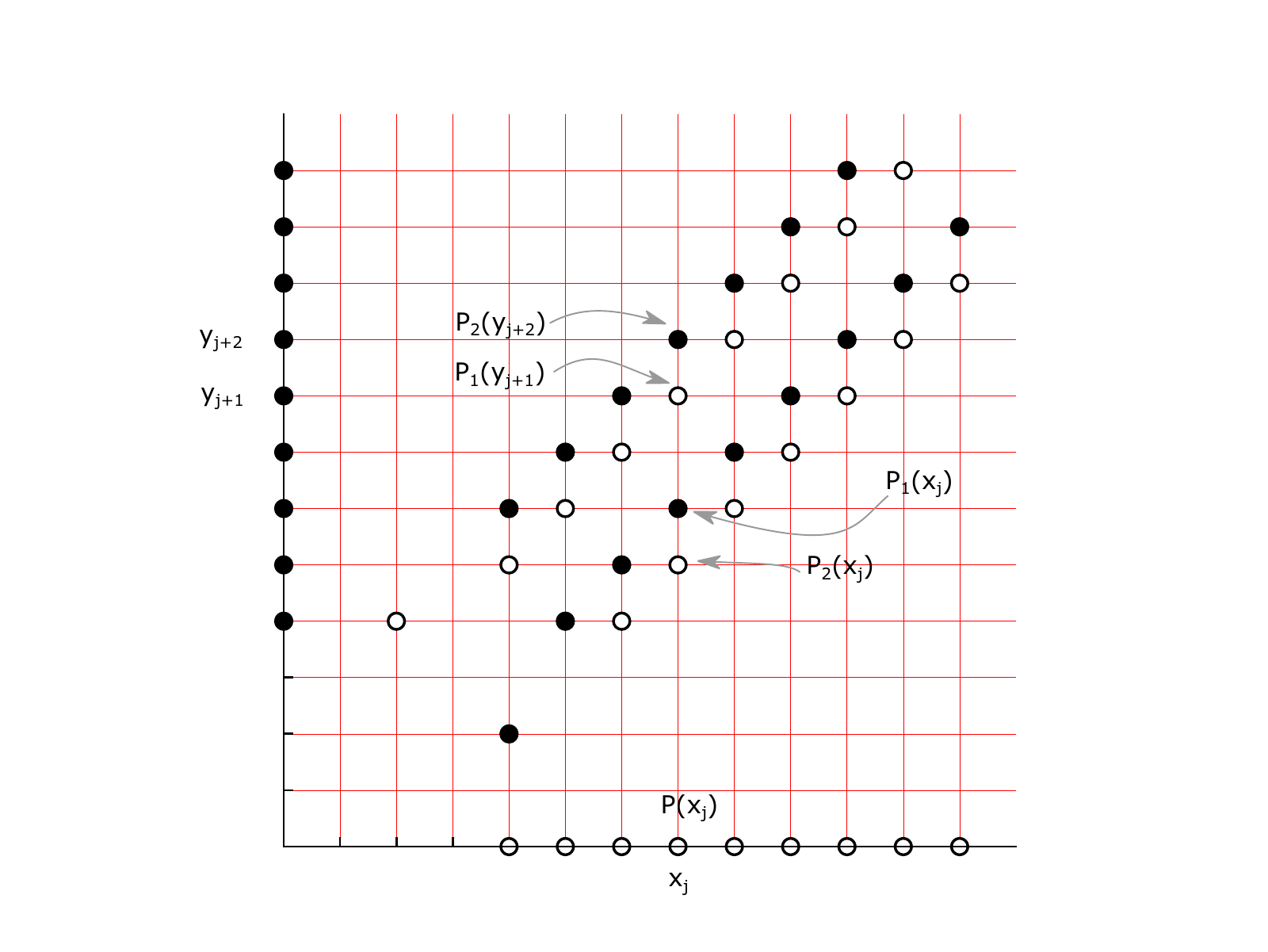}
\par\end{centering}
\caption{$1/10$ game: The $3\to2$ move based TIPG for bias $1/10$\label{fig:1by10correct}}
\end{figure}

With respect to \Figref{1by10correct} we use Mochon's assignment
with $f(y_{i})=(y_{-2}-y_{i})\left(\Gamma_{1}-y_{i}\right)(\Gamma_{2}-y_{i})$
as 
\[
\frac{f(y_{j})c'(x_{j})}{\prod_{k\neq j}(y_{k}-y_{j})}.
\]
Following the scheme as described above the probabilities become
\begin{align*}
P_{2}(y_{j+2}) & =\frac{-f(y_{j+2})c(x_{j})}{4.3(\delta y)^{2}y_{j+2}}\\
P_{1}(y_{j+1}) & =\frac{-f(y_{j+1})c(x_{j})}{3.2(\delta y)^{2}y_{j+1}}\\
P_{1}(x_{j}) & =\frac{-f(y_{j-1})c(x_{j})}{3.2(\delta y)^{2}y_{j-1}}\\
P_{2}(x_{j}) & =\frac{-f(y_{j-2})c(x_{j})}{4.3(\delta y)^{2}y_{j-2}}\\
P(x_{j}) & =\frac{f(0)c(x_{j})\delta y}{y_{j+2}y_{j+1}y_{j-1}y_{j-2}}
\end{align*}
where we added the minus sign to account for the fact that $f$ is
negative for coordinates between $y_{-2}$ and $\Gamma_{1}$. Imposing
the symmetry constraints $P_{1}(y_{j})=P_{1}(x_{j})$ we get 
\[
\frac{f(y_{j})c(x_{j-1})}{\cancel{3.2(\delta y)^{2}}y_{j}}=\frac{f(y_{j-1})c(x_{j})}{\cancel{3.2(\delta y)^{2}}y_{j-1}}
\]
which means 
\[
c(x_{j})=\frac{c_{0}f(x_{j})}{x_{j}}
\]
where $c_{0}$ is a constant. This also entails that $P_{2}(y_{j})=P_{2}(x_{j})$,
viz. it satisfies the second symmetry constraint. Finally we can evaluate
\[
P(x_{j})=\frac{f(0)f(x_{j})\delta x}{x_{j+2}x_{j+1}x_{j}x_{j-1}x_{j-2}}=\frac{c_{0}x_{0}(x_{0}-x_{j})}{x_{j}^{5}}\delta x+\mathcal{O}(\delta x^{2})
\]
which means that 
\[
\sum P(x_{j})=\frac{1}{2}=\sum\frac{P(x_{j})}{x_{j}}\to\int_{x_{0}}^{\Gamma}\frac{(x_{0}-x)dx}{x^{5}}=\int_{x_{0}}^{\Gamma}\frac{(x_{0}-x)dx}{x^{6}}.
\]
We can evaluate this as 
\begin{align*}
x_{0}\int_{x_{0}}^{\Gamma}\left(\frac{1}{x^{5}}-\frac{1}{x^{6}}\right)dx & =\int_{x_{0}}^{\Gamma}\left(\frac{1}{x^{4}}-\frac{1}{x^{5}}\right)dx\\
\left[\frac{1}{4x_{0}^{3}}-\frac{1}{5x_{0}^{4}}\right] & =\left[\frac{1}{3x_{0}^{3}}-\frac{1}{4x_{0}^{4}}\right]\\
\left[\frac{1}{4}-\frac{1}{3}\right] & =\left[\frac{1}{5}-\frac{1}{4}\right]\frac{1}{x_{0}}\\
x_{0}\frac{3-4}{\cancel{4}.3} & =\frac{4-5}{5.\cancel{4}}\\
x_{0} & =\frac{3}{5}\implies\epsilon=\frac{3}{5}-\frac{1}{2}=\frac{1}{10}.
\end{align*}

\subsection{Bias $1/10$ Protocol\label{subsec:Bias1by10protocol}}

\subsubsection{The $3\to2$ Move}

In this section we introduce as many parameters as possible within
the TEF to implement the largest class of $3\to2$ moves. However,
we use our insight to choose an appropriate basis so that the parameters
are small which in turn simplifies the analysis.

Recall that 
\[
\left|v\right\rangle =\frac{\sqrt{p_{g_{1}}}\left|g_{1}\right\rangle +\sqrt{p_{g_{2}}}\left|g_{2}\right\rangle +\sqrt{p_{g_{3}}}\left|g_{3}\right\rangle }{N_{g}}
\]
and let 
\begin{align*}
\left|v_{1}\right\rangle  & =\frac{\sqrt{p_{g_{3}}}\left|g_{2}\right\rangle -\sqrt{p_{g_{2}}}\left|g_{3}\right\rangle }{N_{v_{1}}}\\
\left|v_{2}\right\rangle  & =\frac{-\frac{(p_{g_{2}}+p_{g_{3}})}{\sqrt{p_{g_{1}}}}\left|g_{1}\right\rangle +\sqrt{p_{g_{2}}}\left|g_{2}\right\rangle +\sqrt{p_{g_{3}}}\left|g_{3}\right\rangle }{N_{v_{2}}}
\end{align*}
where $N_{v_{1}}^{2}=p_{g_{3}}+p_{g_{2}}$ and $N_{v_{2}}^{2}=\frac{(p_{g_{2}}+p_{g_{3}})^{2}}{p_{g_{1}}}+p_{g_{2}}+p_{g_{3}}$.
Recall that
\begin{align*}
\left|w\right\rangle  & =\frac{\sqrt{p_{h_{1}}}\left|h_{1}\right\rangle +\sqrt{p_{h_{2}}}\left|h_{2}\right\rangle }{N_{h}}\\
\left|w_{1}\right\rangle  & =\frac{\sqrt{p_{h_{_{2}}}}\left|h_{1}\right\rangle -\sqrt{p_{h_{1}}}\left|h_{2}\right\rangle }{N_{h}}.
\end{align*}
Now we define 
\begin{align*}
\left|v'_{1}\right\rangle  & =\cos\theta\left|v_{1}\right\rangle +\sin\theta\left|v_{2}\right\rangle \\
\left|v_{2}'\right\rangle  & =\sin\theta\left|v_{1}\right\rangle -\cos\theta\left|v_{2}\right\rangle 
\end{align*}
where we know (from hindsight) that $\cos\theta\approx1$. The full
unitary (which is manifestly unitary) we define to be
\[
U=\left|w\right\rangle \left\langle v\right|+\left(\alpha\left|v_{1}'\right\rangle +\beta\left|w_{1}\right\rangle \right)\left\langle v_{1}'\right|+\left|v_{2}'\right\rangle \left\langle v_{2}'\right|+\left(\beta\left|v_{1}'\right\rangle -\alpha\left|w_{1}\right\rangle \right)\left\langle w_{1}\right|+\left|v\right\rangle \left\langle w\right|
\]
where $\left|\alpha\right|^{2}+\left|\beta\right|^{2}=1$ for $\alpha,\beta\in\mathbb{C}$.
There is some freedom in choosing $U$ in the sense that $\alpha\left|v\right\rangle +\beta\left|w_{1}\right\rangle $
would also work (then $\left|v\right\rangle \left\langle w\right|\to\left|v_{1}\right\rangle \left\langle w\right|$)
because these do not influence the constraint equation. That is what
we evaluate now. We need terms of the form $EU\left|g_{i}\right\rangle $
with $E=\mathbb{I}^{\{h_{i}\}}$. This entails that on the $\{\left|g_{i}\right\rangle \}$
space
\begin{align*}
E_{h}UE_{g} & =\left|w\right\rangle \left\langle v\right|+\beta\left|w_{1}\right\rangle \left\langle v_{1}'\right|\\
 & =\left|w\right\rangle \left\langle v\right|+\beta\left|w_{1}\right\rangle \left(\cos\theta\left\langle v_{1}\right|+\sin\theta\left\langle v_{2}\right|\right).
\end{align*}
Consequently we have
\begin{align*}
E_{h}U\left|g_{11}\right\rangle  & =\frac{\sqrt{p_{g_{1}}}}{N_{g}}\left|w\right\rangle +\left[\cos\theta.0-\sin\theta\frac{p_{g_{2}}+p_{g_{3}}}{\sqrt{p_{g_{1}}}N_{v_{2}}}\right]\beta\left|w_{1}\right\rangle \\
E_{h}U\left|g_{22}\right\rangle  & =\frac{\sqrt{p_{g_{2}}}}{N_{g}}\left|w\right\rangle +\left[\cos\theta\frac{\sqrt{p_{g_{3}}}}{N_{v_{1}}}+\sin\theta\frac{\sqrt{p_{g_{2}}}}{N_{v_{2}}}\right]\beta\left|w_{1}\right\rangle \\
E_{h}U\left|g_{33}\right\rangle  & =\frac{\sqrt{p_{g_{3}}}}{N_{g}}\left|w\right\rangle +\left[-\cos\theta\frac{\sqrt{p_{g_{2}}}}{N_{v_{1}}}+\sin\theta\frac{\sqrt{p_{g_{3}}}}{N_{v_{2}}}\right]\beta\left|w_{1}\right\rangle .
\end{align*}
Recall that the constraint equation was
\[
\sum x_{h_{i}}\left|h_{i}\right\rangle \left\langle h_{i}\right|-\sum x_{g_{i}}E_{h}U\left|g_{i}\right\rangle \left\langle g_{i}\right|U^{\dagger}E_{h}\ge0
\]
where the first sum becomes 
\[
\left[\begin{array}{cc}
\left\langle x_{h}\right\rangle  & \frac{\sqrt{p_{h_{1}}p_{h_{2}}}}{N_{h}^{2}}(x_{h_{1}}-x_{h_{2}})\\
\text{h.c.} & \frac{p_{h_{2}}x_{h_{1}}+p_{h_{1}}x_{h_{2}}}{N_{h}^{2}}
\end{array}\right]
\]
in the $\left|w\right\rangle ,\left|w_{1}\right\rangle $ basis. Since
we plan to use the $3\to2$ move with one point on the axis, we take
$x_{g_{1}}=0$. Consequently we need only evaluate 
\begin{align*}
x_{g_{2}}E_{h}U\left|g_{2}\right\rangle \left\langle g_{2}\right|U^{\dagger}E_{h}\dot{=} & x_{g_{2}}\left[\begin{array}{cc}
\frac{p_{g_{2}}}{N_{g}^{2}} & \beta\left(\cos\theta\frac{\sqrt{p_{g_{3}}p_{g_{2}}}}{N_{g}N_{v_{1}}}+\sin\theta\frac{p_{g_{2}}}{N_{g}N_{v_{2}}}\right)\\
\text{h.c.} & \left(\cos\frac{\sqrt{p_{g_{3}}}}{N_{v_{1}}}+\sin\theta\frac{\sqrt{p_{g_{2}}}}{N_{v_{2}}}\right)^{2}\left|\beta\right|^{2}
\end{array}\right]\\
x_{g_{3}}E_{h}U\left|g_{3}\right\rangle \left\langle g_{3}\right|U^{\dagger}\dot{E_{h}=} & x_{g_{3}}\left[\begin{array}{cc}
\frac{p_{g_{3}}}{N_{g}^{2}} & \beta\left(-\cos\theta\frac{\sqrt{p_{g_{2}}p_{g_{3}}}}{N_{g}N_{v_{1}}}+\sin\theta\frac{p_{g_{3}}}{N_{g}N_{v_{2}}}\right)\\
\text{h.c.} & \left(-\cos\frac{\sqrt{p_{g_{2}}}}{N_{v_{1}}}+\sin\frac{\sqrt{p_{g_{3}}}}{N_{v_{2}}}\right)^{2}\left|\beta\right|^{2}
\end{array}\right]
\end{align*}
which means that the constraint equation becomes {\tiny
\[
\left[\begin{array}{cc}
\left\langle x_{h}\right\rangle -\left\langle x_{g}\right\rangle  & \frac{\sqrt{p_{h_{1}}p_{h_{2}}}}{N_{h}^{2}}(x_{h_{1}}-x_{h_{2}})-\beta\cos\theta\frac{\sqrt{p_{g_{2}}p_{g_{3}}}}{N_{g}N_{v_{1}}}(x_{g_{2}}-x_{g_{3}})-\beta\sin\theta\left\langle x_{g}\right\rangle \frac{N_{g}}{N_{v_{2}}}\\
\text{h.c.} & \frac{p_{h_{2}}x_{h_{1}}+p_{h_{1}}x_{h_{2}}}{N_{h}^{2}}-\left|\beta\right|^{2}\left[\frac{\cos^{2}\theta}{N_{v_{1}}^{2}}(p_{g_{3}}x_{g_{2}}+p_{g_{2}}x_{g_{3}})+\frac{\sin^{2}\theta}{\left(N_{v_{2}}^{2}/N_{g}^{2}\right)}\left\langle x_{g}\right\rangle +\frac{2\cos\theta\sin\theta\sqrt{p_{g_{3}}p_{g_{2}}}}{N_{v_{1}}N_{v_{2}}}\left(x_{g_{2}}-x_{g_{3}}\right)\right]
\end{array}\right]\ge0.
\]
}We already showed that Mochon's transition is average non-decreasing
viz. $\left\langle x_{h}\right\rangle -\left\langle x_{g}\right\rangle \ge0$.
We set the off-diagonal elements of the matrix above to zero and show
that the second diagonal element, the second eigenvalue therefore,
is positive. 

Setting the off-diagonal to zero one can obtain $\theta$ by solving
the quadratic in terms of $\beta$ although the expression will not
be particularly pretty. To establish existence and positivity we need
to simplify our expressions.

So far everything was exact even though the basis and techniques were
chosen based on experience. Now we claim that $\theta\frac{N_{g}}{N_{v_{2}}}=\mathcal{O}(\delta y)$
at most (where $\delta y=\delta x$ is the lattice spacing) and since
$\delta y$ will be taken to be small we can take the small $\theta\frac{N_{g}}{N_{v_{2}}}$
limit and to first order in it the constraints become
\[
\frac{\frac{\sqrt{p_{h_{1}}p_{h_{2}}}}{N_{h}^{2}}(x_{h_{1}}-x_{h_{2}})-\beta\frac{\sqrt{p_{g_{2}}p_{g_{3}}}}{N_{g}N_{v_{1}}}(x_{g_{2}}-x_{g_{3}})}{\beta\left\langle x_{g}\right\rangle }=\theta\frac{N_{g}}{N_{v_{2}}}+\mathcal{O}(\delta y^{2})
\]
and 
\[
\frac{p_{h_{2}}x_{h_{1}}+p_{h_{1}}x_{h_{2}}}{N_{h}^{2}}-\left|\beta\right|^{2}\left[\frac{p_{g_{3}}x_{g_{2}}+p_{g_{2}}x_{g_{3}}}{N_{v_{1}}^{2}}+2\theta\frac{N_{g}}{N_{v_{2}}}\frac{\sqrt{p_{g_{3}}p_{g_{2}}}}{N_{g}N_{v_{1}}}(x_{g_{2}}-x_{g_{3}})\right]+\mathcal{O}(\delta y^{2})\ge0.
\]
If our claim is wrong when we evaluate $\theta\frac{N_{g}}{N_{v_{2}}}$
we will get zero order terms but as we show in the following section
$\theta\frac{N_{g}}{N_{v_{2}}}=0.\delta y+\mathcal{O}(\delta y^{2})$
in fact.

\subsubsection{Validity of the $3\to2$ Move}

With respect to \Figref{1by10correct} we have 
\begin{align*}
P_{2}(y_{j+2}) & =p_{h_{2}}=\frac{-f(y_{j+2})}{4.3\delta y^{2}y_{j+2}}\\
P_{1}(y_{j+1}) & =p_{g_{3}}=\frac{-f(y_{j+1})}{3.2\delta y^{2}y_{j+1}}\\
P_{1}(x_{j}) & =p_{h_{1}}=\frac{-f(y_{j-1})}{3.2\delta y^{2}y_{j-1}}\\
P_{2}(x_{j}) & =p_{g_{2}}=\frac{-f(y_{j-2})}{4.3\delta y^{2}y_{j-2}}\\
P(x_{j}) & =p_{g_{1}}=\frac{f(0)\delta y}{y_{j+2}y_{j+1}y_{j-1}y_{j-2}}
\end{align*}
where we assumed $f(0)>0$ and $f(y)<0$ for $y>y_{0}'$, $y_{0}'=y_{0}+\delta y$.
We also scaled by $\delta y$ to make $P(x_{j})$ into a nice density.
So far everything is exact. We now convert all expressions to first
order in $\delta y$. To this end we note 
\begin{align*}
f(y_{j+m}) & =f(y_{j})+\frac{\partial f}{\partial y}m\delta y+\mathcal{O}(\delta y^{2})\\
\frac{1}{y_{j+m}} & =(y_{j}+m\delta y)^{-1}=\frac{1}{y_{j}}\left(1+m\frac{\delta y}{y_{j}}\right)^{-1}=\frac{1}{y_{j}}-m\frac{\delta y}{y_{j}^{2}}+\mathcal{O}(\delta y^{2})
\end{align*}
where $\frac{\partial f}{\partial y}$ refers to $\frac{\partial f(y)}{\partial y}|_{y_{j}}$.
We define and evaluate 
\begin{align*}
P_{k}^{m} & =\frac{-f(y_{j+m})}{k\delta y^{2}y_{j+m}}\\
 & =\frac{1}{k\delta y^{2}}\left[-f(y_{j})-\frac{\partial f}{\partial y}m\delta y+\mathcal{O}(\delta y^{2})\right]\left[\frac{1}{y_{j}}-m\frac{\delta y}{y_{j}^{2}}+\mathcal{O}(\delta y^{2})\right]\\
 & =\frac{1}{k\delta y^{2}}\left[-\frac{f}{y_{j}}-m\frac{\delta y}{y_{j}}\left(\frac{\partial f}{\partial y}-\frac{f}{y_{j}}\right)+\mathcal{O}(\delta y^{2})\right]\\
 & =\frac{1}{ky_{j}\delta y^{2}}\left[-f-m\delta y\left(\frac{\partial f}{\partial y}-\frac{f}{y_{j}}\right)+\mathcal{O}(\delta y^{2})\right]
\end{align*}
where $f$ means $f(y_{j})$. In this notation 
\begin{align*}
p_{h_{2}} & =P_{12}^{2},\,p_{h_{1}}=P_{6}^{-1}\\
p_{g_{2}} & =P_{12}^{-2},\,p_{g_{3}}=P_{6}^{1}.
\end{align*}
With an eye at the off-diagonal condition we evaluate 
\[
P_{k_{1}}^{m_{1}}P_{k_{2}}^{m_{2}}=\frac{1}{k_{1}k_{2}}\left(\frac{1}{y_{j}\delta y^{2}}\right)^{2}\left[f^{2}+f\delta y\left(\frac{\partial f}{\partial y}-\frac{f}{y_{j}}\right)\left(m_{1}+m_{2}\right)+\mathcal{O}(\delta y^{2})\right]
\]
and 
\[
P_{k_{1}}^{m_{1}}+P_{k_{2}}^{m_{2}}=\frac{1}{y_{j}\delta y^{2}}\left[-\left(\frac{1}{k_{1}}+\frac{1}{k_{2}}\right)f-\left(\frac{m_{1}}{k_{1}}+\frac{m_{2}}{k_{2}}\right)\delta y\left(\frac{\partial f}{\partial y}-\frac{f}{y_{j}}\right)+\mathcal{O}(\delta y^{2})\right].
\]
We now evaluate 
\begin{align*}
\sqrt{p_{h_{1}}p_{h_{2}}} & =\sqrt{P_{12}^{2}P_{6}^{-1}}=\frac{1}{y_{j}\delta y^{2}}\sqrt{\frac{1}{12.6}\left[f^{2}+f\delta y\left(\frac{\partial f}{\partial y}-\frac{f}{y_{j}}\right)+\mathcal{O}(\delta y^{2})\right]}\\
N_{h}^{2} & =P_{12}^{2}+P_{6}^{-1}=\frac{1}{y_{j}\delta y^{2}}\left[-\left(\frac{1}{12}+\frac{1}{6}\right)f-\cancelto{0}{\left(\frac{2}{12}-\frac{1}{6}\right)}\delta y\left(\frac{\partial f}{\partial y}-\frac{f}{y_{j}}\right)+\mathcal{O}(\delta y^{2})\right]\\
 & =\frac{1}{4y_{j}\delta y^{2}}\left[-f+\mathcal{O}(\delta y^{2})\right]
\end{align*}
and similarly 
\begin{align*}
\sqrt{p_{g_{2}}p_{g_{3}}} & =\sqrt{P_{12}^{-2}P_{6}^{1}}=\frac{1}{y_{j}\delta y^{2}}\sqrt{\frac{1}{12.6}\left[f^{2}-f\delta y\left(\frac{\partial f}{\partial y}-\frac{f}{y_{j}}\right)+\mathcal{O}(\delta y^{2})\right]}\\
N_{g}^{2} & =P_{12}^{-2}+P_{6}^{1}+p_{g_{1}}=\frac{1}{4y_{j}\delta y^{2}}\left[-f+\mathcal{O}(\delta y^{2})\right]+\left[\frac{f(0)\delta y}{y_{j}^{4}}+\mathcal{O}(\delta y^{2})\right]\\
 & =\frac{1}{4y_{j}\delta y^{2}}\left[-f+\mathcal{O}(\delta y^{2})\right]\\
N_{v_{1}}^{2} & =\frac{1}{4y_{j}\delta y^{2}}\left[-f+\mathcal{O}(\delta y^{2})\right]
\end{align*}
where even though it seems like we have neglected $p_{g_{1}}$ when
we take the ratios the meaning of keeping first order in $\delta y$
would become precise. We can actually take $\beta=1$ and obtain 
\begin{align*}
\theta\frac{N_{g}}{N_{v_{2}}} & =\frac{4\sqrt{\frac{1}{12.6}}(-3\delta y)\left[f.(\cancel{1}+\frac{\delta y}{2f}\left(\frac{\partial f}{\partial y}-\frac{f}{y_{j}}\right))-f.(\cancel{1}-\frac{\delta y}{2f}\left(\frac{\partial f}{\partial y}-\frac{f}{y_{j}}\right))+\mathcal{O}(\delta y^{2})\right]}{\left\langle x_{g}\right\rangle }\\
 & =0+\mathcal{O}(\delta y^{2}).
\end{align*}
This shows that to first order the off-diagonal term is zero for $\theta=0$.

Now we show that the second diagonal element is positive to first
order in $\delta y$. Using the fact that $\theta\frac{N_{g}}{N_{v_{2}}}=\mathcal{O}(\delta y^{2})$
we have 
\[
\frac{p_{h_{2}}x_{h_{1}}+p_{h_{1}}x_{h_{2}}}{N_{h}^{2}}-\frac{p_{g_{3}}x_{g_{2}}+p_{g_{2}}x_{g_{3}}}{N_{v_{1}}^{2}}+\mathcal{O}(\delta y^{2})\ge0
\]
 as the positivity condition. This becomes 
\begin{align*}
= & \frac{P_{12}^{2}y_{j-1}+P_{6}^{-1}y_{j+2}}{N_{h}^{2}}-\frac{P_{6}^{1}y_{j-2}+P_{12}^{-2}y_{j+1}}{N_{v_{1}}^{2}}+\mathcal{O}(\delta y^{2})\\
= & \left(\frac{4y_{j}\delta y^{2}}{-f}\right)\frac{1}{y_{j}\delta y^{2}}\\
 & \left\{ \frac{1}{12}\left[-f-2\delta y\gamma\right](y_{j}-\delta y)+\frac{1}{6}\left[-f+\gamma\delta y\right](y_{j}+2\delta y)-\left(\frac{1}{6}\left[-f-\delta y\gamma\right](y_{j}-2\delta y)+\frac{1}{12}\left[-f+2\delta y\gamma\right](y_{j}+\delta y\right)\right\} \\
 & +\mathcal{O}(\delta y^{2})\\
= & \frac{-2}{3f}\left\{ \frac{1}{2}(\cancel{-fy}+f\delta y-2y\delta y\gamma)+(\cancel{-fy}-2f\delta y+y\delta y\gamma)-\left(\left(\cancel{-fy}+2f\delta y-y\delta y\gamma\right)+\frac{1}{2}\left(\cancel{-fy}-f\delta y+2y\delta y\gamma\right)\right)\right\} \\
 & +\mathcal{O}(\delta y^{2})\\
= & \frac{-2}{3f}\left\{ (f\delta y-2y\delta y\gamma)+2(-2f\delta y+y\delta y\gamma)\right\} +\mathcal{O}(\delta y^{2})\\
= & \frac{-2}{3f}\left\{ -3f\delta y\right\} +\mathcal{O}(\delta y^{2})=2\delta y+\mathcal{O}(\delta y^{2})\ge0
\end{align*}
where $\gamma=\left(\frac{\partial f}{\partial y}-\frac{f}{y_{j}}\right)$
and we suppressed the index $j$ in $y_{j}$ for simplicity. This
establishes the validity of the $3\to2$ transition for a closely
spaced lattice. 

Note that only the proof of validity was done perturbatively to first
order in $\delta y$. The unitary itself is known exactly ($\theta$
can be obtained by solving the quadratic).\\
Using $f(y)=(y_{0}'-y)(\Gamma_{1}-y)(\Gamma_{2}-y)$ we can implement
the last two moves in \Figref{1by10correct} as they form a $3\to1$
merge and a $2\to1$ merge (possibly followed by a raise). The only
remaining task is implementing the $2\to2$ move in the last step
because we assumed here that $\sqrt{p_{g_{2}}}\neq0$ (else the vectors
which we assumed are orthonormal, cease to be so).

\subsubsection{The $2\to2$ Move and its validity}

\begin{figure}
\begin{centering}
\includegraphics[width=14cm]{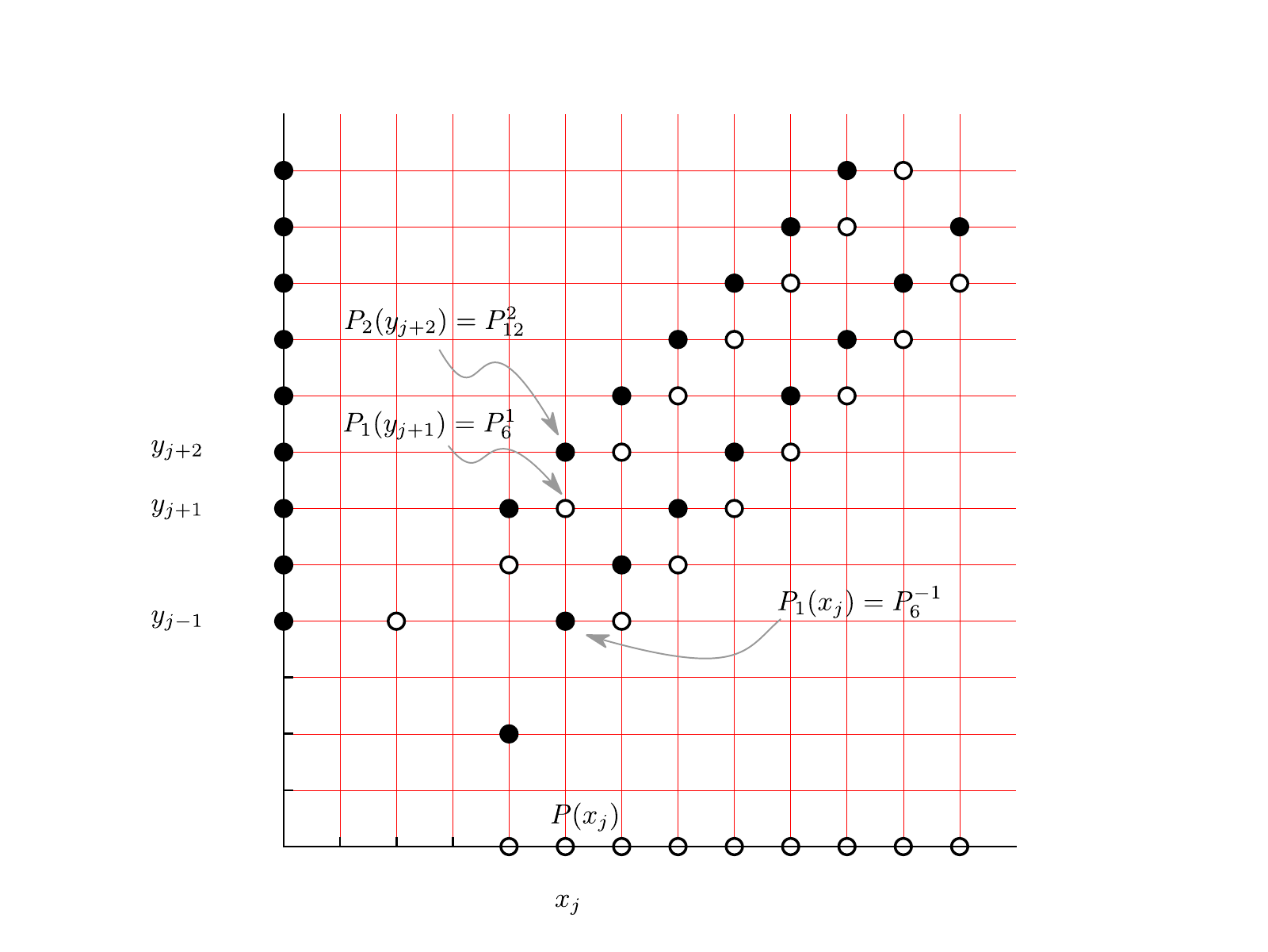}
\par\end{centering}
\caption{First $2\to2$ Transition\label{fig:2to2special}}
\end{figure}

We claim that the $2\to2$ move can be implemented using
\[
U=\left|w\right\rangle \left\langle v\right|+\left(\alpha\left|v\right\rangle +\beta\left|w_{1}\right\rangle \right)\left\langle v_{1}\right|+\left|v\right\rangle \left\langle w\right|+\left(\beta\left|v\right\rangle -\alpha\left|w_{1}\right\rangle \right)\left\langle w_{1}\right|
\]
where as before $\left|\alpha\right|^{2}+\left|\beta\right|^{2}=1$,
\[
\left|v\right\rangle =\frac{1}{N_{g}}\left(\sqrt{p_{g_{1}}}\left|g_{1}\right\rangle +\sqrt{p_{g_{2}}}\left|g_{2}\right\rangle \right),
\]
\[
\left|w\right\rangle =\frac{1}{N_{h}}\left(\sqrt{p_{h_{1}}}\left|h_{1}\right\rangle +\sqrt{p_{h_{2}}}\left|h_{2}\right\rangle \right),
\]
\[
\left|v_{1}\right\rangle =\frac{1}{N_{g}}\left(\sqrt{p_{g_{2}}}\left|g_{1}\right\rangle -\sqrt{p_{g_{1}}}\left|g_{2}\right\rangle \right),
\]
 and 
\[
\left|w_{1}\right\rangle =\frac{1}{N_{h}}\left(\sqrt{p_{h_{2}}}\left|h_{1}\right\rangle -\sqrt{p_{h_{1}}}\left|h_{2}\right\rangle \right).
\]
We evaluate the constraint equation using 
\begin{align*}
E_{h}U\left|g_{11}\right\rangle  & =\frac{\sqrt{p_{g_{1}}}\left|w\right\rangle +\beta e^{-i\phi_{g}}e^{i\phi_{h}}\sqrt{p_{g_{2}}}\left|w_{1}\right\rangle }{N_{g}}\\
E_{h}U\left|g_{22}\right\rangle  & =\frac{\sqrt{p_{g_{2}}}\left|w\right\rangle -\beta e^{-i\phi_{g}}e^{i\phi_{h}}\sqrt{p_{g_{1}}}\left|w_{1}\right\rangle }{N_{g}}
\end{align*}
and
\[
E_{h}U\left|g_{11}\right\rangle \left\langle g_{11}\right|U^{\dagger}E_{h}=\frac{1}{N_{g}^{2}}\begin{array}{c|cc}
 & \left\langle w\right| & \left\langle w_{1}\right|\\
\hline \left|w\right\rangle  & p_{g_{1}} & \beta e^{i(\phi_{h}-\phi_{g})}\sqrt{p_{g_{2}}p_{g_{1}}}\\
\left|w_{1}\right\rangle  & \text{h.c.} & \left|\beta\right|^{2}p_{g_{2}}
\end{array}
\]
(similarly for $L\left|g_{22}\right\rangle \left\langle g_{22}\right|L^{\dagger}$)
as 
\[
\left[\begin{array}{cc}
\left\langle x_{h}\right\rangle -\left\langle x_{g}\right\rangle  & \frac{1}{N_{g}^{2}}\left[\sqrt{p_{h_{1}}p_{h_{2}}}(x_{h_{1}}-x_{h_{2}})-\beta\sqrt{p_{g_{1}}p_{g_{2}}}(x_{g_{1}}-x_{g_{2}})\right]\\
\text{h.c.} & \frac{1}{N_{g}^{2}}\left[p_{h_{2}}x_{h_{1}}+p_{h_{1}}x_{h_{2}}-\left|\beta\right|^{2}(p_{g_{2}}x_{g_{1}}+p_{g_{1}}x_{g_{2}})\right]
\end{array}\right]\ge0
\]
where we absorbed the phase freedom in $\beta$, a free parameter,
which will be fixed shortly. We use the same strategy as above and
take the first diagonal element to be zero. Our burden would be to
first show that 
\[
\sqrt{\frac{p_{h_{1}}p_{h_{2}}}{p_{g_{1}}p_{g_{2}}}}\frac{(x_{h_{1}}-x_{h_{2}})}{(x_{g_{1}}-x_{g_{2}})}=\beta\le1
\]
and subsequently 
\[
\frac{1}{N_{g}^{2}}\left[p_{h_{2}}x_{h_{1}}+p_{h_{1}}x_{h_{2}}-\left|\beta\right|^{2}(p_{g_{2}}x_{g_{1}}+p_{g_{1}}x_{g_{2}})\right]\ge0.
\]
What makes this situation special (compared to the $3\to2$ merge)
is that $f(y_{j-2})=0$ which we use to write 
\[
f(y_{j+k})=\left.\frac{\partial f}{\partial y}\right|_{y_{j-2}}(k+2)\delta y=-(k+2)\alpha\delta y
\]
where 
\[
\alpha=-\left.\frac{\partial f}{\partial y}\right|_{y_{j-2}}=(\Gamma_{1}-y_{j-2})(\Gamma_{2}-y_{j-2}).
\]
Using the axis situation as depicted in \Figref{2to2special} 
 we note that
\begin{align*}
p_{h_{1}} & =P_{1}(x_{j})=\frac{-f(y_{j-1})}{3.2\delta y^{2}y_{j-1}}=\frac{\alpha+\mathcal{O}(\delta y)}{6\delta yy_{j}}\\
p_{h_{2}} & =P_{2}(y_{j+2})=\frac{-f(y_{j+2})}{4.3\delta y^{2}y_{j+2}}=\frac{\alpha+\mathcal{O}(\delta y)}{3\delta yy_{j}}\\
x_{h_{1}} & =y_{j-1},\,x_{h_{2}}=y_{j+2}
\end{align*}
while 
\begin{align*}
p_{g_{1}} & =P(x_{j})=\frac{f(0)\delta y}{y_{j+2}y_{j+1}y_{j-1}y_{j-2}}=\frac{f(0)\delta y+\mathcal{O}(\delta y^{2})}{y_{j}^{4}}\\
p_{g_{2}} & =P_{1}(y_{j+1})=\frac{-f(y_{j+1})}{3.2\delta y^{2}y_{j+1}}=\frac{\alpha+\mathcal{O}(\delta y)}{2\delta yy_{j}}\\
x_{g_{1}} & =0,\,x_{g_{2}}=y_{j+1}.
\end{align*}
This entails 
\begin{align*}
\beta & =\sqrt{\frac{p_{h_{1}}p_{h_{2}}}{p_{g_{1}}p_{g_{2}}}}\frac{(x_{h_{1}}-x_{h_{2}})}{(x_{g_{1}}-x_{g_{2}})}=\sqrt{\frac{\alpha^{\cancel{2}}+\mathcal{O}(\delta y)}{\cancel{\cancelto{3}{6}.3\delta y^{2}}\cancel{y_{j}^{2}}}\frac{\cancel{2}\cancel{\delta y}\cancel{y_{j}^{4}}y_{j}}{\cancel{\delta y}(f(0)\cancel{\alpha}+\mathcal{O}(\delta y))}\frac{\cancel{\left(3\delta y\right)^{2}}}{\cancel{y_{j}^{2}+\mathcal{O}(\delta y)}}}\\
 & =\sqrt{\frac{y_{0}'\alpha+\mathcal{O}(\delta y)}{f(0)}}=\sqrt{\frac{(\Gamma_{1}-y_{j-2})(\Gamma_{2}-y_{j-2})+\mathcal{O}(\delta y)}{\Gamma_{1}\Gamma_{2}}}\le1
\end{align*}
where we used $f(0)=y_{0}'\Gamma_{1}\Gamma_{2}$ and assumed $\delta y$
is small compared $\Gamma$s (which is the case) for the inequality
in the last step to hold.

The second condition can be proven similarly 
\begin{align*}
 & \frac{1}{N_{g}^{2}}\left[p_{h_{2}}x_{h_{1}}+p_{h_{1}}x_{h_{2}}-\left|\beta\right|^{2}(p_{g_{2}}x_{g_{1}}+p_{g_{1}}x_{g_{2}})\right]\\
 & \ge\frac{1}{N_{g}^{2}}\left[p_{h_{2}}x_{h_{1}}+p_{h_{1}}x_{h_{2}}-p_{g_{2}}x_{g_{1}}\right]\\
 & =\frac{1}{N_{g}^{2}}\left[\frac{\alpha+\mathcal{O}(\delta y)}{3\delta yy_{j}}y_{j-1}+\frac{\alpha+\mathcal{O}(\delta y)}{6\delta yy_{j}}y_{j+2}-\frac{f(0)\delta y+\mathcal{O}(\delta y^{2})}{y_{j}^{4}}y_{j+1}\right]\\
 & =\frac{1}{3\delta yN_{g}^{2}}\left[\left(\alpha+\mathcal{O}(\delta y)\right)\left(\frac{3}{2}\right)-\underbrace{\frac{f(0)\delta y^{2}+\mathcal{O}(\delta y^{3})}{y_{j}^{3}}}_{\in\mathcal{O}(\delta y^{2})}\right]\\
 & =\frac{1}{2\delta yN_{g}^{2}}\left[(\Gamma_{1}-y_{j-2})(\Gamma_{2}-y_{j-2})+\mathcal{O}(\delta y)\right]\ge0
\end{align*}
where the last step holds for $\delta y$ small enough.
\begin{figure}[h]
\begin{centering}
\includegraphics[width=14cm]{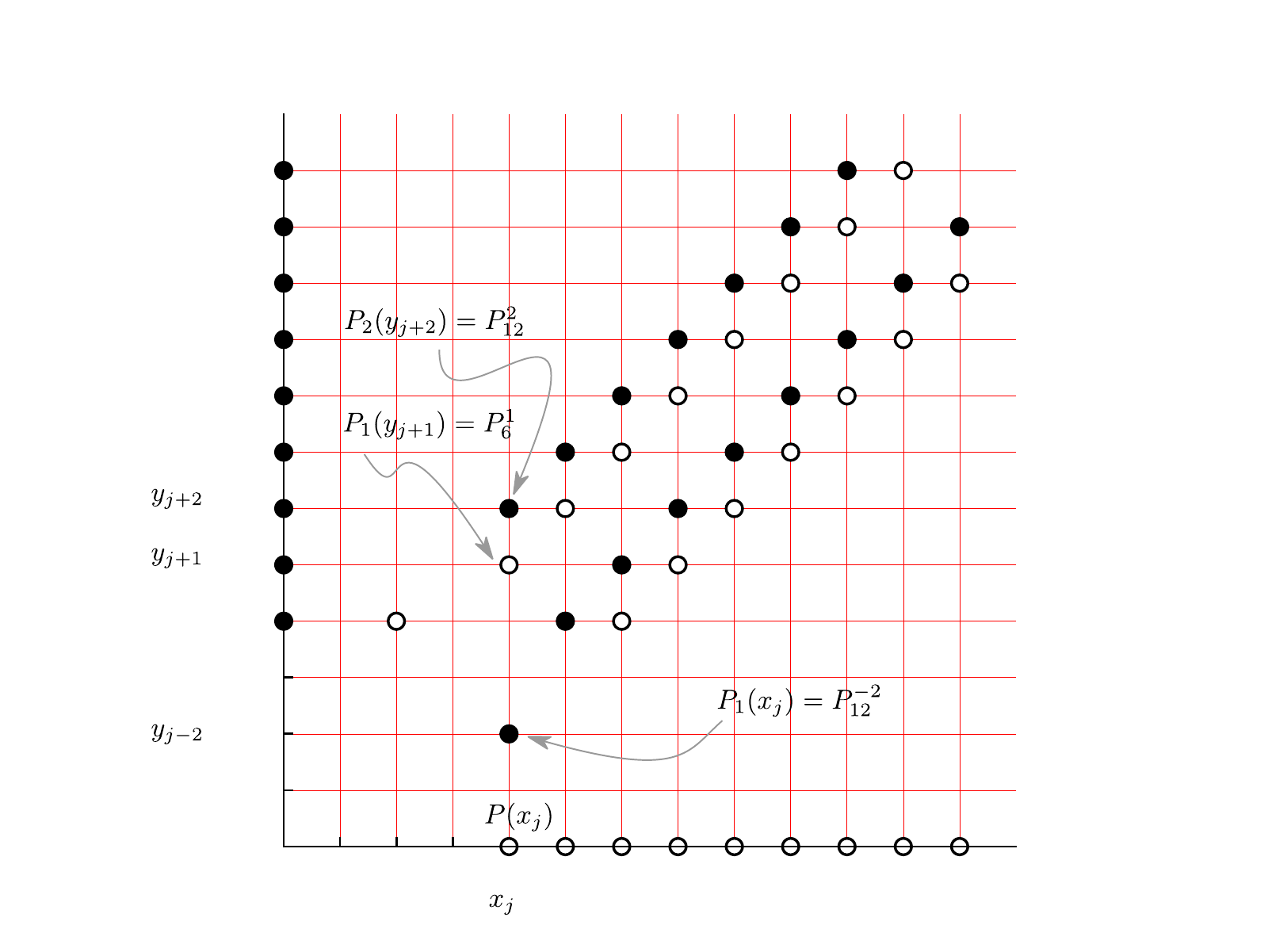}
\par\end{centering}
\caption{Final $2\to2$ Transition. \label{fig:Final-2to2}}
\end{figure}

The $2\to2$ move corresponding to the leftmost (see \Figref{Final-2to2})
and bottommost set of points can be shown to be implementable similarly.

\clearpage{}

\part{Elliptic Monotone Algorithm (EMA)}

\section{Canonical Forms Revisited\label{sec:Canonical-Forms-Revisited}}

\textcolor{purple}{We note that to construct the unitaries involved
in the bias $1/10$ protocol we did not follow any systematic recipe.
We now switch gears and construct an algorithm that can generate the
required unitary for any given $\Lambda$-valid function (see \Defref{lambdaValid}).
Note that corresponding to any WCF protocol with valid functions,
one can find a WCF protocol with strictly valid functions (see \Lemref{ValidToStrictlyValidGame}).
All strictly valid functions are $\Lambda$-valid for some finite
$\Lambda$ (see \Lemref{strictlyValidIsEBM}, \Corref{EBMandLambdaValid}).
Thus we do not lose generality by restricting to $\Lambda$-valid
functions.}

\textcolor{purple}{In this section we formalise the non-trivial constraint
\Eqref{MainConstraintInequality} into two forms which we call the
Canonical Projective Form (CPF) and the Canonical Orthogonal Form
(COF). The CPF is always well defined but the corresponding COF may
contain diverging eigenvalues. However since we restrict to $\Lambda$-valid
functions, as we will see, the COF will also always be well defined.
We need the COF as it is this that we use in the Elliptic Monotone
Algorithm (EMA) algorithm.}

\textcolor{purple}{We continue to use purple for the intuitive parts
and start using }\textcolor{blue}{blue for the proofs}\textcolor{purple}{.}

\subsection{The Canonical Projective Form (CPF) and the Canonical Orthogonal
Form (COF)\label{subsec:CPFandCOF}}

We always use the convention $p_{g_{i}},p_{h_{i}}>0$. \textcolor{purple}{This is important else in some of the statements
one can find trivial counter-examples. Recall \Thmref{TEFconstraint}
which formally states the main result of \Secref{TEF}. }\textcolor{purple}{Note that the number of points initially, $n_{g}$,
and finally, $n_{h}$, may differ. To facilitate further discussion
we formalise the aforesaid condition into an object and its property.
First, however, we define the following notation.}
\begin{defn}[Transition]
 Consider two finitely supported functions $g,h:\mathbb{R}_{\ge}\to\mathbb{R}_{\ge}$.
A transition is defined as $g=\sum_{i=1}^{n_{g}}p_{g_{i}}[x_{g_{i}}]\to h=\sum_{i=1}^{n_{h}}p_{h_{i}}[x_{h_{i}}]$
where $[y](x):=\delta_{xy}$ and $p_{g_{i}}>0$, $p_{h_{i}}>0$. \label{def:transition}
\end{defn}

\begin{defn}[Canonical Projective Form (CPF) for a give transition]

For a given transition the \emph{Canonical Projective Form (CPF)}
is given by the set of $m\times m$ matrices $X_{h}$, $X_{g}$, $U$,
$D$ and $m$ dimensional vectors $\left|v\right\rangle $, $\left|w\right\rangle $
where 
\[
X_{h}:=\sum_{i=1}^{n_{h}}x_{h_{i}}\left|h_{i}\right\rangle \left\langle h_{i}\right|,\,X_{g}:=\sum_{i=1}^{n_{g}}x_{g_{i}}\left|g_{i}\right\rangle \left\langle g_{i}\right|,
\]
\[
\left|w\right\rangle :=\sum_{i=1}^{n_{h}}\sqrt{p_{h_{i}}}\left|h_{i}\right\rangle ,\,\left|v\right\rangle :=\sum_{i=1}^{n_{g}}\sqrt{p_{g_{i}}}\left|g_{i}\right\rangle ,
\]
\[
D:=X_{h}-E_{h}UX_{g}U^{\dagger}E_{h}
\]
and $U$ is a unitary which satisfies 
\[
U\left|v\right\rangle =\left|w\right\rangle 
\]
 for $E_{h}=\sum\left|h_{i}\right\rangle \left\langle h_{i}\right|$,
orthonormal basis vectors $\left\{ \left|g_{1}\right\rangle ,\left|g_{2}\right\rangle \dots\left|g_{n_{g}}\right\rangle ,\left|h_{1}\right\rangle ,\left|h_{2}\right\rangle \dots\left|h_{n_{h}}\right\rangle \right\} $,
$m=n_{g}+n_{h}$.
\end{defn}

\begin{defn}[legal CPF]
 A CPF is \emph{legal} if $D\ge0$.
\end{defn}

In this language then our objective is to find a legal CPF for a given
transition.

\textcolor{purple}{Surprising as it may seem it }\textcolor{purple}{\emph{suffices
to restrict to real unitaries viz. orthogonal matrices}}\textcolor{purple}{.
This will be justified in the next section but we already make this
restriction in everything that follows (unless stated otherwise).
In this section we try to reach an equivalence between a legal CPF
and what we call the legal Canonical Orthogonal Form (COF). }

\textcolor{purple}{The latter will be, roughly speaking, an inequality
of the form $X_{h}-OX_{g}O^{T}\ge0$ where $X_{h}=\text{diag}\{x_{h_{1}},x_{h_{2}}\dots,x_{h_{n_{h}}},\xi,\xi\dots\}$
and $X_{g}=\text{diag}\{x_{g_{1}},x_{g_{2}}\dots,x_{g_{n_{g}}},0,0\dots\}$
for a large $\xi$. It is easy to see that if we can find an $O$
that satisfies the COF for a given transition then the same $O$ would
satisfy the TEF inequality. It is almost trivial to note that a $\Lambda$
valid function admit matrices of the COF form but we will show this
later. Proving the other way, i.e. every legal CPF entails the corresponding
COF must also be legal, is more non-trivial. Doing this requires handling
the infinities and the matrix sizes more carefully. We only sketch
an argument for this as we do not use it in the algorithm.}

\begin{defn}[($n,\xi$) Canonical Orthogonal Form (COF) for a transition, $\xi$
COF for a transition]

For a given transition and two numbers $n\ge\max(n_{h},n_{g})$, $\xi\ge\max(x_{h_{1}},x_{h_{2}}\dots x_{h_{n_{h}}})$
an \emph{$(n,\xi)$ Canonical Orthogonal Form (COF)} is given by the
set of $n\times n$ matrices $X_{h}$, $X_{g}$, $O$, $D$ and vectors
$\left|v\right\rangle $, $\left|w\right\rangle $ where 
\[
X_{h}:=\text{diag}\{x_{h_{1}},x_{h_{2}}\dots,x_{h_{n_{h}}},\xi,\xi\dots\},
\]
\[
X_{g}:=\text{diag}\{x_{g_{1}},x_{g_{2}}\dots,x_{g_{n_{g}}},0,0\dots\},
\]
\[
\left|v\right\rangle :=\sum_{i=1}^{n_{g}}\sqrt{p_{g_{i}}}\left|i\right\rangle ,
\]
\[
\left|w\right\rangle :=\sum_{i=1}^{n_{h}}\sqrt{p_{h_{i}}}\left|i\right\rangle ,
\]
\[
D:=X_{h}-OX_{g}O^{T}
\]
and the matrix $O$ is orthogonal which satisfies
\[
O\left|v\right\rangle =\left|w\right\rangle .
\]

A \emph{$\xi$ Canonical Orthogonal Form (COF)} is an $(n,\xi)$ COF
with $n=n_{h}+n_{g}-1$.
\end{defn}

\begin{defn}[$n$ legal COF, legal COF]
 An $(n,\xi)$ COF is an \emph{$n$ legal COF} if $D\ge0$ in the
limit $\xi\to\infty$. A \emph{legal COF} is a $\xi$ COF such that
$D\ge0$ in the limit $\xi\to\infty$.
\end{defn}

\textcolor{purple}{Imagine you found a legal COF corresponding to
some transition. One can then sandwich $D$ between a positive matrix
as $EDE$ to get
\[
\left[\begin{array}{c|ccc}
X_{h}\\
\hline  & 1\\
 &  & \ddots\\
 &  &  & 1
\end{array}\right]-\underbrace{\left[\begin{array}{ccc|ccc}
1 &  & \\
 & \ddots & \\
 &  & 1\\
\hline  &  &  & \xi^{-1/2}\\
 &  &  &  & \ddots\\
 &  &  &  &  & \xi^{-1/2}
\end{array}\right]}_{:=E}UX_{g}U^{\dagger}\left[\begin{array}{ccc|ccc}
1 &  & \\
 & \ddots & \\
 &  & 1\\
\hline  &  &  & \xi^{-1/2}\\
 &  &  &  & \ddots\\
 &  &  &  &  & \xi^{-1/2}
\end{array}\right].
\]
Note that $D\ge0\iff EDE\ge0$ because $E$ is diagonal (which means
one can write $EDE=(E\sqrt{D})(\sqrt{D}E)$ which in turn is of the
$A^{T}A$ form). From the legality of the COF, $D\ge0$ in the limit
$\xi\to\infty$ and in this limit $E$ becomes a projector. After
some relabelling (and appropriately expanding the space to $m=n_{g}+n_{h}$
dimensions) the inequality reduces to a CPF. This observation readily
extends to the $n$ legal case where $n\le n_{g}+n_{h}$. It turns
out that one can, and we show this later, always express an $n'$
legal COF as an $n$ legal COF with $n\le n_{g}+n_{h}$ (in fact we
can prove that $n\le n_{g}+n_{h}-1$). We have established the following
statement.}

\begin{prop}
Consider a transition. If there exists an $n$ legal COF corresponding
to it then there exists a legal CPF for the said transition.
\end{prop}

\textcolor{purple}{How about the reverse? Given a legal CPF can we
find the corresponding $n$ legal COF? We are given 
\[
D=\left[\begin{array}{c|ccc}
X_{h}\\
\hline  & 0\\
 &  & \ddots\\
 &  &  & 0
\end{array}\right]-\underbrace{\left[\begin{array}{ccc|ccc}
1 &  & \\
 & \ddots & \\
 &  & 1\\
\hline  &  &  & 0\\
 &  &  &  & \ddots\\
 &  &  &  &  & 0
\end{array}\right]}_{=E_{h}}U\left[\begin{array}{c|c}
0\\
\hline  & X_{g}
\end{array}\right]U^{\dagger}\left[\begin{array}{ccc|ccc}
1 &  & \\
 & \ddots & \\
 &  & 1\\
\hline  &  &  & 0\\
 &  &  &  & \ddots\\
 &  &  &  &  & 0
\end{array}\right]\ge0.
\]
Replacing the appended diagonal zeros in the first matrix (one containing
$X_{h}$) with $1$s yields an equivalent inequality. Next note that
we can conjugate by a permutation matrix to get 
\[
\left[\begin{array}{c|c}
0\\
\hline  & X_{g}
\end{array}\right]=\tilde{U}\left[\begin{array}{c|c}
X_{g}\\
\hline  & 0
\end{array}\right]\tilde{U}.
\]
Finally we write the diagonal zeros in $E_{h}$ as $1/\xi^{1/2}$
and use the reverse of the trick above to recover an $m$ legal COF
where recall $m:=n_{g}+n_{h}$. This sketches the proof of the following
statement.}

\begin{prop}
Consider a transition. If there exists legal CPF corresponding to
it then there exists an $m$ legal COF for the said transition (where
recall $m:=n_{g}+n_{h}$).
\end{prop}

\subsection{From EBM to EBRM to COF\label{subsec:EBMtoEBRMtoCOF}}

\textcolor{purple}{We briefly summarise, at the cost being redundant,
how Aharonov et al. prove that valid functions are equivalent to the
EBM functions (assuming the operator monotones are on/the spectrum
of the matrices is in $[0,\Lambda]$). They do this by showing that
the set of EBM functions forms a convex cone $K$. Then they take
the dual of this cone to get $K^{*}$. }\textcolor{purple}{\emph{This
dual happens to be the set of operator monotone functions.}}\textcolor{purple}{{}
Then they find the bi-dual $K^{**}$ and define the objects in this
to be valid functions. They then show that $K=K^{**}$ which is to
say that valid functions are equivalent to EBM functions. Note that
all of this is assuming the aforesaid $[0,\Lambda]$ condition.}

\textcolor{purple}{This is an extremely useful result because checking
if a function is EBM is hard. Checking if a function is valid is a
piece of cake because mathematical wizards have neatly characterised
the set of operator monotone functions. }

\textcolor{purple}{One can do even better. Instead of EBM functions,
consider EBRM functions where the matrices are additionally restricted
to be real. Let this set be given by $K'$. It turns out that its
dual $K'^{*}$ is also the set of operator monotone functions \parencite{Tobias2018}
viz. $K'^{*}=K^{*}$. Aharonov et al's proof for $K=K^{**}$ can be
applied to the real case as is to get $K'=K^{**}$ (granted we assume
the same $[0,\Lambda]$ condition). }

\textcolor{purple}{Since Mochon's point games (and even the ones built
later) use valid functions, the aforesaid simplification justifies
why it suffices to restrict to real matrices.}

We use the definition of Prob (\Defref{prob}), EBM line transition
(\Defref{EBMlineTransition}), EBM function (\Defref{EBMfunctions},
\Defref{lambdaEBMfunctions}), Operator Monotone functions (\Defref{operatorMonotone},
\Defref{operatorMonotoneLambda}) and their characterisation (\Lemref{opMonChar},
\Lemref{opMonCharLambda}), $\Lambda$ valid functions (\Defref{lambdaValid})
and finally its equivalence with EBM functions (\Corref{EBMandLambdaValid}).

\subsection*{Equivalence of EBM and EBRM}

\textcolor{purple}{First we define EBRM transitions and EBRM functions
similar to their EBM analogues except with the further restriction
that the matrices and vectors involved are real.}
\begin{defn}[EBRM transitions]
 Let $g,h:[0,\infty)\to[0,\infty)$ be two functions with finite
supports. The transition $g\to h$ is EBRM if there exist two real
matrices $0\le G\le H$ and a (not necessarily normalised) vector
$\left|\psi\right\rangle $ such that $g=\text{prob}[G,\psi]$ and
$h=\text{prob}[H,\psi]$.
\end{defn}

\begin{defn}[{$K'$, EBRM functions; $K'_{\Lambda}$, EBRM functions on $[0,\Lambda]$ }]
 A function $a:[0,\infty)\to\mathbb{R}$ with finite support is an
EBRM function if the transition $a^{-}\to a^{+}$ is EBRM, where $a^{+}:\mathbb{R}_{\ge0}\to\mathbb{R}_{\ge0}$
and $a^{-}:\mathbb{R}_{\ge0}\to\mathbb{R}_{\ge0}$ denote, respectively,
the positive and the negative part of $a$ ($a=a^{+}-a^{-}).$ We
denote by $K'$ the set of EBRM functions.

For any finite $\Lambda\in(0,\infty)$, a function $a:[0,\Lambda)\to\mathbb{R}$
with finite support is an EBRM function with support on $[0,\Lambda]$
if the transition $a^{-}\to a^{+}$ is EBRM with its spectrum in $[0,\Lambda]$,
where $a^{+}$ and $a^{-}$ denote, respectively, the positive and
the negative part of $a$ (again, $a=a^{+}-a^{-}$). We denote by
$K_{\Lambda}'$ the set of EBRM functions with support on $[0,\Lambda]$.
\end{defn}

\begin{defn}[Real operator monotone functions]
 A function $f:(0,\infty)\to\mathbb{R}$ is a real operator monotone
if for all real matrices $0\le A\le B$ we have $f(A)\le f(B)$.

A function $f:(0,\Lambda)\to\mathbb{R}$ is a real operator monotone
on $[0,\Lambda]$ if for all real matrices $0\le A\le B$ with spectrum
in $[0,\Lambda]$ we have $f(A)\le f(B)$.
\end{defn}

\begin{lem*}
$K_{\Lambda}^{\prime*}$ is the set of real operator monotones on
$[0,\Lambda]$.
\end{lem*}
\begin{proof}[Proof sketch.]
Aharonov et al. showed that $K_{\Lambda}^{*}$ (which is, recall,
the dual of the set of EBM functions on $[0,\Lambda]$) is the set
of operator monotone functions on $[0,\Lambda]$ (see Lemma 3.9 of
\parencite{ACG+14}). Their proof can be adapted here by restricting
to real matrices which entails that $K_{\Lambda}^{\prime*}$ is the
set of real operator monotone functions on $[0,\Lambda]$. 
\end{proof}
\begin{lem}
$K_{\Lambda}^{*}=K_{\Lambda}^{\prime*}$ and $K^{*}=K^{\prime*}$,
i.e. the set of operator monotones on $[0,\Lambda]$ equals the set
of real operator monotones on $[0,\Lambda]$ and the set of operator
monotones equals the set of real operator monotones.\label{lem:opMonotoneEQrealOpMonotone}
\end{lem}

\begin{cor}
$K'_{\Lambda}=K_{\Lambda}^{\prime**}=K_{\Lambda}^{**}=K_{\Lambda}$,
i.e. the set of EBRM functions on $[0,\Lambda]$ = the set of $\Lambda$
valid functions (dual of EBRM functions) = the set of $\Lambda$ valid
functions (dual of EBM functions) = the set of EBM functions on $[0,\Lambda]$.\label{cor:EBMisEBRMisValid}
\end{cor}

\begin{cor}
Any strictly valid function is EBRM.
\end{cor}

\textcolor{purple}{We now sketch the proof of \Lemref{opMonotoneEQrealOpMonotone}.
It is clear that the set of real operator monotones must contain the
set of operator monotones because operator monotones are by definition
required to work in the restricted real case as well. The set of real
operator monotones might be bigger but that does not happen to be
the case. This is because we can encode an $n\times n$ hermitian
matrix into a $2n\times2n$ real symmetric matrix. This is achieved
by replacing each complex number $\alpha+i\beta$ with the matrix
\[
\alpha\left[\begin{array}{cc}
1\\
 & 1
\end{array}\right]+\beta\left[\begin{array}{cc}
 & -1\\
1
\end{array}\right].
\]
Note that the matrices have the exact same properties as $1$ and
$i$ respectively. This corresponds to (after some permutation) writing
a complex matrix $W=W_{\Re}+iW_{\Im}$ as a real symmetric 
\[
W'=\left[\begin{array}{cc}
W_{\Re} & -W_{\Im}\\
W_{\Im} & W_{\Re}
\end{array}\right]
\]
where $W_{\Re}$ and $W_{\Im}$ are real. For a Hermitian $W^{\dagger}=W$
we must have $W_{\Re}=W_{\Re}^{T}$ and $W_{\Im}=-W_{\Im}^{T}$ which
makes $W'=W'^{T}$ a symmetric matrix. The spectra of $W$ and $W'$
are the same. This is established most easily by looking at the diagonal
decomposition. $W=USU^{\dagger}$ which would get transformed to $W'=U'S'U'^{\dagger}$.
Since $S$ is real $S'$ is also real with doubly degenerate eigenvalues
(except for the degeneracy already present in $S$). Thus if we have
$0\le A\le B$ then we would also have $0\le A'\le B'$ as $A-B$
and $A'-B'$ would have the same spectrum where we used $A'$ and
$B'$ to represent real symmetric analogues of the hermitian matrices
$A$ and $B$. The other way is trivial. Hence we have an equivalence
which means that requiring a function to be operator monotone on complex
matrices is the same as requiring it to be operator monotone on real
symmetric matrices of twice the size (at most). This means that the
set of real operator monotones is the same as the set of operator
monotones.}

\subsection*{EBRM to COF | Mochon's Variant}

\textcolor{purple}{We just saw how we can reduce our problem from
the set of EBM transitions to the set of EBRM transitions. We now
show that each EBRM transition can be written in a standard form,
which we call the Canonical Orthogonal Form (COF). The following is
actually due to Mochon/Kitaev \cite{Mochon07} but there was a minor
mistake in one of the arguments which we have corrected. The interesting
part is showing that one can always restrict the matrices to a certain
size which in turn depends on the number of points involved in the
transition. }
\begin{lem}
\begin{comment}
If we are given $G\le H$ with their spectrum in $[a,b]$ and a $\left|\psi\right\rangle $
such that 
\[
g=\text{Prob}[G,\left|\psi\right\rangle ]=\sum_{i=1}^{n_{g}}p_{g_{i}}[x_{g_{i}}]
\]
 and 
\[
h=\text{Prob}[H,\left|\psi\right\rangle ]=\sum_{i=1}^{n_{h}}p_{h_{i}}[x_{h_{i}}]
\]
with $p_{g_{i}},p_{h_{i}}>0$ and $x_{g_{i}}\neq x_{g_{j}}$, $x_{h_{i}}\neq x_{h_{j}}$
for $i\neq j$ then it is possible to find an orthogonal matrix $O$,
diagonal matrices $X_{h}$, $X_{g}$ of size at most $n_{g}+n_{h}-1$
such that 
\[
O\underbrace{\left[\begin{array}{ccc|cc}
x_{g_{1}} &  & \\
 & \ddots & \\
 &  & x_{g_{n_{g}}}\\
\hline  &  &  & a\\
 &  &  &  & \ddots
\end{array}\right]}_{:=X_{g}}O^{T}\le\left[\begin{array}{ccc|cc}
x_{h_{1}} &  & \\
 & \ddots & \\
 &  & x_{h_{n_{h}}}\\
\hline  &  &  & b\\
 &  &  &  & \ddots
\end{array}\right]:=X_{h},
\]
and the vector $\left|\psi\right\rangle =\sum_{i=1}^{n_{h}}\sqrt{p_{h_{i}}}\left|i\right\rangle =\sum_{i=1}^{n_{g}}\sqrt{p_{g_{i}}}O\left|i\right\rangle $.
\end{comment}
For every EBRM transition $g\to h$ with spectrum in $[a,b]$ there
exists an orthogonal matrix $O$, diagonal matrices $X_{h}$, $X_{g}$
(with no multiplicities except possibly those of $a$ and $b$) of
size at most $n_{g}+n_{h}-1$ such that 
\[
O\underbrace{\left[\begin{array}{ccccc}
x_{g_{1}}\\
 & \ddots\\
 &  & x_{g_{n_{g}}}\\
 &  &  & a\\
 &  &  &  & \ddots
\end{array}\right]}_{:=X_{g}}O^{T}\le\left[\begin{array}{ccccc}
x_{h_{1}}\\
 & \ddots\\
 &  & x_{h_{n_{h}}}\\
 &  &  & b\\
 &  &  &  & \ddots
\end{array}\right]=X_{h},
\]
and the vector $\left|\psi\right\rangle :=\sum_{i=1}^{n_{h}}\sqrt{p_{h_{i}}}\left|i\right\rangle =\sum_{i=1}^{n_{g}}\sqrt{p_{g_{i}}}O\left|i\right\rangle $.\label{lem:EBRMisCOF}
\end{lem}

\begin{proof}
An EBRM entails that we are given $G\le H$ with their spectrum in
$[a,b]$ and a $\left|\psi\right\rangle $ such that 
\[
g=\text{Prob}[G,\left|\psi\right\rangle ]=\sum_{i=1}^{n_{g}}p_{g_{i}}[x_{g_{i}}]
\]
 and 
\[
h=\text{Prob}[H,\left|\psi\right\rangle ]=\sum_{i=1}^{n_{h}}p_{h_{i}}[x_{h_{i}}]
\]
with $p_{g_{i}},p_{h_{i}}>0$ and $x_{g_{i}}\neq x_{g_{j}}$, $x_{h_{i}}\neq x_{h_{j}}$
for $i\neq j$ but the dimension and multiplicities can be arbitrary.
First we show that one can always choose the eigenvectors $\left|g_{i}\right\rangle $
of $G$ with eigenvalue $x_{g_{i}}$ such that 
\[
\left|\psi\right\rangle =\sum_{i=1}^{n_{g}}\sqrt{p_{g_{i}}}\left|g_{i}\right\rangle .
\]
Consider $P_{g_{i}}$ to be the projector on the eigenspace with eigenvalue
$x_{g_{i}}$. Note that 
\[
\left|g_{i}\right\rangle :=\frac{P_{g_{i}}\left|\psi\right\rangle }{\sqrt{\left\langle \psi\right|P_{g_{i}}\left|\psi\right\rangle }}
\]
fits the bill. Similarly we choose/define $\left|h_{i}\right\rangle $
so that 
\[
\left|\psi\right\rangle =\sum_{i=1}^{n_{h}}\sqrt{p_{h_{i}}}\left|h_{i}\right\rangle .
\]
Consider now the projector onto the $\{\left|g_{i}\right\rangle \}$
space 
\[
\Pi_{g}=\sum_{i=1}^{n_{g}}\left|g_{i}\right\rangle \left\langle g_{i}\right|.
\]
Note that this will not have all eigenvectors with eigenvalues $\in\{x_{g_{i}}\}$.
Similarly we define 
\[
\Pi_{h}=\sum_{i=1}^{n_{h}}\left|h_{i}\right\rangle \left\langle h_{i}\right|.
\]
We further define $G':=\Pi_{g}G\Pi_{g}+a(\mathbb{I}-\Pi_{g})$ and
$H':=\Pi_{h}H\Pi_{h}+b(\mathbb{I}-\Pi_{h})$. These definitions are
useful as we can show 
\[
G'\le H'.
\]
From $G=\Pi_{g}G\Pi_{g}+(\mathbb{I}-\Pi_{g})G(\mathbb{I}-\Pi_{g})$
we can conclude that $\Pi_{g}G\Pi_{g}+a(\mathbb{I}-\Pi_{g})\le G$.
This entails $G'\le G$. Using a similar argument one can also establish
that $H\le H'$. Combining these we get $G'\le H'$. \\
Consider the projector 
\[
\Pi:=\text{projector on span}\{\{\left|g_{i}\right\rangle \}_{i=1}^{n_{g}},\{\left|h_{i}\right\rangle \}_{i=1}^{n_{h}}\}
\]
and note that this has at most $n_{g}+n_{h}-1$ dimension because
$\left|\psi\right\rangle $ lives in the span of $\{\left|g_{i}\right\rangle \}$
and in the span of $\{\left|h_{i}\right\rangle \}$ so one of the
basis vectors at least is not independent. Now note that 
\[
G'':=\Pi G'\Pi\le\Pi H'\Pi=:H''
\]
because we can always conjugate an inequality by a positive semi-definite
matrix on both sides. Note also that $\Pi\left|\psi\right\rangle =\left|\psi\right\rangle $
which means the matrices and the vectors have the claimed dimension.
We now establish that $\text{Prob}[H'',\left|\psi\right\rangle ]=h$
and $\text{Prob}[G'',\left|\psi\right\rangle ]=g$. For this we first
write the projector tailored to the $g$ basis as $\Pi=\Pi_{g}+\Pi_{g_{\perp}}$
where $\Pi_{g_{\perp}}$ is meant to enlarge the space to the $\text{span}\{h_{i}\}_{i=1}^{n_{h}}$.
With this we evaluate 
\begin{align*}
G'' & =\left(\Pi_{g}+\Pi_{g_{\perp}}\right)\left[\Pi_{g}G\Pi_{g}+a(\mathbb{I}-\Pi_{g})\right]\left(\Pi_{g}+\Pi_{g_{\perp}}\right)\\
 & =\Pi_{g}G\Pi_{g}+a\Pi_{g_{\perp}}.
\end{align*}
Manifestly then $\text{Prob}[G'',\left|\psi\right\rangle ]=g$. By
a similar argument one can establish the $h$ claim. Note that that
$G''$ and $H''$ have no multiplicities except possibly in $a$ and
$b$ respectively. Thus we conclude we can always restrict to the
claimed dimension and form.
\end{proof}

\begin{cor}
For every EBRM transition the corresponding COF is legal. 
\end{cor}

\textcolor{purple}{The COF is of interest because we can use it to
interpret our inequalities geometrically and use the tools thereof.
We study this connection next.}

\section{Ellipsoids\label{sec:Ellipsoids}}

\subsection{The inequality as containment of ellipsoids\label{subsec:inequalityAsEllipsoids}}

\textcolor{purple}{We try to show that the matrix inequality of the
form $0\le G\le H$ can be geometrically viewed as the containment
of a smaller ellipsoid inside a larger one.}

Consider an unnormalised vector $\left|u\right\rangle =\sum_{j}u_{j}\left|h_{j}\right\rangle $
with $u_{j}\in\mathbb{R}$. Note that the set 
\[
\{\left|u\right\rangle |\left\langle u\right|X_{h}\left|u\right\rangle =1\}
\]
 describes the surface of an ellipsoid where $X_{h}=\diag(x_{h_{1}},x_{h_{2}}\dots)$.
This is easy to see as the constraint corresponds to 
\[
x_{h_{1}}u_{1}^{2}+x_{h_{2}}u_{2}^{2}+\dots=1
\]
which is of the form 
\[
\frac{u_{1}^{2}}{a_{1}^{2}}+\frac{u_{2}^{2}}{a_{2}^{2}}+\dots=1
\]
which, in turn, is the equation of an ellipsoid in the variables $\{u_{i}\}$
with axes $a_{1}=1/\sqrt{x_{h_{1}}},a_{2}=1/\sqrt{x_{h_{2}}}\dots$.
An inequality would correspond to points inside or outside the ellipsoid.
It is also useful to note that if we start with some arbitrary (even
unnormalised) vector $\left|u\right\rangle $ then the point on the
ellipse along this direction are given by 
\[
\mathcal{E}_{h}(\left|u\right\rangle )=\frac{\left|u\right\rangle }{\sqrt{\left\langle u\right|X_{h}\left|u\right\rangle }}.
\]
Finally, note that the set $\left\{ \left|u\right\rangle |\left\langle u\right|UX_{g}U^{\dagger}\left|u\right\rangle =1\right\} $
also corresponds to the equation of an ellipsoid with axes $\left\{ 1/\sqrt{x_{g_{i}}}\right\} $
except that it is rotated. This follows from the fact that if we use
$\left|u'\right\rangle =U\left|u\right\rangle $ then the equation
reduces to the standard form in the $u'_{i}$ variables which can
then be used to obtain $u_{i}$s by the aforesaid relations which
is a rotation. We can define a similar map from a vector $\left|u\right\rangle $
to a point on the rotated ellipse as 
\[
\mathcal{E}_{g}(\left|u\right\rangle )=\frac{\left|u\right\rangle }{\sqrt{\left\langle u\right|UX_{g}U^{\dagger}\left|u\right\rangle }}.
\]
With this understanding in place we are ready to get a visual interpretation
of our equation. The statement that 
\begin{align*}
 & X_{h}-UX_{g}U^{\dagger}\ge0\\
\iff & \left\langle u\right|X_{h}\left|u\right\rangle -\left\langle u\right|UX_{g}U^{\dagger}\left|u\right\rangle \ge0 & \forall\left|u\right\rangle \\
\iff & \left\langle u\right|UX_{g}U^{\dagger}\left|u\right\rangle \le1 & \forall\left\{ \left|u\right\rangle |\left\langle u\right|X_{h}\left|u\right\rangle =1\right\} 
\end{align*}
which in turn corresponds to the statement that every point denoted
by $\left|u\right\rangle $ that is on the $h$ ellipse must be on
or inside the $g$ ellipse. Note that if $\left\langle x_{h}\right\rangle -\left\langle x_{g}\right\rangle =0$
then for $\left|u\right\rangle =\left|w\right\rangle $ the inequality
saturates. This in turn means that even for $\mathcal{E}_{h}(\left|w\right\rangle )$
the inequality is saturated as it is the same vector up to a scaling.
The difference is that $\mathcal{E}_{h}(\left|w\right\rangle )$ represents
a point on the $h$ ellipsoid. Since the inequality is saturated it
means that the ellipsoids must touch at this point. Thus $\mathcal{E}_{g}(\left|w\right\rangle )=\mathcal{E}_{h}(\left|w\right\rangle )$
which one can check explicitly as well.

\textcolor{purple}{The discussion so far can only give some intuition
about the visualisation of our constraint equation. This intuition,
as was explained in \Subsecref{EMA-Algorithm}, can be efficiently
used but it requires us to precisely specify the notion of curvature. }

\subsection{Convex Geometry Tools | Weingarten Map and the Support Function\label{subsec:Weingarten}}

\textcolor{purple}{Consider a curve in the plane. One can easily guess
that the curvature must be related to the rate of change of tangents.
This means we must use the second derivative. This can be generalised
to arbitrary dimensions and in this case we obtain a matrix of the
form $\partial_{i}\partial_{j}f$ for some function $f$ which describes
the curve. The eigenvalues of this matrix would tell us the curvature
along the principle directions of curvature, given by the corresponding
eigenvectors. It is possible to follow this idea through for an ellipsoid
but the result becomes rather cumbersome as one must choose a coordinate
system with its origin at the point of interest, aligned along the
normal and re-express all the quantities of interest. }

\textcolor{purple}{A concise way of evaluating the same is based on
a mathematically sophisticated method applicable to all convex bodies.
We state the result for the convex body of interest, an ellipsoid.}

For a normalised direction vector $\left|u\right\rangle $ the support
function corresponding to an ellipsoid $X$ is given by 
\begin{equation}
h(u)=\sqrt{\left\langle u\right|X^{-1}\left|u\right\rangle }=\sqrt{\sum x_{i}^{-1}u_{i}^{2}}.\label{eq:SuppFun}
\end{equation}
The derivative of the support function yields the point on the ellipsoid
where the tangent plane corresponding to the direction $\left|u\right\rangle $
touches the said ellipsoid. It is 
\[
\partial_{i}h(u)=\frac{x_{i}^{-1}u_{i}}{h(u)}.
\]
The most remarkable of all these is the fact is that 
\begin{equation}
h\partial_{j}\partial_{i}h(u)=\left(-\frac{x_{j}^{-1}x_{i}^{-1}u_{i}u_{j}}{h^{2}}+x_{i}^{-1}\delta_{ij}\right)\label{eq:ReverseWeingarten}
\end{equation}
contains as eigenvalues the radii of curvature at the aforesaid point
and as eigenvectors the directions of principle curvature. If instead
of the normal you know the point at which you would like to evaluate
this object then one can use the gradient to first find this normal
and then apply the aforesaid. The normal at a point of contact $\left|c\right\rangle =\sum c_{i}\left|i\right\rangle $
is $\left|u(c)\right\rangle =\mathcal{N}\left(\sum x_{i}c_{i}\left|i\right\rangle \right).$
The results discussed here were deduced as special cases of those
discussed in Section 2.5 of the book on convex bodies by R. Schneider
\cite{Schneider2009}.

\textcolor{purple}{We have stated the basic results needed to proceed
with the description of our algorithm.}

\section{Elliptic Monotone Align (EMA) Algorithm\label{sec:EMAalgorithm}}

\textcolor{purple}{Solving the weak coin flipping (WCF) problem can
be reduced to finding explicit matrices for a given EBM transition
$g=\sum_{i=1}^{n_{g}}p_{g_{i}}[x_{g_{i}}]\to h=\sum_{i=1}^{n_{h}}p_{h_{i}}[x_{h_{i}}]$
where $g$ and $h$ have disjoint support or, equivalently, for a
given EBM function $a=h-g=\sum_{i=1}^{n_{h}}p_{h_{i}}[x_{h_{i}}]-\sum_{i=1}^{n_{g}}p_{g_{i}}[x_{g_{i}}]$.
Here we describe our elliptic monotone align (EMA) algorithm, which
runs by converting the given problem into the same problem of one
less dimension iteratively until it is solved. }

\subsection{Notation}

\textcolor{purple}{This subsection might appear to be particularly
dry as we almost exclusively only introduce definitions; but it is
a necessary evil. We try to motivate the definitions as we move along
but things would not make perfect sense until one reaches the description
and analysis of the algorithm itself.}

At step $k$ of the iteration, the transition $g\to h$ and the associated
function $a=h-g$ used below are given by $g^{(k)}\to h^{(k)}$ and
$a^{(k)}$ respectively. It remains fixed for the said step. We therefore
do not write an explicit dependence on it in the following definitions.
\textcolor{purple}{This is to facilitate the discussion of the iterative
algorithm.} We consider the extended real line $\bar{\mathbb{R}}=\mathbb{R}\cup\{\infty,-\infty\}$
with $1/\infty=-1/\infty:=0$. We also need the extended half line
$\mathbb{\bar{R}}_{\ge}:=\mathbb{R}_{\ge}\cup\{\infty\}$ and $\bar{\mathbb{R}}_{>}:=\mathbb{R}_{>}\cup\{\infty\}$.
\textcolor{purple}{These situations appear unavoidably in the analysis
of certain transitions and correspond to one of the directions of
the ellipsoids having infinite curvature.} We use $\mathcal{N}\left(\left|\psi\right\rangle \right):=\left|\psi\right\rangle =\sqrt{\left\langle \psi|\psi\right\rangle }$.
We usually denote by $[x_{\min},x_{\max}]$ the smallest interval
that contains $\text{supp}(a)$. We call this interval the \emph{support
domain} for $a$. Similarly, we would refer to the smallest interval
containing $\text{supp}(g)\cup\text{supp}(h)$ as the \emph{transition
support domain} for (the transition) $g\to h$\emph{. }We use the
variables $\chi,\xi\in\mathbb{R}$ to be such that they denote an
interval $[\chi,\xi]\supseteq[x_{\min},x_{\max}]$. As these $\chi$
and $\xi$ would later be associated with an interval containing the
spectrum of relevant matrices, we would refer to this interval as
the \emph{spectral domain}. \textcolor{purple}{The need for distinguishing the three is not hard
to justify. A transition $g\to h$ can be such that both $g$ and
$h$ have a term $p_{k}[x_{k}]$ for some $p_{k}>0$ and $x_{k}$.
This term would be absent from $a=h-g$. Thus the transition support
domain and the support domain would be different in general. One might
object to this as we started with the assumption that $g$ and $h$
have disjoint support. The issue is that this assumption does not
necessarily hold once the problem is reduced to a smaller instance
of itself. As for the spectral domain, this is defined from hindsight,
as we know we need to use COFs which (see \Lemref{EBRMisCOF}) use
matrices with spectra that would usually be larger than the transition
support domain.}

\textcolor{purple}{Recall from the discussion of \Subsecref{EMA-Algorithm}
that we intend to use operator monotone functions to make the ellipsoids
touch along a known direction. We already have a characterisation
of operator monotone functions (see \Lemref{opMonCharLambda}). The
function $\lambda x/(\lambda+x)$ can be turned into $-1/(\lambda+x)$
by adding a constant (we will do this carefully shortly). Further,
the characterisation we have expects the input matrices to have their
spectrum in the range $[0,\Lambda]$. We must generalise this as this
assumption can not be made for smaller instances of the same problem
which appear in subsequent iterations. This motivates the following
definitions.}
\begin{defn}[$f_{\lambda}$ on $(\alpha,\beta)$]
 $f_{\lambda}:(\alpha,\beta)\to\mathbb{R}$ is defined for $\lambda\in\mathbb{R}\backslash[-\beta,-\alpha]$
as 
\[
f_{\lambda}(x):=\frac{-1}{\lambda+x}.
\]
\\
\end{defn}

\begin{defn}[{$f_{\lambda}$ on $[\alpha,\beta]$}]
 $f_{\lambda}:[\alpha,\beta]\to\bar{\mathbb{R}}$ is defined for
$\lambda\in\mathbb{R}\backslash(-\beta,-\alpha)$. For $\lambda\in\mathbb{R}\backslash[-\beta,-\alpha]$
we define 
\[
f_{\lambda}(x):=\frac{-1}{\lambda+x}.
\]
For $\lambda=-\beta$ and $-\alpha$ we retain the same definition
as above except when $x=\beta$ and $\alpha$ respectively in which
case we define
\begin{align*}
f_{-\beta}(\beta) & :=\infty\\
f_{-\alpha}(\alpha) & :=-\infty.
\end{align*}
\label{def:f_lambda}

\begin{rem}
Values for $f_{-\beta}(\beta)$ and $f_{-\alpha}(\alpha)$ are obtained
by taking for $x$, respectively, the left limit (approaching from
the left to $\beta$) and right limit (approaching from the right
to $\alpha$). Also note that the operator monotone $f(x)=x$ is not
included in the aforesaid family of functions.
\end{rem}

\textcolor{purple}{We had to define $f_{\lambda}$ on the two intervals
for technical reasons which we can't quite motivate here. We explicitly
defined $f_{\lambda}$ to be $\infty$ or $-\infty$ where it would
be otherwise undefined (division by zero). These infinities, however,
will only appear in the denominator in our algorithm.}

\textcolor{purple}{Again, from the discussion of \Subsecref{EMA-Algorithm},
we recall that we have to expand the smaller ellipsoid until it touches
the larger ellipsoid. From \Subsecref{inequalityAsEllipsoids} we
can see that the ellipsoid corresponding $X_{h}$, a positive diagonal
matrix, is smaller than the one corresponding to $\gamma X_{h}$ for
$0<\gamma<1$. The $X_{h}$ matrix would correspond to a function
$h$. What would the corresponding function be for $\gamma X_{h}$?
The following definition of $h_{\gamma}$ formalises the answer. We
also introduce $l_{\gamma}$ which helps us check the validity condition
for a transition (similar to \Defref{lambdaValid} and \Corref{EBMandLambdaValid}).
The basic idea is to take the inner product (sum over the finite support
of $a$) of the function $a$ with a given operator monotone. If this
is positive for every operator monotone, then the function $a$ is
valid. From hindsight, since we already know the characterisation
of these operator monotones, we define $l_{\gamma}(\lambda)$ to be
this inner product which must be positive, labelling the operator
monotone by $\lambda$ and encoding the stretching of the $h$ ellipsoid
into $\gamma$. This plays a crucial role in our algorithm as we have
to make sure we use the right stretching, $\gamma$, without actually
knowing the ellipsoids completely. We do not expect the details of
these statements to be clear just yet but we hope the following definitions
appear reasonable.}
\end{defn}

\begin{defn}[$l_{\gamma},l_{\gamma}^{1},a_{\gamma}$]
 Consider the transition $g\to h$ and let $a=h-g$. For $\gamma\in(0,1]$
we define the finitely supported functions $h_{\gamma}:\mathbb{R\to\mathbb{R}_{\ge}}$
and $a_{\gamma}(x):\mathbb{R}\to\mathbb{R}$ as 
\begin{align*}
h_{\gamma}(x) & :=h(x/\gamma)\\
a_{\gamma}(x) & :=h_{\gamma}(x)-g(x).
\end{align*}
Let $S_{\gamma}=[x_{\min}(\gamma),x_{\max}(\gamma)]$ be the support
domain of $a_{\gamma}$. We define $l_{\gamma}:\mathbb{R}\backslash[-x_{\max}(\gamma),-x_{\min}(\gamma)]\to\mathbb{R}$
\[
l_{\gamma}(\lambda):=\sum_{x\in\text{supp}(a_{\gamma})}a_{\gamma}(x)f_{\lambda}(x)
\]
where $f_{\lambda}$ is defined on $S_{\gamma}$.

We define 
\[
l_{\gamma}^{1}:=\sum_{x\in\text{supp}(a_{\gamma})}a_{\gamma}(x)x.
\]

\textcolor{gray}{\label{def:lANDa}}
\end{defn}

\begin{rem}
$h_{\gamma}$ and $g$ might have overlapping support for certain
values of $\gamma$ which justifies the terminology distinguishing
support domain and spectral support domain (introduced at the beginning
of the section).
\end{rem}

\textcolor{purple}{We now define a sort of indicator function, $m$,
which tells us, given the transition $g\to h$, if the transition
corresponding to the scaled ellipsoid $g\to h_{\gamma}$ is valid.
There are some extra parameters this function needs. Consider the
spectrum of the matrices which make this transition EBRM (they must
be EBRM if they are valid, similar to \Corref{EBMandLambdaValid}).
These parameters encode the interval in which this spectrum must be
contained.}

\begin{defn}[$m(\gamma,\chi,\xi)$]
 We define $m:((0,1],\mathbb{R},\mathbb{R})\to\{0,1\}$ to be 
\[
m(\gamma,\chi,\xi):=\begin{cases}
0 & \text{if any of the following root conditions hold}\\
1 & \text{else}.
\end{cases}
\]
where the first root condition is satisfied if there exists a $\lambda\in\mathbb{R}\backslash(-\xi,-\chi)$
such that $l_{\gamma}(\lambda)=0$, and the second root condition
is satisfied if $l_{\gamma}^{1}=0$.\label{def:m}
\end{defn}

\textcolor{purple}{As we are dealing with different representations
of the same object, we define a relation between the matrix instance
of the problem (which involves matrices) and the function instance
thereof (which involves transitions and functions). The matrix instance
contains all the information needed and so in the discussion of the
algorithm we pack everything into a matrix instance to keep things
palpable. The reader can glance through the following and later refer
to them when they are used.}
\begin{defn}[Matrix Instance, $\text{\ensuremath{\underbar{X}}}$ $\to$ Function
Instance, $\text{\ensuremath{\underbar{x}}}$]
 For a \emph{Matrix Instance} defined to be the tuple $\text{\ensuremath{\underbar{X}}}:=(X_{h},X_{g},\left|w\right\rangle ,\left|v\right\rangle )$
where $X_{h},X_{g}$ are diagonal matrices and $\left|w\right\rangle ,\left|v\right\rangle $
are vectors on $\mathbb{R}^{n}$ for some $n$ with equal norm, i.e.
$\left\langle w|w\right\rangle =\left\langle v|v\right\rangle $,
we define the \emph{Function Instance} to be the tuple $\text{\ensuremath{\underbar{x}}}:(g,h,a)$
where $h=\text{Prob}[X_{h},\left|w\right\rangle ]$, $g=\text{Prob}[X_{g},\left|v\right\rangle ]$
and $a=h-g$. \label{def:Xtox}
\end{defn}

\begin{defn}[Attributes of the Function Instance, $\text{\ensuremath{\underbar{x}}}$]
 For a given tuple $\text{\ensuremath{\underbar{x}}}:=(g,h,a)$ as
\Defref{Xtox} we define the attributes $n_{h},n_{g},\{p_{g_{i}}\},\{p_{h_{i}}\},\{x_{g_{i}}\},\{x_{h_{i}}\}$
as they appear by declaring $g\to h$ to be a transition, i.e., 
\begin{itemize}
\item $n_{h}$ as the number of times $h$ is non-zero, 
\item $n_{g}$ as the number of times $g$ is non-zero, 
\item $\{p_{h_{i}}\},\{x_{h_{i}}\},\{p_{g_{i}}\},\{x_{g_{i}}\}$ implicitly
as
\[
h=\sum_{i=1}^{n_{h}}p_{h_{i}}[x_{h_{i}}],\,g=\sum_{i=1}^{n_{g}}p_{g_{i}}[x_{g_{i}}]
\]
(for $p_{h_{i}},p_{g_{i}}>0$). 
\end{itemize}
The \emph{support domain} for $a$ is denoted by $[x_{\min},x_{\max}]$,
i.e., the attributes $x_{\min},x_{\max}$ are defined to be such that
$[x_{\min},x_{\max}]$ is the smallest interval containing $\text{supp}(a)$.

\end{defn}

\begin{rem}
Note that $x_{\min}$ and $x_{\max}$ may not be $x_{\min}=\min\{\{x_{h_{i}}\},\{x_{g_{i}}\}\}$
and $x_{\max}=\max\{\{x_{h_{i}}\},\{x_{g_{i}}\}\}$ respectively because
there can be cancellations in the evaluation of $h-g=a$.
\end{rem}

\begin{defn}[Attributes of the Matrix Instance, $\text{\ensuremath{\underbar{X}}}$]
 We associate the following with a matrix instance.
\begin{itemize}
\item \emph{Spectral domain}: For a tuple $\text{\ensuremath{\underbar{X}}}$
as defined in \Defref{Xtox} we denote the \emph{spectral domain}
by $[\chi,\xi]$ where the attributes $\chi,\xi$ are such that $[\chi,\xi]$
is the smallest interval containing $\text{spec}\{X_{g}\oplus X_{h}\}$. 
\item \emph{Solution}: We say that $O$ is a \emph{solution} to the matrix
instance $\text{\ensuremath{\underbar{X}}}=\left(X_{h},X_{g},\left|w\right\rangle ,\left|v\right\rangle \right)$
if $X_{h}\ge OX_{g}O^{T}$ and $O\left|v\right\rangle =\left|w\right\rangle $. 
\item \emph{Notation}: With respect to a standard orthonormal basis $\{\left|i\right\rangle \}$,
we use the \emph{notation} $X_{h}:=\sum_{i=1}^{k}y_{h_{i}}\left|i\right\rangle \left\langle i\right|$,
$X_{g}:=\sum_{i=1}^{k}y_{g_{i}}\left|i\right\rangle \left\langle i\right|$,
$\left|w\right\rangle :=\sum_{i=1}^{k}\sqrt{q_{h_{i}}}\left|i\right\rangle $,
$\left|v\right\rangle :=\sum_{i=1}^{k}\sqrt{q_{g_{i}}}\left|i\right\rangle $.
\end{itemize}
\end{defn}

\begin{rem}
We index the Matrix Instance and the corresponding function instance
as $\text{\ensuremath{\underbar{X}}}^{(k)}=\left(X_{h}^{(k)},X_{g}^{(k)},\left|w^{(k)}\right\rangle ,\left|v^{(k)}\right\rangle \right)$
and $\text{\ensuremath{\underbar{X}}}^{(k)}\to\text{\ensuremath{\underbar{x}}}^{(k)}=\left(h^{(k)},g^{(k)},a^{(k)}\right)$
respectively. The associated attributes are implicitly assumed to
be correspondingly indexed, e.g., as $\chi^{(k)},\xi^{(k)}$ and $n_{h}^{(k)},n_{g}^{(k)},x_{\min}^{(k)},x_{\max}^{(k)}$.
\end{rem}

\begin{rem}
We introduce two different symbol sets $p,x$ and $q,y$ as it allows
us to describe the proof more neatly by allowing two ways of indexing
the same object. We use $p,x$ for $\text{\ensuremath{\underbar{x}}}$
and $q,y$ for $\text{\ensuremath{\underbar{X}}}$ which are essentially
the same. 
\end{rem}

\subsection{Lemmas for EMA}

\textcolor{purple}{With the notation in place, we can now state and
prove some results which we would need in our algorithm. We do this
in three steps. First, we generalise the results obtained by Aharonov
et al. about operator monotones and their relation with EBM functions.
This is the workhorse of our algorithm. Second, we prove some results
which formalise our intuitive notion of tightening\textemdash stretching
the smaller $h$ ellipsoid until it touches the larger $g$ ellipsoid.
Finally, we prove a generalisation thereof in the case where the curvature
of the smaller $h$ ellipsoid becomes infinite.}

\textcolor{purple}{For a first reading, it might be better to focus
on the statements, and come back to the proofs after reading the algorithm.}

\subsubsection{Generalisations\label{subsec:EMA-Lemmas-Generalisations}}

\begin{figure}
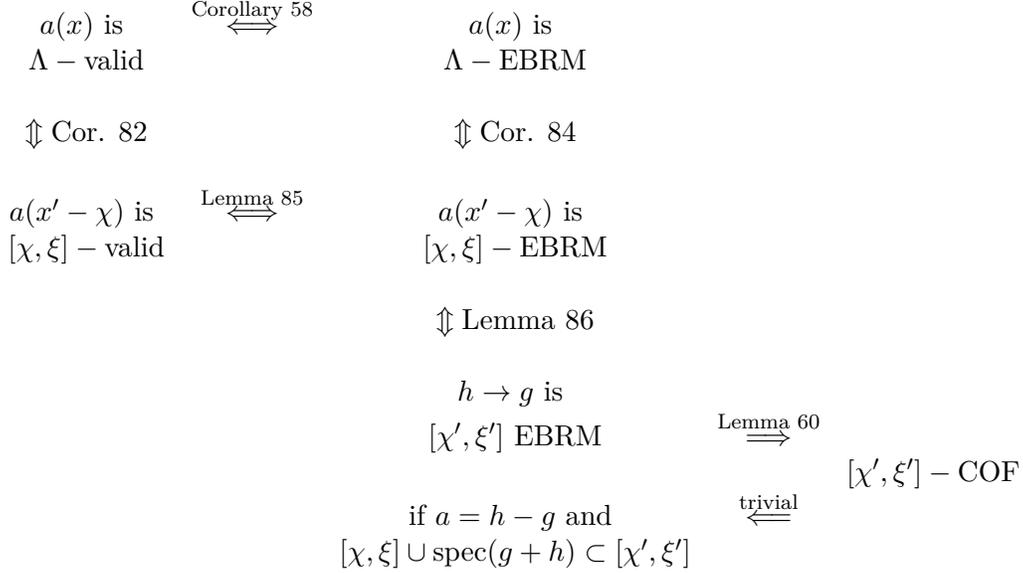

\[
\begin{array}{ccccc}
a(x)\text{ is } & \overset{\text{Corollary \prettyref{cor:EBMisEBRMisValid}}}{\iff} & a(x)\text{ is }\\
\Lambda-\text{valid} &  & \Lambda-\text{EBRM}\\
\\
\Updownarrow\text{Cor. \prettyref{cor:moveValidFns}} &  & \Updownarrow\text{Cor. \prettyref{cor:moveEBRMfns}}\\
\\
a(x'-\chi)\text{ is } & \overset{\text{\prettyref{lem:GenValidIsEBRM}}}{\iff} & a(x'-\chi)\text{ is }\\{}
[\chi,\xi]-\text{valid} &  & \left[\chi,\xi\right]-\text{EBRM}\\
\\
 &  & \Updownarrow\text{\prettyref{lem:EBRMfunIsEBRMtrans_evenWithCommon}}\\
\\
 &  & h\to g\text{ is }\\
 &  & [\chi',\xi']\text{ EBRM} & \overset{\text{\prettyref{lem:EBRMisCOF}}}{\implies}\\
 &  &  &  & \text{[\ensuremath{\chi',\xi']-\text{COF}}}\\
 &  & \text{if }a=h-g\text{ and } & \overset{\text{trivial}}{\impliedby}\\
 &  & [\chi,\xi]\cup\text{spec}(g+h)\subset[\chi',\xi']
\end{array}
\]

\caption{Generalisation schematized.\label{fig:Generalisation-schematized}}
\end{figure}

Keep the bigger picture, \Figref{Generalisation-schematized}, in
mind to retain a sense of direction. \textcolor{purple}{Our main objective here would be twofold. First,
we wish to generalise \Corref{EBMisEBRMisValid} from being restricted
to matrices with their spectrum in $[0,\Lambda]$ to being applicable
for matrices with their spectrum in $[\chi,\xi]$. Second, we wish
to extend the result from valid functions to valid transitions, including
the case of overlapping support. }

\textcolor{purple}{To establish the first, our strategy would be to
find a relation between $[0,\Lambda]$ valid functions and $[\chi,\xi]$
valid functions (which we will define carefully soon) and then a relation
between $[0,\Lambda]$ EBRM functions and $[\chi,\xi]$ EBRM functions.
Then we use the link between $[0,\Lambda]$ valid and $[0,\Lambda]$
EBRM functions to establish the equivalence of $[\chi,\xi]$ valid
functions and $[\chi,\xi]$ EBRM functions. Along the way we sharpen
our understanding of operator monotone functions which should make
the definitions of $f_{\lambda}$, $l$ and $m$ (see \Defref{lANDa}
and \Defref{m}) obvious. The second objective can be met with by
a single, albeit, slightly long argument.}

\textcolor{purple}{Let us start with extending our definition of the
Canonical Orthogonal Form to accommodate matrices with their spectrum
in $[\chi,\xi]$.}

\begin{defn}[{Canonical Orthogonal Form (COF) with spectrum in $[\chi,\xi]$}]
 For a given transition $g\to h$ let $[\chi,\xi]$ be such that
it contains $\text{supp}(g)$ and $\text{supp}(h)$. We define the
Canonical Orthogonal Form (COF) with its spectrum in $[\chi,\xi]$
by the set of $n\times n$ matrices $X_{h},X_{g},O,D$ and vectors
$\left|v\right\rangle ,\left|w\right\rangle $ where 
\[
X_{h}:=\text{diag}\{x_{h_{1}},x_{h_{2}}\dots,x_{h_{n_{h}}},\xi,\xi\dots\},
\]
\[
X_{g}:=\text{diag}\{x_{g_{1}},x_{g_{2}}\dots,x_{g_{n_{g}}},\chi,\chi\dots\},
\]
\[
\left|v\right\rangle :=\sum_{i=1}^{n_{g}}\sqrt{p_{g_{i}}}\left|i\right\rangle ,
\]
\[
\left|w\right\rangle :=\sum_{i=1}^{n_{h}}\sqrt{p_{h_{i}}}\left|i\right\rangle ,
\]
\[
D:=X_{h}-OX_{g}O^{\dagger},
\]
the matrix $O$ is orthogonal which satisfies
\[
\left|v\right\rangle =O\left|w\right\rangle 
\]
and $n=n_{g}+n_{h}-1$.
\end{defn}

\begin{defn}[{Legal COF with spectrum in $[\chi,\xi]$}]
 A COF with spectrum in $[\chi,\xi]$ is legal if $D\ge0$.
\end{defn}

\textcolor{purple}{We obviously need to generalise the notion of operator
monotone functions to the range $[\chi,\xi]$ as well to achieve our
first objective.}

\begin{defn}[{Operator monotone functions on $[\chi,\xi]$}]
 A function $f:[\chi,\xi]\to\mathbb{R}$ is operator monotone on
$[\chi,\xi]$ if for all real symmetric matrices $H,G$ with $\text{spec}(H\oplus G)\in[\chi,\xi]$
and $H\ge G$ we have $f(H)\ge f(G)$.
\end{defn}

\textcolor{purple}{What happens if we try to shift/translate the interval
on which an operator monotone is defined? This is a natural question
to ask, an answer to which would also directly relate our new definition
to the previous one. }
\begin{claim}
$f(x)$ is an operator monotone function on $[\chi,\xi]$ if and only
if $f'(x')=f(x'-x_{0})$ is an operator monotone function on $[\chi+x_{0},\xi+x_{0}]$.\label{claim:moveOpMon}
\end{claim}

\begin{proof}
\textcolor{blue}{Consider real symmetric matrices $H\ge G$ with $\text{spec}(H\oplus G)\in[\chi,\xi]$
and let $f(x)$ be operator monotone on $[\chi,\xi]$. We must consider
$f'(x')=f(x=x'-x_{0})$ which is the same as $f'(x+x_{0})=f(x)$.
We show that $f'$ is an operator monotone on $[\chi+x_{0},\xi+x_{0}]$.
Note that $H':=H+x_{0}\mathbb{I}$ and $G':=G+x_{0}\mathbb{I}$ are
such that $H'\ge G'$ and $\text{spec}(H'\oplus G')\in[\chi+x_{0},\xi+x_{0}]$.
Note that $f'(H')=f(H)$ and $f'(G')=f(G)$ because
\begin{align*}
f'(H') & =f'(H+x_{0}\mathbb{I})\\
 & =O_{h}f'(H_{d}+x_{0}\mathbb{I})O_{h}^{T}\\
 & =O_{h}f(H_{d})O_{h}^{T}\\
 & =f(H)
\end{align*}
and similarly for $G$ where $H=O_{h}H_{d}O_{h}^{T}$ for $O_{h}$
orthogonal and $H_{d}$ diagonal. Since $f$ is operator monotone
on $[\chi,\xi]$ we have $f(H)\ge f(G)$ which entails $f'(H')\ge f'(G')$.
Since this holds for all $H',G'$ with their $\text{spec}(H'\oplus G')\in[\chi+x_{0},\xi+x_{0}]$
we can conclude that $f'$ is an operator monotone on $[\chi+x_{0},\xi+x_{0}]$.
The other way follows by setting $\chi+x_{0}$ to $\chi$, $\xi+x_{0}$
to $\xi$, $x_{0}$ to $-x_{0}$ but since all these were arbitrary
to start with, the reasoning goes through unchanged.}
\end{proof}

\textcolor{purple}{We now note that from the characterisation of operator
monotone functions we initially had (see \Lemref{opMonCharLambda}),
we can construct one which is easier to shift/translate (in the aforesaid
sense).}

\begin{cor}[{Characterisation of operator monotone functions on $[0,\Lambda]$}]
 Any operator monotone function $f:[0,\Lambda]\to\mathbb{R}$ can
be written as 
\[
f(x)=c_{0}+c_{1}x-\int\frac{1}{\lambda+x}d\tilde{\omega}(\lambda)
\]
with the integral ranging over $\lambda\in(-\infty,-\Lambda)\cup(0,\infty)$
satisfying $\int\frac{1}{\lambda(1+\lambda)}d\tilde{\omega}(\lambda)<\infty$.\label{cor:opMonLambdaCharV2}
\end{cor}

\begin{proof}
\textcolor{blue}{Consider the characterisation given in \Lemref{opMonCharLambda}
according to which we had $f(x)=c_{0}'+c_{1}x+\int\frac{\lambda x}{\lambda+x}d\omega(\lambda)$
with $\int\frac{\lambda}{1+\lambda}d\omega(\lambda)<\infty$. We can
write 
\begin{align*}
f(x) & =c_{0}'+c_{1}x+\int\left(\lambda-\frac{\lambda^{2}}{\lambda+x}\right)d\omega(\lambda)\\
 & =c_{0}+c_{1}x-\int\frac{\lambda^{2}d\omega(\lambda)}{\lambda+x}
\end{align*}
where with $d\tilde{\omega}=\lambda^{2}d\omega(\lambda)$ we obtain
the claimed form. Note that the finiteness of $\int\frac{\lambda}{1+\lambda}d\omega$
is necessary to conclude that $c_{0}=c_{0}'+\int\frac{\lambda}{1+\lambda}d\omega$
is also finite.}
\end{proof}

\textcolor{purple}{Note that this form also makes it easier for us
to handle any divergences as there is only the denominator one has
to deal with. }

\textcolor{purple}{This can now be shifted/translated to allow for
a characterisation of our shifted/translated operator monotones.}

\begin{cor}[{Characterisation of operator monotone functions on $[\chi,\xi]$}]
 Any operator monotone function $f':[\chi,\xi]\to\mathbb{R}$ can
be written as 
\[
f'(x')=c_{0}'+c_{1}'x'-\int\frac{1}{\lambda'+x'}d\tilde{\omega}'(\lambda')
\]
with the integral ranging over $\lambda'\in(-\infty,-\xi)\cup(-\chi,\infty)$
satisfying $\int\frac{1}{(\lambda'+\chi)\left(1+\lambda'+\chi\right)}d\tilde{\omega}'(\lambda')<\infty$.\label{cor:opMonGenChar}
\end{cor}

\begin{proof}
\textcolor{blue}{We follow the convention that $x'\in[\chi,\xi]$
while the unprimed $x\in[0,\xi-\chi]$. From \Claimref{moveOpMon}
we know that $f(x)$ is operator monotone on $[0,\xi-\chi]$ if and
only if $f'(x')=f(x'-\chi)$ is operator monotone on $[\chi,\xi]$
where $x'=x+\chi$. Since we already have a characterisation for $f(x)$
we can characterise $f'(x')$ as $f(x'-\chi)$. From \Corref{opMonLambdaCharV2}
we have 
\begin{align*}
f'(x') & =c_{0}+c_{1}(x'-\chi)-\int\frac{d\tilde{\omega}(\lambda)}{\lambda+x'-\chi}\\
 & =c_{0}'+c_{1}x'-\int\frac{d\tilde{\omega}'(\lambda')}{\lambda'+x'}
\end{align*}
where $\lambda'=\lambda-\chi$. Since we had $\lambda\in(-\infty,-(\xi-\chi))\cup(0,\infty)$
it entails $\lambda'\in(-\infty,-\xi)\cup(-\chi,\infty)$. The condition
on the integral $\int\frac{d\tilde{\omega}(\lambda)}{\lambda\left(\lambda+x\right)}<\infty$
can be expressed in terms of $\lambda'$ as $\int\frac{d\tilde{\omega}'(\lambda')}{\left(\lambda'+\chi\right)(1+\lambda'+\chi)}<\infty$
with $d\tilde{\omega}'(\lambda')=d\tilde{\omega}(\lambda'+\chi)$.
With $c_{1}=c_{1}'$ and $c_{0}'=c_{0}-c_{1}\chi$ we obtain the claimed
form.}
\end{proof}

\textcolor{purple}{We now generalise \Defref{lambdaValid} to $(\chi,\xi)$
valid functions (we intended $(\chi,\xi)$ to indicate a tuple and
not an open set so do not read too much into this notation).}

\begin{defn}[$(\chi,\xi)$ valid function]
 A finitely supported function $a:\mathbb{R}\to\mathbb{R}$ with
$\text{supp}(a)\in[\chi,\xi]$ is $(\chi,\xi)$ valid if for every
operator monotone function $f$ on $[\chi,\xi]$ we have $\sum_{x\in\text{supp}(a)}a(x)f(x)\ge0$.
\end{defn}

\begin{rem}
Since in \Corref{opMonGenChar} $d\tilde{\omega}'$ is a measure,
to establish $(\chi,\xi)$ validity of functions, it would suffice
to restrict our attention to operator monotones $f'(x')=x'$, $f'(x')=-\frac{1}{\lambda'+x'}$
with $x'\in[\chi,\xi]$, $\lambda'\in(-\infty,-\xi)\cup(-\chi,\infty)$.
\end{rem}

\textcolor{purple}{By shifting/translating the characterisation of
operator monotone functions we can shift/translate valid functions
as well.}
\begin{cor}[$a(x)$ is $(\chi,\xi)$ valid $\iff$ $a(x'-x_{0})$ is $(\chi+x_{0},\xi+x_{0})$
valid]
 A finitely supported function $a:\mathbb{R}\to\mathbb{R}$ with
$\text{supp}(a)\in[\chi,\xi]$ is $(\chi,\xi)$ valid if and only
if the function $a'(x'):=a(x'-x_{0}):\mathbb{R}_{\ge}\to\mathbb{R}$
is $(\chi-x_{0},\xi-x_{0})$ valid.\label{cor:moveValidFns}
\end{cor}

\begin{proof}
\textit{\textcolor{blue}{$a$ is $(\chi,\xi)$ valid entails $\sum_{x\in\text{supp}(a)}a(x)f(x)\ge0$
for all $f$ operator monotone on $[\chi,\xi]$. We can write the
sum as $\sum a(x'-x_{0})f(x'-x_{0})\ge0$. Using \Claimref{moveOpMon}
we note that $f'(x')=f(x'-x_{0})$ is operator monotone on $[\chi+x_{0},\xi+x_{0}]$.
For $a'(x')=a(x'-x_{0})$ we thus have $\sum a'(x')f'(x')\ge0$ which
means $a'(x')$ is a $(\chi+x_{0},\xi+x_{0})$ valid function. The
other way follows similarly.}}
\end{proof}

\textcolor{purple}{In accordance with our strategy, we have established
a relation between $(0,\Lambda)$ valid functions and $(\chi,\xi)$
valid functions (in fact we have a more general result). We now proceed
with establishing its analogue for EBRM functions.}

\begin{defn}[{EBRM on $[\chi,\xi]$}]
 A finitely supported function $a:\mathbb{R}\to\mathbb{R}$ is EBRM
on $[\chi,\xi]$ if there exist real symmetric matrices $H\ge G$
with their spectrum in $[\chi,\xi]$ and a vector $\left|w\right\rangle $
such that $a=\text{Prob}[H,\left|w\right\rangle ]-\text{Prob}[G,\left|w\right\rangle ]$.
\end{defn}

\begin{cor}[{$a(x)$ is EBRM on $[\chi,\xi]$ $\iff$ $a(x+\chi)$ is EBRM on $[0,\xi-\chi]$}]
 A finitely supported function $a:\mathbb{R}\to\mathbb{R}$ with
$\text{supp}(a)\in[\chi,\xi]$ is EBRM on $[\chi,\xi]$ if and only
if the function $a'(x):=(x+\chi):\mathbb{R}_{\ge}\to\mathbb{R}$ is
EBRM on $[0,\xi-\chi]$.\label{cor:moveEBRMfns}
\end{cor}

\begin{proof}
\textcolor{blue}{If $a$ is EBRM on $[\chi,\xi]$ it follows that
there exist real symmetric matrices with $H\ge G$ and a vector $\left|w\right\rangle $
such that $\text{spec}[H\oplus G]\in[\chi,\xi]$ and $a=\text{Prob}[H,\left|w\right\rangle ]-\text{Prob}[G,\left|w\right\rangle ]$.
Clearly, $H':=H-\chi\mathbb{I}\ge G-\chi\mathbb{I}=:G'$ and $a'(x)=\text{Prob}[H',\left|w\right\rangle ]-\text{Prob}[G',\left|w\right\rangle ]=a(x+\chi)$
with $\text{spec}[H'\oplus G']\in[0,\xi-\chi]$. This means $a'$
is EBRM on $[0,\xi-\chi]$. The other way follows similarly.}
\end{proof}

\textcolor{purple}{We have done all the hard work for meeting the
first objective. We now simply combine our results so far to prove
the desired equivalence.}

\begin{lem}[{$a(x)$ is $(\chi,\xi)$ valid function $\iff$ $a(x)$ is EBRM on
$[\chi,\xi]$}]
 A finitely supported function $a:\mathbb{R}\to\mathbb{R}$ with
$\text{supp}(a)\in[\chi,\xi]$ being $(\chi,\xi)$ valid is equivalent
to it being EBRM on $[\chi,\xi]$.\label{lem:GenValidIsEBRM}
\end{lem}

\begin{proof}
\textcolor{blue}{From \Corref{moveValidFns} we know that $a(x)$
being $(\chi,\xi)$ valid is equivalent to $a(x+\chi)$ being $\Lambda=\xi-\chi$
valid. From \Corref{EBMisEBRMisValid} we know that $a(x+\chi)$ is
equivalently EBRM on $[0,\xi-\chi]$. Finally using \Corref{moveEBRMfns}
we know that $a(x+\chi)$ being EBRM on $[0,\xi-\chi]$ is equivalent
to $a(x)$ being EBRM on $[\chi,\xi]$. }
\end{proof}

\textcolor{purple}{The second objective will be achieved in a single
shot.}

\begin{lem}[EBRM function $\iff$ EBRM transition even with common support]

If we write an EBRM function $a$ with spectrum in $[\chi',\xi']$
as $a=h-g$ with $h,g:\mathbb{R}_{\ge}\to\mathbb{R}_{\ge}$ which
may have common support then $g\to h$ is an EBRM transition with
spectrum in $[\chi,\xi]$ and with (the smallest) matrix size (at
most) $n_{g}+n_{h}-1$ where $[\chi,\xi]$ is the smallest interval
containing $[\chi',\xi']$ and $\text{supp}(h)\cup\text{supp}(g)$.

Conversely, if $g\to h$ is an EBRM transition with spectrum in $[\chi,\xi]$
with $h,g:\mathbb{R}_{\ge}\to\mathbb{R}_{\ge}$ which may have common
support then $a=h-g$ is an EBRM function with its spectrum in $[\chi,\xi]$
(the smallest) matrix size at most $n_{g}+n_{h}-1$.\label{lem:EBRMfunIsEBRMtrans_evenWithCommon}
\end{lem}

\begin{proof}
\textcolor{blue}{To prove the first statement we write $a=a^{+}-a^{-}$
where $a^{+}=\sum_{i=1}^{n'_{h}}p'_{h_{i}}[x_{h_{i}}],a^{-}=\sum_{i=1}^{n'_{g}}p'_{g_{i}}[x_{g_{i}}]$,
for $a^{+},a^{-}:\mathbb{R}_{\ge}\to\mathbb{R}_{\ge}$, represent
the positive and the negative parts of $a$. Note that $a^{+}$ and
$a^{-}$ by virtue of this definition can't have any common support.
Consider $\Delta=\sum_{i=1}^{n_{\Delta}}c_{i}[x_{i}]:\mathbb{R}_{\ge}\to\mathbb{R}_{\ge}$
to be such that $h=a^{+}+\Delta$ and $g=a^{-}+\Delta$. This is always
the case because $h-g=a$. Consider the case where $\text{supp}(\Delta)\cap\text{supp}(a)=\emptyset$.
In this case $n_{g}=n_{g}'+n_{\Delta}$ and $n_{h}=n_{h}'+n_{\Delta}$.
Since $a$ is an EBRM function we have a legal COF, viz $O'X_{g}'O^{\prime T}\le X_{h}'$
and $\left|w'\right\rangle =O'\left|v'\right\rangle $, of dimension
$(n'=n_{g}'+n_{h}'-1)$ from \Lemref{EBRMisCOF}. To obtain the matrices
corresponding to $g\to h$ we expand the space to $n=n_{g}+n_{h}-1$
dimensions and define $X_{g}=X'_{g}\oplus X$, $X_{h}=X'_{h}\oplus X$,
$O=O'\oplus\mathbb{I}$, $\left|v\right\rangle =\left|v'\right\rangle +\sum_{i=n'}^{n}\sqrt{c_{i+1-n'}}\left|i\right\rangle $
where $X=\text{diag}\{x_{1},x_{2}\dots x_{n_{\Delta}}\}$. This is
just an elaborate way of adding the points in $\Delta$ to the matrices
and the vectors in such a way that the part corresponding to $\Delta$
remains unchanged. The other cases can be similarly demonstrated with
the only difference being in the relation between $n_{g},n'_{g}$
and $n_{h},n_{h}'$. Suppose $\Delta$ is non-zero only at one point.
If $\Delta$ adds a point where $a^{-}$ had a point then it does
not contribute to increasing the number of points in $g$ that is
$n_{g}=n_{g}'$ but it does increase the number in $h$ that is $n_{h}=n_{h}'+1$.
This means that we have one extra dimension to find the matrices certifying
$g\to h$ is EBRM which is precisely what is needed to append that
extra idle point as described above. Similarly one can reason for
adding a point where $a^{+}$ had a point and finally extend it to
the most general case of $\text{supp}(\Delta)\cap\text{supp}(a)\neq\emptyset$
which may involve multiple points.}

\textcolor{blue}{We now prove the converse. Since $g\to h$ is an
EBRM transition from \Lemref{EBRMisCOF} we know that it admits a
legal COF, that is $OX_{g}O^{T}\le X_{h}$ and $O\left|v\right\rangle =\left|w\right\rangle $
with dimension $n_{g}+n_{h}-1$. To be able to show that $a=h-g=a^{+}-a^{-}$
(where $a^{+}$ and $a^{-}$ are again the positive and negative part
of $a$) is an EBRM function it suffices to show that $a$ is a valid
function. This follows directly from the COF and operator monotones
as $Of(X_{g})O^{T}\le f(X_{h})$ implies $\left\langle v\right|f(X_{g})\left|v\right\rangle \le\left\langle w\right|f(X_{h})\left|w\right\rangle $
which in turn is $\sum h(x)f(x)-\sum g(x)f(x)\ge0$ and that is the
same as $\sum a(x)f(x)\ge0$ for all $f$ operator monotone on the
spectrum of $X_{h}\oplus X_{g}$, viz. $a$ is valid. From \Lemref{GenValidIsEBRM}
we conclude that $a$ is also EBRM with size at most $n_{g}+n_{h}-1$
(actually we can make a stronger statement by saying the size should
be at most $n_{g}'+n_{h}'-1$ where $\left|\text{supp}(a^{+})\right|=n_{h}'$
and $\left|\text{supp}(a^{-})\right|=n_{g}'$).}
\end{proof}

\textcolor{purple}{This completes the first, and longest, step of
our groundwork for discussing the algorithm. }Our achievement so far has been schematized in \Figref{Generalisation-schematized}.

\subsubsection{For the finite part}

\textcolor{purple}{For the second step, we state the following fact
(see \cite{Bhatia2013}). The proof of this statement is interesting
in its own right but here we only state it and use it for terminating
our recursive algorithm. }

\begin{fact}[Weyl's Monotonicity Theorem]
 If $H$ is positive semi-definite and $A$ is Hermitian then $\lambda_{j}^{\downarrow}(A+H)\ge\lambda_{j}^{\downarrow}(A)$
for all $j$ where $\lambda_{j}^{\downarrow}(M)$ represents the $j^{\text{th}}$
largest eigenvalue of the Hermitian matrix $M$.
\end{fact}

\begin{cor}
If $H\ge G$ then $\lambda_{j}^{\downarrow}(H)\ge\lambda_{j}^{\downarrow}(G)$
for all $j$.\label{cor:WeylCorollary}
\end{cor}

\textcolor{purple}{At some point, if the algorithm reaches a point
where there is no vector constraint (that is the vector $\left|v\right\rangle =\left|w\right\rangle =0$)
then one can use the aforesaid result to conclude that the solution,
orthogonal matrix $O$, must be a permutation matrix (this will become
clear later, we mention it to motivate the relevance of the result). }

\textcolor{purple}{We now state a continuity condition which we subsequently
use to establish that when we stretch the $h$ ellipsoid, there would
always exist the perfect amount of stretching that makes the $h$
ellipsoid just touch the $g$ ellipsoid. The non-triviality here is
that we have to conclude this without fully knowing the ellipsoids.}

\begin{claim}[Continuity of $l$]
 Let $[x_{\min},x_{\max}]$ be the smallest interval containing $\text{supp}(a)$.
$l(\lambda)$ is continuous in the intervals $\lambda\in(-x_{\min},\infty]$
and $\lambda\in[-\infty,-x_{\max})$ (see \Defref{lANDa}).\label{lem:continuityL}
\end{claim}

\begin{proof}
\textcolor{blue}{Since $l(\lambda)$ is just a rational function of
$\lambda$ it suffices to show that the denominator doesn't become
zero in the said range. The roots of the denominator are of the form
$\lambda+x=0$ for $x\in\{\{x_{g_{i}}\},\{x_{h_{i}}\}\}$. Hence the
largest root will be $\lambda=-x_{\min}$ and the smallest $\lambda=-x_{\max}$.
Neither of the intervals defined in the statement contain any roots
and therefore we can conclude that $l(\lambda)$ will be continuous
therein. Note that the function $f_{\lambda}$ on $[x_{\min},x_{\max}]$
is not even defined for $\lambda$ in $(-x_{\max},-x_{\min})$.}
\end{proof}

\begin{sloppy}
\begin{lem}[Tightening with the matrix spectrum unknown]
 Consider a finitely supported valid function $a$. Let $[x_{\min}(\gamma),x_{\max}(\gamma)]$
be the smallest interval containing $\text{supp}(a_{\gamma})$. Consider
$m(\gamma,x_{\min}(\gamma),x_{\max}(\gamma))$ as a function of $\gamma$
(see \Defref{m}). $m$ has at least one root in the interval $(0,1]$.\label{lem:tighteningNoSpect}
\end{lem}

\end{sloppy}

\begin{proof}
\textcolor{blue}{To prove the claim it suffices to show that $l_{\gamma}(\lambda)$
has a root in the range $(0,\infty)$ for some $\gamma\in(0,1]$.
Note that we are given a valid function $a$ which means $\text{supp}(a)\in\mathbb{R}_{\ge}$.}

\textcolor{blue}{We assume that $l_{\gamma=1}(\lambda)>0$ for all
$\lambda\in(0,\infty)$ because if this was not the case then we trivially
have $\gamma=1$ as a root, i.e. $m(1,x_{\min}(1),x_{\max}(1))=0$. }

\textcolor{blue}{Notice that since $\sum h(x)=\sum g(x)$ we have
\begin{align*}
\lambda l(\lambda) & =\sum h(x)(\lambda f_{\lambda}(x)+1)-\sum g(x)(\lambda f_{\lambda}(x)+1)\\
 & =\sum h(x)\frac{x}{\lambda+x}-\sum g(x)\frac{x}{\lambda+x}.
\end{align*}
Therefore for the remainder of this proof we redefine $f_{\lambda}=\frac{1}{\lambda}\frac{x}{\lambda+x}$
without changing the value of $l$ or by extension $l_{\gamma}$ (the
$1/\lambda$ factor is partly why we restricted $\lambda$ to $(0,\infty)$
instead of the more general $(-x_{\min},\infty)$). Note that $\lim_{\gamma\to0^{+}}l_{\gamma}(\lambda)<0$
for all $\lambda\in(0,\infty)$ because $h_{\gamma}(x)=h(x/\gamma)$
which means $\lim_{\gamma\to0}\sum h_{\gamma}(x)f_{\lambda}(x)=\lim_{\gamma\to0}\sum h(x)f_{\lambda}(\gamma x)=0$
since $\lim_{x\to0}f_{\lambda}(x)=0$. This in turn means $\lim_{\gamma\to0^{+}}l_{\gamma}(\lambda)=-\sum g(x)f_{\lambda}(x)<0$.}

\textcolor{blue}{Further, each term constituting $l_{\gamma}(\lambda)$
is finite for $\lambda\in(0,\infty)$ since  for $\lambda>0$ the
denominators are of the form $\lambda+x$ which are always positive.
Hence  $l_{\gamma}(\lambda)$ as a function of $\lambda\in[0,\infty)$
and $\gamma\in(0,1]$ is continuous. By continuity then between $\gamma=0^{+}$
and $\gamma=1$ there should be a root.}

\textcolor{blue}{It remains to justify why we extended the range of
$\lambda$ from $(0,\infty)$ to $(-\infty,-x_{\max})\cup(-x_{\min},\infty)$
in the definition of $m$ (see \Defref{m}) as it appears in the statement
of the lemma. This is due to the fact that $l_{\gamma}(\lambda)$
is continuous for $\lambda$ in the stated range, see \Lemref{continuityL},
and so there might be a root which appears in the extended range.
If this is the case we would like to use this possibly higher value
of $\gamma$ (because for a small enough value a non-negative root
must appear due to the aforesaid reasoning). This can help us avoid
infinities (we will explain this later).}
\end{proof}

\textcolor{purple}{Once we are guaranteed that there is at least one
perfect stretching amount, we want to know the spectrum of the matrices.
We state a slightly more general result which is a direct consequence
of the results from the previous subsubsection. The difference is
that it is stated in a form that would be useful for the algorithm.}

\begin{lem}[Matrix spectrum from a valid function]
 Consider a valid function $a$, i.e. an $a$ such that $l(\lambda)\ge0$
and $l^{1}\ge0$ for $\lambda\in[0,\infty)$ (see \Defref{lANDa})
and let $[\chi,\xi]$ be such that for $\lambda\in[-\infty,-\xi)\cup(-\chi,\infty]$
we have $l(\lambda)\ge0$. 

There exists a legal COF, corresponding to the function $a$, with
its spectrum contained in $[\chi,\xi]$. \label{lem:MatSpecFromValidFun}
\end{lem}

\begin{proof}
\textcolor{blue}{Since $l(\lambda)\ge0$ for $\lambda\in(-\infty,-\xi)\cup(-\chi,\infty)$
and $l^{1}\ge0$ we know from \Corref{opMonGenChar} that $a$ is
$(\chi,\xi)$ valid. From \Lemref{GenValidIsEBRM} we know that $a$
is EBRM on $[\chi,\xi]$. Finally from \Lemref{EBRMisCOF} we know
that there exists a legal COF with spectrum in $[\chi,\xi]$.}
\end{proof}

\textcolor{purple}{Recall that in \Subsecref{EMA-Algorithm} we said
that we focus on operator monotone functions $f$ with the special
property that $f^{-1}$ is also an operator monotone. We now establish
that $f_{\lambda}$s (see \Defref{f_lambda}) have this property.}

\begin{lem}[$H\ge G\iff f_{\lambda}(H)\ge f_{\lambda}(G)$]
 Let $H,G$ be real symmetric matrices and $[\chi,\xi]$ be the smallest
interval containing $\text{spec}[H\oplus G]$ and $f_{\lambda}$ be
on $(\chi,\xi)$. $H\ge G$ if and only if $f_{\lambda}(H)\ge f_{\lambda}(G)$.\label{lem:extrOpMonAreInvertible}
\end{lem}

\begin{proof}
\textcolor{blue}{$H\ge G\implies f_{\lambda}(H)\ge f_{\lambda}(G)$
because $f_{\lambda}$ is an operator monotone function for matrices
with spectrum in $[\chi,\xi]$. We prove the converse. We find the
inverse function of $f_{\lambda}$ and show that it is also an operator
monotone. Start with recalling that for $x\in[\chi,\xi]$ we have
\[
y=f_{\lambda}(x)=\frac{-1}{\lambda+x}\implies x=-\frac{1}{y}-\lambda
\]
 where $\lambda\in\mathbb{R}\backslash[\chi,\xi]$. Thus $f_{\lambda}^{-1}(y)=-\frac{1}{y}-\lambda$.
For a given $\lambda$ either $f_{\lambda}(\chi)$ and $f_{\lambda}(\xi)$
are both greater than zero or both less than zero. Hence the operator
monotones $f'_{\lambda'}(y)$ on $[f_{\lambda}(\chi),f_{\lambda}(\xi)]$
permit $\lambda'=0$. Consequently $f'_{\lambda'=0}(y)=\frac{-1}{y}$
is an operator monotone on $[f_{\lambda}(\chi),f_{\lambda}(\xi)]$.
A constant is also an operator monotone. Thus we conclude $f_{\lambda}^{-1}(y)$
is an operator monotone on the required interval establishing the
converse.}
\end{proof}

\textcolor{purple}{This completes the second step of the groundwork.}

\subsubsection{For Wiggle-v; the infinite part\label{subsec:lemmas-for-Wiggle-v}}

\textcolor{purple}{For the final step of the groundwork, we need to
establish a result which lets us tackle the divergences head-on. What
is this divergence issue? Recall form our discussion in \Subsecref{EMA-Algorithm},
our strategy was to tighten (we saw hints of how that can be done
in the previous sections) and then find the operator monotone $f_{\lambda}$
for which the ellipsoids touch along the $\left|w\right\rangle $
direction. What happens if one of the ellipsoids under the action
of this operator monotone has infinite curvature along some directions?
One can in fact show that there are cases where this would necessarily
happen. Having infinite curvature means that the corresponding matrix
has a divergence. It is like an ellipse gets mapped to a line. Our
algorithm will fail in this situation because the normal at the tip
of a line is ill defined.}

\textcolor{purple}{Our strategy is to show that tightness is preserved
under the action of $f_{\lambda}$. This means that if for some $\lambda'$
we consider the ellipsoids obtained by applying $f_{\lambda'}$ and
we find that they touch along $\left|w\right\rangle $ then for some
other $\lambda''$ ($\neq\lambda'$) they would continue to touch
but along some other direction. This allows us to consider the sequence
leading to the divergence. We use this sequence in the analysis of
the algorithm.}

\textcolor{purple}{Here we start by showing this result in the case
where everything is well defined and then extend it to the divergent
case. }

\begin{lem}[Strict inequality under $f_{\lambda}$]
 $H>G$ if and only if $f_{\lambda}(H)>f_{\lambda}(G)$ where $f_{\lambda}$
is on $(\chi,\xi)\supset\text{spec}[H\oplus G]$. \label{lem:StrictInequality}
\end{lem}

\begin{proof}
\textcolor{blue}{Note that $H>G\iff H':=H+\lambda\mathbb{I}>G+\lambda\mathbb{I}=:G'$
where $\lambda\in\bar{\mathbb{R}}\backslash[-\xi,-\chi]$ (by definition
of $f_{\lambda}$ on $(\chi,\xi)$). There can be two cases, either
both the matrices are strictly positive or both are strictly negative.
Let us assume the former (the other follows similarly). We have 
\begin{align*}
 & H'>G'>0\\
\iff & \mathbb{I}>H^{\prime-1/2}G'H^{\prime-1/2}\\
\iff & \mathbb{I}<H^{\prime1/2}G{}^{\prime-1}H^{\prime1/2}\\
\iff & H^{\prime-1}<G^{\prime-1}
\end{align*}
where the first inequality follows from the fact that multiplication
by a positive matrix doesn't affect the inequality (hint: it only
changes the vectors $\left|w\right\rangle $ we use to show $\left\langle w\right|\left(H'-G'\right)\left|w\right\rangle >0$
but maps the set of rays to themselves as the norm of the vector might
change), the second follows from the fact that one can diagonalise
the matrices (identity stays the same) and then it is just a set of
inequalities involving real numbers, and the third follows from again
multiplication by a positive matrix. The last one is the same as $f_{\lambda}(H)>f_{\lambda}(G)$.}
\end{proof}

\begin{cor}[Tightness preservation under $f_{\lambda}$]
Let $H\ge G$ and $f_{\lambda}$ be on $(\chi,\xi)\supset\text{spec}[H\oplus G]$.
There exists a $\left|w\right\rangle $ such that $\left\langle w\right|\left(H-G\right)\left|w\right\rangle =0$
if and only if there exists a $\left|w_{\lambda}\right\rangle $ such
that $\left\langle w_{\lambda}\right|\left(f_{\lambda}(H)-f_{\lambda}(G)\right)\left|w_{\lambda}\right\rangle =0$.\label{cor:tightnessResult}
\end{cor}

\begin{proof}
\textcolor{blue}{The contrapositive of the aforesaid condition is
that $f_{\lambda}(H)>f_{\lambda}(G)$ if and only if $H>G$ which
holds due to \Lemref{StrictInequality}.}
\end{proof}

\begin{lem}[Extending tightness preservation under $f_{\lambda}$ to apparently
divergent situations]
\begin{sloppy}Let $X_{h},X_{g}$ be diagonal matrices with $\text{spec}[X_{h}]\in(\chi,\xi]$,
$\text{spec}[X_{g}]\in[\chi,\xi)$ and let $f_{\lambda}$ be on $[\chi,\xi]$.
Let, further, $O$ be an orthogonal matrix such that $X_{h}\ge OX_{g}O^{T}$. 

There exists a vector $\left|w\right\rangle $ such that $\left\langle w\right|\left(f_{-\xi}(X_{h})-Of_{-\xi}(X_{g})O^{T}\right)\left|w\right\rangle =0$
if and only if there exists a $\left|w_{\lambda}\right\rangle $ such
that $\left\langle w_{\lambda}\right|\left(f_{\lambda}(X_{h})-Of_{\lambda}(X_{g})O^{T}\right)\left|w_{\lambda}\right\rangle =0$
for $\lambda\in\mathbb{R}\backslash[\chi,\xi]$. 

Similarly, there exists a vector $\left|w\right\rangle $ such that
$\left\langle w\right|\left(f_{-\chi}(X_{h})-Of_{-\chi}(X_{g})O^{T}\right)\left|w\right\rangle =0$
if and only if there exists a $\left|w_{\lambda}\right\rangle $ such
that $\left\langle w_{\lambda}\right|\left(f_{\lambda}(X_{h})\allowbreak-Of_{\lambda}(X_{g})O^{T}\right)\left|w_{\lambda}\right\rangle =0$
for $\lambda\in\mathbb{R}\backslash[\chi,\xi]$. 

\end{sloppy}\label{lem:tightnessPreserveInfAlso}
\end{lem}

\begin{proof}
\textcolor{blue}{The trouble with this version of the tightness statement
is that $X_{h}$ has an eigenvalue $\xi$ (if it doesn't then it reduces
to the previous statement) which means that $f_{-\xi}(X_{h})$ is
not well defined. We assume that $X_{h}$ can be expressed as 
\[
X_{h}=\left[\begin{array}{cc}
X_{h}'\\
 & \xi\mathbb{I}''
\end{array}\right]
\]
where $X_{h}'$ has no eigenvalue equal to $\xi$ and $\mathbb{I}''$
is the identity matrix in the subspace. We can write 
\begin{align*}
 & X_{h}>OX_{g}O^{T}\\
 & \iff\left[\begin{array}{cc}
f_{\lambda}(X_{h}')\\
 & f_{\lambda}(\xi\mathbb{I}'')
\end{array}\right]>Of_{\lambda}(X_{g})O^{T}\,\text{for}\,\lambda\in\mathbb{R}\backslash[-\xi,-\chi]\\
 & \iff\left[\begin{array}{cc}
f_{\lambda}(X_{h}')\\
 & \mathbb{I}''
\end{array}\right]>\left[\begin{array}{cc}
\mathbb{I}'\\
 & f_{\lambda}(\xi\mathbb{I}'')^{-1/2}
\end{array}\right]Of_{\lambda}(X_{g})O^{T}\left[\begin{array}{cc}
\mathbb{I}'\\
 & f_{\lambda}(\xi\mathbb{I}'')^{-1/2}
\end{array}\right]\,\text{for}\,\lambda\in\mathbb{R}\backslash[-\xi,-\chi]
\end{align*}
where in the last line the expression has a well defined limit for
$\lambda=-\xi$. This establishes the contrapositive variant of the
statement we wanted to prove (similar to the strategy used for proving
\Corref{tightnessResult}) once we note the following. If $\left\langle w\right|\left(f_{-\xi}(X_{h})-Of_{-\xi}(X_{g})O^{T}\right)\left|w\right\rangle =0$
it is easy to see that $\left[\begin{array}{cc}
0\\
 & \mathbb{I}''
\end{array}\right]\left|w\right\rangle =0$ otherwise due to the constraint on the spectrum of $X_{g}$ the aforesaid
expression would be $\infty$. This entails that 
\[
\left\langle w\right|\left(\left[\begin{array}{cc}
f_{-\xi}(X_{h}')\\
 & \mathbb{I}''
\end{array}\right]-\left[\begin{array}{cc}
\mathbb{I}'\\
 & f_{-\xi}(\xi\mathbb{I}'')^{-1/2}
\end{array}\right]Of_{-\xi}(X_{g})O^{T}\left[\begin{array}{cc}
\mathbb{I}'\\
 & f_{-\xi}(\xi\mathbb{I}'')^{-1/2}
\end{array}\right]\right)\left|w\right\rangle =0.
\]
One can similarly prove the case for $f_{-\chi}(X_{g})$.}
\end{proof}

\subsection{The Algorithm\label{subsec:EMA-The-Algorithm-itself}}

\textcolor{purple}{We now state our algorithm and formally state its
correctness. Thereafter, we motivate each step of the algorithm and
prove its correctness. }

\begin{sloppy}
\begin{defn}[EMA Algorithm]
 \label{def:EMAalgorithm} Given a finitely supported function $a$
(we assume it is $\Lambda$-valid) proceed in the following three
phases.\\
~~\\
\textbf{PHASE 1: INITIALISATION}
\begin{itemize}
\item \textbf{Tightening procedure}: Let $[x_{\min}(\gamma'),x_{\max}(\gamma')]$
be the support domain for $a_{\gamma'}$. Let $\gamma\in(0,1]$ be
the largest root of $m(\gamma',x_{\min}(\gamma'),x_{\max}(\gamma'))$.
Let $x_{\max}:=x_{\max}(\gamma)$ and $x_{\min}:=x_{\min}(\gamma)$. 
\item \textbf{Spectral domain for the representation}: Find the smallest
interval $[\chi,\xi]$ such that $l_{\gamma}(\lambda)\ge0$ for $\lambda\in\mathbb{\bar{R}}\backslash[\chi,\xi]$.
If $\text{supp}(g),\text{supp}(h)$ is not contained in $[\chi,\xi]$
then from all expansions of $[\chi,\xi]$ that contain the aforesaid
sets, pick the smallest. Relabel this interval to $[\chi,\xi]$.
\item \textbf{Shift}: Transform 
\[
a(x)\to a'(x'):=a(x'+\chi-1)
\]
where instead of $1$ any positive constant would do (justified by
\Corref{moveEBRMfns}). Similarly transform 
\begin{align*}
g(x) & \to g'(x'):=g(x'+\chi-1)\\
h(x) & \to h'(x'):=h(x'+\chi-1).
\end{align*}
Relabel $a'$ to be $a$, $g'$ to be $g$ and $h'$ to be $h$. (Remark:
We do not deduce $h$ and $g$ from $a$ as its positive and negative
part because they might now have common support due to the tightening
procedure.)
\item \textbf{The matrices}: For $n:=n_{g}+n_{h}-1$ we define $n\times n$
matrices with spectrum in $[\chi,\xi]$ and $n$ dimensional vectors
as 
\begin{align*}
X_{g}^{(n)} & =\text{diag}[\chi,\chi,\dots x_{g_{1}},x_{g_{2}}\dots,x_{g_{n_{g}}}],\\
X_{h_{\gamma}}^{(n)} & =\text{diag}[\gamma x_{h_{1}},\gamma x_{h_{2}},\dots,\gamma x_{h_{n_{h}}},\xi,\xi,\dots],\\
\left|v^{(n)}\right\rangle  & \doteq\left[0,0\dots,\sqrt{p_{g_{1}}},\sqrt{p_{g_{2}}},\dots,\sqrt{p_{g_{n_{g}}}}\right],\\
\left|w^{(n)}\right\rangle  & \doteq\left[\sqrt{p_{h_{1}}},\sqrt{p_{h_{2}}},\dots,\sqrt{p_{h_{n_{h}}}},0,0\dots\right]
\end{align*}
where $g=\sum_{i=1}^{n_{g}}p_{g_{i}}[x_{g_{i}}]$ and $h=\sum_{i=1}^{n_{h}}p_{h_{i}}[x_{h_{i}}]$.
Note that $n_{g}$ and $n_{h}$ may be different. 
\item \textbf{Bootstrapping the iteration}: 
\begin{itemize}
\item Basis: $\left\{ \left|t_{h_{i}}^{(n+1)}\right\rangle \right\} $ where
$\left|t_{h_{i}}^{(n+1)}\right\rangle :=\left|i\right\rangle $ for
$i=1,2\dots n$ where $\left|i\right\rangle $ refers to the standard
basis in which the matrices and the vectors were originally written.
\item Matrix Instance: $\text{\ensuremath{\underbar{X}}}^{(n)}=\{X_{h}^{(n)},X_{g}^{(n)},\left|w^{(n)}\right\rangle ,\left|v^{(n)}\right\rangle \}$.
\end{itemize}
\end{itemize}
\textbf{PHASE 2: ITERATION}
\begin{itemize}
\item Objective: Find the objects $\left|u_{h}^{(k)}\right\rangle ,\bar{O}_{g}^{(k)},\bar{O}_{h}^{(k)}$
and $s^{(k)}$ (which together relate $O^{(k)}$ to $O^{(k-1)}$ where
$O^{(k)}$ solves $\text{\ensuremath{\underbar{X}}}^{(k)}$ and $O^{(k-1)}$
solves $\text{\ensuremath{\underbar{X}}}^{(k-1)}$ that is yet to
be defined)
\item Input: We will assume we are given
\begin{itemize}
\item Basis: $\left\{ \left|t_{h_{i}}^{(k+1)}\right\rangle \right\} $
\item Matrix Instance: $\text{\ensuremath{\underbar{X}}}^{(k)}=\left(X_{h}^{(k)},X_{g}^{(k)},\left|w^{(k)}\right\rangle ,\left|v^{(k)}\right\rangle \right)$
with attribute $\chi^{(k)}>0$
\item Function Instance: $\text{\ensuremath{\underbar{X}}}^{(k)}\to\text{\ensuremath{\underbar{x}}}^{(k)}=\left(h^{(k)},g^{(k)},a^{(k)}\right)$
\end{itemize}
\item Output: 
\begin{itemize}
\item Basis: $\left\{ \left|u_{h}^{(k)}\right\rangle ,\left|t_{h_{i}}^{(k)}\right\rangle \right\} $
\item Matrix Instance: $\text{\ensuremath{\underbar{X}}}^{(k-1)}=\left(X_{h}^{(k-1)},X_{g}^{(k-1)},\left|w^{(k-1)}\right\rangle ,\left|v^{(k-1)}\right\rangle \right)$
with attribute $\chi^{(k-1)}>0$
\item Function Instance: $\text{\ensuremath{\underbar{X}}}^{(k-1)}\to\text{\ensuremath{\underbar{x}}}^{(k-1)}=\left(h^{(k-1)},g^{(k-1)},a^{(k-1)}\right)$
\item Unitary Constructors: Either $\bar{O}_{g}^{(k)}$ and $\bar{O}_{h}^{(k)}$
are returned or $\bar{O}^{(k)}$ is returned. If $\bar{O}^{(k)}$
is returned, set $\bar{O}_{g}^{(k)}:=\bar{O}^{(k)}$ and $\bar{O}_{h}^{(k)}=\mathbb{I}$.
\item Relation: If $s^{(k)}$ is not specified, define $s^{(k)}:=1$.\\
If $s^{(k)}=1$ then use 
\[
O^{(k)}:=\bar{O}_{h}^{(k)}\left(\left|u_{h}^{(k)}\right\rangle \left\langle u_{h}^{(k)}\right|+O^{(k-1)}\right)\bar{O}_{g}^{(k)}
\]
else use 
\[
O^{(k)}:=\left[\bar{O}_{h}^{(k)}\left(\left|u_{h}^{(k)}\right\rangle \left\langle u_{h}^{(k)}\right|+O^{(k-1)}\right)\bar{O}_{g}^{(k)}\right]^{T}.
\]
\end{itemize}
\item Algorithm:
\begin{itemize}
\item \textbf{Boundary condition:} \textbf{If} $n_{g}=0$ and $n_{h}=0$
\textbf{then} set $k_{0}=k$ and \textbf{jump to} phase 3.
\item \textbf{Tighten}: Define $X_{h_{\gamma'}}^{(k)}:=\gamma'X^{(k)}$.
Let $\gamma$ be the largest root of $m(\gamma',\chi_{\gamma'}^{(k)},\xi_{\gamma'}^{(k)})$
for $a^{(k)}$ where $\chi_{\gamma'}^{(k)},\xi_{\gamma'}^{(k)}$ are
such that $[\chi_{\gamma'}^{(k)},\xi_{\gamma'}^{(k)}]$ is the smallest
interval containing $\text{spec}[X_{h_{\gamma'}}^{(k)}\oplus X_{g}^{(k)}]$.
Relabel $X_{h_{\gamma}}^{(k)}$ to $X_{h}^{(k)}$, $\chi_{\gamma}^{(k)}$
to $\chi^{(k)}$ and $\xi_{\gamma}^{(k)}$ to $\xi^{(k)}$ for notational
ease. Similarly relabel $a_{\gamma}^{(k)}$ to $a^{(k)}$, $h_{\gamma}^{(k)}$
to $h^{(k)}$, $l_{\gamma}^{(k)}$ to $l^{(k)}$. Update $x_{\min}$
and $x_{\max}$ to be such that $\text{supp}(a^{(k)})\in[x_{\min}^{(k)},x_{\max}^{(k)}]$
is the smallest such interval. Define $s^{(k)}:=1$.
\item \textbf{Honest align}: \textbf{If} $l^{1(k)}=0$ \textbf{then} define
$\eta=-\chi^{(k)}+1$
\[
X_{h}^{\prime(k)}:=X_{h}^{(k)}+\eta,\quad X_{g}^{\prime(k)}:=X_{g}+\eta.
\]
\textbf{Else}: Pick a root $\lambda$ of the function $l^{(k)}(\lambda')$
in the domain $\mathbb{R}\backslash(-\xi^{(k)},-\chi^{(k)})$. In
the following two cases we consider the function $f_{\lambda}$ on
$[\chi^{(k)},\xi^{(k)}]$.
\begin{itemize}
\item If $\lambda\neq-\chi^{(k)}$ then: Let $\eta=-f_{\lambda}(\chi^{(k)})+1$
where any positive constant could be chosen instead of $1$. Define
\[
X_{h}^{\prime(k)}:=f_{\lambda}(X_{h}^{(k)})+\eta,\quad X_{g}^{\prime(k)}:=f_{\lambda}(X_{g}^{(k)})+\eta.
\]
\item If $\lambda=-\chi^{(k)}$ then: Update $s^{(k)}=-1$. Let $\eta=-f_{\lambda}(\xi^{(k)})-1$
where any positive constant could be chosen instead of 1. Define 
\[
X_{h}^{\prime(k)}:=X_{g}^{\prime\prime(k)},\quad X_{g}^{\prime(k)}:=X_{h}^{\prime\prime(k)},
\]
where
\[
X_{h}^{\prime\prime(k)}:=-f_{\lambda}(X_{h}^{(k)})-\eta,\quad X_{g}^{\prime\prime(k)}:=-f_{\lambda}(X_{g}^{(k)})-\eta
\]
and make the replacement 
\begin{align*}
\left|v^{(k)}\right\rangle  & \to\left|w^{(k)}\right\rangle \\
\left|w^{(k)}\right\rangle  & \to\left|v^{(k)}\right\rangle .
\end{align*}
\end{itemize}
\item \textbf{\textcolor{black}{Remove spectral collision: If}}\textcolor{black}{{}
$\lambda=-\chi^{(k)}$ or $\lambda=-\xi^{(k)}$ }\textbf{\textcolor{black}{then}}\textcolor{black}{{} }
\begin{enumerate}
\item \textbf{Idle point:} \textbf{If} for some $j',j$, we have $q_{g_{j'}}^{(k)}=q_{h_{j}}^{(k)}$
and $y_{g_{j'}}^{(k)}=y_{h_{j}}^{(k)}$ \textbf{then} the solution
is given by \Defref{idlePoint} \textbf{}\\
\textbf{Jump} to \textbf{End}.
\item \textbf{\textcolor{black}{Final Extra}}\textcolor{black}{: }\textbf{\textcolor{black}{If}}\textcolor{black}{{}
for some $j,j'$ we have $q_{g_{j'}}^{(k)}>q_{h_{j}}^{(k)}$ and $y_{g_{j'}}^{(k)}=y_{h_{j}}^{(k)}$
}\textbf{\textcolor{black}{then}}\textcolor{black}{{} the solution is
given by \Defref{finalExtra}} \textbf{}\\
\textbf{Jump} to \textbf{End}.
\item \textbf{Initial Extra}: \textbf{If} for some $j,j'$ we have $q_{g_{j'}}^{(k)}<q_{h_{j}}^{(k)}$
and $y_{g_{j'}}^{(k)}=y_{h_{j}}^{(k)}$ \textbf{then} the solution
is given by \Defref{initialExtra}\textbf{}\\
\textbf{Jump} to \textbf{End}.
\end{enumerate}
\item \textbf{Evaluate the Reverse Weingarten Map}: 
\begin{enumerate}
\item Consider the point $\left|w^{(k)}\right\rangle /\sqrt{\left\langle w^{(k)}\right|X_{h}^{\prime(k)}\left|w^{(k)}\right\rangle }$
on the ellipsoid $X_{h}^{\prime(k)}$. Evaluate the normal at this
point as $\left|u_{h}^{(k)}\right\rangle =\mathcal{N}\left(\sum_{i=1}^{n_{h}^{(k)}}\sqrt{p_{h_{i}}^{(k)}}x_{h_{i}}^{\prime(k)}\left|t_{h_{i}}^{(k+1)}\right\rangle \right)$.
Similarly evaluate $\left|u_{g}^{(k)}\right\rangle $, the normal
at the point $\left|v^{(k)}\right\rangle /\sqrt{\left\langle w^{(k)}\right|X_{g}^{\prime(k)}\left|w^{(k)}\right\rangle }$
on the ellipsoid $X_{g}^{\prime(k)}$. 
\item Recall that for a given diagonal matrix $X=\sum_{i}y_{i}\left|i\right\rangle \left\langle i\right|>0$
and normal vector $\left|u\right\rangle =\sum_{i}u_{i}\left|i\right\rangle $
the Reverse Weingarten map is given by $W_{ij}=\left(-\frac{y_{j}^{-1}y_{i}^{-1}u_{i}u_{j}}{r^{2}}+y_{i}^{-1}\delta_{ij}\right)$
where $r=\sqrt{\sum y_{i}^{-1}u_{i}^{2}}$. Evaluate the Reverse Weingarten
maps $W_{h}^{\prime(k)}$ and $W_{g}^{\prime(k)}$ along $\left|u_{h}^{(k)}\right\rangle $
and $\left|u_{g}^{(k)}\right\rangle $ respectively.\textcolor{blue}{{}
}
\item Find the eigenvectors and eigenvalues of the Reverse Weingarten maps.
The eigenvectors of $W_{h}'$ form the $h$ tangent (and normal) vectors
$\left\{ \left\{ \left|t_{h_{i}}^{(k)}\right\rangle \right\} ,\left|u_{h}^{(k)}\right\rangle \right\} $.
The corresponding radii of curvature are obtained from the eigenvalues
$\left\{ \{r_{h_{i}}^{(k)}\},0\right\} =\left\{ \{c_{h_{i}}^{(k)-1}\},0\right\} $
which are inverses of the curvature values. The tangents are labelled
in the decreasing order of radii of curvature (increasing order of
curvature). Similarly for the $g$ tangent (and normal) vectors. Fix
the sign freedom in the eigenvectors by requiring $\left\langle t_{h_{i}}^{(k)}|w^{(k)}\right\rangle \ge0$
and $\left\langle t_{g_{i}}^{(k)}|v^{(k)}\right\rangle \ge0$. 
\end{enumerate}
\item \textbf{Finite Method: If }$\lambda\neq-\xi^{(k)}$ and $\lambda\neq-\chi^{(k)}$,
i.e. if it is the finite case \textbf{then}
\begin{enumerate}
\item $\bar{O}^{(k)}:=\left|u_{h}^{(k)}\right\rangle \left\langle u_{g}^{(k)}\right|+\sum_{i=1}^{k-1}\left|t_{h_{i}}^{(k)}\right\rangle \left\langle t_{g_{i}}^{(k)}\right|$
\item $\left|v^{(k-1)}\right\rangle :=\bar{O}^{(k)}\left|v^{(k)}\right\rangle -\left\langle u_{h}^{(k)}\right|\bar{O}^{(k)}\left|v^{(k)}\right\rangle \left|u_{h}^{(k)}\right\rangle $
and $\left|w^{(k-1)}\right\rangle :=\left|w^{(k)}\right\rangle -\left\langle u_{h}^{(k)}|w^{(k)}\right\rangle \left|u_{h}^{(k)}\right\rangle $.
\item Define $X_{h}^{(k-1)}:=\text{diag}\{c_{h_{1}}^{(k)},c_{h_{2}}^{(k)}\dots,c_{h_{k-1}}^{(k)}\}$,
$X_{g}^{(k-1)}:=\text{diag}\{c_{g_{1}}^{(k)},c_{g_{2}}^{(k)}\dots c_{g_{k-1}}^{(k)}\}$.
\item \textbf{Jump} to \textbf{End}.
\end{enumerate}
\item \textbf{Wiggle-v Method: If }$\lambda=-\xi^{(k)}$ or $\lambda=-\chi^{(k)}$
\textbf{then}
\begin{enumerate}
\item $\left|u_{h}^{(k)}\right\rangle $ is renamed to $\left|\bar{u}_{h}^{(k)}\right\rangle $,
$\left|u_{g}^{(k)}\right\rangle $ remains the same.
\item Let $\tau=\cos\theta:=\left\langle u_{g}^{(k)}|v^{(k)}\right\rangle /\left\langle \bar{u}_{h}^{(k)}|w^{(k)}\right\rangle $.
Let $\left|\bar{t}_{h}^{(k)}\right\rangle $ be an eigenvector of
$X_{h}^{\prime(k)-1}$ with zero eigenvalue (comment: this is also
perpendicular to $\left|w^{(k)}\right\rangle $). Redefine 
\[
\left|u_{h}^{(k)}\right\rangle :=\cos\theta\left|\bar{u}_{h}^{(k)}\right\rangle +\sin\theta\left|\bar{t}_{h}^{(k)}\right\rangle ,
\]
\[
\left|t_{h_{k}}^{(k)}\right\rangle =s\left(-\sin\theta\left|\bar{u}_{h}^{(k)}\right\rangle +\cos\theta\left|\bar{t}_{h}^{(k)}\right\rangle \right)
\]
 where the sign $s\in\{1,-1\}$ is fixed by demanding $\left\langle t_{h_{k}}^{(k)}|w^{(k)}\right\rangle \ge0$. 
\item $\bar{O}^{(k)}$ and $\left|v^{(k-1)}\right\rangle ,\left|w^{(k-1)}\right\rangle $
are evaluated as step i and ii of the finite case (a).
\item Define 
\[
X_{h}^{\prime(k-1)}:=\text{diag}\{c_{h_{1}}^{(k)},c_{h_{2}}^{(k)},\dots,c_{h_{k-1}}^{(k)}\},\qquad X_{g}^{\prime(k-1)}:=\text{diag}\{c_{g_{1}}^{(k)},c_{g_{2}}^{(k)},\dots,c_{g_{k-1}}^{(k)}\}.
\]
Let $[\chi^{\prime(k-1)},\xi^{\prime(k-1)}]$ denote the smallest
interval containing $\text{spec}[X_{h}^{\prime(k-1)}\oplus X_{g}^{\prime(k-1)}]$.
Let $\lambda'=-\chi'^{(k-1)}+1$ where instead of $1$ any positive
number would also work. Consider $f_{\lambda''}$ on $[\chi^{\prime(k-1)},\xi^{\prime(k-1)}]$.
Let $\eta=-f_{\lambda'}(\chi{}^{\prime(k-1)})+1$. Define 
\[
X_{h}^{(k-1)}:=f_{\lambda'}(X_{h}^{\prime(k-1)})+\eta,\qquad X_{g}^{(k-1)}:=f_{\lambda'}(X_{g}^{\prime(k-1)})+\eta.
\]
\item \textbf{Jump} to \textbf{End}.
\end{enumerate}
\item \textbf{End}: Restart the current phase (phase 2) with the newly obtained
$(k-1)$ sized objects.
\end{itemize}
~~
\end{itemize}
\textbf{PHASE 3: RECONSTRUCTION}~

Let $k_{0}$ be the iteration at which the algorithm stops. Using
the relation 
\[
O^{(k)}=\bar{O}_{g}^{(k)}\left(\left|u_{h}^{(k)}\right\rangle \left\langle u_{h}^{(k)}\right|+O^{(k-1)}\right)\bar{O}_{h}^{(k)}
\]
 (or its transpose if $s^{(k)}=-1$), evaluate $O^{(k_{1})}$ from
$O^{(k_{0})}:=\mathbb{I}_{k_{0}}$, then $O^{(k_{2})}$ from $O^{(k_{1})}$,
then $O^{(k_{3})}$ from $O^{(k_{2})}$ and so on until $O^{(n)}$
is obtained which solves the matrix instance $\text{\ensuremath{\underbar{X}}}^{(n)}$
we started with. In terms of EBRM matrices, the solution is given
by $H=X_{h}^{(n)},$ $G=O^{(n)}X_{g}O^{(n)T}$, and $\left|w\right\rangle =\left|w^{(n)}\right\rangle $. 
\end{defn}

\end{sloppy}
\begin{thm}[Correctness of the EMA Algorithm]
 Given a $\Lambda$-valid function, the EMA algorithm (see \Defref{EMAalgorithm})
always finds an orthogonal matrix $O$ of size at most $n\times n$
where $n=n_{g}+n_{h}$, such that the constraints on $O$ stated in
\Thmref{TEFconstraint-inf} corresponding to the function $a$, are
satisfied.\label{thm:EMAcorrectnessFormal}
\end{thm}

\begin{defn}[Spectral Collision: Case Idle Point]
 ~~

\begin{align*}
\left\{ \left|u_{h}^{(k)}\right\rangle ,\left|t_{h_{1}}^{(k)}\right\rangle ,\left|t_{h_{2}}^{(k)}\right\rangle ,\dots\left|t_{h_{k-1}}^{(k)}\right\rangle \right\} \overset{\text{componentwise}}{:=}\\
\left\{ \left|t_{h_{j}}^{(k+1)}\right\rangle ,\left|t_{h_{1}}^{(k+1)}\right\rangle ,\left|t_{h_{2}}^{(k+1)}\right\rangle ,\dots\left|t_{h_{j-1}}^{(k+1)}\right\rangle ,\left|t_{h_{j+1}}^{(k+1)}\right\rangle ,\dots\left|t_{h_{k}}^{(k+1)}\right\rangle \right\} ,
\end{align*}
\[
\bar{O}^{(k)}:=\sum_{i=1}^{k}\left|a_{i}\right\rangle \left\langle t_{h_{i}}^{(k+1)}\right|,
\]
where
\[
\left\{ \left|a_{1}\right\rangle ,\left|a_{2}\right\rangle \dots\left|a_{k}\right\rangle \right\} \overset{\text{componentwise}}{:=}
\]
\[
\begin{cases}
\begin{array}{c}
\Big\{\left|t_{h_{1}}^{(k+1)}\right\rangle ,\left|t_{h_{2}}^{(k+1)}\right\rangle ,\dots\left|t_{h_{j-1}}^{(k+1)}\right\rangle ,\left|t_{h_{j'}}^{(k+1)}\right\rangle ,\left|t_{h_{j}}^{(k+1)}\right\rangle ,\left|t_{h_{j+1}}^{(k+1)}\right\rangle ,\\
\dots,\left|t_{h_{j'-1}}^{(k+1)}\right\rangle ,\left|t_{h_{j'+1}}^{(k+1)}\right\rangle \dots\left|t_{h_{k}}^{(k+1)}\right\rangle \Big\}\\
\\
\end{array} & j<j'\\
\begin{array}{c}
\Big\{\left|t_{h_{1}}^{(k+1)}\right\rangle ,\left|t_{h_{2}}^{(k+1)}\right\rangle ,\dots\left|t_{h_{j'-1}}^{(k+1)}\right\rangle ,\left|t_{h_{j'+1}}^{(k+1)}\right\rangle \dots\\
\left|t_{h_{j-1}}^{(k+1)}\right\rangle ,\left|t_{h_{j'}}^{(k+1)}\right\rangle ,\left|t_{h_{j}}^{(k+1)}\right\rangle ,\left|t_{h_{j+1}}^{(k+1)}\right\rangle \dots\left|t_{h_{k}}^{(k+1)}\right\rangle \Big\}\\
\\
\end{array} & j>j'\\
\left\{ \left|t_{h_{1}}^{(k+1)}\right\rangle ,\left|t_{h_{2}}^{(k+1)}\right\rangle ,\dots\left|t_{h_{k}}^{(k+1)}\right\rangle \right\}  & j=j',
\end{cases}
\]
and  
\[
X_{h}^{(k-1)}:=\sum_{i\neq j}y_{h_{i}}^{(k)}\left|t_{h_{i}}^{(k+1)}\right\rangle \left\langle t_{h_{i}}^{(k+1)}\right|,
\]
\[
X_{g}^{(k-1)}:=\bar{O}^{(k)}X_{g}^{(k)}\bar{O}^{(k)T}-y_{h_{j}}\left|t_{h_{j}}^{(k+1)}\right\rangle \left\langle t_{h_{j}}^{(k+1)}\right|,
\]
 
\[
\left|w^{(k-1)}\right\rangle =\mathcal{N}\left[\left|w^{(k)}\right\rangle -\sqrt{p_{h_{j}}}\left|t_{h_{j}}^{(k+1)}\right\rangle \right],\,\left|v^{(k-1)}\right\rangle =\mathcal{N}\left[\bar{O}^{(k)}\left|v^{(k)}\right\rangle -\sqrt{p_{h_{j}}}\left|t_{h_{j}}^{(k+1)}\right\rangle \right].
\]
\textcolor{gray}{}\\
\textcolor{gray}{(This specifies $\text{\ensuremath{\underbar{X}}}^{(k-1)}:=\{X_{h}^{(k-1)},X_{g}^{(k-1)},\left|w^{(k-1)}\right\rangle ,\left|v^{(k-1)}\right\rangle \}$.)} \label{def:idlePoint}
\end{defn}

\begin{defn}[Spectral Collision: Case Final Extra]
 ~~

\textcolor{black}{{} }$\text{\ensuremath{\underbar{X}}}^{(k-1)}:=(X_{h}^{(k-1)},X_{g}^{(k-1)},\left|w^{(k-1)}\right\rangle ,\left|v^{(k-1)}\right\rangle )$
where $X_{h}^{(k-1)}=\sum_{i=1}^{k-1}y_{h_{i}}^{(k-1)}\left|t_{h_{i}}^{(k)}\right\rangle \left\langle t_{h_{i}}^{(k)}\right|$,
$X_{g}^{(k-1)}=\sum_{i=1}^{k-1}y_{g_{i}}^{(k-1)}\left|t_{h_{i}}^{(k)}\right\rangle \left\langle t_{h_{i}}^{(k)}\right|$,
$\left|v^{(k-1)}\right\rangle =\mathcal{N}\left[\sum_{i=1}^{k-1}\sqrt{q_{g_{i}}^{(k-1)}}\left|t_{h_{i}}^{(k)}\right\rangle \right]$,
$\left|w^{(k-1)}\right\rangle =\mathcal{N}\left[\sum_{i=1}^{k-1}\sqrt{q_{h_{i}}^{(k-1)}}\left|t_{h_{i}}^{(k)}\right\rangle \right]$
where the coordinates and weights are given by
\begin{align*}
\left\{ q_{h_{1}}^{(k-1)},\dots q_{h_{k-1}}^{(k-1)}\right\} \overset{\text{componentwise}}{=} & \left\{ q_{h_{1}}^{(k)},q_{h_{2}}^{(k)}\dots,q_{h_{j-1}}^{(k)},q_{h_{j+1}}^{(k)},\dots q_{h_{k}}^{(k)}\right\} \\
\left\{ q_{g_{1}}^{(k-1)},\dots q_{g_{k-1}}^{(k-1)}\right\} \overset{\text{componentwise}}{=} & \left\{ q_{g_{2}}^{(k)}\dots,q_{g_{j'-1}}^{(k)},q_{g_{j'}}^{(k)}-q_{h_{j}}^{(k)},q_{g_{j'+1}}^{(k)},q_{g_{j'+2}}^{(k)}\dots q_{g_{k}}^{(k)}\right\} \\
\left\{ y_{g_{1}}^{(k-1)},\dots y_{g_{k-1}}^{(k-1)}\right\} \overset{\text{componentwise}}{=} & \left\{ y_{g_{2}}^{(k)},\dots y_{g_{k}}^{(k)}\right\} \\
\left\{ y_{h_{1}}^{(k-1)},\dots y_{h_{k-1}}^{(k-1)}\right\} \overset{\text{componentwise}}{=} & \left\{ y_{h_{1}}^{(k)},\dots y_{h_{j-1}}^{(k)},y_{h_{j+1}}^{(k)}\dots,y_{h_{k}}^{(k)}\right\} ,
\end{align*}
the basis is given by 
\begin{align*}
\left\{ \left|u_{h}^{(k)}\right\rangle ,\left|t_{h_{1}}^{(k)}\right\rangle \dots\left|t_{h_{k-1}}^{(k)}\right\rangle \right\} \overset{\text{componentwise}}{=}\\
\left\{ \left|t_{h_{j}}^{(k+1)}\right\rangle ,\left|t_{h_{1}}^{(k+1)}\right\rangle ,\left|t_{h_{2}}^{(k+1)}\right\rangle ,\dots\left|t_{h_{j-1}}^{(k+1)}\right\rangle ,\left|t_{h_{j+1}}^{(k+1)}\right\rangle ,\left|t_{h_{j+2}}^{(k+1)}\right\rangle \dots\left|t_{h_{k}}^{(k+1)}\right\rangle \right\} .
\end{align*}
The orthogonal matrices are given by $\bar{O}_{h}^{(k)}:=\sum\left|t_{h_{i}}^{(k+1)}\right\rangle \left\langle a_{i}\right|$
where 
\[
\left\{ \left|a_{1}\right\rangle ,\dots\left|a_{k}\right\rangle \right\} \to\left\{ \left|u_{h}^{(k)}\right\rangle ,\left|t_{h_{1}}^{(k)}\right\rangle \dots\left|t_{h_{k-1}}^{(k)}\right\rangle \right\} ,
\]
$\bar{O}_{g}^{(k)}:=\tilde{O}^{(k)}\bar{O}_{h}^{(k)}$ where 
\begin{align*}
\tilde{O}^{(k)}:= & \mathcal{N}\left[\sqrt{q_{h_{j}}^{(k)}}\left|u_{h}^{(k)}\right\rangle +\sqrt{q_{g_{j'}}^{(k)}-q_{h_{j}}^{(k)}}\left|t_{h_{j'}}^{(k)}\right\rangle \right]\mathcal{N}\left[\sqrt{q_{g_{1}}^{(k)}}\left\langle u_{h}^{(k)}\right|+\sqrt{q_{g_{j'}}^{(k)}}\left\langle t_{h_{j'}}^{(k)}\right|\right]\\
 & +\mathcal{N}\left[\sqrt{q_{g_{j'}}^{(k)}-q_{h_{j}}^{(k)}}\left|u_{h}^{(k)}\right\rangle -\sqrt{q_{h_{j}}^{(k)}}\left|t_{h_{j'}}^{(k)}\right\rangle \right]\mathcal{N}\left[\sqrt{q_{g_{j'}}^{(k)}}\left\langle u_{h}^{(k)}\right|-\sqrt{q_{g_{1}}^{(k)}}\left\langle t_{h_{j'}}^{(k)}\right|\right]\\
 & +\sum_{i\in\{1,\dots k\}\backslash j'}\left|t_{h_{i}}^{(k)}\right\rangle \left\langle t_{h_{i}}^{(k)}\right|.
\end{align*}
\label{def:finalExtra}
\end{defn}

\begin{defn}[Spectral Collision: Case Initial Extra]
 ~~

$\text{\ensuremath{\underbar{X}}}^{(k-1)}:=(X_{h}^{(k-1)},X_{g}^{(k-1)},\left|w^{(k-1)}\right\rangle ,\left|v^{(k-1)}\right\rangle )$
where $X_{h}^{(k-1)}=\sum_{i=1}^{k-1}y_{h_{i}}^{(k-1)}\left|t_{h_{i}}^{(k)}\right\rangle \left\langle t_{h_{i}}^{(k)}\right|$,
$X_{g}^{(k-1)}=\sum_{i=1}^{k-1}y_{g_{i}}^{(k-1)}\left|t_{h_{i}}^{(k)}\right\rangle \left\langle t_{h_{i}}^{(k)}\right|$,
$\left|v^{(k-1)}\right\rangle =\mathcal{N}\left[\sum_{i=1}^{k-1}\sqrt{q_{g_{i}}^{(k-1)}}\left|t_{h_{i}}^{(k)}\right\rangle \right]$,
$\left|w^{(k-1)}\right\rangle =\mathcal{N}\left[\sum_{i=1}^{k-1}\sqrt{q_{h_{i}}^{(k-1)}}\left|t_{h_{i}}^{(k)}\right\rangle \right]$
where the coordinates and weights are given by
\begin{align*}
\left\{ q_{h_{1}}^{(k-1)},\dots q_{h_{k-1}}^{(k-1)}\right\} \overset{\text{componentwise}}{=} & \left\{ q_{h_{1}}^{(k)}\dots,q_{h_{j-1}}^{(k)},q_{h_{j}}^{(k)}-q_{g_{j'}}^{(k)},q_{h_{j+1}}^{(k)},q_{h_{j+2}}^{(k)}\dots q_{h_{k-1}}^{(k)}\right\} \\
\left\{ q_{g_{1}}^{(k-1)},\dots q_{g_{k-1}}^{(k-1)}\right\} \overset{\text{componentwise}}{=} & \left\{ q_{g_{1}}^{(k)},q_{g_{2}}^{(k)}\dots,q_{g_{j'-1}}^{(k)},q_{g_{j'+1}}^{(k)},\dots q_{g_{k}}^{(k)}\right\} \\
\left\{ y_{g_{1}}^{(k-1)},\dots y_{g_{k-1}}^{(k-1)}\right\} \overset{\text{componentwise}}{=} & \left\{ y_{g_{1}}^{(k)},\dots y_{g_{j'-1}}^{(k)},y_{g_{j'+1}}^{(k)}\dots,y_{g_{k}}^{(k)}\right\} \\
\left\{ y_{h_{1}}^{(k-1)},\dots y_{h_{k-1}}^{(k-1)}\right\} \overset{\text{componentwise}}{=} & \left\{ y_{h_{1}}^{(k)},\dots y_{h_{k-1}}^{(k)}\right\} ,
\end{align*}
the basis is given by 
\begin{align*}
\left\{ \left|u_{h}^{(k)}\right\rangle ,\left|t_{h_{1}}^{(k)}\right\rangle \dots\left|t_{h_{k-1}}^{(k)}\right\rangle \right\} \overset{\text{componentwise}}{=}\\
\left\{ \left|t_{h_{j}}^{(k+1)}\right\rangle ,\left|t_{h_{1}}^{(k+1)}\right\rangle ,\left|t_{h_{2}}^{(k+1)}\right\rangle ,\dots\left|t_{h_{j-1}}^{(k+1)}\right\rangle ,\left|t_{h_{j+1}}^{(k+1)}\right\rangle ,\left|t_{h_{j+2}}^{(k+1)}\right\rangle \dots\left|t_{h_{k}}^{(k+1)}\right\rangle \right\} .
\end{align*}
The orthogonal matrices are given by $\bar{O}_{h}^{(k)}:=\tilde{O}^{(k)}\sum\left|a_{i}\right\rangle \left\langle t_{h_{i}}^{(k+1)}\right|$
where 
\[
\left\{ \left|a_{1}\right\rangle ,\dots\left|a_{k}\right\rangle \right\} \overset{\text{componentwise}}{=}\left\{ \left|t_{h_{1}}^{(k)}\right\rangle ,\left|t_{h_{2}}^{(k)}\right\rangle \dots\left|t_{h_{k-1}}^{(k)}\right\rangle ,\left|u_{h}^{(k)}\right\rangle \right\} .
\]
 
\begin{align*}
\tilde{O}^{(k)}:= & \mathcal{N}\left[\sqrt{q_{g_{j'}}^{(k)}}\left|u_{h}^{(k)}\right\rangle +\sqrt{q_{h_{j}}^{(k)}-q_{g_{j'}}^{(k)}}\left|t_{h_{j}}^{(k)}\right\rangle \right]\mathcal{N}\left[\sqrt{q_{h_{k}}^{(k)}}\left\langle u_{h}^{(k)}\right|+\sqrt{q_{g_{j}}^{(k)}}\left\langle t_{h_{j}}^{(k)}\right|\right]\\
 & +\mathcal{N}\left[\sqrt{q_{h_{j}}^{(k)}-q_{g_{j'}}^{(k)}}\left|u_{h}^{(k)}\right\rangle -\sqrt{q_{g_{j'}}^{(k)}}\left|t_{h_{j}}^{(k)}\right\rangle \right]\mathcal{N}\left[\sqrt{q_{g_{j}}^{(k)}}\left\langle u_{h}^{(k)}\right|-\sqrt{q_{h_{k}}^{(k)}}\left\langle t_{h_{j}}^{(k)}\right|\right]\\
 & +\sum_{i\in\{1,\dots k\}\backslash j}\left|t_{h_{i}}^{(k)}\right\rangle \left\langle t_{h_{i}}^{(k)}\right|
\end{align*}
and $\bar{O}_{h}^{(k)}$ is given by the basis change $\left\{ \left|t_{h_{1}}^{(k+1)}\right\rangle ,\dots\left|t_{h_{k}}^{(k+1)}\right\rangle \right\} \to\left\{ \left|u_{h}^{(k)}\right\rangle ,\left|t_{h_{1}}^{(k)}\right\rangle \dots\left|t_{h_{k-1}}^{(k)}\right\rangle \right\} $.\label{def:initialExtra}
\end{defn}

\textcolor{purple}{We start with motivating} the exact step of the
algorithm \textcolor{blue}{and then provide a proof or justification
for the claims made in that step.}

\subsubsection{Phase 1: Initialisation}

We are given a $\Lambda$-valid transition $g\to h$ and the EBRM
function $a=h-g$%
\begin{comment}
 which we can write as $a=h-g$ where $g,h:\mathbb{R}_{\ge}\to\mathbb{R}_{\ge}$
with no common support. 
\end{comment}
. (Remark: We use below the notation used in the definition of a transition.)

\smallskip{}

\textcolor{purple}{Since the function is EBRM we know there are matrices
$H\ge G$ and a vector $\left|\psi\right\rangle $ such that $a=\text{Prob}[H,\left|\psi\right\rangle ]-\text{Prob}[G,\left|\psi\right\rangle ]$.
We also know that the maximum matrix size we need to consider is $n_{g}+n_{h}-1$.
We want to know the spectrum of the matrices involved to proceed. }

\textcolor{purple}{The picture we have in mind is the following. We
know that $H\ge G$ in terms of ellipsoids means that the $H$ ellipsoid
is inside the $G$ ellipsoid (the order gets reversed). We try to
expand the $H$ ellipsoid (which means scaling down the matrix $H$)
until it touches the $G$ ellipsoid. When they touch we know that
the corresponding spectrum of the matrices is optimal in some sense.
This would be trivial if we already knew $H$ and $G$ but it serves
as a good picture nonetheless.}

\textcolor{purple}{What we do know is the function $a=h-g$. We use
the equivalence between EBRM and valid functions to perform the aforesaid
tightening procedure even without knowing the matrices. We use $a_{\gamma}=h_{\gamma}-g$
where $h_{\gamma}(x)=h(x/\gamma)$ and check if $a_{\gamma}$ stays
valid as we shrink $\gamma$ from one to zero. We stop the moment
we see any signature of tightness. Using this $a_{\gamma}$ we determine
the spectrum of the matrices certifying the EBRM claim.}

\bigskip{}

\textcolor{purple}{We start with tightening till we find some operator
monotone labelled by $\lambda$ for which $l_{\gamma'}(\lambda)$
disappears. This captures the notion of the ellipsoids touching as
after applying this operator monotone, along the $\left|w\right\rangle $
direction, the ellipsoids must touch.}\smallskip{}

\textbf{Tightening procedure}: Let $[x_{\min}(\gamma'),x_{\max}(\gamma')]$
be the support domain for $a_{\gamma'}$. Let $\gamma\in(0,1]$ be
the largest root of $m(\gamma',x_{\min}(\gamma'),x_{\max}(\gamma'))$.
Let $x_{\max}:=x_{\max}(\gamma)$ and $x_{\min}:=x_{\min}(\gamma)$. 

\smallskip{}

\textcolor{blue}{\textcolor{blue}{First we must show that there would indeed be a root
of $m$ as a function of $\gamma'$ in the range $(0,1]$. This is
a direct consequence of \Lemref{tighteningNoSpect}. Second we must
show that if we can find the matrices certifying $a_{\gamma}$ is
EBRM we can find the matrices certifying $a$ is EBRM. This follows
from the observation that $\gamma X_{h}\ge OX_{g}O^{T}$ implies that
$X_{h}\ge\gamma X_{h}\ge OX_{g}O^{T}$.}}

\bigskip{}

\textcolor{purple}{We found a signature of tightness. Now we find
the spectrum of the matrices involved.}\smallskip{}

\textbf{Spectral domain for the representation}: Find the smallest
interval $[\chi,\xi]$ such that $l_{\gamma}(\lambda)\ge0$ for $\lambda\in\mathbb{\bar{R}}\backslash[\chi,\xi]$.
If $\text{supp}(g),\text{supp}(h)$ is not contained in $[\chi,\xi]$
then from all expansions of $[\chi,\xi]$ that contain the aforesaid
sets, pick the smallest. Relabel this interval to $[\chi,\xi]$.  

\smallskip{}

\textcolor{blue}{\textcolor{blue}{The interval so obtained will contain the spectrum
of the matrices that certify $a_{\gamma}$ is EBRM. This is justified
by \Lemref{MatSpecFromValidFun} using the fact that $l_{\gamma}^{1}\ge0$
due to the previous step.}}\bigskip{}

\textcolor{purple}{We need our matrices to be positive to be able
to use the elliptic picture. We therefore shift the spectrum of the
matrices so that the smallest eigenvalue required is one (where we
could have used any positive number).}\smallskip{}

\textbf{Shift}: Transform 
\[
a(x)\to a'(x'):=a(x'+\chi-1)
\]
where instead of $1$ any positive constant would do (justified by
\Corref{moveEBRMfns}). Similarly transform 
\begin{align*}
g(x) & \to g'(x'):=g(x'+\chi-1)\\
h(x) & \to h'(x'):=h(x'+\chi-1).
\end{align*}
Relabel $a'$ to be $a$, $g'$ to be $g$ and $h'$ to be $h$. (Remark:
We do not deduce $h$ and $g$ from $a$ as its positive and negative
part because they might now have common support due to the tightening
procedure.)

\smallskip{}

\textcolor{blue}{\textcolor{blue}{We use }\Corref{moveEBRMfns}\textcolor{blue}{{} to
deduce that if $a(x)$ is EBRM with spectrum in $[\chi,\xi]$ then
$a'(x')=a(x'+\chi-1)$ is EBRM with spectrum in $[1,\xi-\chi+1]$.
We must also show that if we can find the matrices certifying $a'$
is EBRM then we can find the matrices certifying $a$ is EBRM. This
is a direct consequence of the fact that $X_{h}'\ge OX_{g}'O^{T}\iff X_{h}-(\chi-1)\mathbb{I}\ge O(X_{g}-(\chi-1)\mathbb{I})O^{T}$.
The orthogonal matrix, $O$, which is of primary interest remains
unchanged. }}\bigskip{}

\textcolor{purple}{With the spectrum determined and adjusted to the
elliptic picture, which we put to use soon, we fix everything except
the orthogonal matrix by using the Canonical Orthogonal Form (up to
a permutation).}\smallskip{}

\textbf{The matrices}: For $n:=n_{g}+n_{h}-1$ we define $n\times n$
matrices with spectrum in $[\chi,\xi]$ and $n$ dimensional vectors
as 
\begin{align*}
X_{g}^{(n)} & =\text{diag}[\chi,\chi,\dots x_{g_{1}},x_{g_{2}}\dots,x_{g_{n_{g}}}],\\
X_{h_{\gamma}}^{(n)} & =\text{diag}[\gamma x_{h_{1}},\gamma x_{h_{2}},\dots,\gamma x_{h_{n_{h}}},\xi,\xi,\dots],\\
\left|v^{(n)}\right\rangle  & \doteq\left[0,0\dots,\sqrt{p_{g_{1}}},\sqrt{p_{g_{2}}},\dots,\sqrt{p_{g_{n_{g}}}}\right],\\
\left|w^{(n)}\right\rangle  & \doteq\left[\sqrt{p_{h_{1}}},\sqrt{p_{h_{2}}},\dots,\sqrt{p_{h_{n_{h}}}},0,0\dots\right]
\end{align*}
where $g=\sum_{i=1}^{n_{g}}p_{g_{i}}[x_{g_{i}}]$ and $h=\sum_{i=1}^{n_{h}}p_{h_{i}}[x_{h_{i}}]$.
Note that $n_{g}$ and $n_{h}$ may be different. 

\smallskip{}

\textcolor{blue}{\textcolor{blue}{We use \Lemref{EBRMfunIsEBRMtrans_evenWithCommon}
to deduce that $g\to h$ is a valid transition from the validity of
$a$. Then we use \Lemref{EBRMisCOF} to write the diagonal matrices
as described above given the valid transition $g\to h$, upto a permutation.
Our objective is to find a matrix $O^{(n)}$ such that $O^{(n)}\left|v^{(n)}\right\rangle =\left|w^{(n)}\right\rangle $
while satisfying the inequality $X_{h}^{(n)}\ge O^{(n)}X_{g}^{(n)}O^{(n)T}$. }}\bigskip{}

\textcolor{purple}{We now remove all the redundant information and
pack it into a form which we can iteratively reduce to a simpler form.
}\smallskip{}

\textbf{Bootstrapping the iteration}: 
\begin{itemize}
\item[\textendash{}] Basis: $\left\{ \left|t_{h_{i}}^{(n+1)}\right\rangle \right\} $
where $\left|t_{h_{i}}^{(n+1)}\right\rangle :=\left|i\right\rangle $
for $i=1,2\dots n$ where $\left|i\right\rangle $ refers to the standard
basis in which the matrices and the vectors were originally written.
\item[\textendash{}] Matrix Instance: $\text{\ensuremath{\underbar{X}}}^{(n)}=\{X_{h}^{(n)},X_{g}^{(n)},\left|w^{(n)}\right\rangle ,\left|v^{(n)}\right\rangle \}$.
\end{itemize}

\subsubsection{Phase 2: Iteration}

\textcolor{purple}{}
\begin{figure}
\begin{centering}
\includegraphics[width=12cm]{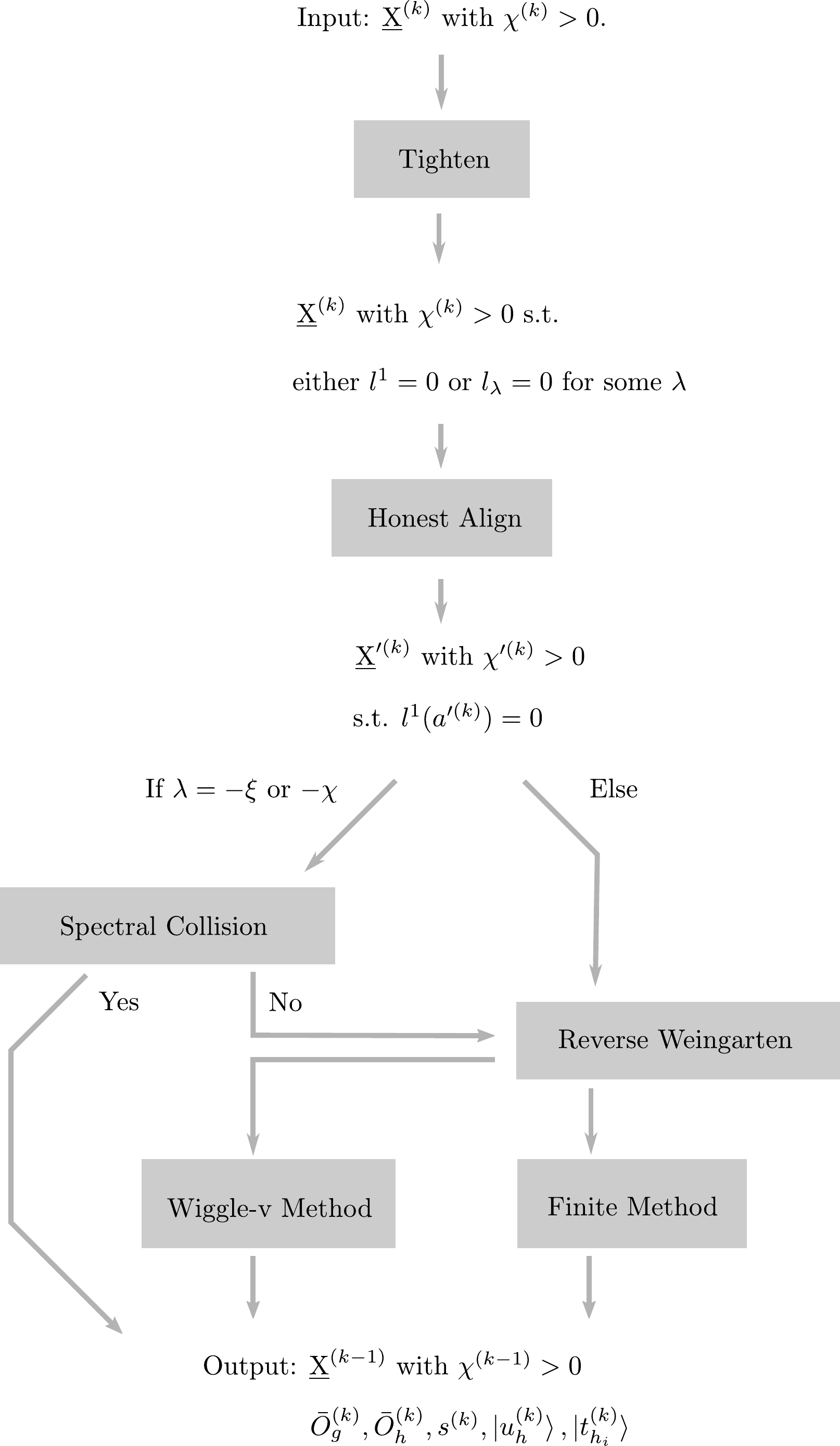}
\par\end{centering}
\textcolor{purple}{\caption{Overview of the main step, the iteration, of the algorithm (excluding
the boundary condition).\label{fig:Overview-of-Iteration}}
}
\end{figure}

\bigskip{}
\textcolor{purple}{We take as input the matrices $X_{g},X_{h}$ together
with the vectors $\left|w\right\rangle ,\left|v\right\rangle $ and
churn out the same objects with one less dimension. We also output
objects that, once we have iteratively reduced the problem to triviality,
can be put together to find the orthogonal matrix $O$. See \Figref{Overview-of-Iteration}
for a schematic reference. }\smallskip{}

\begin{itemize}
\item Objective: Find the objects $\left|u_{h}^{(k)}\right\rangle ,\bar{O}_{g}^{(k)},\bar{O}_{h}^{(k)}$
and $s^{(k)}$ (which together relate $O^{(k)}$ to $O^{(k-1)}$ where
$O^{(k)}$ solves $\text{\ensuremath{\underbar{X}}}^{(k)}$ and $O^{(k-1)}$
solves $\text{\ensuremath{\underbar{X}}}^{(k-1)}$ that is yet to
be defined)
\item Input: We assume we are given
\begin{itemize}
\item Basis: $\left\{ \left|t_{h_{i}}^{(k+1)}\right\rangle \right\} $
\item Matrix Instance: $\text{\ensuremath{\underbar{X}}}^{(k)}=\left(X_{h}^{(k)},X_{g}^{(k)},\left|w^{(k)}\right\rangle ,\left|v^{(k)}\right\rangle \right)$
with attribute $\chi^{(k)}>0$
\item Function Instance: $\text{\ensuremath{\underbar{X}}}^{(k)}\to\text{\ensuremath{\underbar{x}}}^{(k)}=\left(h^{(k)},g^{(k)},a^{(k)}\right)$
\end{itemize}
\item Output: 
\begin{itemize}
\item Basis: $\left\{ \left|u_{h}^{(k)}\right\rangle ,\left|t_{h_{i}}^{(k)}\right\rangle \right\} $
\item Matrix Instance: $\text{\ensuremath{\underbar{X}}}^{(k-1)}=\left(X_{h}^{(k-1)},X_{g}^{(k-1)},\left|w^{(k-1)}\right\rangle ,\left|v^{(k-1)}\right\rangle \right)$
with attribute $\chi^{(k-1)}>0$
\item Function Instance: $\text{\ensuremath{\underbar{X}}}^{(k-1)}\to\text{\ensuremath{\underbar{x}}}^{(k-1)}=\left(h^{(k-1)},g^{(k-1)},a^{(k-1)}\right)$
\item Unitary Constructors: Either $\bar{O}_{g}^{(k)}$ and $\bar{O}_{h}^{(k)}$
are returned or $\bar{O}^{(k)}$ is returned. If $\bar{O}^{(k)}$
is returned, set $\bar{O}_{g}^{(k)}:=\bar{O}^{(k)}$ and $\bar{O}_{h}^{(k)}=\mathbb{I}$.
\item Relation: If $s^{(k)}$ is not specified, define $s^{(k)}:=1$.\\
If $s^{(k)}=1$ then use 
\[
O^{(k)}:=\bar{O}_{h}^{(k)}\left(\left|u_{h}^{(k)}\right\rangle \left\langle u_{h}^{(k)}\right|+O^{(k-1)}\right)\bar{O}_{g}^{(k)}
\]
else use 
\[
O^{(k)}:=\left[\bar{O}_{h}^{(k)}\left(\left|u_{h}^{(k)}\right\rangle \left\langle u_{h}^{(k)}\right|+O^{(k-1)}\right)\bar{O}_{g}^{(k)}\right]^{T}.
\]
\smallskip{}
\textcolor{blue}{Our task is to solve the matrix instance $\text{\ensuremath{\underbar{X}}}^{(k)}$,
i.e. find a real unitary $O^{(k)}$ such that $X_{h}^{(k)}\ge O^{(k)}X_{g}^{(k)}O^{(k)T}$
and $O^{(k)}\left|v^{(k)}\right\rangle =\left|w^{(k)}\right\rangle $.
We assume that the solution exists and show that the solution to the
smaller instance, denoted by $\text{\ensuremath{\underbar{X}}}^{(k-1)}$
must also exist. More precisely, we show that $O^{(k)}$ must have
the form $O^{(k)}=\left(\left|u_{h}^{(k)}\right\rangle \left\langle u_{h}^{(k)}\right|+O^{(k-1)}\right)\bar{O}^{(k)}$
(for a solution to exist) which satisfies the aforesaid constraints
granted we can find $O^{(k-1)}$ which acts on a $k-1$ dimensional
Hilbert space orthogonal to $\left|u_{h}^{(k)}\right\rangle $ and
satisfies constraints of the same form in the smaller dimension, viz.
$X_{h}^{(k-1)}\ge O^{(k-1)}X_{g}^{(k-1)}O^{(k-1)T}$ and $O^{(k-1)}\left|v^{(k-1)}\right\rangle =\left|w^{(k-1)}\right\rangle $.
Hence the assumption that $O^{(k)}$ has a solution allows us to deduce
that $O^{(k-1)}$ must also have a solution. This allow us to iteratively
solve the problem.}

\textcolor{blue}{In certain trivial cases, where a point appears both
before and after a transition viz. $X_{g}^{(k)}$ and $X_{h}^{(k)}$
have a common eigenvalue, the solution has the form $O^{(k)}=\bar{O}_{h}^{(k)}\left(\left|u_{h}^{(k)}\right\rangle \left\langle u_{h}^{(k)}\right|+O^{(k-1)}\right)\bar{O}_{g}^{(k)}$.
Finally, in one of the ``infinite'' cases denoted by the ``Wiggle-v
method'' the solution will have the form $O^{(k)}=\left[\left(\left|u_{h}^{(k)}\right\rangle \left\langle u_{h}^{(k)}\right|+O^{(k-1)}\right)\bar{O}^{(k)}\right]^{T}$.} \bigskip{}
\end{itemize}
\item Algorithm:\\
\bigskip{}

\textcolor{purple}{If we reach a stage where the vector constraints
have disappeared then we can simply stop.}\\
\smallskip{}

\begin{itemize}
\item \textbf{Boundary condition:} \textbf{If} $n_{g}=0$ and $n_{h}=0$
\textbf{then} set $k_{0}=k$ and \textbf{jump to} phase 3.\\
\smallskip{}
\\
\textcolor{blue}{\textcolor{blue}{We assumed that an $O^{(k)}$ satisfying the constraints
(listed right after the input/output section) exists. In this case
it means that there exists an $O^{(k)}$ such that $X_{h}^{(k)}\ge O^{(k)}X_{g}^{(k)}O^{(k)T}$
as there is no vector $\left|v^{(k)}\right\rangle ,\left|w^{(k)}\right\rangle $
to impose further constraints. Using \Corref{WeylCorollary} with
$H=X_{h}^{(k)}$ and $G=O^{(k)}X_{g}^{(k)}O^{(k)T}$ we conclude that
$O^{(k)}$ need only be a permutation matrix. Note that this step
can never be entered right after the $\text{\ensuremath{\underbar{X}}}^{(n)}$
instance as we start with assuming $n_{g},n_{h}>0$. Further, since
the protocol by construction always returns $X_{h}$ and $X_{g}$
in the ascending order the permutation matrix will be $\mathbb{I}$. }}\\
\bigskip{}
\end{itemize}
\textcolor{purple}{Finally, we deal with the interesting case. We
again use the picture where the $H$ ellipsoid is contained inside
the $G$ ellipsoid. We expand the $H$ ellipsoid (which corresponds
to shrinking the $H$ matrix) until it touches the $G$ ellipsoid
as before by working with the function $a$.}
\begin{itemize}
\item \textbf{Tighten}: Define $X_{h_{\gamma'}}^{(k)}:=\gamma'X^{(k)}$.
Let $\gamma$ be the largest root of $m(\gamma',\chi_{\gamma'}^{(k)},\xi_{\gamma'}^{(k)})$
for $a^{(k)}$ where $\chi_{\gamma'}^{(k)},\xi_{\gamma'}^{(k)}$ are
such that $[\chi_{\gamma'}^{(k)},\xi_{\gamma'}^{(k)}]$ is the smallest
interval containing $\text{spec}[X_{h_{\gamma'}}^{(k)}\oplus X_{g}^{(k)}]$.
Relabel $X_{h_{\gamma}}^{(k)}$ to $X_{h}^{(k)}$, $\chi_{\gamma}^{(k)}$
to $\chi^{(k)}$ and $\xi_{\gamma}^{(k)}$ to $\xi^{(k)}$ for notational
ease. Similarly relabel $a_{\gamma}^{(k)}$ to $a^{(k)}$, $h_{\gamma}^{(k)}$
to $h^{(k)}$, $l_{\gamma}^{(k)}$ to $l^{(k)}$. Update $x_{\min}$
and $x_{\max}$ to be such that $\text{supp}(a^{(k)})\in[x_{\min}^{(k)},x_{\max}^{(k)}]$
is the smallest such interval. Define $s^{(k)}:=1$.

\textcolor{blue}{\textcolor{blue}{Our burden again is to show that $m$ as a function
of $\gamma'$ has a root. Unlike the first tightening procedure this
time we know the spectrum of the matrices involved. Since we are given
(by assumption) that the matrix instance has a solution we know that
$l_{\gamma'=1}(\lambda)\ge0$ and $l_{\gamma'=1}^{1}\ge0$ for $\lambda\in\bar{\mathbb{R}}\backslash[\chi_{\gamma'=1}^{(k)},\xi_{\gamma'=1}^{(k)}]$
using \Lemref{GenValidIsEBRM}. We also know that $\chi_{\gamma'}^{(k)}>0$
which means that $a^{(k)}$ (as deduced from the function instance
of $\text{\ensuremath{\underbar{X}}}^{(k)}$) is a valid function.
This observation lets us conclude that $m$ as a function of $\gamma'$
has a root in the required range because the reasoning behind a similar
claim proved in \Lemref{tighteningNoSpect} goes through unchanged.}}

\textcolor{purple}{The tightening procedure guarantees we will be
able to find a $\lambda$ which corresponds to an operator monotone
such that after applying this function the ellipsoids, which we do
not even know completely yet, must touch along the $\left|w\right\rangle $
direction. This piece of information is key to reducing the problem
to a smaller instance of itself. Recall the picture with the $H$
ellipsoid contained inside the $G$ ellipsoid. If we know that they,
in addition, touch at some known point then it must be so that the
inner ellipsoid is more curved than the outer ellipsoid. When expressed
algebraically, this condition essentially becomes that requirement
that an ellipsoid $H^{(k-1)}$ that encodes the curvature of the ellipsoid
$H^{(k)}$ at the point of contact must be contained inside the corresponding
$G^{(k-1)}$ ellipsoid which encodes the curvature of the $G^{(k)}$
ellipsoid. The vector condition also reduces similarly. Subtleties
arise when $\lambda$ happens to have boundary values in its allowed
range as this yields infinities and this has an interesting consequence. }
\item \textbf{Honest align}: \textbf{If} $l^{1(k)}=0$ \textbf{then} define
$\eta=-\chi^{(k)}+1$
\[
X_{h}^{\prime(k)}:=X_{h}^{(k)}+\eta,\quad X_{g}^{\prime(k)}:=X_{g}+\eta.
\]
\textbf{Else}: Pick a root $\lambda$ of the function $l^{(k)}(\lambda')$
in the domain $\mathbb{R}\backslash(-\xi^{(k)},-\chi^{(k)})$. In
the following two cases we consider the function $f_{\lambda}$ on
$[\chi^{(k)},\xi^{(k)}]$.
\begin{itemize}
\item If $\lambda\neq-\chi^{(k)}$ then: Let $\eta=-f_{\lambda}(\chi^{(k)})+1$
where any positive constant could be chosen instead of $1$. Define
\[
X_{h}^{\prime(k)}:=f_{\lambda}(X_{h}^{(k)})+\eta,\quad X_{g}^{\prime(k)}:=f_{\lambda}(X_{g}^{(k)})+\eta.
\]
\item If $\lambda=-\chi^{(k)}$ then: Update $s^{(k)}=-1$. Let $\eta=-f_{\lambda}(\xi^{(k)})-1$
where any positive constant could be chosen instead of 1. Define 
\[
X_{h}^{\prime(k)}:=X_{g}^{\prime\prime(k)},\quad X_{g}^{\prime(k)}:=X_{h}^{\prime\prime(k)},
\]
where
\[
X_{h}^{\prime\prime(k)}:=-f_{\lambda}(X_{h}^{(k)})-\eta,\quad X_{g}^{\prime\prime(k)}:=-f_{\lambda}(X_{g}^{(k)})-\eta
\]
and make the replacement 
\begin{align*}
\left|v^{(k)}\right\rangle  & \to\left|w^{(k)}\right\rangle \\
\left|w^{(k)}\right\rangle  & \to\left|v^{(k)}\right\rangle .
\end{align*}
 \textcolor{gray}{}
\end{itemize}
\textcolor{purple}{If we have $\lambda=-\chi^{(k)}$ or $-\xi^{(k)}$
it means that at least one of the matrices (among $X_{g}^{(k)}$ and
$X_{h}^{(k)}$ under $f_{\lambda}$) would diverge. We must remove
eigenvalues common to both matrices as isolating the divergence makes
it easier to handle.}
\item \textbf{\textcolor{black}{Remove spectral collision: If}}\textcolor{black}{{}
$\lambda=-\chi^{(k)}$ or $\lambda=-\xi^{(k)}$ }\textbf{\textcolor{black}{then}}\textcolor{black}{{} }

\textcolor{purple}{If it so happens that the coordinate and the probability
associated is the same we must leave the associated vector unchanged
(up to a relabelling). The following simply formalises this procedure
and encodes the remaining non-trivial part into a problem of one less
dimension. }\begin{sloppy}\smallskip{}

\begin{enumerate}
\item \textbf{Idle point:} \textbf{If} for some $j',j$, we have $q_{g_{j'}}^{(k)}=q_{h_{j}}^{(k)}$
and $y_{g_{j'}}^{(k)}=y_{h_{j}}^{(k)}$ \textbf{then} the solution
is given by 
\begin{align*}
\left\{ \left|u_{h}^{(k)}\right\rangle ,\left|t_{h_{1}}^{(k)}\right\rangle ,\left|t_{h_{2}}^{(k)}\right\rangle ,\dots\left|t_{h_{k-1}}^{(k)}\right\rangle \right\} \overset{\text{componentwise}}{:=}\\
\left\{ \left|t_{h_{j}}^{(k+1)}\right\rangle ,\left|t_{h_{1}}^{(k+1)}\right\rangle ,\left|t_{h_{2}}^{(k+1)}\right\rangle ,\dots\left|t_{h_{j-1}}^{(k+1)}\right\rangle ,\left|t_{h_{j+1}}^{(k+1)}\right\rangle ,\dots\left|t_{h_{k}}^{(k+1)}\right\rangle \right\} ,
\end{align*}
\[
\bar{O}^{(k)}:=\sum_{i=1}^{k}\left|a_{i}\right\rangle \left\langle t_{h_{i}}^{(k+1)}\right|,
\]
where
\[
\left\{ \left|a_{1}\right\rangle ,\left|a_{2}\right\rangle \dots\left|a_{k}\right\rangle \right\} \overset{\text{componentwise}}{:=}
\]
\[
\begin{cases}
\begin{array}{c}
\Big\{\left|t_{h_{1}}^{(k+1)}\right\rangle ,\left|t_{h_{2}}^{(k+1)}\right\rangle ,\dots\left|t_{h_{j-1}}^{(k+1)}\right\rangle ,\left|t_{h_{j'}}^{(k+1)}\right\rangle ,\left|t_{h_{j}}^{(k+1)}\right\rangle ,\left|t_{h_{j+1}}^{(k+1)}\right\rangle ,\\
\dots,\left|t_{h_{j'-1}}^{(k+1)}\right\rangle ,\left|t_{h_{j'+1}}^{(k+1)}\right\rangle \dots\left|t_{h_{k}}^{(k+1)}\right\rangle \Big\}\\
\\
\end{array} & j<j'\\
\begin{array}{c}
\Big\{\left|t_{h_{1}}^{(k+1)}\right\rangle ,\left|t_{h_{2}}^{(k+1)}\right\rangle ,\dots\left|t_{h_{j'-1}}^{(k+1)}\right\rangle ,\left|t_{h_{j'+1}}^{(k+1)}\right\rangle \dots\\
\left|t_{h_{j-1}}^{(k+1)}\right\rangle ,\left|t_{h_{j'}}^{(k+1)}\right\rangle ,\left|t_{h_{j}}^{(k+1)}\right\rangle ,\left|t_{h_{j+1}}^{(k+1)}\right\rangle \dots\left|t_{h_{k}}^{(k+1)}\right\rangle \Big\}\\
\\
\end{array} & j>j'\\
\left\{ \left|t_{h_{1}}^{(k+1)}\right\rangle ,\left|t_{h_{2}}^{(k+1)}\right\rangle ,\dots\left|t_{h_{k}}^{(k+1)}\right\rangle \right\}  & j=j',
\end{cases}
\]
and  
\[
X_{h}^{(k-1)}:=\sum_{i\neq j}y_{h_{i}}^{(k)}\left|t_{h_{i}}^{(k+1)}\right\rangle \left\langle t_{h_{i}}^{(k+1)}\right|,
\]
\[
X_{g}^{(k-1)}:=\bar{O}^{(k)}X_{g}^{(k)}\bar{O}^{(k)T}-y_{h_{j}}\left|t_{h_{j}}^{(k+1)}\right\rangle \left\langle t_{h_{j}}^{(k+1)}\right|,
\]
 
\[
\left|w^{(k-1)}\right\rangle =\mathcal{N}\left[\left|w^{(k)}\right\rangle -\sqrt{p_{h_{j}}}\left|t_{h_{j}}^{(k+1)}\right\rangle \right],\,\left|v^{(k-1)}\right\rangle =\mathcal{N}\left[\bar{O}^{(k)}\left|v^{(k)}\right\rangle -\sqrt{p_{h_{j}}}\left|t_{h_{j}}^{(k+1)}\right\rangle \right].
\]
\textcolor{gray}{}\\
\textcolor{gray}{(This specifies $\text{\ensuremath{\underbar{X}}}^{(k-1)}:=\{X_{h}^{(k-1)},X_{g}^{(k-1)},\left|w^{(k-1)}\right\rangle ,\left|v^{(k-1)}\right\rangle \}$.)} \textbf{}\\
\textbf{Jump} to \textbf{End}.

\smallskip{}
\textcolor{blue}{\begin{sloppy}In this proof by $x_{h_{i}}$ we mean $y_{h_{i}}$,
similarly by $x_{g_{i}}$ we mean $y_{h_{i}}$; we apologise for the
inconvenience. We want to find an $O^{(k)}$ such that $X_{h}^{(k)}\ge O^{(k)}X_{g}^{(k)}O^{(k)T}$
and $O^{(k)}\left|v^{(k)}\right\rangle =\left|w^{(k)}\right\rangle $.
We do this in two stages. First, we re-arrange the entries of $X_{g}^{(k)}$
as $X_{g}^{\prime(k)}:=O_{p}^{(k)}X_{g}^{(k)}O_{p}^{(k)T}$ and define
$\left|v_{p}^{(k)}\right\rangle :=O_{p}^{(k)}\left|v\right\rangle $
for an $O_{p}^{(k)}$ to be specified later. The re-arrangement will
be such that $x_{g_{j'}}$ sits at the $j,j$ location while the rest
of the elements of $X_{g}^{\prime(k)}$ are arranged in the increasing
order. Second, we solve our initial problem under the assumption that
$j=j'$. The non-trivial part here would be showing that we can take
$O^{(k)}$ to have the form $\left(\left|j\right\rangle \left\langle j\right|+O^{(k-1)}\right)\bar{O}^{(k)}$
without loss of generality.\\
Let us start with the first step. We denote the orthogonal matrix
$O=\sum_{i}\left|b_{i}\right\rangle \left\langle a_{i}\right|$ by
$\left\{ \left|a_{1}\right\rangle ,\left|a_{2}\right\rangle ,\dots\left|a_{k}\right\rangle \right\} \to\left\{ \left|b_{1}\right\rangle ,\left|b_{2}\right\rangle ,\dots\left|b_{k}\right\rangle \right\} $
where $\left\{ \left|b_{i}\right\rangle \right\} $ and $\left\{ \left|a_{i}\right\rangle \right\} $
each constitute an orthonormal basis. Using this notation then for
the case $j<j'$, we define $O_{p}^{(k)}$ by 
\begin{align*}
\left\{ \left|t_{h_{1}}^{(k+1)}\right\rangle ,\left|t_{h_{2}}^{(k+1)}\right\rangle ,\dots\left|t_{h_{k}}^{(k+1)}\right\rangle \right\} \to\\
\left\{ \left|t_{h_{1}}^{(k+1)}\right\rangle ,\left|t_{h_{2}}^{(k+1)}\right\rangle ,\dots\left|t_{h_{j-1}}^{(k+1)}\right\rangle ,\left|t_{h_{j'}}^{(k+1)}\right\rangle ,\left|t_{h_{j}}^{(k+1)}\right\rangle ,\left|t_{h_{j+1}}^{(k+1)}\right\rangle ,\dots,\left|t_{h_{j'-1}}^{(k+1)}\right\rangle ,\left|t_{h_{j'+1}}^{(k+1)}\right\rangle \dots\left|t_{h_{k}}^{(k+1)}\right\rangle \right\}  & ,
\end{align*}
for $j'<j$ we define it by
\begin{align*}
\left\{ \left|t_{h_{1}}^{(k+1)}\right\rangle ,\left|t_{h_{2}}^{(k+1)}\right\rangle ,\dots\left|t_{h_{k}}^{(k+1)}\right\rangle \right\} \to\\
\left\{ \left|t_{h_{1}}^{(k+1)}\right\rangle ,\left|t_{h_{2}}^{(k+1)}\right\rangle ,\dots\left|t_{h_{j'-1}}^{(k+1)}\right\rangle ,\left|t_{h_{j'+1}}^{(k+1)}\right\rangle \dots\left|t_{h_{j-1}}^{(k+1)}\right\rangle ,\left|t_{h_{j'}}^{(k+1)}\right\rangle ,\left|t_{h_{j}}^{(k+1)}\right\rangle ,\left|t_{h_{j+1}}^{(k+1)}\right\rangle \dots\left|t_{h_{k}}^{(k+1)}\right\rangle \right\} 
\end{align*}
and if $j'=j$ we set $O_{p}^{(k)}=\mathbb{I}^{(k)}$. \\
For the second step, we solve the main problem under the assumption
that $j'=j$. We are given $X_{g}^{'(k)}=\text{diag}\{x_{g_{1}}',x_{g_{2}}'\dots x_{g_{k}}'\}$
and $X_{h}^{(k)}=\text{diag}\{x_{h_{1}},x_{h_{2}}\dots x_{h_{k}}\}$
which are such that $x_{h_{j}}=x_{g_{j}}^{\prime}$; $\left|v^{\prime(k)}\right\rangle \doteq(\sqrt{q'_{g_{1}}},\sqrt{q'_{g_{2}}},\dots\sqrt{q'_{g_{k}}})^{T}$
and $\left|w^{(k)}\right\rangle \doteq(\sqrt{q_{h_{1}}},\sqrt{q_{h_{2}}},\dots\sqrt{q_{h_{k}}})^{T}$
are such that $q_{h_{j}}=q_{g_{j}}'$. Let us define the matrix instance
to be $\text{\ensuremath{\underbar{X}}}^{\prime(k)}=\{X_{h}^{(k)},X_{g}^{\prime(k)},\left|v^{\prime(k)}\right\rangle ,\left|w^{(k)}\right\rangle \}$.
We have to find an $O^{\prime(k)}$ such that $X_{h}^{(k)}\ge O^{\prime(k)}X_{g}^{\prime(k)}O^{\prime(k)T}$
and $O^{\prime(k)}\left|v^{\prime(k)}\right\rangle =\left|w\right\rangle $.
Let $\text{\ensuremath{\underbar{X}}}^{\prime(k-1)}=\left\{ X_{h}^{(k-1)},X_{g}^{\prime(k-1)},\left|v^{\prime(k-1)}\right\rangle ,\left|w^{(k-1)}\right\rangle \right\} $
be the matrix instance obtained after removing the $j^{\text{th}}$
entry from the vectors, viz. $\left|v^{\prime(k-1)}\right\rangle :=\sum_{i\neq j}\sqrt{q_{g_{i}}'}\left|t_{h_{i}}^{(k+1)}\right\rangle $,
$\left|w^{(k-1)}\right\rangle :=\sum_{i\neq j}\sqrt{q_{h_{i}}}\left|t_{h_{i}}^{(k+1)}\right\rangle $
and similarly defining $X_{g}^{\prime(k-1)}=\text{diag}\{x_{g_{1}}^{\prime},x_{g_{2}}^{\prime}\dots x_{g_{j-1}}^{\prime},x'_{g_{j+1}},\dots x_{g_{k}}\}$,
$X_{h}^{(k-1)}=\text{diag}\left\{ x_{h_{1}},x_{h_{2}}\dots x_{h_{j-1}},x_{h_{j+1}},\dots x_{h_{k}}\right\} $.
Note that $a^{(k)}=a^{(k-1)}$ as the $j^{\text{th}}$ point gets
cancelled. This means that if there is an $O^{\prime(k)}$ satisfying
the aforementioned constraints $a^{(k)}$ is EBRM on the spectral
domain of $\text{\ensuremath{\underbar{X}}}^{(k)}$. Since $a^{(k)}=a^{(k-1)}$
we know that $a^{(k-1)}$ is also EBRM on the same domain. From \Lemref{EBRMfunIsEBRMtrans_evenWithCommon}
(we will justify that $k$ is large enough separately) we conclude
that there must also exist an $O^{\prime(k-1)}$ which satisfies $X_{h}^{(k-1)}\ge O^{\prime(k-1)}X_{g}^{\prime(k-1)}O^{\prime(k-1)T}$
and $O^{\prime(k-1)}\left|v^{\prime(k-1)}\right\rangle =\left|w^{(k)}\right\rangle $.
\\
With all this in place we can claim that without loss of generality
we can write $O^{\prime(k)}=\left|t_{h_{j}}\right\rangle \left\langle t_{h_{j}}\right|+O^{\prime(k-1)}$
because if we can find some other $\tilde{O}^{\prime(k)}$ which satisfies
the required constraints then there exists an $O^{\prime(k-1)}$ which
satisfies the corresponding constraints in the smaller dimension and
that means we can show $O^{\prime(k)}$ also satisfies the required
constraints,
\begin{align*}
X_{h}^{(k)}=x_{h_{j}}\left|t_{h_{j}}^{(k+1)}\right\rangle \left\langle t_{h_{j}}^{(k+1)}\right|+X_{h}^{(k-1)} & \ge\\
x_{g_{j}}\left|t_{h_{j}}^{(k+1)}\right\rangle \left\langle t_{h_{j}}^{(k+1)}\right|+O^{\prime(k-1)}X_{g}^{\prime(k-1)}O^{\prime(k-1)} & =\\
\left(\left|t_{h_{j}}^{(k+1)}\right\rangle \left\langle t_{h_{j}}^{(k+1)}\right|+O^{\prime(k-1)}\right)X_{g}^{\prime(k)}\left(\left|t_{h_{j}}^{(k+1)}\right\rangle \left\langle t_{h_{j}}^{(k+1)}\right|+O^{\prime(k-1)}\right)^{T} & =\\
O^{\prime(k)}X_{g}^{\prime(k)}O^{\prime(k)T},
\end{align*}
along with 
\[
O^{\prime(k)}\left|v^{\prime(k)}\right\rangle =\sqrt{q'_{g_{j}}}\left|t_{h_{j}}^{(k+1)}\right\rangle +O^{\prime(k-1)}\left|v^{\prime(k-1)}\right\rangle =\sqrt{q'_{g_{j}}}\left|t_{h_{j}}^{(k+1)}\right\rangle +\left|w^{(k-1)}\right\rangle =\left|w^{(k-1)}\right\rangle .
\]
\\
It remains to combine the two steps to produce the matrix $\bar{O}^{(k)}$,
the vectors $\left\{ \left|n_{h}^{(k)}\right\rangle ,\left\{ \left|t_{h_{i}}^{(k)}\right\rangle \right\} \right\} $,
along with $\text{\ensuremath{\underbar{X}}}^{(k-1)}$. We use $X_{g}^{\prime(k)}=O_{p}^{(k)}X_{g}^{(k)}O_{p}^{(k)T}$
from the first step and substitute it in the inequality which we showed
would hold, i.e. 
\[
X_{h}^{(k)}\ge O^{\prime(k)}X_{g}^{\prime(k)}O^{\prime(k)T}=O^{\prime(k)}O_{p}^{(k)}X_{g}O_{p}^{(k)T}O^{\prime(k)T}
\]
and using $O_{p}^{(k)}\left|v^{(k)}\right\rangle =\left|v^{\prime(k)}\right\rangle $
we have 
\[
O^{\prime(k)}\left|v^{\prime(k)}\right\rangle =O^{\prime(k)}O_{p}^{(k)}\left|v^{(k)}\right\rangle =\left|w^{(k)}\right\rangle .
\]
Comparing the inequality to the form $X_{h}^{(k)}\ge O^{(k)}X_{g}^{(k)}O^{(k)T}$,
$O^{(k)}\left|v^{(k)}\right\rangle =\left|w^{(k)}\right\rangle $
for 
\[
O^{(k)}=\left(\left|n_{h}^{(k)}\right\rangle \left\langle n_{h}^{(k)}\right|+O^{(k-1)}\right)\bar{O}^{(k)}
\]
 we get $\bar{O}^{(k)}=O_{p}^{(k)}$, $\left|n_{h}^{(k)}\right\rangle =\left|t_{h_{j}}^{(k+1)}\right\rangle $
and $O^{(k-1)}=O^{\prime(k-1)}$. Note that this $O^{(k)}$ is consistent
with comparing the equality with $O^{(k)}\left|v^{(k)}\right\rangle =\left|w^{(k)}\right\rangle $.
The basis for the sub-problem, i.e. the $(k-1)$ dimensional problem,
was the same as before except for the fact that we removed $\left|t_{h_{j}}^{(k+1)}\right\rangle $.
Thus we define $\left\{ \left|t_{h_{1}}^{(k)}\right\rangle ,\left|t_{h_{2}}^{(k)}\right\rangle \dots\left|t_{h_{k-1}}^{(k)}\right\rangle \right\} =\left\{ t_{h_{1}}^{(k+1)},t_{h_{2}}^{(k+1)}\dots t_{h_{j-1}}^{(k+1)},t_{h_{j+1}}^{(k+1)},\dots t_{h_{k}}^{(k+1)}\right\} $.
Identifying 
\[
\text{\ensuremath{\underbar{X}}}^{(k-1)}=\left\{ X_{h}^{(k-1)},X_{g}^{(k-1)},\left|v^{(k-1)}\right\rangle ,\left|w^{(k-1)}\right\rangle \right\} 
\]
 with 
\[
\text{\ensuremath{\underbar{X}}}^{\prime(k-1)}=\left\{ X_{h}^{(k-1)},X_{g}^{\prime(k-1)},\left|v^{\prime(k-1)}\right\rangle ,\left|w^{(k-1)}\right\rangle \right\} 
\]
 completes the argument since $O^{(k-1)}$ was already identified
with $O^{\prime(k-1)}$ so we are just labelling here.\end{sloppy}}\bigskip{}

\item \textbf{\textcolor{black}{Final Extra}}\textcolor{black}{: }\textbf{\textcolor{black}{If}}\textcolor{black}{{}
for some $j,j'$ we have $q_{g_{j'}}^{(k)}>q_{h_{j}}^{(k)}$ and $y_{g_{j'}}^{(k)}=y_{h_{j}}^{(k)}$
}\textbf{\textcolor{black}{then}}\textcolor{black}{{} the solution is
given by} \textcolor{black}{{} }$\text{\ensuremath{\underbar{X}}}^{(k-1)}:=(X_{h}^{(k-1)},X_{g}^{(k-1)},\left|w^{(k-1)}\right\rangle ,\left|v^{(k-1)}\right\rangle )$
where $X_{h}^{(k-1)}=\sum_{i=1}^{k-1}y_{h_{i}}^{(k-1)}\left|t_{h_{i}}^{(k)}\right\rangle \left\langle t_{h_{i}}^{(k)}\right|$,
$X_{g}^{(k-1)}=\sum_{i=1}^{k-1}y_{g_{i}}^{(k-1)}\left|t_{h_{i}}^{(k)}\right\rangle \left\langle t_{h_{i}}^{(k)}\right|$,
$\left|v^{(k-1)}\right\rangle =\mathcal{N}\left[\sum_{i=1}^{k-1}\sqrt{q_{g_{i}}^{(k-1)}}\left|t_{h_{i}}^{(k)}\right\rangle \right]$,
$\left|w^{(k-1)}\right\rangle =\mathcal{N}\left[\sum_{i=1}^{k-1}\sqrt{q_{h_{i}}^{(k-1)}}\left|t_{h_{i}}^{(k)}\right\rangle \right]$
where the coordinates and weights are given by
\begin{align*}
\left\{ q_{h_{1}}^{(k-1)},\dots q_{h_{k-1}}^{(k-1)}\right\} \overset{\text{componentwise}}{=} & \left\{ q_{h_{1}}^{(k)},q_{h_{2}}^{(k)}\dots,q_{h_{j-1}}^{(k)},q_{h_{j+1}}^{(k)},\dots q_{h_{k}}^{(k)}\right\} \\
\left\{ q_{g_{1}}^{(k-1)},\dots q_{g_{k-1}}^{(k-1)}\right\} \overset{\text{componentwise}}{=} & \left\{ q_{g_{2}}^{(k)}\dots,q_{g_{j'-1}}^{(k)},q_{g_{j'}}^{(k)}-q_{h_{j}}^{(k)},q_{g_{j'+1}}^{(k)},q_{g_{j'+2}}^{(k)}\dots q_{g_{k}}^{(k)}\right\} \\
\left\{ y_{g_{1}}^{(k-1)},\dots y_{g_{k-1}}^{(k-1)}\right\} \overset{\text{componentwise}}{=} & \left\{ y_{g_{2}}^{(k)},\dots y_{g_{k}}^{(k)}\right\} \\
\left\{ y_{h_{1}}^{(k-1)},\dots y_{h_{k-1}}^{(k-1)}\right\} \overset{\text{componentwise}}{=} & \left\{ y_{h_{1}}^{(k)},\dots y_{h_{j-1}}^{(k)},y_{h_{j+1}}^{(k)}\dots,y_{h_{k}}^{(k)}\right\} ,
\end{align*}
the basis is given by 
\begin{align*}
\left\{ \left|u_{h}^{(k)}\right\rangle ,\left|t_{h_{1}}^{(k)}\right\rangle \dots\left|t_{h_{k-1}}^{(k)}\right\rangle \right\} \overset{\text{componentwise}}{=}\\
\left\{ \left|t_{h_{j}}^{(k+1)}\right\rangle ,\left|t_{h_{1}}^{(k+1)}\right\rangle ,\left|t_{h_{2}}^{(k+1)}\right\rangle ,\dots\left|t_{h_{j-1}}^{(k+1)}\right\rangle ,\left|t_{h_{j+1}}^{(k+1)}\right\rangle ,\left|t_{h_{j+2}}^{(k+1)}\right\rangle \dots\left|t_{h_{k}}^{(k+1)}\right\rangle \right\} .
\end{align*}
The orthogonal matrices are given by $\bar{O}_{h}^{(k)}:=\sum\left|t_{h_{i}}^{(k+1)}\right\rangle \left\langle a_{i}\right|$
where 
\[
\left\{ \left|a_{1}\right\rangle ,\dots\left|a_{k}\right\rangle \right\} \to\left\{ \left|u_{h}^{(k)}\right\rangle ,\left|t_{h_{1}}^{(k)}\right\rangle \dots\left|t_{h_{k-1}}^{(k)}\right\rangle \right\} ,
\]
$\bar{O}_{g}^{(k)}:=\tilde{O}^{(k)}\bar{O}_{h}^{(k)}$ where 
\begin{align*}
\tilde{O}^{(k)}:= & \mathcal{N}\left[\sqrt{q_{h_{j}}^{(k)}}\left|u_{h}^{(k)}\right\rangle +\sqrt{q_{g_{j'}}^{(k)}-q_{h_{j}}^{(k)}}\left|t_{h_{j'}}^{(k)}\right\rangle \right]\mathcal{N}\left[\sqrt{q_{g_{1}}^{(k)}}\left\langle u_{h}^{(k)}\right|+\sqrt{q_{g_{j'}}^{(k)}}\left\langle t_{h_{j'}}^{(k)}\right|\right]\\
 & +\mathcal{N}\left[\sqrt{q_{g_{j'}}^{(k)}-q_{h_{j}}^{(k)}}\left|u_{h}^{(k)}\right\rangle -\sqrt{q_{h_{j}}^{(k)}}\left|t_{h_{j'}}^{(k)}\right\rangle \right]\mathcal{N}\left[\sqrt{q_{g_{j'}}^{(k)}}\left\langle u_{h}^{(k)}\right|-\sqrt{q_{g_{1}}^{(k)}}\left\langle t_{h_{j'}}^{(k)}\right|\right]\\
 & +\sum_{i\in\{1,\dots k\}\backslash j'}\left|t_{h_{i}}^{(k)}\right\rangle \left\langle t_{h_{i}}^{(k)}\right|.
\end{align*}
\\
\textbf{Jump} to \textbf{End}.

\smallskip{}
\textcolor{blue}{\begin{sloppy} We are given $\text{\ensuremath{\underbar{X}}}^{(k)}=(X_{h}^{(k)},X_{g}^{(k)},\left|w^{(k)}\right\rangle ,\left|v^{(k)}\right\rangle )$
where $X_{h}^{(k)}=\sum_{i=1}^{k}y_{h_{i}}^{(k)}\left|t_{h_{i}}^{(k+1)}\right\rangle \left\langle t_{h_{i}}^{(k+1)}\right|$,
$X_{g}^{(k)}=\sum_{i=1}^{k}y_{g_{i}}^{(k)}\left|t_{h_{i}}^{(k+1)}\right\rangle \left\langle t_{h_{i}}^{(k+1)}\right|$,
$\left|v^{(k)}\right\rangle =\sum_{i=1}^{k}q_{g_{i}}^{(k)}\left|t_{h_{i}}^{(k+1)}\right\rangle $,
$\left|w^{(k)}\right\rangle =\sum_{i=1}^{k}q_{h_{i}}^{(k)}\left|t_{h_{i}}^{(k+1)}\right\rangle $
which means the corresponding function instance $\text{\ensuremath{\underbar{x}}}^{(k)}=(h^{(k)},g^{(k)},a^{(k)})$
where, in particular we have, $a^{(k)}=\sum_{i\in\{1,\dots k\}\backslash j}q_{h_{i}}^{(k)}[y_{h_{i}}]-\sum_{i\in\{1,\dots k\}\backslash j'}q_{g_{i}}^{(k)}[y_{g_{i}}]-(q_{g_{j'}}^{(k)}-q_{h_{j}}^{(k)})[y_{h_{j}}]$.
Since we assume $\text{\ensuremath{\underbar{X}}}^{(k)}$ has a solution
it follows that $a^{(k)}$ is $[\chi,\xi]$ valid. Thus the transition
$g^{(k-1)}:=a_{-}^{(k)}\to a_{+}^{(k)}=:h^{(k-1)}$ is also $[\chi,\xi]$
valid where $g^{(k-1)}$ comprises $n_{g}^{(k-1)}=n_{g}^{(k)}$ points
and $h^{(k-1)}$ comprises $n_{h}^{(k-1)}=n_{h}^{(k)}-1$ points (using
the attributes corresponding to the function instance $(h^{(k-1)},g^{(k-1)},h^{(k-1)}-g^{(k-1)})$;
The notation would be of the form $g=\sum_{i=1}^{n_{g}}p_{g_{i}}[x_{g_{i}}]$
and $h=\sum_{i=1}^{n_{h}}p_{h_{i}}[x_{h_{i}}]$). Since $k=n_{g}^{(k)}+n_{h}^{(k)}-1$
the aforesaid relation yields $k-1=n_{g}^{(k-1)}+n_{h}^{(k-1)}-1$.
We conclude that $\text{\ensuremath{\underbar{X}}}^{(k-1)}:=(X_{h}^{(k-1)},X_{g}^{(k-1)},\left|w^{(k-1)}\right\rangle ,\left|v^{(k-1)}\right\rangle )$
where $X_{h}^{(k-1)}=\sum_{i=1}^{k-1}y_{h_{i}}^{(k-1)}\left|t_{h_{i}}^{(k)}\right\rangle \left\langle t_{h_{i}}^{(k)}\right|$,
$X_{g}^{(k-1)}=\sum_{i=1}^{k-1}y_{g_{i}}^{(k-1)}\left|t_{h_{i}}^{(k)}\right\rangle \left\langle t_{h_{i}}^{(k)}\right|$,
$\left|v^{(k-1)}\right\rangle =\mathcal{N}\left[\sum_{i=1}^{k-1}\sqrt{q_{g_{i}}^{(k-1)}}\left|t_{h_{i}}^{(k)}\right\rangle \right]$,
$\left|w^{(k-1)}\right\rangle =\mathcal{N}\left[\sum_{i=1}^{k-1}\sqrt{q_{h_{i}}^{(k-1)}}\left|t_{h_{i}}^{(k)}\right\rangle \right]$
has a solution for 
\begin{align*}
\left\{ q_{h_{1}}^{(k-1)},\dots q_{h_{k-1}}^{(k-1)}\right\} \overset{\text{componentwise}}{=} & \left\{ q_{h_{1}}^{(k)},q_{h_{2}}^{(k)}\dots,q_{h_{j-1}}^{(k)},q_{h_{j+1}}^{(k)},\dots q_{h_{k}}^{(k)}\right\} \\
\left\{ q_{g_{1}}^{(k-1)},\dots q_{g_{k-1}}^{(k-1)}\right\} \overset{\text{componentwise}}{=} & \left\{ q_{g_{2}}^{(k)}\dots,q_{g_{j'-1}}^{(k)},q_{g_{j'}}^{(k)}-q_{h_{j}}^{(k)},q_{g_{j'+1}}^{(k)},q_{g_{j'+2}}^{(k)}\dots q_{g_{k}}^{(k)}\right\} \\
\left\{ y_{g_{1}}^{(k-1)},\dots y_{g_{k-1}}^{(k-1)}\right\} \overset{\text{componentwise}}{=} & \left\{ y_{g_{2}}^{(k)},\dots y_{g_{k}}^{(k)}\right\} \\
\left\{ y_{h_{1}}^{(k-1)},\dots y_{h_{k-1}}^{(k-1)}\right\} \overset{\text{componentwise}}{=} & \left\{ y_{h_{1}}^{(k)},\dots y_{h_{j-1}}^{(k)},y_{h_{j+1}}^{(k)}\dots,y_{h_{k}}^{(k)}\right\} 
\end{align*}
as the corresponding function instance $\text{\ensuremath{\underbar{x}}}^{(k-1)}$
is indeed given by $(h^{(k-1)},g^{(k-1)},a^{(k-1)}=a^{(k)})$. Here
$\{\left|t_{h_{i}}^{(k)}\right\rangle \}$ constitute an orthonormal
basis which we relate to $\left|t_{h_{i}}^{(k+1)}\right\rangle $
shortly. We used the fact that $q_{g_{1}}^{(k)}=0$ as $y_{g_{1}}^{(k)}=\chi$
(To see this note that $k-1>n_{g}^{(k-1)}$ which means that many
$q_{g_{i}}$ are zero; by convention we write the smallest eigenvalue,
$\chi$ first to increase the matrix size so the first $i=1,2\dots\left(k-1-n_{g}^{(k-1)}\right)$
$q_{i}$s are zero.). This means that there must exist an $O^{(k-1)}$
which solves $\text{\ensuremath{\underbar{X}}}^{(k-1)}$.\\
Let us take a moment to note the following basis change manoeuvre.
Note that $X'_{h}\ge O'X_{g}'O^{\prime T}$ with $O'\left|v'\right\rangle =\left|w'\right\rangle $
is equivalent to $X_{h}\ge OX_{g}O^{T}$ with $O\left|v\right\rangle =\left|w\right\rangle $
where $O=\bar{O}_{h}^{T}O'\bar{O}_{g}$, $\bar{O}_{g}\left|v\right\rangle =\left|v'\right\rangle $,
$\bar{O}_{h}\left|w\right\rangle =\left|w'\right\rangle $, $\bar{O}_{h}X_{h}\bar{O}_{h}^{T}=X_{h}'$,
$\bar{O}X_{g}\bar{O}_{g}^{T}=X_{g}'$ which is easy to see by a simple
substitution.\\
We first expand the matrix $\text{\ensuremath{\underbar{X}}}^{(k-1)}$
to $k$ dimensions as follows. We already had $X_{h}^{(k-1)}\ge O^{(k-1)}X_{g}^{(k-1)}O^{(k-1)T}$
with $O^{(k-1)}\left|v^{(k-1)}\right\rangle =\left|w^{(k-1)}\right\rangle $
which we expand as 
\begin{align*}
\underbrace{y_{h_{j}}^{(k)}\left|u_{h}^{(k)}\right\rangle \left\langle u_{h}^{(k)}\right|+X_{h}^{(k-1)}}_{:=X_{h}^{\prime(k)}} & \ge\\
\underbrace{\left(\left|u_{h}^{(k)}\right\rangle \left\langle u_{h}^{(k)}\right|+O^{(k-1)}\right)}_{:=O^{\prime(k)}}\underbrace{\left(y_{h_{j}}^{(k)}\left|u_{h}^{(k)}\right\rangle \left\langle u_{h}^{(k)}\right|+X_{g}^{(k-1)}\right)}_{:=X_{g}^{\prime(k)}}\left(\left|u_{h}^{(k)}\right\rangle \left\langle u_{h}^{(k)}\right|+O^{(k-1)}\right)^{T}
\end{align*}
 with $\left|v^{\prime(k)}\right\rangle =\mathcal{N}\left[\sqrt{q_{h_{j}}^{(k)}}\left|u_{h}^{(k)}\right\rangle +\left|v^{(k-1)}\right\rangle \right]$
and $\left|w'^{(k)}\right\rangle =\mathcal{N}\left[\sqrt{q_{h_{j}}^{(k)}}\left|u_{h}^{(k)}\right\rangle +\left|w^{(k-1)}\right\rangle \right]$.
Note that the matrix instance $\text{\ensuremath{\underbar{X}}}^{\prime(k)}:=(X_{h}^{\prime(k)},X_{g}^{\prime(k)},\left|v^{\prime(k)}\right\rangle ,\left|w^{\prime(k)}\right\rangle )$
yields $\text{\ensuremath{\underbar{x}}}^{\prime(k)}=\text{\ensuremath{\underbar{x}}}^{(k)}$.
We can now use the equivalence we pointed out above to establish a
relation between $X_{h}^{(k)}\ge O^{(k)}X_{g}^{(k)}O^{(k)T}$ and
$X_{h}^{\prime(k)}\ge O^{\prime(k)}X_{g}^{\prime(k)}O^{\prime(k)T}$
by finding $\bar{O}_{g}$ and $\bar{O}_{h}$. We define, somewhat
arbitrarily, 
\begin{align*}
\left\{ \left|u_{h}^{(k)}\right\rangle ,\left|t_{h_{1}}^{(k)}\right\rangle \dots\left|t_{h_{k-1}}^{(k)}\right\rangle \right\} \overset{\text{componentwise}}{=}\\
\left\{ \left|t_{h_{j}}^{(k+1)}\right\rangle ,\left|t_{h_{1}}^{(k+1)}\right\rangle ,\left|t_{h_{2}}^{(k+1)}\right\rangle ,\dots\left|t_{h_{j-1}}^{(k+1)}\right\rangle ,\left|t_{h_{j+1}}^{(k+1)}\right\rangle ,\left|t_{h_{j+2}}^{(k+1)}\right\rangle \dots\left|t_{h_{k}}^{(k+1)}\right\rangle \right\} .
\end{align*}
We require $\bar{O}_{h}^{(k)}\left|w^{(k)}\right\rangle $ to be $\left|w^{\prime(k)}\right\rangle $.
This is simply a permutation matrix given by $\left\{ \left|t_{h_{1}}^{(k+1)}\right\rangle ,\dots\left|t_{h_{k}}^{(k+1)}\right\rangle \right\} \to\left\{ \left|u_{h}^{(k)}\right\rangle ,\left|t_{h_{1}}^{(k)}\right\rangle \dots\left|t_{h_{k-1}}^{(k)}\right\rangle \right\} $.
Note that this yields $\bar{O}_{h}^{(k)T}X_{h}^{\prime(k)}\bar{O}_{h}^{(k)}=X_{h}^{(k)}$.
It remains to find $\bar{O}_{g}^{(k)}$ which we demand must satisfy
$\bar{O}_{g}^{(k)}\left|v^{(k)}\right\rangle =\left|v^{\prime(k)}\right\rangle $.
Observe first that $\bar{O}_{h}^{(k)}\left|v^{(k)}\right\rangle =\sqrt{q_{g_{1}}^{(k)}}\left|u_{h}^{(k)}\right\rangle +\sum_{i=2}^{k}\sqrt{q_{g_{i}}^{(k)}}\left|t_{h_{i-1}}^{(k)}\right\rangle $.
We must now apply 
\begin{align*}
\tilde{O}^{(k)}:= & \mathcal{N}\left[\sqrt{q_{h_{j}}^{(k)}}\left|u_{h}^{(k)}\right\rangle +\sqrt{q_{g_{j'}}^{(k)}-q_{h_{j}}^{(k)}}\left|t_{h_{j'}}^{(k)}\right\rangle \right]\mathcal{N}\left[\sqrt{q_{g_{1}}^{(k)}}\left\langle u_{h}^{(k)}\right|+\sqrt{q_{g_{j'}}^{(k)}}\left\langle t_{h_{j'}}^{(k)}\right|\right]\\
 & +\mathcal{N}\left[\sqrt{q_{g_{j'}}^{(k)}-q_{h_{j}}^{(k)}}\left|u_{h}^{(k)}\right\rangle -\sqrt{q_{h_{j}}^{(k)}}\left|t_{h_{j'}}^{(k)}\right\rangle \right]\mathcal{N}\left[\sqrt{q_{g_{j'}}^{(k)}}\left\langle u_{h}^{(k)}\right|-\sqrt{q_{g_{1}}^{(k)}}\left\langle t_{h_{j'}}^{(k)}\right|\right]\\
 & +\sum_{i\in\{1,\dots k\}\backslash j'}\left|t_{h_{i}}^{(k)}\right\rangle \left\langle t_{h_{i}}^{(k)}\right|
\end{align*}
to get $\bar{O}_{g}^{(k)}\left|v^{(k)}\right\rangle =\left|v^{\prime(k)}\right\rangle $
where we defined $\bar{O}_{g}^{(k)}:=\tilde{O}^{(k)}\bar{O}_{h}^{(k)}$.
(Note the expression could be simplified by using $q_{g_{1}}=0$ which
in fact is necessary for probability conservation.) Using $y_{h_{j}}^{(k)}=y_{g_{j'}}^{(k)}$
we can also see that $\bar{O}_{g}^{(k)T}X_{g}^{\prime(k)}\bar{O}_{g}^{(k)}$
is essentially $X_{g}^{(k)}$ with $\chi^{(k)}$ at $\left|t_{h_{1}}^{(k+1)}\right\rangle $
replaced by $y_{g_{j'}}(=y_{h_{j}})$. One can conclude therefore
that $X_{g}^{\prime(k)}\ge\bar{O}_{g}^{(k)}X_{g}^{(k)}\bar{O}_{g}^{(k)T}$.
Following the substitution manoeuvre we have
\begin{align*}
X_{h}^{\prime(k)} & \ge O^{\prime(k)}X_{g}^{\prime(k)}O^{\prime(k)T}\ge O^{\prime(k)}\bar{O}_{g}^{(k)}X_{g}^{(k)}\bar{O}_{g}^{(k)T}O^{\prime(k)T}\\
\iff\bar{O}_{h}^{(k)T}X_{h}^{\prime(k)}\bar{O}_{h}^{(k)} & \ge\underbrace{\bar{O}_{h}^{(k)T}O^{\prime(k)}\bar{O}_{g}^{(k)}}_{:=O^{(k)}}X_{g}^{(k)}\bar{O}_{g}^{(k)T}O^{\prime(k)T}\bar{O}_{h}^{(k)}\\
\iff X_{h}^{(k)} & \ge O^{(k)}X_{g}^{(k)}O^{(k)T}
\end{align*}
and similarly 
\begin{align*}
O^{\prime(k)}\left|v^{\prime(k)}\right\rangle  & =\left|w^{\prime(k)}\right\rangle \\
\iff O^{\prime(k)}\bar{O}_{g}^{(k)}\left|v^{(k)}\right\rangle  & =\bar{O}_{h}^{(k)}\left|w^{(k)}\right\rangle \\
\iff O^{(k)}\left|v^{(k)}\right\rangle  & =\left|w^{(k)}\right\rangle .
\end{align*}
This completes the proof.\end{sloppy}}\bigskip{}

\item \textbf{Initial Extra}: \textbf{If} for some $j,j'$ we have $q_{g_{j'}}^{(k)}<q_{h_{j}}^{(k)}$
and $y_{g_{j'}}^{(k)}=y_{h_{j}}^{(k)}$ \textbf{then} the solution
is given by $\text{\ensuremath{\underbar{X}}}^{(k-1)}:=(X_{h}^{(k-1)},X_{g}^{(k-1)},\left|w^{(k-1)}\right\rangle ,\left|v^{(k-1)}\right\rangle )$
where $X_{h}^{(k-1)}=\sum_{i=1}^{k-1}y_{h_{i}}^{(k-1)}\left|t_{h_{i}}^{(k)}\right\rangle \left\langle t_{h_{i}}^{(k)}\right|$,
$X_{g}^{(k-1)}=\sum_{i=1}^{k-1}y_{g_{i}}^{(k-1)}\left|t_{h_{i}}^{(k)}\right\rangle \left\langle t_{h_{i}}^{(k)}\right|$,
$\left|v^{(k-1)}\right\rangle =\mathcal{N}\left[\sum_{i=1}^{k-1}\sqrt{q_{g_{i}}^{(k-1)}}\left|t_{h_{i}}^{(k)}\right\rangle \right]$,
$\left|w^{(k-1)}\right\rangle =\mathcal{N}\left[\sum_{i=1}^{k-1}\sqrt{q_{h_{i}}^{(k-1)}}\left|t_{h_{i}}^{(k)}\right\rangle \right]$
where the coordinates and weights are given by
\begin{align*}
\left\{ q_{h_{1}}^{(k-1)},\dots q_{h_{k-1}}^{(k-1)}\right\} \overset{\text{componentwise}}{=} & \left\{ q_{h_{1}}^{(k)}\dots,q_{h_{j-1}}^{(k)},q_{h_{j}}^{(k)}-q_{g_{j'}}^{(k)},q_{h_{j+1}}^{(k)},q_{h_{j+2}}^{(k)}\dots q_{h_{k-1}}^{(k)}\right\} \\
\left\{ q_{g_{1}}^{(k-1)},\dots q_{g_{k-1}}^{(k-1)}\right\} \overset{\text{componentwise}}{=} & \left\{ q_{g_{1}}^{(k)},q_{g_{2}}^{(k)}\dots,q_{g_{j'-1}}^{(k)},q_{g_{j'+1}}^{(k)},\dots q_{g_{k}}^{(k)}\right\} \\
\left\{ y_{g_{1}}^{(k-1)},\dots y_{g_{k-1}}^{(k-1)}\right\} \overset{\text{componentwise}}{=} & \left\{ y_{g_{1}}^{(k)},\dots y_{g_{j'-1}}^{(k)},y_{g_{j'+1}}^{(k)}\dots,y_{g_{k}}^{(k)}\right\} \\
\left\{ y_{h_{1}}^{(k-1)},\dots y_{h_{k-1}}^{(k-1)}\right\} \overset{\text{componentwise}}{=} & \left\{ y_{h_{1}}^{(k)},\dots y_{h_{k-1}}^{(k)}\right\} ,
\end{align*}
the basis is given by 
\begin{align*}
\left\{ \left|u_{h}^{(k)}\right\rangle ,\left|t_{h_{1}}^{(k)}\right\rangle \dots\left|t_{h_{k-1}}^{(k)}\right\rangle \right\} \overset{\text{componentwise}}{=}\\
\left\{ \left|t_{h_{j}}^{(k+1)}\right\rangle ,\left|t_{h_{1}}^{(k+1)}\right\rangle ,\left|t_{h_{2}}^{(k+1)}\right\rangle ,\dots\left|t_{h_{j-1}}^{(k+1)}\right\rangle ,\left|t_{h_{j+1}}^{(k+1)}\right\rangle ,\left|t_{h_{j+2}}^{(k+1)}\right\rangle \dots\left|t_{h_{k}}^{(k+1)}\right\rangle \right\} .
\end{align*}
The orthogonal matrices are given by $\bar{O}_{h}^{(k)}:=\tilde{O}^{(k)}\sum\left|a_{i}\right\rangle \left\langle t_{h_{i}}^{(k+1)}\right|$
where 
\[
\left\{ \left|a_{1}\right\rangle ,\dots\left|a_{k}\right\rangle \right\} \overset{\text{componentwise}}{=}\left\{ \left|t_{h_{1}}^{(k)}\right\rangle ,\left|t_{h_{2}}^{(k)}\right\rangle \dots\left|t_{h_{k-1}}^{(k)}\right\rangle ,\left|u_{h}^{(k)}\right\rangle \right\} .
\]
 
\begin{align*}
\tilde{O}^{(k)}:= & \mathcal{N}\left[\sqrt{q_{g_{j'}}^{(k)}}\left|u_{h}^{(k)}\right\rangle +\sqrt{q_{h_{j}}^{(k)}-q_{g_{j'}}^{(k)}}\left|t_{h_{j}}^{(k)}\right\rangle \right]\mathcal{N}\left[\sqrt{q_{h_{k}}^{(k)}}\left\langle u_{h}^{(k)}\right|+\sqrt{q_{g_{j}}^{(k)}}\left\langle t_{h_{j}}^{(k)}\right|\right]\\
 & +\mathcal{N}\left[\sqrt{q_{h_{j}}^{(k)}-q_{g_{j'}}^{(k)}}\left|u_{h}^{(k)}\right\rangle -\sqrt{q_{g_{j'}}^{(k)}}\left|t_{h_{j}}^{(k)}\right\rangle \right]\mathcal{N}\left[\sqrt{q_{g_{j}}^{(k)}}\left\langle u_{h}^{(k)}\right|-\sqrt{q_{h_{k}}^{(k)}}\left\langle t_{h_{j}}^{(k)}\right|\right]\\
 & +\sum_{i\in\{1,\dots k\}\backslash j}\left|t_{h_{i}}^{(k)}\right\rangle \left\langle t_{h_{i}}^{(k)}\right|
\end{align*}
and $\bar{O}_{h}^{(k)}$ is given by the basis change $\left\{ \left|t_{h_{1}}^{(k+1)}\right\rangle ,\dots\left|t_{h_{k}}^{(k+1)}\right\rangle \right\} \to\left\{ \left|u_{h}^{(k)}\right\rangle ,\left|t_{h_{1}}^{(k)}\right\rangle \dots\left|t_{h_{k-1}}^{(k)}\right\rangle \right\} $.\textbf{}\\
\textbf{Jump} to \textbf{End}.
\end{enumerate}
\smallskip{}

\textcolor{blue}{\begin{sloppy} This proof will be very similar to the previous one.
We are given $\text{\ensuremath{\underbar{X}}}^{(k)}=(X_{h}^{(k)},X_{g}^{(k)},\left|w^{(k)}\right\rangle ,\left|v^{(k)}\right\rangle )$
where $X_{h}^{(k)}=\sum_{i=1}^{k}y_{h_{i}}^{(k)}\left|t_{h_{i}}^{(k+1)}\right\rangle \left\langle t_{h_{i}}^{(k+1)}\right|$,
$X_{g}^{(k)}=\sum_{i=1}^{k}y_{g_{i}}^{(k)}\left|t_{h_{i}}^{(k+1)}\right\rangle \left\langle t_{h_{i}}^{(k+1)}\right|$,
$\left|v^{(k)}\right\rangle =\sum_{i=1}^{k}q_{g_{i}}^{(k)}\left|t_{h_{i}}^{(k+1)}\right\rangle $,
$\left|w^{(k)}\right\rangle =\sum_{i=1}^{k}q_{h_{i}}^{(k)}\left|t_{h_{i}}^{(k+1)}\right\rangle $
which means the corresponding function instance $\text{\ensuremath{\underbar{x}}}^{(k)}=(h^{(k)},g^{(k)},a^{(k)})$
where, in particular we have, 
\[
a^{(k)}=\sum_{i\in\{1,\dots k\}\backslash j}q_{h_{i}}^{(k)}[y_{h_{i}}]+(q_{h_{j}}^{(k)}-q_{g_{j'}}^{(k)})[y_{h_{j}}]-\sum_{i\in\{1,\dots k\}\backslash j'}q_{g_{i}}^{(k)}[y_{g_{i}}].
\]
Since we assume $\text{\ensuremath{\underbar{X}}}^{(k)}$ has a solution
it follows that $a^{(k)}$ is $[\chi,\xi]$ valid. Thus the transition
$g^{(k-1)}:=a_{-}^{(k)}\to a_{+}^{(k)}=:h^{(k-1)}$ is also $[\chi,\xi]$
valid where $g^{(k-1)}$ comprises $n_{g}^{(k-1)}=n_{g}^{(k)}-1$
points and $h^{(k-1)}$ comprises $n_{h}^{(k-1)}=n_{h}^{(k)}$ points
(using the attributes corresponding to the function instance $(h^{(k-1)},g^{(k-1)},h^{(k-1)}-g^{(k-1)})$;
The notation would be of the form $g=\sum_{i=1}^{n_{g}}p_{g_{i}}[x_{g_{i}}]$
and $h=\sum_{i=1}^{n_{h}}p_{h_{i}}[x_{h_{i}}]$). Since $k=n_{g}^{(k)}+n_{h}^{(k)}-1$
the aforesaid relation yields $n_{g}^{(k-1)}+n_{h}^{(k-1)}-1=k-1$.
We conclude that $\text{\ensuremath{\underbar{X}}}^{(k-1)}:=(X_{h}^{(k-1)},X_{g}^{(k-1)},\left|w^{(k-1)}\right\rangle ,\left|v^{(k-1)}\right\rangle )$
where $X_{h}^{(k-1)}=\sum_{i=1}^{k-1}y_{h_{i}}^{(k-1)}\left|t_{h_{i}}^{(k)}\right\rangle \left\langle t_{h_{i}}^{(k)}\right|$,
$X_{g}^{(k-1)}=\sum_{i=1}^{k-1}y_{g_{i}}^{(k-1)}\left|t_{h_{i}}^{(k)}\right\rangle \left\langle t_{h_{i}}^{(k)}\right|$,
$\left|v^{(k-1)}\right\rangle =\mathcal{N}\left[\sum_{i=1}^{k-1}\sqrt{q_{g_{i}}^{(k-1)}}\left|t_{h_{i}}^{(k)}\right\rangle \right]$,
$\left|w^{(k-1)}\right\rangle =\mathcal{N}\left[\sum_{i=1}^{k-1}\sqrt{q_{h_{i}}^{(k-1)}}\left|t_{h_{i}}^{(k)}\right\rangle \right]$
have a solution for 
\begin{align*}
\left\{ q_{h_{1}}^{(k-1)},\dots q_{h_{k-1}}^{(k-1)}\right\} \overset{\text{componentwise}}{=} & \left\{ q_{h_{1}}^{(k)}\dots,q_{h_{j-1}}^{(k)},q_{h_{j}}^{(k)}-q_{g_{j'}}^{(k)},q_{h_{j+1}}^{(k)},q_{h_{j+2}}^{(k)}\dots q_{h_{k-1}}^{(k)}\right\} \\
\left\{ q_{g_{1}}^{(k-1)},\dots q_{g_{k-1}}^{(k-1)}\right\} \overset{\text{componentwise}}{=} & \left\{ q_{g_{1}}^{(k)},q_{g_{2}}^{(k)}\dots,q_{g_{j'-1}}^{(k)},q_{g_{j'+1}}^{(k)},\dots q_{g_{k}}^{(k)}\right\} \\
\left\{ y_{g_{1}}^{(k-1)},\dots y_{g_{k-1}}^{(k-1)}\right\} \overset{\text{componentwise}}{=} & \left\{ y_{g_{1}}^{(k)},\dots y_{g_{j'-1}}^{(k)},y_{g_{j'+1}}^{(k)}\dots,y_{g_{k}}^{(k)}\right\} \\
\left\{ y_{h_{1}}^{(k-1)},\dots y_{h_{k-1}}^{(k-1)}\right\} \overset{\text{componentwise}}{=} & \left\{ y_{h_{1}}^{(k)},\dots y_{h_{k-1}}^{(k)}\right\} ,
\end{align*}
as the corresponding function instance $\text{\ensuremath{\underbar{x}}}^{(k-1)}$
is indeed given by $(h^{(k-1)},g^{(k-1)},a^{(k-1)}=a^{(k)})$. Here
$\{\left|t_{h_{i}}^{(k)}\right\rangle \}$ constitute an orthonormal
basis which we relate to $\left|t_{h_{i}}^{(k+1)}\right\rangle $
shortly. We used the fact that $q_{h_{k}}^{(k)}=0$ as $y_{h_{k}}^{(k)}=\xi$.
(To see this note that $k-1>n_{h}^{(k-1)}$ which means that many
$q_{h_{i}}$ are zero; by convention we write the smallest eigenvalue,
$x_{h_{1}}$ first all the way till $x_{h_{n_{h}}}$ and then to increase
the matrix size we append zeros so the $i=n_{h},n_{h}+1\dots k$ yield
$q_{h_{i}}=0$.) This means that there must exist an $O^{(k-1)}$
which solves $\text{\ensuremath{\underbar{X}}}^{(k-1)}$. \\
Let us take a moment to note the following basis change manoeuvre.
$X'_{h}\ge O'X_{g}'O^{\prime T}$ with $O'\left|v'\right\rangle =\left|w'\right\rangle $
is equivalent to $X_{h}\ge OX_{g}O^{T}$ with $O\left|v\right\rangle =\left|w\right\rangle $
where $O=\bar{O}_{h}^{T}O'\bar{O}_{g}$, $\bar{O}_{g}\left|v\right\rangle =\left|v'\right\rangle $,
$\bar{O}_{h}\left|w\right\rangle =\left|w'\right\rangle $, $\bar{O}_{h}X_{h}\bar{O}_{h}^{T}=X_{h}'$,
$\bar{O}X_{g}\bar{O}_{g}^{T}=X_{g}'$ which is easy to see by a simple
substitution.\\
We first expand the matrix $\text{\ensuremath{\underbar{X}}}^{(k-1)}$
to $k$ dimensions as follows. We already had $X_{h}^{(k-1)}\ge O^{(k-1)}X_{g}^{(k-1)}O^{(k-1)T}$
with $O^{(k-1)}\left|v^{(k-1)}\right\rangle =\left|w^{(k-1)}\right\rangle $
which we expand as 
\begin{align*}
\underbrace{y_{h_{j}}^{(k)}\left|u_{h}^{(k)}\right\rangle \left\langle u_{h}^{(k)}\right|+X_{h}^{(k-1)}}_{:=X_{h}^{\prime(k)}}\ge\\
\underbrace{\left(\left|u_{h}^{(k)}\right\rangle \left\langle u_{h}^{(k)}\right|+O^{(k-1)}\right)}_{:=O^{\prime(k)}}\underbrace{\left(y_{h_{j}}^{(k)}\left|u_{h}^{(k)}\right\rangle \left\langle u_{h}^{(k)}\right|+X_{g}^{(k-1)}\right)}_{:=X_{g}^{\prime(k)}}\left(\left|u_{h}^{(k)}\right\rangle \left\langle u_{h}^{(k)}\right|+O^{(k-1)}\right)^{T}
\end{align*}
 with $\left|v^{\prime(k)}\right\rangle =\mathcal{N}\left[\sqrt{q_{g_{j'}}^{(k)}}\left|u_{h}^{(k)}\right\rangle +\left|v^{(k-1)}\right\rangle \right]$
and $\left|w'^{(k)}\right\rangle =\mathcal{N}\left[\sqrt{q_{g_{j'}}^{(k)}}\left|u_{h}^{(k)}\right\rangle +\left|w^{(k-1)}\right\rangle \right]$.
Note that the matrix instance $\text{\ensuremath{\underbar{X}}}^{\prime(k)}:=(X_{h}^{\prime(k)},X_{g}^{\prime(k)},\left|v^{\prime(k)}\right\rangle ,\left|w^{\prime(k)}\right\rangle )$
yields $\text{\ensuremath{\underbar{x}}}^{\prime(k)}=\text{\ensuremath{\underbar{x}}}^{(k)}$.
We can now use the equivalence we pointed out above to establish a
relation between $X_{h}^{(k)}\ge O^{(k)}X_{g}^{(k)}O^{(k)T}$ and
$X_{h}^{\prime(k)}\ge O^{\prime(k)}X_{g}^{\prime(k)}O^{\prime(k)T}$
by finding $\bar{O}_{g}$ and $\bar{O}_{h}$. We define, somewhat
arbitrarily, 
\begin{align*}
\left\{ \left|u_{h}^{(k)}\right\rangle ,\left|t_{h_{1}}^{(k)}\right\rangle \dots\left|t_{h_{k-1}}^{(k)}\right\rangle \right\} \overset{\text{componentwise}}{=}\\
\left\{ \left|t_{h_{j}}^{(k+1)}\right\rangle ,\left|t_{h_{1}}^{(k+1)}\right\rangle ,\left|t_{h_{2}}^{(k+1)}\right\rangle ,\dots\left|t_{h_{j-1}}^{(k+1)}\right\rangle ,\left|t_{h_{j+1}}^{(k+1)}\right\rangle ,\left|t_{h_{j+2}}^{(k+1)}\right\rangle \dots\left|t_{h_{k}}^{(k+1)}\right\rangle \right\} .
\end{align*}
We require $\bar{O}_{g}^{(k)}\left|v^{(k)}\right\rangle $ to be $\left|v^{\prime(k)}\right\rangle $.
This is simply a permutation matrix given by $\left\{ \left|t_{h_{1}}^{(k+1)}\right\rangle ,\dots\left|t_{h_{k}}^{(k+1)}\right\rangle \right\} \to\left\{ \left|u_{h}^{(k)}\right\rangle ,\left|t_{h_{1}}^{(k)}\right\rangle \dots\left|t_{h_{k-1}}^{(k)}\right\rangle \right\} $.
Note that this yields $\bar{O}_{g}^{(k)T}X_{g}^{\prime(k)}\bar{O}_{g}^{(k)}=X_{g}^{(k)}$
as $y_{h_{j}}^{(k)}=y_{g_{j'}}^{(k)}$. It remains to find $\bar{O}_{h}^{(k)}$
which we demand must satisfy $\bar{O}_{h}^{(k)}\left|w^{(k)}\right\rangle =\left|w^{\prime(k)}\right\rangle $.
Let us define $\bar{O}_{h}^{(k)}=\tilde{O}^{(k)}\left(\sum_{i=1}^{k}\left|a_{i}\right\rangle \left\langle t_{h_{i}}^{(k+1)}\right|\right)$.
Observe that for $\tilde{O}^{(k)}=\mathbb{I}$ we have $\bar{O}_{h}^{(k)}\left|w^{(k)}\right\rangle =q_{h_{k}}^{(k)}\left|u_{h}^{(k)}\right\rangle +\sum_{i=1}^{k-1}q_{h_{i}}^{(k)}\left|t_{h_{i}}^{(k)}\right\rangle $
where 
\[
\left\{ \left|a_{1}\right\rangle ,\dots\left|a_{k}\right\rangle \right\} \overset{\text{componentwise}}{=}\left\{ \left|t_{h_{1}}^{(k)}\right\rangle ,\left|t_{h_{2}}^{(k)}\right\rangle \dots\left|t_{h_{k-1}}^{(k)}\right\rangle ,\left|u_{h}^{(k)}\right\rangle \right\} .
\]
If we define 
\begin{align*}
\tilde{O}^{(k)}:= & \mathcal{N}\left[\sqrt{q_{g_{j'}}^{(k)}}\left|u_{h}^{(k)}\right\rangle +\sqrt{q_{h_{j}}^{(k)}-q_{g_{j'}}^{(k)}}\left|t_{h_{j}}^{(k)}\right\rangle \right]\mathcal{N}\left[\sqrt{q_{h_{k}}^{(k)}}\left\langle u_{h}^{(k)}\right|+\sqrt{q_{g_{j}}^{(k)}}\left\langle t_{h_{j}}^{(k)}\right|\right]\\
 & +\mathcal{N}\left[\sqrt{q_{h_{j}}^{(k)}-q_{g_{j'}}^{(k)}}\left|u_{h}^{(k)}\right\rangle -\sqrt{q_{g_{j'}}^{(k)}}\left|t_{h_{j}}^{(k)}\right\rangle \right]\mathcal{N}\left[\sqrt{q_{g_{j}}^{(k)}}\left\langle u_{h}^{(k)}\right|-\sqrt{q_{h_{k}}^{(k)}}\left\langle t_{h_{j}}^{(k)}\right|\right]\\
 & +\sum_{i\in\{1,\dots k\}\backslash j}\left|t_{h_{i}}^{(k)}\right\rangle \left\langle t_{h_{i}}^{(k)}\right|
\end{align*}
we get $\bar{O}_{h}^{(k)}\left|w^{(k)}\right\rangle =\left|w^{\prime(k)}\right\rangle $
as desired. We can also see that $\bar{O}_{h}^{(k)T}X_{g}^{\prime(k)}\bar{O}_{h}^{(k)}$
is essentially $X_{g}$ with $\xi^{(k)}$ at $\left|t_{h_{k}}^{(k+1)}\right\rangle $
replaced by $y_{h_{j}}$. We therefore conclude that $X_{g}^{\prime(k)}\ge\bar{O}_{g}^{(k)}X_{g}^{(k)}\bar{O}_{g}^{\prime(k)}$.
Following the substitution manoeuvre we have
\begin{align*}
X_{h}^{\prime(k)} & \ge O^{\prime(k)}X_{g}^{\prime(k)}O^{\prime(k)T}\ge O^{\prime(k)}\bar{O}_{g}^{(k)}X_{g}^{(k)}\bar{O}_{g}^{(k)T}O^{\prime(k)T}\\
\iff\bar{O}_{h}^{(k)T}X_{h}^{\prime(k)}\bar{O}_{h}^{(k)} & \ge\underbrace{\bar{O}_{h}^{(k)T}O^{\prime(k)}\bar{O}_{g}^{(k)}}_{:=O^{(k)}}X_{g}^{(k)}\bar{O}_{g}^{(k)T}O^{\prime(k)T}\bar{O}_{h}^{(k)}\\
\iff X_{h}^{(k)} & \ge O^{(k)}X_{g}^{(k)}O^{(k)T}
\end{align*}
and similarly 
\begin{align*}
O^{\prime(k)}\left|v^{\prime(k)}\right\rangle  & =\left|w^{\prime(k)}\right\rangle \\
\iff O^{\prime(k)}\bar{O}_{g}^{(k)}\left|v^{(k)}\right\rangle  & =\bar{O}_{h}^{(k)}\left|w^{(k)}\right\rangle \\
\iff O^{(k)}\left|v^{(k)}\right\rangle  & =\left|w^{(k)}\right\rangle .
\end{align*}
This completes the proof.\end{sloppy}}\end{sloppy}\bigskip{}

\item \textbf{Evaluate the Reverse Weingarten Map}: 
\begin{enumerate}
\item Consider the point $\left|w^{(k)}\right\rangle /\sqrt{\left\langle w^{(k)}\right|X_{h}^{\prime(k)}\left|w^{(k)}\right\rangle }$
on the ellipsoid $X_{h}^{\prime(k)}$. Evaluate the normal at this
point as $\left|u_{h}^{(k)}\right\rangle =\mathcal{N}\left(\sum_{i=1}^{n_{h}^{(k)}}\sqrt{p_{h_{i}}^{(k)}}x_{h_{i}}^{\prime(k)}\left|t_{h_{i}}^{(k+1)}\right\rangle \right)$.
Similarly evaluate $\left|u_{g}^{(k)}\right\rangle $, the normal
at the point $\left|v^{(k)}\right\rangle /\sqrt{\left\langle w^{(k)}\right|X_{g}^{\prime(k)}\left|w^{(k)}\right\rangle }$
on the ellipsoid $X_{g}^{\prime(k)}$. 
\item Recall that for a given diagonal matrix $X=\sum_{i}y_{i}\left|i\right\rangle \left\langle i\right|>0$
and normal vector $\left|u\right\rangle =\sum_{i}u_{i}\left|i\right\rangle $
the Reverse Weingarten map is given by $W_{ij}=\left(-\frac{y_{j}^{-1}y_{i}^{-1}u_{i}u_{j}}{r^{2}}+y_{i}^{-1}\delta_{ij}\right)$
where $r=\sqrt{\sum y_{i}^{-1}u_{i}^{2}}$. Evaluate the Reverse Weingarten
maps $W_{h}^{\prime(k)}$ and $W_{g}^{\prime(k)}$ along $\left|u_{h}^{(k)}\right\rangle $
and $\left|u_{g}^{(k)}\right\rangle $ respectively.\textcolor{blue}{{}
}
\item Find the eigenvectors and eigenvalues of the Reverse Weingarten maps.
The eigenvectors of $W_{h}'$ form the $h$ tangent (and normal) vectors
$\left\{ \left\{ \left|t_{h_{i}}^{(k)}\right\rangle \right\} ,\left|u_{h}^{(k)}\right\rangle \right\} $.
The corresponding radii of curvature are obtained from the eigenvalues
$\left\{ \{r_{h_{i}}^{(k)}\},0\right\} =\left\{ \{c_{h_{i}}^{(k)-1}\},0\right\} $
which are inverses of the curvature values. The tangents are labelled
in the decreasing order of radii of curvature (increasing order of
curvature). Similarly for the $g$ tangent (and normal) vectors. Fix
the sign freedom in the eigenvectors by requiring $\left\langle t_{h_{i}}^{(k)}|w^{(k)}\right\rangle \ge0$
and $\left\langle t_{g_{i}}^{(k)}|v^{(k)}\right\rangle \ge0$. 
\end{enumerate}
\item \textbf{Finite Method: If }$\lambda\neq-\xi^{(k)}$ and $\lambda\neq-\chi^{(k)}$,
i.e. if it is the finite case \textbf{then}
\begin{enumerate}
\item $\bar{O}^{(k)}:=\left|u_{h}^{(k)}\right\rangle \left\langle u_{g}^{(k)}\right|+\sum_{i=1}^{k-1}\left|t_{h_{i}}^{(k)}\right\rangle \left\langle t_{g_{i}}^{(k)}\right|$
\item $\left|v^{(k-1)}\right\rangle :=\bar{O}^{(k)}\left|v^{(k)}\right\rangle -\left\langle u_{h}^{(k)}\right|\bar{O}^{(k)}\left|v^{(k)}\right\rangle \left|u_{h}^{(k)}\right\rangle $
and $\left|w^{(k-1)}\right\rangle :=\left|w^{(k)}\right\rangle -\left\langle u_{h}^{(k)}|w^{(k)}\right\rangle \left|u_{h}^{(k)}\right\rangle $.
\item Define $X_{h}^{(k-1)}:=\text{diag}\{c_{h_{1}}^{(k)},c_{h_{2}}^{(k)}\dots,c_{h_{k-1}}^{(k)}\}$,
$X_{g}^{(k-1)}:=\text{diag}\{c_{g_{1}}^{(k)},c_{g_{2}}^{(k)}\dots c_{g_{k-1}}^{(k)}\}$.
\item \textbf{Jump} to \textbf{End}.

\begin{sloppy}

\textcolor{blue}{Our first burden is to prove that $O^{(k)}$ must
have the form $\left(\left|u_{h}^{(k)}\right\rangle \left\langle u_{h}^{(k)}\right|+O^{(k-1)}\right)\bar{O}^{(k)}$
for $\bar{O}^{(k)}:=\left|u_{h}^{(k)}\right\rangle \left\langle u_{g}^{(k)}\right|+\sum_{i=1}^{k-1}\left|t_{h_{i}}^{(k)}\right\rangle \left\langle t_{g_{i}}^{(k)}\right|$
if $O^{(k)}$ is to be a solution of the matrix instance $\text{\ensuremath{\underbar{X}}}^{(k)}$.
This is best explained by imagining that Arthur is trying to find
the orthogonal matrix and Merlin already knows the orthogonal matrix
but has still been following the steps performed so far. Recall that
we are now at a point where
\begin{align*}
\sum a'(x)x & =\left\langle w\right|X_{h}'\left|w\right\rangle -\left\langle v\right|X_{g}'\left|v\right\rangle \\
 & =\left\langle w\right|X_{h}'\left|w\right\rangle -\left\langle w\right|OX_{g}'O^{T}\left|w\right\rangle \\
 & =0.
\end{align*}
From Merlin's point of view along the $\left|w\right\rangle $ direction
the ellipsoids $X_{h}'$ and $OX_{g}'O^{T}$ touch. Suppose he started
with the ellipsoids $X'_{h},X_{g}'$ and only subsequently rotated
the second one. He can mark the point along the direction $\left|v\right\rangle $
on the $X_{g}'$ ellipsoid as the point that would after rotation
touch the $X_{h}'$ ellipsoid because as $X'_{g}\to OX_{g}'O^{T}$
the point along the $\left|v\right\rangle $ direction would get mapped
to the point along the direction $O\left|v\right\rangle =\left|w\right\rangle $.
Now, since the ellipsoids touch it must be so, Merlin deduces, that
the normal of the ellipsoid $X_{g}'$ at the point $\left|v\right\rangle /\sqrt{\left\langle v\right|X_{g}'\left|v\right\rangle }$
is mapped to the normal of the ellipsoid $X_{h}'$ at the point $\left|w\right\rangle /\sqrt{\left\langle w\right|X_{h}'\left|w\right\rangle }$
when $X_{g}'$ is rotated to $OX_{g}'O^{T}$, i.e. $O\left|u_{g}\right\rangle =\left|u_{h}\right\rangle $. }

\textcolor{blue}{From Arthur's point of view, who has been following
Merlin's reasoning, in addition to knowing that $O$ must satisfy
$O\left|v\right\rangle =\left|w\right\rangle $ he now knows that
it must also satisfy $O\left|u_{g}\right\rangle =\left|u_{h}\right\rangle $. }

\textcolor{blue}{Merlin further concludes that the curvature of the
$X_{g}'$ ellipsoid at the point $\left|v\right\rangle /\sqrt{\left\langle v\right|X'_{g}\left|v\right\rangle }$
must be more than the curvature of the $X_{h}'$ ellipsoid at the
point $\left|w\right\rangle /\sqrt{\left\langle w\right|X_{h}'\left|w\right\rangle }$.
To be precise, he needs to find a method for evaluating this curvature.
He knows that the brute-force way of doing this is to find a coordinate
system with its origin on the said point and then imagining the manifold,
locally, as a function from $n-1$ coordinates to one coordinate,
call it $x_{n}(x_{1},x_{2}\dots x_{n-1})$ (think of a sphere centred
at the origin; it can be thought of, locally, as a function from $x$
and $y$ to $z$ given by $z=\sqrt{x^{2}+y^{2}}$). The curvature
of this object is a generalisation of the second derivative which
forms a matrix with its elements given by $\partial_{x_{i}}\partial_{x_{j}}x_{n}$.
Since this matrix is symmetric he knows it can be diagonalised. The
directions of the eigenvectors of this matrix he calls the principle
directions of curvature where the curvature values are the corresponding
eigenvalues. He recalls that there is a simpler way of evaluating
these principle directions and curvatures which uses the Weingarten
map. The eigenvectors of the Reverse Weingarten map $W_{h}'$, evaluated
for $X_{h}'$ at $\left|w\right\rangle $, yield the normal and tangent
vectors with the corresponding eigenvalues zero and radii of curvature
respectively. Curvature is the inverse of the radius of curvature.
Similarly for the Reverse Weingarten map $W'_{g}$ evaluated for $X_{g}'$
at $\left|v\right\rangle $.}

\textcolor{blue}{With this knowledge Merlin deduces that he can write,
for some $\tilde{O}_{ij}\in\mathbb{R}$ such that $\sum_{j}\tilde{O}_{ij}\tilde{O}_{jk}=\delta_{ik}$,
\begin{align*}
O^{(k)} & =\left|u_{h}\right\rangle \left\langle u_{g}\right|+\sum_{i,j}\tilde{O}_{ij}\left|t_{h_{i}}\right\rangle \left\langle t_{g_{j}}\right|\\
 & =\left(\left|u_{h}\right\rangle \left\langle u_{h}\right|+\underbrace{\sum_{i,j}\tilde{O}_{ij}\left|t_{h_{i}}\right\rangle \left\langle t_{h_{j}}\right|}_{=O^{(k-1)}}\right)\left(\underbrace{\left|u_{h}\right\rangle \left\langle u_{g}\right|+\sum_{i}\left|t_{h_{i}}\right\rangle \left\langle t_{g_{i}}\right|}_{=\bar{O}^{(k)}}\right)
\end{align*}
where he re-introduced the superscript in the orthogonal operators.
He then turns to his intuition about the curvature of the smaller
ellipsoid being more than that of the larger ellipsoid. He observes
that equivalently, the radius of curvature of the smaller ellipsoid
must be smaller than that of the larger ellipsoid. To make this precise
he first notes that the Weingarten map $W_{g}'$ gets transformed
to $OW'_{g}O^{T}$ when $X_{g}'$ is rotated as $OX_{g}'O^{T}$. He
considers the point $\left|w\right\rangle /\sqrt{\left\langle w\right|X_{h}'\left|w\right\rangle }$,
which is shared by both the $X_{h}'$ and the $OX'_{g}O^{T}$ ellipsoid.
It must be so, he reasons, that along all directions in the tangent
plane, the $X_{h}'$ ellipsoid (the smaller one, remember larger $X_{h}'$
means smaller ellipsoid) must have a smaller radius of curvature than
the $OX_{g}'O^{T}$ ellipsoid, i.e. for all $\left|t\right\rangle \in\text{span}\{\left|t_{h_{i}}\right\rangle \}$,
$\left\langle t\right|W'_{h}\left|t\right\rangle \le\left\langle t\right|OW'_{g}O^{T}\left|t\right\rangle $.
Restricting his attention to the tangent space he deduces the statement
is equivalent to $W'_{h}\le OW'_{g}O^{T}$. He writes this out explicitly
as $\sum c_{h_{i}}^{-1}\left|t_{h_{i}}\right\rangle \left\langle t_{h_{i}}\right|\le\sum c_{g_{i}}^{-1}O\left|t_{g_{i}}\right\rangle \left\langle t_{g_{i}}\right|O^{T}$.
Now he uses the form of $O$ he had deduced to obtain $\sum c_{h_{i}}^{-1}\left|t_{h_{i}}\right\rangle \left\langle t_{h_{i}}\right|\le\sum c_{g_{i}}^{-1}O^{(k-1)}\left|t_{h_{i}}\right\rangle \left\langle t_{h_{i}}\right|O^{(k-1)T}$.
From this he is able to deduce that the inequality $X_{h}^{(k-1)}\ge O^{(k-1)}X_{g}^{(k-1)}O^{(k-1)T}$
must hold. }

\textcolor{blue}{Merlin's reasoning entails, Arthur summarises, that
$O^{(k)}$ must always have the form 
\[
O^{(k)}=\left(\left|u_{h}^{(k)}\right\rangle \left\langle u_{h}^{(k)}\right|+O^{(k-1)}\right)\bar{O}^{(k)}
\]
and that $O^{(k-1)}$ must satisfy the constraint 
\[
X_{h}^{(k-1)}\ge O^{(k-1)}X_{g}^{(k-1)}O^{(k-1)T}.
\]
Merlin, surprised by the similarity of the constraint he obtained
with the one he started with, extends his reasoning to the vector
itself. He knows that $O^{(k)}\left|v^{(k)}\right\rangle =\left|w^{(k)}\right\rangle $
but now he substitutes for $O^{(k)}$ to obtain $\left(\left|u_{h}^{(k)}\right\rangle \left\langle u_{h}^{(k)}\right|+O^{(k-1)}\right)\bar{O}^{(k)}\left|v^{(k)}\right\rangle =\left|w^{(k)}\right\rangle $.
He observes that $O^{(k-1)}$ can not influence the $\left|u_{h}^{(k)}\right\rangle $
component of the vector $\bar{O}^{(k)}\left|v^{(k)}\right\rangle $.
He thus projects out the $\left|u_{h}^{(k)}\right\rangle $ component
to obtain 
\[
O^{(k-1)}\underbrace{\left(\bar{O}^{(k)}\left|v^{(k)}\right\rangle -\left\langle u_{h}\right|\bar{O}^{(k)}\left|v^{(k)}\right\rangle \left|u_{h}\right\rangle \right)}_{=\left|v^{(k-1)}\right\rangle }=\underbrace{\left|w^{(k)}\right\rangle -\left\langle u_{h}^{(k)}|w^{(k)}\right\rangle \left|u_{h}^{(k)}\right\rangle }_{=\left|w^{(k-1)}\right\rangle }.
\]
With this, Arthur realises, he can reduce his problem involving a
$k$-dimensional orthogonal matrix into a smaller problem in $k-1$
dimensions with exactly the same form. Since Merlin's orthogonal matrix
was any arbitrary solution, and since the constraints involved do
not depend explicitly on the solution (only on the initial problem),
Arthur concludes that this reduction must hold for all possible solutions.}

\end{sloppy}
\end{enumerate}
\item \textbf{Wiggle-v Method: If }$\lambda=-\xi^{(k)}$ or $\lambda=-\chi^{(k)}$
\textbf{then}\\
\textcolor{purple}{The aforesaid method relies on matching the normals.
It works well so long as the correct operator monotone (the monotone
that yields $X'_{h}$ and $X'_{g}$ for which $\left|w\right\rangle /\sqrt{\left\langle w\right|X'_{h}\left|w\right\rangle }$
is a point on both $X_{h}'$ and $OX_{g}'O^{T}$) doesn't yield infinities.
If the operator monotone yields infinities it means that one of the
directions involved has infinite curvature which in turn means that
the component of the normal along this direction can be arbitrary.
To see this, imagine having a line contained inside an ellipsoid (both
centred at the origin) touching its boundaries. The line can be thought
of as an ellipse with infinite curvature along one of the directions.
The normal of the line at its tip is arbitrary and therefore we can't
require the usual condition that normals of the two curves must coincide.
The solution is to consider the sequence leading to the aforesaid
situation.}
\begin{enumerate}
\item $\left|u_{h}^{(k)}\right\rangle $ is renamed to $\left|\bar{u}_{h}^{(k)}\right\rangle $,
$\left|u_{g}^{(k)}\right\rangle $ remains the same.
\item Let $\tau=\cos\theta:=\left\langle u_{g}^{(k)}|v^{(k)}\right\rangle /\left\langle \bar{u}_{h}^{(k)}|w^{(k)}\right\rangle $.
Let $\left|\bar{t}_{h}^{(k)}\right\rangle $ be an eigenvector of
$X_{h}^{\prime(k)-1}$ with zero eigenvalue (comment: this is also
perpendicular to $\left|w^{(k)}\right\rangle $). Redefine 
\[
\left|u_{h}^{(k)}\right\rangle :=\cos\theta\left|\bar{u}_{h}^{(k)}\right\rangle +\sin\theta\left|\bar{t}_{h}^{(k)}\right\rangle ,
\]
\[
\left|t_{h_{k}}^{(k)}\right\rangle =s\left(-\sin\theta\left|\bar{u}_{h}^{(k)}\right\rangle +\cos\theta\left|\bar{t}_{h}^{(k)}\right\rangle \right)
\]
 where the sign $s\in\{1,-1\}$ is fixed by demanding $\left\langle t_{h_{k}}^{(k)}|w^{(k)}\right\rangle \ge0$. 
\item $\bar{O}^{(k)}$ and $\left|v^{(k-1)}\right\rangle ,\left|w^{(k-1)}\right\rangle $
are evaluated as step 1 and 2 of the finite case.
\item Define 
\[
X_{h}^{\prime(k-1)}:=\text{diag}\{c_{h_{1}}^{(k)},c_{h_{2}}^{(k)},\dots,c_{h_{k-1}}^{(k)}\},\qquad X_{g}^{\prime(k-1)}:=\text{diag}\{c_{g_{1}}^{(k)},c_{g_{2}}^{(k)},\dots,c_{g_{k-1}}^{(k)}\}.
\]
Let $[\chi^{\prime(k-1)},\xi^{\prime(k-1)}]$ denote the smallest
interval containing $\text{spec}[X_{h}^{\prime(k-1)}\oplus X_{g}^{\prime(k-1)}]$.
Let $\lambda'=-\chi'^{(k-1)}+1$ where instead of $1$ any positive
number would also work. Consider $f_{\lambda''}$ on $[\chi^{\prime(k-1)},\xi^{\prime(k-1)}]$.
Let $\eta=-f_{\lambda'}(\chi{}^{\prime(k-1)})+1$. Define 
\[
X_{h}^{(k-1)}:=f_{\lambda'}(X_{h}^{\prime(k-1)})+\eta,\qquad X_{g}^{(k-1)}:=f_{\lambda'}(X_{g}^{\prime(k-1)})+\eta.
\]
\item \textbf{Jump} to \textbf{End}.

\textcolor{blue}{We start with the case $\lambda=-\xi^{(k)}$. The
other case with $\lambda=-\chi^{(k)}$ follows analogously. For the
moment just imagine $\eta=0$ for simplicity; for $\eta\neq0$ the
argument goes through essentially unchanged. Note that because $\left\langle w\right|f_{-\xi}(X_{h})\left|w\right\rangle -\left\langle v\right|f_{-\xi}(X_{g})\left|v\right\rangle $
is zero we can conclude that $y_{h_{i}}^{(k)}=\xi$ implies $q_{h_{i}}=0$.
After the application of the map $f_{-\xi}$ these $y_{h_{i}}^{(k)}$s
and $y_{g_{i}}^{(k)}$s would become infinities but $\left\langle t_{h_{i}}^{(k+1)}|w\right\rangle $
and $\left\langle t_{g_{i}}^{(k+1)}|v\right\rangle $ would be zero
where we suppressed the superscripts for $\left|v^{(k)}\right\rangle $
and $\left|w^{(k)}\right\rangle $. Since the eigenvalues are arranged
in the ascending order in $X_{h}^{(k)}$ (in the $\left\{ \left|t_{h_{i}}^{(k+1)}\right\rangle \right\} $
basis) we have $y_{h_{k}}^{(k)}=\xi$ and the corresponding vector
is $\left|t_{h_{k}}^{(k+1)}\right\rangle =:\left|\bar{t}_{h}\right\rangle $.
It would be useful to define $\left|\tilde{t}_{h_{i}}\right\rangle =\left|t_{h_{i}}\right\rangle $
for $i=1,2,\dots j-1$ and $\left|\bar{t}_{h_{l}}\right\rangle =\left|t_{h_{i}}\right\rangle $
for $i=j,j+1,\dots k$, $l=\left(i-j\right)+1$ where $j$ is the
smallest $i$ for which $x_{h_{i}}=\xi$ (their existence is a straight
forward consequence of dimension counting, $k\ge n_{g}+n_{h}-1$).
This allows us to speak of the subspace with eigenvalue $\xi$ of
$X_{h}^{(k)}$ easily. We focus on the two dimensional plane spanned
by $\left|w\right\rangle $ and $\left|\bar{t}_{h}\right\rangle $.}

\textcolor{blue}{Consider the M-view (Merlin's point of view). Since
Merlin has a solution $O^{(k)}$ to the matrix instance 
\[
\text{\ensuremath{\underbar{X}}}^{(k)}=\{X_{h}^{(k)},X_{g}^{(k)},\left|w^{(k)}\right\rangle ,\left|v^{(k)}\right\rangle \}
\]
 his solution is also a solution to the matrix instance 
\[
\text{\ensuremath{\underbar{X}}}^{(k)}(\lambda):=\left\{ f_{\lambda}(X_{h}^{(k)}),f_{\lambda}(X_{g}^{(k)}),\left|w^{(k)}\right\rangle ,\left|v^{(k)}\right\rangle \right\} 
\]
 for $\lambda\le-\xi$ but close enough to $-\xi$ such that $f_{\lambda}(X_{h}),f_{\lambda}(X_{g})>0$.
This is a consequence of $f_{\lambda}$ being operator monotone. Using
\Corref{tightnessResult} and \Lemref{tightnessPreserveInfAlso} we
know that since the ellipsoids corresponding to the matrix instance
$\text{\ensuremath{\underbar{X}}}(-\xi)$ touch along $\left|w\right\rangle $
(as we are given that $\left\langle w\right|f_{-\xi}(X_{h})\left|w\right\rangle -\left\langle w\right|Of_{-\xi}(X_{g})O^{T}\left|w\right\rangle =\left\langle w\right|f_{-\xi}(X_{h})\left|w\right\rangle -\left\langle v\right|f_{-\xi}(X_{g})\left|v\right\rangle =0$)
there must also exist some vector $\left|c(\lambda)\right\rangle $
such that $\left\langle c(\lambda)\right|f_{\lambda}(X_{h})\left|c(\lambda)\right\rangle -\left\langle c(\lambda)\right|Of_{\lambda}(X_{g})O^{T}\left|c(\lambda)\right\rangle =0$
that is the ellipsoids corresponding to the matrix instance $\text{\ensuremath{\underbar{X}}}(\lambda)$
touch along the said direction. (Caution: Do not confuse $\left|c(\lambda)\right\rangle $
with $c_{h_{i}}/c_{g_{i}}$. The latter are used for curvature values
and the former refers to the contact vector just defined.) Note that
to match the other conditions of the lemma it suffices to assume that
$X_{h}$ and $X_{g}$ do not have a common eigenvalue which in turn
is guaranteed by the ``remove spectral collision'' part.}

\textcolor{blue}{It is easy to convince oneself that $\lim_{\lambda\to-\xi}\left|c(\lambda)\right\rangle =\left|w\right\rangle $
(hint: argue along the lines $f_{\lambda}(X_{h})$ is very close to
$f_{-\xi}(X_{h})$ and so the vectors should also be very close which
satisfy the condition). Note that we can write 
\[
\left|w\right\rangle =\sum_{i=1}^{j-1}q_{h_{i}}\left|\tilde{t}_{h_{i}}\right\rangle 
\]
because $\left\langle \bar{t}_{h_{i}}|w\right\rangle =0$. There is
no such restriction on $\left|c(\lambda)\right\rangle $ which can
have the more general form $\left|c(\lambda)\right\rangle =\sum_{i=1}^{j-1}c(\lambda)_{i}\left|\tilde{t}_{h_{i}}\right\rangle +\sum_{i=j}^{k}c(\lambda)_{i}\left|\bar{t}_{h_{l}}\right\rangle $
where $l=(i-j)+1$. Restating one of the limit conditions, for $i=j,j+1\dots k$,
we must have the $\lim_{\lambda\to-\xi}c(\lambda)_{i}=0$. At this
point we use the fact that if $O$ is a solution it entails that 
\[
\acute{O}(\lambda):=\left(\sum_{i=1}^{j-1}\left|\tilde{t}_{h_{i}}\right\rangle \left\langle \tilde{t}_{h_{i}}\right|+\sum_{i,m=1}^{k-j+1}Q(\lambda)_{im}\left|\bar{t}_{h_{i}}\right\rangle \left\langle \bar{t}_{h_{m}}\right|\right)O
\]
is also a solution, where $Q(\lambda)$ is an orthogonal matrix in
the space spanned by $\left\{ \left|\bar{t}_{h_{i}}\right\rangle \right\} $.
This is a consequence of the fact that $\{\left|\bar{t}_{h_{i}}\right\rangle \}$
spans an eigenspace (with the same eigenvalue, $f_{\lambda}(\xi)$,)
of $f_{\lambda}(X_{h})$. We can use this freedom to ensure that the
point of contact always has the form 
\[
\left|c(\lambda)\right\rangle =\sum_{i=1}^{j-1}c(\lambda)_{i}\left|\tilde{t}_{h_{i}}\right\rangle +\bar{c}(\lambda)\left|\bar{t}_{h}\right\rangle 
\]
where $\bar{c}(\lambda)=\sqrt{\sum_{i=j}^{k}c(\lambda)_{i}^{2}}$
which must vanish in the limit $\lambda\to-\xi$ as its constituents
disappear in the said limit. Similarly $\lim_{\lambda\to-\xi}c(\lambda)_{i}=q_{h_{i}}$.}

\textcolor{blue}{Next we evaluate the normals $\left|u_{h}(\lambda)\right\rangle $
at $\left|c(\lambda)\right\rangle $ for the ellipsoid represented
by $f_{\lambda}(X_{h})$ and similarly the normal $\left|\bar{u}_{h}\right\rangle $
at $\left|w\right\rangle $ for the ellipsoid represented by $f_{-\xi}(X_{h})$
to show that $\lim_{\lambda\to-\xi}\left|u_{h}(\lambda)\right\rangle \neq\left|\bar{u}_{h}\right\rangle $
(see \Figref{infiniteCurvature}). Notice that the right-most term
in $\left|u_{h}(\lambda)\right\rangle =\mathcal{N}\left[\sum_{i=1}^{j-1}f_{\lambda}(y_{h_{i}})c(\lambda)_{i}\left|\tilde{t}_{h_{i}}\right\rangle +f_{\lambda}(\xi)\bar{c}(\lambda)\left|\bar{t}_{h}\right\rangle \right]$
has $f_{\lambda}(\xi)$ approaching infinity and $\bar{c}(\lambda)$
approaching zero as $\lambda$ tends to $-\xi$. This is why it can
have a finite component along $\left|\bar{t}_{h}\right\rangle $.
On the other hand, $\left|\bar{u}_{h}\right\rangle =\mathcal{N}\left[\sum_{i=1}^{j-1}f_{-\xi}(y_{h_{i}})q_{h_{i}}\left|\tilde{t}_{h_{i}}\right\rangle \right]$
which has no component along $\left|\bar{t}_{h}\right\rangle $. Since
$\lim_{\lambda\to-\xi}f_{\lambda}(y_{h_{i}})=f_{-\xi}(y_{h_{i}})$
and $\lim_{\lambda\to-\xi}c(\lambda)_{i}=q_{h_{i}}$ for $i\in\{1,2\dots j-1\}$,
we can write 
\[
\lim_{\lambda\to-\xi}\left|u_{h}(\lambda)\right\rangle =\cos\theta\left|\bar{u}_{h}\right\rangle +\sin\theta\left|\bar{t}_{h}\right\rangle :=\left|u_{h}\right\rangle .
\]
Evidently, we must use $\left|u_{h}\right\rangle $ instead of $\left|\bar{u}_{h}\right\rangle $
to be able to use the reasoning of the finite method. However, we
do not know $\cos\theta$ yet. }

\textcolor{blue}{Our strategy is to proceed as in the finite method
with the assumption that $\left|c(\lambda)\right\rangle $ is known
(which it isn't as we only know it exists and how it behaves in the
limit of $\lambda\to-\xi$) and then use a consistency condition to
find $\cos\theta$ in terms of known quantities. At this point we
re-introduce the superscripts as we will reduce the dimension of the
problem as we proceed. Let the normal and tangents at $O^{T}\left|c(\lambda)\right\rangle $
for $f_{\lambda}(X_{g})$ be given by $\left\{ \left|u_{g}^{(k)}(\lambda)\right\rangle ,\left\{ t_{g_{i}}^{(k)}(\lambda)\right\} \right\} $.
Similarly at $\left|c(\lambda)\right\rangle $ for $f_{\lambda}(X_{h})$
the normal and tangents are $\left\{ \left|u_{h}^{(k)}(\lambda)\right\rangle ,\left\{ t_{h_{i}}^{(k)}(\lambda)\right\} \right\} $.
From the finite method we know that $O^{(k)}(\lambda):=\left(\left|u_{h}(\lambda)\right\rangle \left\langle u_{h}(\lambda)\right|+O^{(k-1)}\right)\bar{O}^{(k)}$
where $\bar{O}^{(k)}=\left|u_{h}^{(k)}(\lambda)\right\rangle \left\langle u_{g}^{(k)}(\lambda)\right|+\sum_{i}\left|t_{h_{i}}^{(k)}\right\rangle \left\langle t_{g_{i}}^{(k)}\right|$
can be used to reduce the problem into a smaller instance of itself.
In particular, we must have $\left\langle u_{h}^{(k)}(\lambda)|w\right\rangle =\left\langle u_{h}^{(k)}(\lambda)\right|O^{(k)}(\lambda)\left|v\right\rangle =\left\langle u_{g}^{(k)}(\lambda)|v\right\rangle $
because $O^{(k-1)}$ can influence only the subspace spanned by $\left\{ \left|t_{h_{i}}^{(k)}\right\rangle \right\} $
and the component of the vectors $\left|w\right\rangle $ and $O^{(k)}\left|v\right\rangle $
along $\left|u_{h}^{(k)}(\lambda)\right\rangle $ must match for consistency.}

\textcolor{blue}{We can determine $\cos\theta$ by taking the limit
of the aforesaid condition as $\left\langle u_{h}|w\right\rangle =\left\langle u_{g}|v\right\rangle $
where we again suppressed the superscripts. Substituting $\left|u_{h}\right\rangle =\cos\theta\left|\bar{u}_{h}\right\rangle +\sin\theta\left|\bar{t}_{h}\right\rangle $
we obtain 
\[
\cos\theta=\frac{\left\langle u_{g}|v\right\rangle }{\left\langle \bar{u}_{h}|w\right\rangle }.
\]
}

\textcolor{blue}{It now remains to find the limit of the reverse Weingarten
maps. The reverse Weingarten map for $f_{\lambda}(X_{g})$ along the
normal $\left|u_{g}(\lambda)\right\rangle $ is not of concern because
it has a well defined limit as $\lambda\to-\xi$. We consider the
case for $f_{\lambda}(X_{h})$ along the normal $\left|u_{h}(\lambda)\right\rangle $.
Note that the support function as defined in \Eqref{SuppFun} is finite
in the limit $\lambda\to-\xi$ (use the definition of the normal to
get $\sum x_{i}^{-1}u_{i}^{2}=\sum x_{i}^{-1}x_{i}^{2}c_{i}^{2}=\sum x_{i}c_{i}^{2}=\left\langle c\right|X\left|c\right\rangle $,
plug in $\left|c\right\rangle =\left|w\right\rangle $, $X=f_{-\xi}(X_{h})$
and then use the fact that $\left\langle w\right|f_{-\xi}(X_{h})\left|w\right\rangle -\left\langle v\right|f_{-\xi}(X_{g})\left|v\right\rangle =0$
which means both must be finite by noting that we already dealt with
the troublesome case of $\infty-\infty$ in the ``remove spectral
collision'' part). Let us denote it by $h(\lambda)$. Now the reverse
Weingarten map as defined in \Eqref{ReverseWeingarten} is given by
\[
\left(W_{h}(\lambda)\right)_{im}=-\frac{1}{h(\lambda)^{2}}\frac{u_{h_{i}}(\lambda)u_{h_{m}}(\lambda)}{f_{\lambda}(y_{h_{i}})f_{\lambda}(y_{h_{m}})}+\frac{\delta_{im}}{f_{\lambda}(x_{h_{i}})}.
\]
Since $\lim_{\lambda\to-\xi}\left|u(\lambda)\right\rangle $ is well
defined, $\lim_{\lambda\to-\xi}h(\lambda)$ is finite, we only need
to show that $\lim_{\lambda\to-\xi}1/f_{\lambda}(y_{h_{i}})$ is well
defined. (We assumed $\eta$ is zero so $f_{-\xi}(y_{h_{i}})\neq0$.
If $\eta$ is not zero we must consider $f_{-\xi}(y_{h_{i}})+\eta$
everywhere but that changes no argument.) For $i=1,2\dots j-1$, $f_{-\xi}(y_{h_{i}})$
is finite but for $i=j,j+1\dots k$, $f_{-\xi}(y_{h_{i}})$ is not
well defined however $1/f_{-\xi}(y_{h_{i}})=0$. We therefore conclude
that 
\[
\lim_{\lambda\to-\xi}\left(W_{h}(\lambda)\right)_{im}=\begin{cases}
-\frac{1}{h^{2}}\frac{u_{h_{i}}u_{h_{m}}}{f_{-\xi}(y_{h_{i}})f_{-\xi}(y_{h_{m}})}+\frac{\delta_{im}}{f_{-\xi}(y_{h_{i}})} & i,m\in\{1,2\dots j-1\}\\
0 & i,m\in\{j,j+1\dots k\}
\end{cases}:=\left(W_{h}\right)_{im}
\]
which is simply the reverse Weingarten map evaluated for $f_{-\xi}(X_{h})$
along $\left|u_{h}\right\rangle =\cos\theta\left|\bar{u}_{h}\right\rangle +\sin\theta\left|\bar{t}_{h}\right\rangle $
and $\cos\theta=\left\langle u_{g}|v\right\rangle /\left\langle \bar{u}_{h}|w\right\rangle $.
It remains to relate $W_{h}$ with the reverse Weingarten map, $\bar{W}_{h}$,
evaluated for $f_{-\xi}(X_{h})$ along $\left|\bar{u}_{h}\right\rangle $.
Surprisingly, it is easy to see that $W_{h}=\bar{W}_{h}$ because
only the $\cos\theta\left|\bar{u}_{h}\right\rangle $ part contributes
to the non-zero portion of $W_{h}$ and the $\cos\theta$ factor gets
cancelled due to the $h^{2}$ term. Further, recall that the normal
vector is always an eigenvector of the reverse Weingarten map evaluated
along it, with eigenvalue zero. This tells us that if there is(are)
tangent(s) with zero radius of curvature then the normal is not uniquely
defined. This confirms what we already knew. Now since both $\left|\bar{u}_{h}\right\rangle $,
$\left|\bar{t}_{h}\right\rangle $ have zero eigenvalues for $\bar{W}_{h}(=W_{h})$
and $\left|u\right\rangle =\cos\theta\left|\bar{u}_{h}\right\rangle +\sin\theta\left|\bar{t}_{h}\right\rangle $
we define $\left|t_{h}\right\rangle :=s\left(\sin\theta\left|\bar{u}_{h}\right\rangle -\cos\theta\left|\bar{t}_{h}\right\rangle \right)$
to span the same space so that $\left|u\right\rangle $ is the correct
normal vector (as we deduced earlier in our discussion) and $\left|t_{h}\right\rangle $
is the correct tangent vector corresponding to the point $\left|w\right\rangle $
of $f_{-\xi}(X_{h})$. }

\textcolor{blue}{The final step is to convert the condition on the
reverse Weingarten map into a condition on the Weingarten map (inverse
of the reverse Weingarten map). After extracting the tangent vectors
appropriately, one simply needs to add a constant before inverting
to obtain the Weingarten map condition. This is done in the last step.
This completes the proof of the wiggle-v method for $\lambda=-\xi$.}

\textcolor{blue}{To see how the same reasoning applies to the $\lambda=-\chi$
case first note that for $\lambda\ge-\chi$ we have $f_{\lambda}(X_{h}),f_{\lambda}(X_{g})<0$
(assuming $\eta=0$ as before). The condition $f_{\lambda}(X_{h})\ge Of_{\lambda}(X_{g})O^{T}$
can then be expressed as $-f_{\lambda}(X_{g})\ge-O^{T}f_{\lambda}(X_{h})O$
with $O^{T}\left|w\right\rangle =\left|v\right\rangle $ which can
now be reasoned analogous to the aforementioned analysis. }

\textcolor{blue}{}
\begin{figure}
\centering{}\textcolor{blue}{\includegraphics[width=14cm]{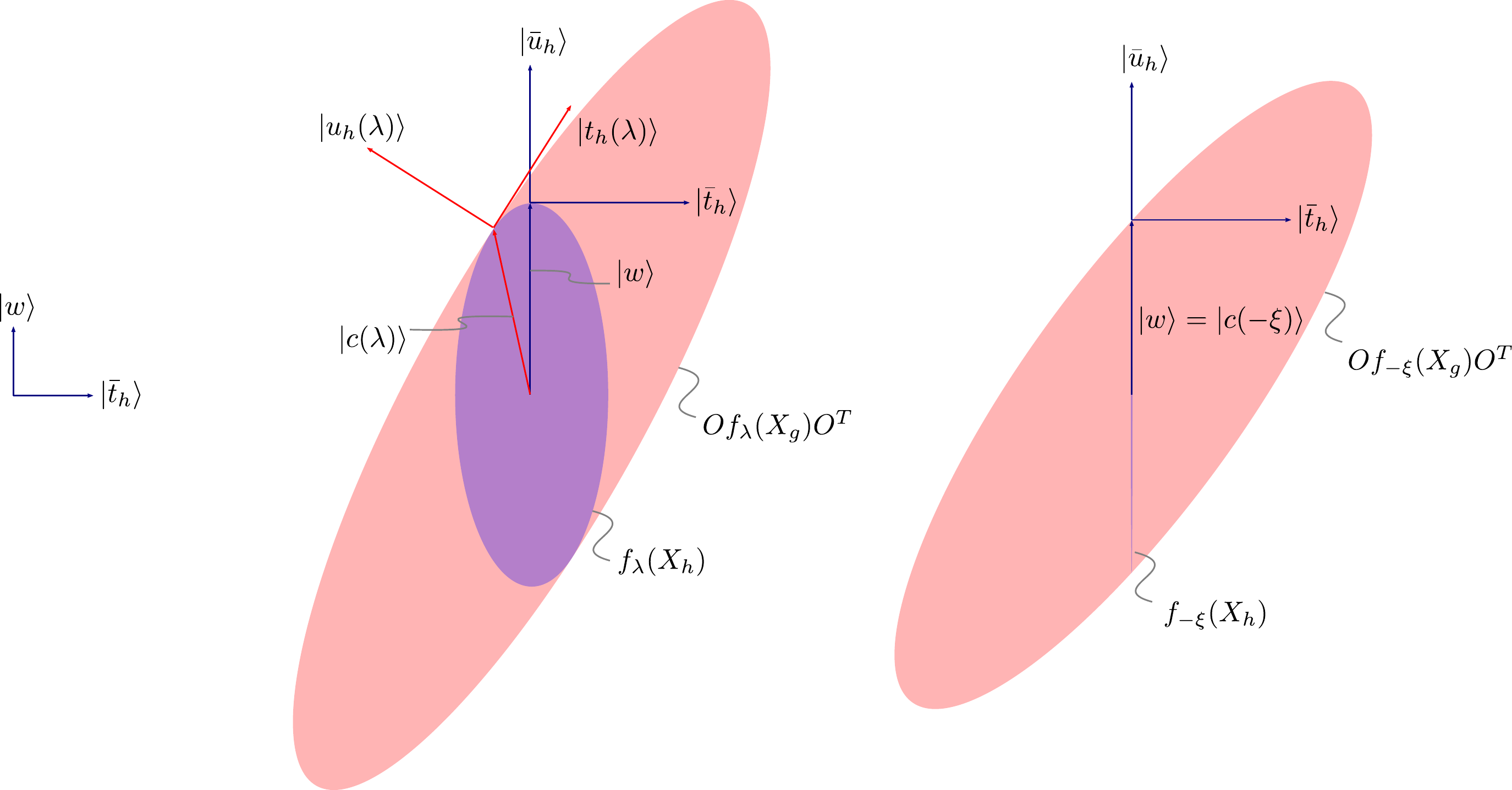}\caption{A sequence leading to infinite curvature.\label{fig:infiniteCurvature}}
}
\end{figure}

\end{enumerate}
\item \textbf{End}: Restart the current phase (phase 2) with the newly obtained
$(k-1)$ sized objects.\\
We end with giving the dimension argument. The dimension after every
iteration is $k-1\ge n_{g}^{(k-1)}+n_{h}^{(k-1)}-1$ if we start with
the assumption that $k\ge n_{g}^{(k)}+n_{h}^{(k)}-1$. \textcolor{blue}{The
reason is that either $n_{g}^{(k-1)}=n_{g}^{(k)}-1$ or $=n_{g}^{(k)}$.
Similarly, either $n_{h}^{(k-1)}=n_{h}^{(k)}-1$ or $=n_{h}^{(k)}$.
Justification of this is simply that we remove at least one component
from the two vectors (from the $n_{g}^{(k)}$ for the usual wiggle-v).
To see this, note that in the finite case we remove one from both
as we write express the vector in a new basis. This new basis is the
space where the vector has finite support. We then remove one of the
components in the sub-problem. In the infinite case, it is possible
that we remove one and add one for $n_{h}^{(k-1)}$, assuming it is
the usual wiggle-v, but we necessarily reduce $n_{g}^{(k-1)}$ as
this is similar to the finite case. For the other wiggle-v, $g$ and
$h$ get swapped but the counting stays the same.}
\end{itemize}
\end{itemize}

\subsubsection{Phase 3: Reconstruction}

Let $k_{0}$ be the iteration at which the algorithm stops. Using
the relation 
\[
O^{(k)}=\bar{O}_{g}^{(k)}\left(\left|u_{h}^{(k)}\right\rangle \left\langle u_{h}^{(k)}\right|+O^{(k-1)}\right)\bar{O}_{h}^{(k)}
\]
 (or its transpose if $s^{(k)}=-1$), evaluate $O^{(k_{1})}$ from
$O^{(k_{0})}:=\mathbb{I}_{k_{0}}$, then $O^{(k_{2})}$ from $O^{(k_{1})}$,
then $O^{(k_{3})}$ from $O^{(k_{2})}$ and so on until $O^{(n)}$
is obtained which solves the matrix instance $\text{\ensuremath{\underbar{X}}}^{(n)}$
we started with. In terms of EBRM matrices, the solution is given
by $H=X_{h}^{(n)},$ $G=O^{(n)}X_{g}O^{(n)T}$, and $\left|w\right\rangle =\left|w^{(n)}\right\rangle $.

\pagebreak{}

\section{Conclusion\label{sec:Conclusion}}

In the first part, we described a framework which we used to construct
the unitaries required to implement the bias $1/10$ protocol. In
the second part, we described the EMA algorithm which allows us to
find the unitaries corresponding to arbitrary $\Lambda$-valid moves,
which combined with the framework, allows us to numerically find quantum
WCF protocols with arbitrarily small bias.

A preliminary implementation of the EMA algorithm on python \parencite{Arora+18},
which is usable but not automated enough for an end-user, has already
yielded the following interesting results and future research directions.
\begin{enumerate}
\item \emph{Mochon's denominator needs neither padding nor operator monotones.}
For assignments given by Mochon's denominator, it is known that $\left\langle x_{h}\right\rangle =\left\langle x_{g}\right\rangle $
which means that for the first iteration of the algorithm, we need
not use any operator monotone function. This was clear. The surprising
result of the numerical implementation was that even for subsequent
iterations one need not use operator monotones, which also explains
why\footnote{To see this, note that the only time we spill over to the extra dimensions,
is when we use the wiggle-v method. Otherwise, we stay inside the
first $\max(n_{g},n_{h})$ dimensions.} we did not need padding, i.e. the (solution) orthogonal matrix has
size $n=n_{g}=n_{h}$. An interesting open question is to analytically
prove this as Mochon's denominator based assignment is very close
to the kind of assignments used in Mochon's protocols. If analytic
expressions for the unitaries corresponding to the former (Mochon's
denominator) can be found, the unitaries corresponding to the assignments
used in Mochon's protocols would be a small perturbation thereof.
Hence, this might lead us to analytic expressions for the unitaries
involved in games with arbitrarily low biases.
\item \emph{Moves in the bias $1/18$ protocol do not need padding (no wiggle-v).}
We already know analytically that there are specific cases where padding
is required. However, when we tried to numerically implement the moves
involved in Mochon's protocols going as low as $\epsilon=1/18$, to
our surprise, we found that in no case was padding necessary (which
means the wiggle-v method was never invoked). It would be interesting
to see if this can be proven to be the case for all of Mochon's moves.
There is another class of games that achieve arbitrarily low bias
due to Pelchat and Høyer \parencite{pelchat13} whose moves are also
worth investigating in this regard, and even otherwise.
\item \emph{Trick to improve the precision of the EMA algorithm.} The algorithm
tries to find a $\lambda$ such that $\left\langle w\right|f_{\lambda}(X_{h})\left|w\right\rangle -\left\langle v\right|f_{\lambda}(X_{g})\left|v\right\rangle =0$.
In the finite case, one must also have for consistency, $\left\langle w|n_{h}(\lambda)\right\rangle =\left\langle v|n_{g}(\lambda)\right\rangle $
(this is because in subsequent steps, the space orthogonal is affected
so if the component of the honest states along the normals is not
mapped correctly, it would not get fixed later; this would mean there
is no solution as we are only imposing necessary conditions). We observed
that, numerically, we get a better precision if we use the latter
condition for fine-tuning the result (after applying the former for
obtaining a more course-grained solution). While analytically, the
first condition implies the latter exactly, this ceases to be the
case numerically due to the finiteness of precision. A careful error
analysis of the EMA algorithm would be required to fully understand
this behaviour. As a first step, we can understand this improvement
as a direct consequence of the fact that the honest state is explicitly
mapped correctly (up to the precision of the machine, which is about
16 floating points for most computers) if we use the method involving
normals while in the latter, this should happen implicitly. 
\end{enumerate}
Limitations of the current implementation.
\begin{enumerate}
\item \emph{Limited wiggle-v.} We have not fully implemented the Wiggle-v
method which means that it would be cumbersome to apply it to the
general merge and split, for instance. However, for them we already
give the explicit Blinkered Unitaries. For the rest, as we already
saw, it does not even seem necessary. 
\item \emph{Minor pending issues.} Sometimes due to noise (arising from
finiteness of the precision) our global minimiser gets trapped into
local minima and has to be guided manually by looking at the graph.
This means that a refined algorithm should also be able to solve the
problem. Further, we did not implement the systematic method defined
by the EMA algorithm for finding the spectrum of the matrices it uses
but it appears that almost any guess works for Mochon's assignments.
\end{enumerate}
An important open problem that remains is to account for noise in
the quantum formalism itself. We assumed all along that the unitary
is applied exactly. Now there are two possible effects of the unitary
being noisy. The first effect is on the coordinates of the points.
This can be accounted for by raising all the points a little bit (proportional
to the noise). Doing this for each move would yield a cumulative effect
on the final point, that is, the bias. This should not be very hard
to evaluate. The second effect is on the weights of the points. This
must be handled with care as most of the games rely on a cancellation
of weight across different moves. One way to proceed would be to imagine,
after a lossy operation, the remaining game/protocol is implemented
with a slightly scaled down weight but with the same relative proportion.
The rest of the weight should be collected in the end and merged with
the final point to get a small raise. This would quantify the effect
of this type of noise on the bias.

\section{Acknowledgements}

We are thankful to Nicolas Cerf, Mathieu Brandeho and Ognyan Oreshkov
for various insightful discussions. We acknowledge support from the
Belgian Fonds de la Recherche Scientifique \textendash{} FNRS under
grants no F.4515.16 (QUICTIME) and R.50.05.18.F (QuantAlgo). The QuantAlgo
project has received funding from the QuantERA ERA-NET Cofund in Quantum
Technologies implemented within the European Union's Horizon 2020
Programme. ASA further acknowledges the FNRS for support through the
FRIA grant, 3/5/5 \textendash{} MCF/XH/FC \textendash{} 16754. 

\pagebreak{}

\printbibliography

\pagebreak{}

\appendix

\section{Blinkered $m\to n$ Transition\label{sec:Blinkered-transition}}

Recall that the unitary we had described was of the form $U=\left|w\right\rangle \left\langle v\right|+\left|v\right\rangle \left\langle w\right|+\sum\left|v_{i}\right\rangle \left\langle v_{i}\right|+\sum\left|w_{i}\right\rangle \left\langle w_{i}\right|$.
It is evident that having a scheme for generating these $\left|v_{i}\right\rangle $
and $\left|w_{i}\right\rangle $ will be useful as we explore more
complicated transitions. More precisely, we need to complete a set
containing one vector into a complete orthonormal basis. Let us do
this first and then return to the analysis of a $3\to2$ merge.

\subsubsection*{Completing an Orthonormal Basis}

Consider an orthonormal complete set of basis vectors $\left\{ \left|g_{i}\right\rangle \right\} $,
and a vector $\left|v\right\rangle =\frac{\sum_{i}\sqrt{p_{i}}\left|g_{i}\right\rangle }{\sqrt{\sum_{i}p_{i}}}$.
We describe a scheme for constructing vectors $\left|v_{i}\right\rangle $
s.t. $\left\{ \left|v\right\rangle ,\left\{ \left|v_{i}\right\rangle \right\} \right\} $
is a complete orthonormal set of basis vectors. Formally, we can do
this inductively. Instead, we do this by examples for that makes it
intuitive and demonstrates the generalisable argument right away.
The first we define to be 
\[
\left|v_{1}\right\rangle =\frac{\sqrt{p_{1}}\left|g_{1}\right\rangle -\frac{p_{1}}{\sqrt{p_{2}}}\left|g_{2}\right\rangle }{\sqrt{p_{1}+\frac{p_{1}^{2}}{p_{2}}}}\left(=\frac{\sqrt{p_{1}}\left|g_{1}\right\rangle -\sqrt{p_{2}}\left|g_{2}\right\rangle }{\sqrt{p_{1}+p_{2}}},\,\text{the familiar one}\right)
\]
which is manifestly normalised and orthogonal to $\left|v\right\rangle $,
i.e. $\left\langle v|v_{1}\right\rangle =p_{1}-p_{1}=0$. The next
vector is 
\[
\left|v_{2}\right\rangle =\frac{\sqrt{p_{1}}\left|g_{1}\right\rangle +\sqrt{p_{2}}\left|g_{2}\right\rangle -\frac{\left(p_{1}+p_{2}\right)}{\sqrt{p_{3}}}\left|g_{3}\right\rangle }{\sqrt{p_{1}+p_{2}+\frac{(p_{1}+p_{2})^{2}}{p_{3}}}}
\]
which is again manifestly normalised and orthogonal to $\left|v_{1}\right\rangle $
because $\left\langle v_{2}|v_{1}\right\rangle =\left\langle v|v_{1}\right\rangle $.
$\left\langle v|v_{2}\right\rangle =p_{1}+p_{2}-(p_{1}+p_{2})=0$.
Similarly one can construct the $\left(k+1\right)^{\text{th}}$ basis
vector as 
\[
\left|v_{k}\right\rangle =\frac{\sum_{i=1}^{k}\sqrt{p_{k}}\left|g_{k}\right\rangle -\frac{\sum_{i=1}^{k}p_{k}}{\sqrt{p_{k+1}}}\left|g_{k+1}\right\rangle }{N_{k}}
\]
where the $N_{k}=\sqrt{\sum_{i=1}^{k}p_{k}+\frac{(\sum_{i=1}^{k}p_{k})^{2}}{p_{k+1}}}$
and obtain the full set.

\subsubsection*{The Analysis}

Back to the analysis. Recall that the constraint equation was 
\[
\underbrace{\sum x_{h_{i}}\left|h_{ii}\right\rangle \left\langle h_{ii}\right|}_{\text{I}}+\underbrace{x\mathbb{I}^{\{g_{ii}\}}}_{\text{II}}\ge\underbrace{\sum x_{g_{i}}U\left|g_{ii}\right\rangle \left\langle g_{ii}\right|U^{\dagger}}_{\text{III}}
\]
where we have introduced the notation $\left|h_{ii}\right\rangle =\left|h_{i}h_{i}\right\rangle $
in the interest of efficiency. The $g_{1},g_{2},g_{3}\to h_{1},h_{2}$
transition requires us to know 
\[
U=\left|v\right\rangle \left\langle w\right|+\left|w\right\rangle \left\langle v\right|+\left|v_{1}\right\rangle \left\langle v_{1}\right|+\left|v_{2}\right\rangle \left\langle v_{2}\right|+\left|w_{1}\right\rangle \left\langle w_{1}\right|.
\]
Using the procedure above we can evaluate the vectors of interest
\begin{align*}
\left|v\right\rangle  & =\frac{\sqrt{p_{g_{1}}}\left|g_{11}\right\rangle +\sqrt{p_{g_{2}}}\left|g_{22}\right\rangle +\sqrt{p_{g_{3}}}\left|g_{33}\right\rangle }{N_{g}}\\
\left|v_{1}\right\rangle  & =\frac{\sqrt{p_{g_{1}}}\left|g_{11}\right\rangle -\frac{p_{g_{1}}}{\sqrt{p_{g_{2}}}}\left|g_{22}\right\rangle }{N_{g_{1}}}\\
\left|v_{2}\right\rangle  & =\frac{\sqrt{p_{g_{1}}}\left|g_{11}\right\rangle +\sqrt{p_{g_{2}}}\left|g_{22}\right\rangle -\frac{\left(p_{g_{1}}+p_{g_{2}}\right)}{\sqrt{p_{g_{3}}}}\left|g_{33}\right\rangle }{N_{g_{2}}}\\
\left|w\right\rangle  & =\frac{\sqrt{p_{h_{1}}}\left|h_{11}\right\rangle +\sqrt{p_{h_{2}}}\left|h_{22}\right\rangle }{N_{h}}\\
\left|w_{1}\right\rangle  & =\frac{\sqrt{p_{h_{2}}}\left|h_{11}\right\rangle -\sqrt{p_{h_{1}}}\left|h_{22}\right\rangle }{N_{h}}
\end{align*}
where $N_{g},\,N_{g_{1}},\,N_{g_{2}},\,N_{h}$ are normalisations.
In fact we want to express the constraints in this basis. To evaluate
the first term we use the above to find 
\begin{align*}
\left|h_{11}\right\rangle  & =\frac{\sqrt{p_{h_{1}}}\left|w\right\rangle +\sqrt{p_{h_{2}}}\left|w_{1}\right\rangle }{N_{h}}\\
\left|h_{22}\right\rangle  & =\frac{\sqrt{p_{h_{2}}}\left|w\right\rangle -\sqrt{p_{h_{1}}}\left|w_{1}\right\rangle }{N_{h}}
\end{align*}
which leads to
\begin{align*}
\text{I} & =x_{h_{1}}\left|h_{11}\right\rangle \left\langle h_{11}\right|+x_{h_{2}}\left|h_{22}\right\rangle \left\langle h_{22}\right|\\
 & =\frac{x_{h_{1}}}{N_{h}^{2}}\left[\begin{array}{c|cc}
 & \left\langle w\right| & \left\langle w_{1}\right|\\
\hline \left|w\right\rangle  & p_{h_{1}} & \sqrt{p_{h_{1}}p_{h_{2}}}\\
\left|w_{1}\right\rangle  & \sqrt{p_{h_{1}}p_{h_{2}}} & p_{h_{2}}
\end{array}\right]+\frac{x_{h_{2}}}{N_{h}^{2}}\left[\begin{array}{c|cc}
 & \left\langle w\right| & \left\langle w_{1}\right|\\
\hline \left|w\right\rangle  & p_{h_{2}} & -\sqrt{p_{h_{1}}p_{h_{2}}}\\
\left|w_{1}\right\rangle  & -\sqrt{p_{h_{1}}p_{h_{2}}} & p_{h_{1}}
\end{array}\right]\\
 & =\frac{1}{N_{h}^{2}}\left[\begin{array}{c|cc}
 & \left\langle w\right| & \left\langle w_{1}\right|\\
\hline \left|w\right\rangle  & p_{h_{1}}x_{h_{1}}+p_{h_{2}}x_{h_{2}} & \sqrt{p_{h_{1}}p_{h_{2}}}(x_{h_{1}}-x_{h_{2}})\\
\left|w_{1}\right\rangle  & \sqrt{p_{h_{1}}p_{h_{2}}}(x_{h_{1}}-x_{h_{2}}) & p_{h_{2}}x_{h_{1}}+p_{h_{1}}x_{h_{2}}
\end{array}\right].
\end{align*}
(Remark: We had made a mistake in this term which was causing the
matrix to sometimes become negative; after correction, the matrix
seems to be positive for Mochon's f-function based construction) Evaluation
of II is nearly trivial for identity can be expressed in any basis
and that yields 
\begin{align*}
\text{II} & =x(\left|v\right\rangle \left\langle v\right|+\left|v_{1}\right\rangle \left\langle v_{1}\right|+\left|v_{2}\right\rangle \left\langle v_{2}\right|)\\
 & =\left[\begin{array}{c|ccc}
 & \left\langle v\right| & \left\langle v_{1}\right| & \left\langle v_{2}\right|\\
\hline \left|v\right\rangle  & x\\
\left|v_{1}\right\rangle  &  & x\\
\left|v_{2}\right\rangle  &  &  & x
\end{array}\right].
\end{align*}
For the last term 
\[
\text{III}=\underbrace{x_{g_{1}}U\left|g_{11}\right\rangle \left\langle g_{11}\right|U^{\dagger}}_{\text{(i)}}+\underbrace{x_{g_{2}}U\left|g_{22}\right\rangle \left\langle g_{22}\right|U^{\dagger}}_{\text{(ii)}}+\underbrace{x_{g_{3}}U\left|g_{33}\right\rangle \left\langle g_{33}\right|U^{\dagger}}_{\text{(iii)}}
\]
 We evaluate 
\begin{align*}
U\left|g_{11}\right\rangle  & =\frac{\sqrt{p_{g_{1}}}}{N_{g}}\left|w\right\rangle +\frac{\sqrt{p_{g_{1}}}}{N_{g_{1}}}\left|v_{1}\right\rangle +\frac{\sqrt{p_{g_{1}}}}{N_{g_{2}}}\left|v_{2}\right\rangle \\
U\left|g_{22}\right\rangle  & =\frac{\sqrt{p_{g_{2}}}}{N_{g}}\left|w\right\rangle +\frac{\left(-\frac{p_{g_{1}}}{\sqrt{p_{g_{2}}}}\right)}{N_{g_{1}}}\left|v_{1}\right\rangle +\frac{\sqrt{p_{g_{2}}}}{N_{g_{2}}}\left|v_{2}\right\rangle \\
U\left|g_{33}\right\rangle  & =\frac{\sqrt{p_{g_{3}}}}{N_{g}}\left|w\right\rangle +0\left|v_{1}\right\rangle +\frac{\left(-\frac{p_{g_{1}}+g_{g_{2}}}{\sqrt{p_{g_{3}}}}\right)}{N_{g_{2}}}\left|v_{2}\right\rangle .
\end{align*}
We must now find each sub term, starting with the most regular
\[
\text{(i)}=x_{g_{1}}p_{g_{1}}\left[\begin{array}{c|ccc}
 & \left\langle v_{1}\right| & \left\langle v_{2}\right| & \left\langle w\right|\\
\hline \left|v_{1}\right\rangle  & \frac{1}{N_{g_{1}}^{2}} & \frac{1}{N_{g_{1}}N_{g_{2}}} & \frac{1}{N_{g_{1}}N_{g}}\\
\left|v_{2}\right\rangle  & \frac{1}{N_{g_{2}}N_{g_{1}}} & \frac{1}{N_{g_{2}}^{2}} & \frac{1}{N_{g_{2}}N_{g}}\\
\left|w\right\rangle  & \frac{1}{N_{g}N_{g_{1}}} & \frac{1}{N_{g}N_{g_{2}}} & \frac{1}{N_{g}^{2}}
\end{array}\right].
\]
For the second term, we re-write $U\left|g_{22}\right\rangle =\sqrt{p_{g_{2}}}\left(\frac{1}{N_{g}}\left|w\right\rangle -\frac{1}{N'_{g_{1}}}\left|v_{1}\right\rangle +\frac{1}{N_{g_{2}}}\left|v_{2}\right\rangle \right)$
where we have defined 
\[
N'_{g_{1}}=\frac{p_{g_{2}}}{p_{g_{1}}}N_{g_{1}}
\]
to obtain 
\[
(ii)=x_{g_{2}}p_{g_{2}}\left[\begin{array}{c|ccc}
 & \left\langle v_{1}\right| & \left\langle v_{2}\right| & \left\langle w\right|\\
\hline \left|v_{1}\right\rangle  & \frac{1}{N_{g_{1}}^{\prime2}} & -\frac{1}{N'_{g_{1}}N_{g_{2}}} & -\frac{1}{N'_{g_{1}}N_{g}}\\
\left|v_{2}\right\rangle  & -\frac{1}{N_{g_{2}}N'_{g_{1}}} & \frac{1}{N_{g_{2}}^{2}} & \frac{1}{N_{g_{2}}N_{g}}\\
\left|w\right\rangle  & -\frac{1}{N_{g}N'_{g_{1}}} & \frac{1}{N_{g}N_{g_{2}}} & \frac{1}{N_{g}^{2}}
\end{array}\right]
\]
and finally $U\left|g_{33}\right\rangle =\sqrt{p_{g_{3}}}\left(\frac{1}{N_{g}}\left|w\right\rangle +0\left|v_{1}\right\rangle -\frac{1}{N'_{g_{2}}}\left|v_{2}\right\rangle \right)$
with 
\[
N'_{g_{2}}=\frac{p_{g_{3}}}{p_{g_{1}}+p_{g_{2}}}
\]
to get 
\[
\text{(iii)}=x_{g_{3}}p_{g_{3}}\left[\begin{array}{c|ccc}
 & \left\langle v_{1}\right| & \left\langle v_{2}\right| & \left\langle w\right|\\
\hline \left|v_{1}\right\rangle \\
\left|v_{2}\right\rangle  &  & \frac{1}{N_{g_{2}}^{\prime2}} & -\frac{1}{N'_{g_{2}}N_{g}}\\
\left|w\right\rangle  &  & -\frac{1}{N_{g}N'_{g_{2}}} & \frac{1}{N_{g}^{2}}
\end{array}\right].
\]
Now we can combine all of these into a single matrix and try to obtain
some simpler constraints.

\noindent
{\tiny%
\noindent\begin{minipage}[t]{1\columnwidth}%
\[
M\overset{\text{def}}{=}\left[\begin{array}{c|ccccc}
 & \left\langle v\right| & \left\langle v_{1}\right| & \left\langle v_{2}\right| & \left\langle w\right| & \left\langle w_{1}\right|\\
\hline \left|v\right\rangle  & x\\
\left|v_{1}\right\rangle  &  & x-\frac{x_{g_{1}}p_{g_{1}}}{N_{8_{1}}^{2}}-\frac{x_{g_{2}}p_{g_{2}}}{N_{g_{1}}^{\prime2}} & -\frac{x_{g_{1}}p_{g_{1}}}{N_{g_{1}}N_{g_{2}}}+\frac{x_{g_{2}}p_{g_{2}}}{N_{g_{1}}'N_{g_{2}}} & -\frac{x_{g_{1}}p_{g_{1}}}{N_{g_{1}}N_{g}}+\frac{x_{g_{2}}p_{g_{2}}}{N_{g_{1}}'N_{g}}\\
\left|v_{2}\right\rangle  &  & -\frac{x_{g_{1}}p_{g_{1}}}{N_{g_{2}}N_{g_{1}}}+\frac{x_{g_{2}}p_{g_{2}}}{N_{g_{2}}N'_{g_{1}}} & x-\frac{x_{g_{1}}p_{g_{1}}}{N_{g_{2}}^{2}}-\frac{x_{g_{2}}p_{g_{2}}}{N_{g_{2}}^{2}}-\frac{x_{g_{3}}p_{g_{3}}}{N_{g_{2}}^{\prime2}} & -\frac{x_{g_{1}}p_{g_{1}}}{N_{g_{2}}N_{g}}-\frac{x_{g_{2}}p_{g_{2}}}{N_{g_{2}}N_{g}}+\frac{x_{g_{3}}p_{g_{3}}}{N_{g_{2}}'N_{g}}\\
\left|w\right\rangle  &  & -\frac{x_{g_{1}}p_{g_{1}}}{N_{g}N_{g_{1}}}+\frac{x_{g_{2}}p_{g_{2}}}{N_{g}N_{g_{1}}'} & -\frac{x_{g_{1}}p_{g_{1}}}{N_{g}N_{g_{2}}}-\frac{x_{g_{2}}p_{g_{2}}}{N_{g}N_{g_{2}}}+\frac{x_{g_{3}}p_{g_{3}}}{N_{g}N'_{g_{2}}} & \frac{p_{h_{1}}x_{h_{1}}+p_{h_{2}}x_{h_{2}}}{N_{h}^{2}}-\frac{1}{N_{g}^{2}}\sum_{i}x_{g_{i}}p_{g_{i}} & \frac{\sqrt{p_{h_{1}}p_{h_{2}}}}{N_{h}^{2}}(x_{h_{1}}-x_{h_{2}})\\
\left|w_{1}\right\rangle  &  &  &  & \frac{\sqrt{p_{h_{1}}p_{h_{2}}}}{N_{h}^{2}}(x_{h_{1}}-x_{h_{2}}) & \frac{p_{h_{2}}x_{h_{1}}+p_{h_{1}}x_{h_{2}}}{N_{h}^{2}}
\end{array}\right]\ge0.
\]
\end{minipage}

}

Despite this appearing to be a complicated expression, we can conclude
that it will always be so that larger the $x$ looser will be the
constraint. To show this and to simplify this calculation, note that
$M$ can be split into a scalar condition, $x\ge0$ (from the $\left|v\right\rangle \left\langle v\right|$
part) and a sub-matrix which we choose to write as 
\[
\begin{array}{c|c|c}
 & \begin{array}{cc}
\left\langle v_{1}\right| & \left\langle v_{2}\right|\end{array} & \begin{array}{cc}
\left\langle w\right| & \left\langle w_{1}\right|\end{array}\\
\hline \begin{array}{c}
\left|v_{1}\right\rangle \\
\left|v_{2}\right\rangle 
\end{array} & C & B^{T}\\
\hline \begin{array}{c}
\left|w\right\rangle \\
\left|w_{1}\right\rangle 
\end{array} & B & A
\end{array}\ge0.
\]
Now since $\left[\begin{array}{cc}
C & B^{T}\\
B & A
\end{array}\right]\ge0\iff\left[\begin{array}{cc}
A & B\\
B^{T} & C
\end{array}\right]\ge0\iff C\ge0,\,A-BC^{-1}B^{T}\ge0,\,(\mathbb{I}-CC^{-1})B^{T}=0$ using Shur's Complement condition for positivity where $C^{-1}$
is supposed to be the generalised inverse. Since $x$ is in our hands,
we can take it to be sufficiently large so that $C>0$ and thereby
make sure that $\mathbb{I}-CC^{-1}=0$. Evidently then, the only condition
of interest is 
\[
A-BC^{-1}B^{T}\ge0.
\]
We can do even better than this actually. Note that if $C>0$ then
$C^{-1}>0$ and that the second term is of the form 
\[
\underbrace{\left[\begin{array}{cc}
a & b\\
0 & 0
\end{array}\right]}_{B}\underbrace{\left[\begin{array}{cc}
\alpha & \gamma\\
\gamma & \beta
\end{array}\right]}_{C^{-1}}\underbrace{\left[\begin{array}{cc}
a & 0\\
b & 0
\end{array}\right]}_{B^{T}}=\left[\begin{array}{cc}
\left[\begin{array}{cc}
a & b\end{array}\right]\left[\begin{array}{cc}
\alpha & \gamma\\
\gamma & \beta
\end{array}\right]\left[\begin{array}{c}
a\\
b
\end{array}\right] & 0\\
0 & 0
\end{array}\right]\ge0
\]
because $C^{-1}>0$. We can therefore write the constraint equation
as
\[
A\ge BC^{-1}B^{T}\ge0
\]
and note that $A\ge0$ is a necessary condition. This also becomes
a sufficient condition in the limit that $x\to\infty$ because $C^{-1}\to0$
in that case. We have thereby reduced the analysis to simply checking
if 
\[
\left[\begin{array}{cc}
\frac{p_{h_{1}}x_{h_{1}}+p_{h_{2}}x_{h_{2}}}{N_{h}^{2}}-\frac{1}{N_{g}^{2}}\sum_{i}x_{g_{i}}p_{g_{i}} & \frac{\sqrt{p_{h_{1}}p_{h_{2}}}}{N_{h}^{2}}(x_{h_{1}}-x_{h_{2}})\\
\frac{\sqrt{p_{h_{1}}p_{h_{2}}}}{N_{h}^{2}}(x_{h_{1}}-x_{h_{2}}) & \frac{p_{h_{2}}x_{h_{1}}+p_{h_{1}}x_{h_{2}}}{N_{h}^{2}}
\end{array}\right]\ge0.
\]
This being a $2\times2$ matrix can be checked for positivity by the
trace and determinant method. Another possibility is the use of Schur's
Complement conditions again. Here, however, we intend to use a more
general technique (similar to the one used in the split analysis).
Let us introduce 
\[
\left\langle x_{g}\right\rangle \overset{\text{def}}{=}\frac{1}{N_{g}^{2}}\sum_{i}x_{g_{i}}p_{g_{i}},\,\left\langle \frac{1}{x_{h}}\right\rangle \overset{\text{def}}{=}\frac{1}{N_{h}^{2}}\sum_{i}\frac{p_{h_{i}}}{x_{h_{i}}}
\]
and recall/note that term (I) and one element from term (III) constitute
matrix $A$, which can also be written as 
\begin{align*}
A & =x_{h_{1}}\left|h_{11}\right\rangle \left\langle h_{11}\right|+x_{h_{2}}\left|h_{22}\right\rangle \left\langle h_{22}\right|-\left\langle x_{g}\right\rangle \left|w\right\rangle \left\langle w\right|\\
 & =\begin{array}{c|cc}
 & \left\langle h_{11}\right| & \left\langle h_{22}\right|\\
\hline \left|h_{11}\right\rangle  & x_{h_{1}}\\
\left|h_{22}\right\rangle  &  & x_{h_{2}}
\end{array}-\left\langle x_{g}\right\rangle \left|w\right\rangle \left\langle w\right|
\end{align*}
Note that this now has the exact same form as that of the split constraint
with $x_{g_{1}}\to\left\langle x_{g}\right\rangle $. We use the same
$F-M\ge0\iff\mathbb{I}-\sqrt{F}^{-1}M\sqrt{F}^{-1}\ge0$ for $F>0$
technique to obtain $\mathbb{I}\ge\left\langle x_{g}\right\rangle \left|w''\right\rangle \left\langle w''\right|$
where $\left|w''\right\rangle =\frac{\sqrt{\frac{p_{h_{1}}}{x_{h_{1}}}}\left|h_{11}\right\rangle +\sqrt{\frac{p_{h_{2}}}{x_{h_{2}}}}\left|h_{22}\right\rangle }{N_{h}}$.
Normalising this one gets $\left|w'\right\rangle =\frac{\left|w''\right\rangle }{\sqrt{\left\langle \frac{1}{x_{h}}\right\rangle }}$
which entails $\mathbb{I}\ge\left\langle x_{g}\right\rangle \left\langle \frac{1}{x_{h}}\right\rangle \left|w'\right\rangle \left\langle w'\right|$
and that leads us to the final condition 
\[
\frac{1}{\left\langle x_{g}\right\rangle }\ge\left\langle \frac{1}{x_{h}}\right\rangle .
\]
In fact all the techniques used in reaching this result can be extended
to the $m\to n$ transition case as well and so the aforesaid result
should hold in general.

\subsubsection*{}

\section{Mochon's Assignments\label{sec:Mochons-Assignments}}
\begin{lem*}[Mochon's Denominator]
 $\sum_{i=1}^{n}\frac{1}{\prod_{j\neq i}(x_{j}-x_{i})}=0$ for $n\ge2$.
\end{lem*}
\begin{proof}
We prove this by induction (following Mochon's proof, just optimised
for clarity instead of space). For $n=2$ 
\[
\frac{1}{(x_{2}-x_{1})}+\frac{1}{(x_{1}-x_{2})}=0.
\]
Now we show that if the result holds for $n-1$ and it would also
hold for $n$ which would complete the inductive proof. We start with
noting that 
\[
\frac{1}{(x_{n}-x_{i})(x_{1}-x_{i})}=\frac{1}{x_{n}-x_{1}}\left[\frac{1}{x_{1}-x_{i}}-\frac{1}{x_{n}-x_{i}}\right].
\]
This is useful because it helps breaking the product into a sum. My
strategy would be to pull off one common term so that we can apply
the result to the remaining $n-1$ terms. The expression of interest
is 
\begin{align*}
\sum_{i=1}^{n}\frac{1}{\prod_{j\neq i}(x_{j}-x_{i})} & =\frac{1}{\prod_{j\neq1}(x_{j}-x_{1})}+\sum_{i=2}^{n-1}\frac{1}{\prod_{j\neq i}(x_{j}-x_{i})}+\frac{1}{\prod_{j\neq n}(x_{j}-x_{n})}
\end{align*}
where notice that the $i$th term in the sum (of the second term)
can be written as 
\[
\frac{1}{(x_{n}-x_{i})(x_{1}-x_{i})\prod_{j\neq i,1,n}(x_{j}-x_{i})}=\frac{1}{x_{n}-x_{1}}\left[\frac{1}{\prod_{j\neq i,n}(x_{j}-x_{i})}-\frac{1}{\prod_{j\neq1,i}(x_{j}-x_{i})}\right].
\]
The first term can be written as 
\[
\frac{1}{(x_{n}-x_{1})\prod_{j\neq1,n}(x_{j}-x_{1})}
\]
while the last can be written as 
\[
\frac{-1}{(x_{n}-x_{1})\prod_{j\neq n,1}(x_{j}-x_{n})}.
\]
Putting all these together, we get
\begin{align*}
 & \sum_{i=1}^{n}\frac{1}{\prod_{j\neq i}(x_{j}-x_{i})}\\
 & =\frac{1}{(x_{n}-x_{1})}\left[\underbrace{\frac{1}{\prod_{j\neq1,n}(x_{j}-x_{1})}+\sum_{i=2}^{n-1}\frac{1}{\prod_{j\neq i,n}(x_{j}-x_{i})}}-\underbrace{\sum_{i=2}^{n-1}\frac{1}{\prod_{j\neq1,i}(x_{j}-x_{i})}-\frac{1}{\prod_{j\neq1,n}(x_{j}-x_{n})}}\right]\\
 & =\frac{1}{(x_{n}-x_{1})}\left[\sum_{i=1}^{n-1}\frac{1}{\prod_{j\neq i,n}(x_{j}-x_{i})}-\sum_{i=2}^{n}\frac{1}{\prod_{j\neq1,i}(x_{j}-x_{i})}\right]
\end{align*}
where both sums disappear if the result holds for $n-1$. This completes
the proof.
\end{proof}
\begin{lem*}[Mochon's f-assignment Lemma]
 $\sum_{i=1}^{n}\frac{f(x_{i})}{\prod_{j\neq i}(x_{j}-x_{i})}=0$
where $f(x_{i})$ is of order $k\le n-2$.
\end{lem*}
\begin{proof}
Again we do this by induction on $k$. For $k=0$ the result holds
by the previous result. We assume it holds for order $k-1$ and show
using this that it also holds for order $k$ (this proof is also Mochon's).
Let $g(x_{i})$ be a polynomial of order $k-1$ s.t.
\[
\sum_{i=1}^{n}\frac{f(x_{i})}{\prod_{j\neq i}(x_{j}-x_{i})}=\sum_{i=1}^{n}\frac{(x_{1}-x_{i})(x_{2}-x_{i})\dots(x_{k}-x_{i})-g(x_{i})}{\prod_{j\neq i}(x_{j}-x_{i})}.
\]
Notice that the first part of the sum disappears for all $1\le i\le k$
because of the numerator. Consequently we can write the aforesaid
as
\begin{align*}
 & =\sum_{i=k+1}^{n}\frac{(x_{1}-x_{i})(x_{2}-x_{i})\dots(x_{k}-x_{i})}{\prod_{j\neq i}(x_{j}-x_{i})}-\sum_{i=1}^{n}\frac{g(x_{i})}{\prod_{j\neq i}(x_{j}-x_{i})}\\
 & =\sum_{i=k+1}^{n}\frac{1}{\prod_{j\neq i,1,2\dots,k}(x_{j}-x_{i})}\\
 & =0
\end{align*}
where in the first step, the second term becomes zero by assuming
the result holds for $k-1$ and in the second step the sum disappears
because of the previous result (Mochon's Denominator). Note that $k\le n-2$
for the aforesaid argument to work because otherwise the last step
would become invalid.
\end{proof}
\begin{lem*}
$\sum_{i=1}^{n}\frac{x_{i}^{n-1}}{\prod_{j\neq i}(x_{j}-x_{i})}=(-1)^{n-1}$
for $n\ge2$.
\end{lem*}
\begin{proof}
Let us define $d(n):=\sum_{i=1}^{n}\frac{x_{i}^{n-1}}{\prod_{j\neq i}(x_{j}-x_{i})}$
to proceed inductively. We can then write 
\[
d(2)=\frac{x_{1}}{x_{2}-x_{1}}+\frac{x_{2}}{x_{1}-x_{2}}=\frac{x_{1}(x_{1}-x_{2})+x_{2}(x_{2}-x_{1})}{(x_{2}-x_{1})(x_{1}-x_{2})}=-1.
\]
We assume the result holds for $d(n)$ and write 
\begin{align*}
d(n+1) & =\sum_{i=1}^{n+1}\frac{x_{i}^{n}}{\prod_{j\neq i}(x_{j}-x_{i})}\\
 & =\sum_{i=1}^{n+1}\frac{-(x_{n+1}-x_{i})(x_{i}^{n-1})+x_{n+1}x_{i}^{n-1}}{\prod_{j\neq i}(x_{j}-x_{i})}\\
 & =-\sum_{i=1}^{n+1}(x_{n+1}-x_{i})\frac{x_{i}^{n-1}}{\prod_{j\neq i}(x_{j}-x_{i})}+x_{n+1}\underbrace{\sum_{i=1}^{n+1}\frac{x_{i}^{n-1}}{\prod_{j\neq i}(x_{j}-x_{i})}}_{=0\,(\text{Mochon's Denominator})}\\
 & =-\sum_{i=1}^{n}\cancel{\frac{(x_{n+1}-x_{i})}{(x_{n+1}-x_{i})}}\frac{x_{i}^{n-1}}{\prod_{j\neq i,n+1}(x_{j}-x_{i})}+\cancelto{0}{(x_{n+1}-x_{n+1})}\frac{x_{n+1}^{n-1}}{\prod_{j\neq n+1}(x_{j}-x_{n+1})}\\
 & =-d(n).
\end{align*}
\end{proof}
\begin{prop*}
$\left\langle x_{h}\right\rangle -\left\langle x_{g}\right\rangle =\frac{1}{N_{h}^{2}}=\frac{1}{N_{g}^{2}}$
for a Mochon's TDPG assignment with $k=n-2$ and coefficient of $x^{n-2}$
$\pm1$ in $f(x)$. As above here $\left\langle x_{h}\right\rangle =\frac{1}{N_{h}^{2}}\sum p_{h_{i}}x_{h_{i}}$
and $\left\langle x_{g}\right\rangle =\frac{1}{N_{g}^{2}}\sum p_{g_{i}}x_{g_{i}}$.
\end{prop*}
\begin{proof}
Note, to start with, that the coefficient of $x^{n-2}$ being $\pm1$
is not an artificial requirement because for killing $n-2$ points
$f(x)$ will have the form 
\[
f(x)=(x_{k_{1}}-x)(x_{k_{2}}-x)\dots(x_{k_{n-2}}-x)=(-1)^{n-2}x^{n-2}+\tilde{f}(x)
\]
 where $\tilde{f}$ is a polynomial of order $n-2$. Observe that
\begin{align*}
N_{h}^{2}\left(\left\langle x_{h}\right\rangle -\left\langle x_{g}\right\rangle \right) & =\sum_{i=1}^{n}p(x_{i})x_{i}=-\sum_{i=1}^{n}\frac{x_{i}f(x_{i})}{\prod_{j\neq i}(x_{j}-x_{i})}\\
 & =-\sum_{i=1}^{n}\frac{x_{i}(-1)^{n-2}x_{i}^{n-2}}{\prod_{j\neq i}(x_{j}-x_{i})}-\sum_{i}\frac{\tilde{f}(x_{i})}{\prod_{j\neq i}(x_{j}-x_{i})}\\
 & =-(-1)^{n-2}\sum_{i=1}^{n}\frac{x_{i}^{n-1}}{\prod_{j\neq i}(x_{j}-x_{i})}\\
 & =1
\end{align*}
where the second term in the second step vanishes because of Mochon's
f-assignment Lemma and the last step follows from the previous result.
\end{proof}
\clearpage{}

\end{document}